\documentclass[useAMS,usenatbib]{mn2e}

\usepackage{epsfig,color}
\usepackage{natbib}
\bibpunct{(}{)}{;}{a}{}{,}
\def\spose#1{\hbox to 0pt{#1\hss}}
\def\ltsimm{\mathrel{\spose{\lower 3pt\hbox{$\sim$}}
        \raise 2.0pt\hbox{$<$}}}
\def\gtsimm{\mathrel{\spose{\lower 3pt\hbox{$\sim$}}
        \raise 2.0pt\hbox{$>$}}}
\def\Mdot{\hbox{${\dot M}$}}

\def\km{{\rm\thinspace km}}
\def\cm{{\rm\thinspace cm}}

\def\s{{\rm\thinspace s}}
\def\yr{{\rm\thinspace yr}}
\def\g{{\rm\thinspace g}}
\def\kmps{\hbox{${\rm\km\s^{-1}\,}$}}

\def\erg{{\rm\thinspace erg}}

\def\Hz{{\rm\thinspace Hz}}

\def\ster{{\rm\thinspace ster}}
\def\ergps{\hbox{${\rm\erg\s^{-1}\,}$}}

\def\Msol{\hbox{${\rm\thinspace M_{\odot}}$}}

\def\Msolpyr{\hbox{${\rm\Msol\yr^{-1}\,}$}}
\def\pcm{\hbox{${\rm\cm^{-1}\,}$}}
\def\pcm2{\hbox{${\rm\cm^{-2}\,}$}}
\def\pcm3{\hbox{${\rm\cm^{-3}\,}$}}
\def\ergpscm3Hz{\hbox{${\rm\ergps\cm^{-3}\Hz^{-1}\,}$}}
\def\ergpscm3Hzster{\hbox{${\rm\ergps\cm^{-3}\Hz^{-1}\ster^{-1}\,}$}}
\def\gpcm3{\hbox{${\rm\g\cm^{-3}\,}$}}
\def\ergpcm2{\hbox{${\rm\erg\cm^{-2}\,}$}}
\def\ergpcm3{\hbox{${\rm\erg\cm^{-3}\,}$}}
\def\phpscm2{\hbox{${\rm photons\s^{-1}\cm^{-2}\,}$}}

\title[The turbulent destruction of clouds]{The turbulent destruction of clouds - I. A $k$-$\epsilon$ treatment of turbulence in 2D models of adiabatic shock-cloud interactions}
\author[J.~M.~Pittard, S.~A.~E.~G.~Falle, T.~W.~Hartquist and J.~E.~Dyson]
{J. M. Pittard$^{1}$\thanks{E-mail: jmp@ast.leeds.ac.uk}, S.~A.~E.~G.~Falle$^{2}$, T.~W.~Hartquist$^{1}$ and J.~E.~Dyson$^{1}$\\
$^{1}$School of Physics and Astronomy, The University of
        Leeds, Leeds LS2 9JT, UK\\
$^{2}$Department of Applied Mathematics, The University of
        Leeds, Leeds LS2 9JT, UK\\
}

\begin{document}

\date{Accepted by MNRAS, 22${\rm nd}$ July, 2008}

\pagerange{\pageref{firstpage}--\pageref{lastpage}} \pubyear{2005}

\maketitle

\label{firstpage}

\begin{abstract}
The interaction of a shock with a cloud has been extensively studied
in the literature, where the effects of magnetic fields, radiative
cooling and thermal conduction have been considered.  In many cases,
the formation of fully developed turbulence has been prevented by the
artificial viscosity inherent in hydrodynamical simulations. This
problem is particularly severe in some recent simulations designed to
investigate the interaction of a flow with multiple clouds, where the
resolution of individual clouds is necessarily poor. Furthermore, the
shocked flow interacting with the cloud has been assumed to be
completely uniform in all previous single-cloud studies. In reality,
the flow behind the shock is also likely to be turbulent, with
non-uniform density, pressure and velocity structure created as the
shock sweeps over inhomogenities upstream of the cloud (as seen in
recent multiple cloud simulations). To address these twin issues we
use a sub-grid compressible $k$-$\epsilon$ turbulence model to
estimate the properties of the turbulence generated in shock-cloud
interactions and the resulting increase in the transport coefficients
that the turbulence brings.  A detailed comparison with the output
from an inviscid hydrodynamical code puts these new results into
context.

Despite the above concerns, we find that cloud destruction in inviscid
and $k$-$\epsilon$ models occurs at roughly the same speed when the
post-shock flow is smooth and when the density contrast between the
cloud and inter-cloud medium, $\chi \ltsimm 100$. However, there are
increasing and significant differences as $\chi$ increases. The
$k$-$\epsilon$ models also demonstrate better convergence in
resolution tests than inviscid models, a feature which is particularly
useful for multiple-cloud simulations.

Clouds which are over-run by a highly turbulent post-shock environment
are destroyed significantly quicker as they are subject to strong
``buffeting'' by the flow. The decreased lifetime and faster
acceleration of the cloud material to the speed of the ambient flow
leads to a reduction in the total amount of circulation (vorticity)
generated in the interaction, so that the amount of vorticity may be
self-limiting.  Additional calculations with an inviscid code where
the post-shock flow is given random, grid-scale, motions confirms the
more rapid destruction of the cloud.

Our results clearly show that turbulence plays an important role in
shock-cloud interactions, and that environmental turbulence adds a new
dimension to the parameter space which has hitherto been studied. 
\end{abstract}


\section{Introduction}
\label{sec:intro}
Circumstellar, interstellar, and intergalactic environments are
inhomogeneous, with clouds of various densities and temperatures
embedded in a hotter, more tenuous, substrate.  This substrate is
often turbulent, in part due to the interaction of shocks, shells,
winds and jets with these clouds.  The nature of such interactions is
interesting, because the evolution and morphology of large-scale flows
can ultimately be regulated by objects of much smaller size.

In the interstellar medium (ISM), for instance, the interaction of
supernova shock waves with interstellar clouds creates a continuous
interchange of mass and energy between various thermal phases
\citep{Cox:1974,McKee:1977}. Several outcomes are possible.  The
shocks may destroy the clouds, and mix their material into their
surroundings.  Alternatively, the shocks may trigger the collapse of
the clouds and the formation of new stars, thereby removing material
(at least temporarily) from the ISM \citep{Elmegreen:1977}.  The
shocks themselves will slow and material behind them will cool to form
thin dense shells. These shells may then fragment and form new clouds.
Clouds which survive the passage of the shell subsequently find
themselves either inside a hot, low density bubble (if the bubble is
energy-conserving), or exposed to a fierce, high Mach number wind (if
the shell is momentum-driven). An important process is the generation
of vorticity as the diffuse medium flows around the
clouds. Understanding the interaction between shocks/shells/winds/jets
and interstellar clouds is therefore a key step in studies of the
structure and evolution of the ISM \citep[see the recent reviews
by][]{Elmegreen:2004,Scalo:2004}.

A similar interaction occurs in massive, early-type, stellar systems
where each star blows a powerful, clumpy, wind which collides with the
other.  The impact of the clumps causes the wind-wind collision region
to become highly turbulent \citep{Pittard:2007a}, with implications
for particle acceleration, the timescales for equilibrium ionization
and electron heating, and the physical mixing of the winds.

In this work we consider the adiabatic interation of a shock with a
cloud.  An extensive literature of analytical and numerical
investigations of shock-cloud interactions now exists \citep[see,
e.g.,][and references therein]{Nakamura:2006}. The effects of ordered
magnetic fields, radiative cooling, and thermal conduction have all
been considered, but not simultaneously until recently
\citep{Orlando:2008}.  High-power laser experiments of shock-cloud
interactions \citep[e.g.,][]{Klein:2003} have complemented this
literature. While thermal conduction acts to prevent hydrodynamic
instabilities, radiative cooling enhances them. The effect of ordered
magnetic fields is more complicated: instabilities are prevented in
cases where a magnetic field provides a high tension at the surface of
the cloud, but the effects depend on the strength, orientation and
scale of the magnetic field. Instabilities and vortical motions may be
prevented in some directions, but not necessarily in
others. Interestingly, in three-dimensional simulations a magnetic
field may actually enhance the fragmentation of a cloud
\citep*{Gregori:1999,Shin:2008}, though the actual mixing of the cloud
and ambient medium remains hindered.  Even if the magnetic fields are
not strong enough to directly affect the dynamics, they can
significantly reduce the effects of thermal conduction in directions
normal to the field lines, thus allowing instabilities to develop
\citep{Orlando:2008}. 

Given the fact that in shock-cloud interactions the Reynolds number is
typically high, and the above predisposition for instabilities to
develop, the fluid velocity field around the cloud is expected to vary
significantly and irregularly in both position and time. This
``turbulence'' transports and mixes the cloud material much more
effectively than a comparable laminar flow. Turbulence is also
effective at ``mixing'' the momentum of a fluid (i.e. accelerating
material ripped off the cloud to the ambient flow speed), and at
transferring heat.  In turbulent flows of high Reynolds number there
is a separation of scales - large-scale motions will be strongly
influenced by the geometry of the cloud, and control the transport and
mixing of cloud material, while the behaviour of small-scale motions
is determined almost entirely by the rate at which they receive energy
from the large scales, and by the viscosity.  Hence these small-scale
motions have a universal character, independent of the flow geometry.

While much insight has been gained from previous numerical
investigations of shock-cloud interactions, the artificial viscosity
inherent in all such simulations has the potential to prevent the
formation of fully developed turbulence, to limit the turbulent mixing
of cloud material into the surrounding medium, and to hinder the
destruction of the cloud.  Furthermore, all previous simulations are
highly idealized in the sense that the flow behind the shock is
assumed to be perfectly smooth and uniform (i.e. laminar). In reality,
random inhomogenities in the ambient medium upstream of the cloud will
deform the shock, and will cause velocity, density and pressure
structures to develop in the post-shock flow. The post-shock flow will
then also be ``turbulent'', and it is expected that the destruction of
the cloud will be more rapid in such conditions. 

In this work we address whether fully developed turbulence has been
prevented in all previous numerical works. For high Reynolds number
flows, the only tractable method is a statistical approach i.e. to
describe the turbulent flow, not in terms of a velocity field
${\bf u}(x,t)$, but in terms of some statistics. A model based
on such statistics can lead to a tractable set of equations, because
statistical fields vary smoothly (if at all) in position and time.  To
do this we use a sub-grid turbulent viscosity model, where an attempt
is made to calculate the properties of the turbulence and the
resulting increase in the transport coefficients. The most widely used
is the so-called $k$-$\epsilon$ model, where the properties of the
turbulence are described by two variables, the turbulent energy per
unit mass, $k$, and the turbulent dissipation rate per unit mass,
$\epsilon$. The addition of viscous and diffusive terms in the fluid
equations simulates the turbulent mixing of the cloud and intercloud
medium due to shear instabilities. In this way the subgrid
turbulence model emulates a high Reynolds number flow. Incorporating
a $k$-$\epsilon$ model into shock-cloud simulations should produce
more realistic results than those from inviscid codes, where the
viscosity is purely numerical and the size of shear instabilities is
determined by the resolution of the numerical grid. A turbulent
viscosity model differs from simply adding physical viscosity to the
grid, because the turbulent viscosity is largest in shear layers and
essentially vanishes in regions with little shear, whereas models
with grid viscosity have the same viscosity everywhere.

The structure of this paper is as follows. The key
physics of a shock-cloud interaction is reviewed in
Section~\ref{sec:review}. Section~\ref{sec:setup} introduces the
investigation and the numerical method used. The results are presented
in Section~\ref{sec:results}, where the effects of turbulence on the
cloud evolution are described.  A discussion of the relevance of our
results to shock-cloud and wind-cloud observations is given in
Section~\ref{sec:discussion}. Section~\ref{sec:summary}
summarizes the conclusions of this work, and possible future
work is noted in Section~\ref{sec:future}.

\section{The interaction of a shock with a cloud}
\label{sec:review}
The interaction of a shock with a cloud is a highly non-linear and
complex phenomenon which can be vastly simplified if some assumptions
are made. In this work we assume that the magnetic field is too weak
to be dynamically important (though it must be strong enough to reduce
the thermal conductivity and effective mean-free-path), and that the
interaction is adiabatic.  In the interstellar medium, the typical
magnetic field is $\approx 5\;\mu$G, and the magnetic pressure is
typically a few times higher than the thermal pressure
\citep[e.g.,][]{Cox:2005}.  However, if the shock driven into the
cloud is strong and adiabatic the postshock thermal pressure becomes
much higher than the postshock magnetic pressure, justifying our
assumption that the magnetic field is too weak to be dynamically
important.  We also assume that the magnetic field does not prevent
the full mixing of initially disparate phases (i.e.  that turbulence
drives the reconnection needed to allow this mixing).

The effects of thermal conduction are also ignored in this work. The
efficiency of thermal conduction in magnetized turbulent plasmas
remains highly uncertain \citep[see, e.g.,][]{Pittard:2007b}. X-ray
observations of the hot intracluster medium
\citep{Ettori:2000,Vikhlinin:2001} and theoretical considerations
\citep*{Narayan:2001,Asai:2004,Chandran:2004} suggest that the
coefficient of heat conduction is at least five times lower than the
Spitzer value in the presence of tangled magnetic fields. Furthermore,
if turbulent resistivity can reconnect field lines quickly enough,
turbulent heat transport may be more efficient than thermal conduction
\citep{Cho:2003,Chandran:2004,Lazarian:2006}.

For the interaction to be adiabatic, radiative cooling must be
unimportant. This is often the case for clouds in hot, low density
environments (e.g., in planetary nebulae, in bubbles blown around
individual or groups of massive stars, and in starburst and superwind
environments). The behaviour of the cloud is more likely to be
adiabatic if the cloud is small. Since the assumption of adiabaticity
preserves the scale-free nature of the simulations, the interaction
is then determined by whether there is a sufficient rate of collisions
within the plasma to give it fluid properties, by the dominant
mechanism for the damping of hydromagnetic waves, by the Reynolds
number, and by the way the turbulence is driven. Each of these issues
is discussed below. In order that their importance can readily be
determined, two different examples are considered. In the first
scenario the cloud is ionized, and has a radius $r_{\rm c} = 1\;{\rm
pc}$, density $n_{\rm c} = 0.4\;{\rm cm^{-3}}$, and temperature
$T_{\rm c} = 8000\;{\rm K}$. In the second scenario we consider a
neutral cloud with $r_{\rm c} = 0.46\;{\rm pc}$, $n_{\rm c} = 30\;{\rm
cm^{-3}}$, and $T_{\rm c} = 100\;{\rm K}$. In both cases we imagine
that the clouds are in approximate pressure equilibrium with a
surrounding medium with $n_{\rm ic} = 3 \times 10^{-3}\;{\rm cm^{-3}}$
and $T_{\rm ic} = 10^{6}\;{\rm K}$, and are struck by a high speed
Mach 10 shock with velocity $v_{\rm b} = 1.52\times 10^{8}\;{\rm
cm\;s^{-1}}$.  The shocked intercloud gas has a density $\rho = 2.6
\times 10^{-26}\;{\rm g\;cm^{-3}}$, temperature $T = 3.2 \times
10^{7}\;{\rm K}$, and velocity $u_{\rm ics} = 0.75 v_{\rm b} = 1.1
\times 10^{8}\;{\rm cm\;s^{-1}}$. The parameters noted above are
typical of small interstellar clouds \citep{McKee:1977}.

\subsection{Collisional or collisionless?}
\label{sec:collisions}
The first consideration is whether the mean-free-path is short enough
that a collisional treatment can be adopted (i.e. whether the plasma
has fluid-like properties). If this is the case, complications due to
collisionless wave-particle interactions and their associated effects
such as Landau damping can be ignored \citep{Parker:1979}. The damping
of hydromagnetic waves in a collisionless plasma is stronger than in a
collisional plasma, since collisions serve to suppress the
wave-particle interactions which damp the wave so strongly in the
collisionless case.

The Coulomb collision cross-section for electron or ion scattering is
$\sigma \approx 10^{-12}/T_{\rm eV}^{2}\;{\rm cm^{2}}$, where $T_{\rm
eV}$ is the temperature in electron volts. For neutral material the
gas atomic cross-section is $\sim 10^{-16}\;{\rm cm^{2}}$.  Hence the
mean-free-path within the ionized and neutral clouds is $\lambda =
1/n\sigma \approx 10^{12}\;{\rm cm}$ and $\approx 10^{14}\;{\rm cm}$
respectively. These values are much smaller than the cloud radii, so a
fluid-like treatment is appropriate. On the other hand, the
mean-free-path in the post-shock intercloud medium is $\lambda \approx
6 \times 10^{20}\;{\rm cm}$, which is significantly larger than the
radius of the clouds. However, even a small B-field will significantly
reduce the effective mean-free-path (the mean-free-path is then of
order the gyroradius, which for $B = 3\;\mu$G and $T = 10^{7}\;$K is
$r_{\rm g} \sim 10^{9}\;$cm for protons), so the shocked intercloud
gas can also be considered to be collisional.

\subsection{Reynolds number, eddies and instabilities}
\label{sec:reynoldsnumber}
The Reynolds number of flow past a cloud is ${\rm Re} = u r_{\rm
c}/\nu$, where $u$ is the average flow velocity past the cloud,
$r_{\rm c}$ is the radius of the cloud, and $\nu$ is the kinematic
viscosity. For a fully ionized, non-magnetic gas of density $\rho$ and
temperature $T$,
\begin{equation}
\label{eq:kinviscosity}
\nu = 2.21 \times 10^{-15} \frac{T^{5/2} A^{1/2}}{Z^{4}\rho\;{\rm ln}\;\Lambda}\;\;{\rm cm^{2}\;s^{-1}},
\end{equation}
where $A$ and $Z$ are the atomic weight and charge of the positive
ions, and ${\rm ln}\;\Lambda$ is the coulomb logarithm
\citep{Spitzer:1956}.  In a magnetized plasma, the kinematic viscosity
is of the order of the mean-free-path (i.e. the particle gyroradius)
times the typical thermal velocity.

The characteristic Reynolds number of the interaction is higher in the
cloud material than the surrounding environment, since the cloud is
considerably cooler. The shock driven into the cloud has a speed $v =
v_{\rm b}/\sqrt{\chi}$, where $\chi$ is the density contrast of the
cloud with respect to its surroundings. For the ionized cloud
considered in Section~\ref{sec:review}, the post-shock temperature is
$2.6\times10^{5}\;{\rm K}$ and $\nu \approx 5 \times 10^{20}\;{\rm
cm^{2}\;s^{-1}}$. Hence ${\rm Re} \approx 7 \times 10^{5}$.  If
the cloud is magnetized with a post-shock field $B \sim 10\;\mu$G, the
kinematic viscosity for protons is $\nu \sim 10^{15}\;{\rm
cm^{2}\;s^{-1}}$ and the Reynolds number is proportionally higher. In
the neutral cloud the Reynolds number calculated from the kinematic
molecular viscosity is similarly high. The Reynolds number of the
flow around a cloud is also high. For a flow with $B = 3\;\mu$G and $T
= 10^{7}\;$K, the kinematic viscosity $\nu \sim 10^{17}\;{\rm
cm^{2}\;s^{-1}}$.  In all cases, the viscous stresses acting on the
boundary layer which forms as the shock sweeps over the cloud are
negligible, and a turbulent energy cascade ensues.

The largest eddies have a length scale, $l$, comparable in size to the
cloud, while the smallest eddies where the turbulent energy is
dissipated have a length scale $\eta \sim {\rm Re}^{-3/4}\;l$.
Due to the nonlinear term ${\bf u} \cdot \nabla {\bf u}$ in the
equation of motion, large eddies, created by instabilities in the mean
flow, are themselves subject to inertial instabilities and rapidly
``break-up'' or evolve into yet smaller vortices. Energy is transferred
into vortices of about one-half their size in a time comparable to
their ``turnover time'' \citep[$t = l/u$, where $u$ is the
characteristic velocity of eddies of size $l$,][]{Davidson:2004}.
Provided that energy is constantly injected at large scales,
small-scale eddies are superimposed on larger eddies. The timescale to
set up a turbulent energy cascade is roughly twice the turnover time
of the largest eddies (i.e. $t = 2 l/u$). Since $l/u \sim r_{\rm
c}/u_{\rm ics}$, the setup time is $t \sim 2 r_{\rm c}/v_{\rm b} =
t_{\rm sc}$, where $t_{\rm sc}$ is the timescale for the shock in the
intercloud medium to sweep over the cloud. For dense clouds ($\chi \gg
1$), this setup time is much shorter than the survival time of the
cloud, which is a few times the ``cloud crushing'' timescale,
\begin{equation}
\label{eq:tcc}
t_{\rm cc} \equiv \frac{\chi^{1/2}r_{\rm c}}{v_{\rm b}},
\end{equation}
for the cloud to be crushed by the initial shock that is driven into
it \citep*{Klein:1994}.

Simultaneously with the top-down energy transfer from large to small
eddies, Kelvin-Helmholtz (KH) and Rayleigh-Taylor (RT) instabilities
inject energy from the bottom-up, since the smallest scale
disturbances grow fastest.  The KH and RT growth times are
\begin{equation}
t_{\rm KH} \sim \frac{t_{\rm cc}}{k_{\lambda}r_{\rm c}},\;\;\;\;\;\;\;t_{\rm RT} \sim \frac{t_{\rm cc}}{(k_{\lambda}r_{\rm c})^{1/2}},
\end{equation}
where $k_{\lambda}$ is the wave-number of the perturbation
\citep{Klein:1994}. The smallest scale of the instabilities is set by
the scale at which the damping of hydromagnetic waves occurs, which was
shown in Section~\ref{sec:collisions} to be through particle
collisions rather than wave-particle interactions.

For the unmagnetized ionized cloud the minimum scale due to viscous
damping is $\eta_{\rm vis} \sim {\rm Re}^{-3/4}\;r_{\rm c} \sim 4
\times 10^{-5}\;r_{\rm c}$.  However, in ionized plasmas the thermal
conductivity is even more effective at damping waves with a
significant longitudinal component.  The thermometric conductivity $K
= \kappa T/U$, where $U$ is the thermal energy per unit volume and
$\kappa$ is the thermal conduction coefficient \citep{Parker:1979}.
Here we find that $K \approx 40\;\nu$, and thermal conduction prevents
instabilities with a scale smaller than $\eta_{\rm tc} \sim 6 \times
10^{-4}\;r_{\rm c}$. In magnetized ionized clouds the lengthscale at
which instabilities are damped is smaller if the magnetic field is
sufficiently weak to be dynamically unimportant. On the other hand, if
the magnetic field is strong enough to be dynamically important,
magnetic tension increases the minimum lengthscale at which
instabilities occur.

A disturbance of wavelength $\lambda = 10^{-3}\;r_{\rm c}$ grows
in a timescale of $t_{\rm KH} = 1.6\times10^{-4}\;t_{\rm cc}$ and $t_{\rm
RT} = 1.3\times10^{-2}\;t_{\rm cc}$. With the ionized cloud parameters
given above, $\chi=133$, $t_{\rm cc} = 5.8\; t_{\rm sc}$, and we
obtain $t_{\rm KH} \approx 10^{-3}\;t_{\rm sc}$ and $t_{\rm RT}
\approx 0.1\; t_{\rm sc}$. These timescales are much faster than the
one to setup the turbulent energy cascade (which was shown above to be
$\sim t_{\rm cc}$), and the (ensemble-averaged) turbulent spectrum
will differ from the classical Kolmogorov spectrum because it is
driven by energy input at both large and small scales.

Photoionization at the surface of neutral clouds maintains most of
the C in the form $C^{+}$ \citep[e.g.,][]{Hartquist:1998}. An
upper limit to the ratio of the ionized to neutral mass density,
$\rho^{*}/\rho$, is therefore the fractional abundance by mass of
carbon, which has a cosmic value of about $3 \times 10^{-3}$
\citep[e.g.,][]{Dopita:2003}.  \citet{Kulsrud:1969} note that for
hydromagnetic waves with $\lambda < \lambda_{2}$, the charged
particles move as if the neutrals were absent, while for $\lambda >
\lambda_{1}$, the entire medium moves. In contrast, when $\lambda_{2}
< \lambda < \lambda_{1}$, the neutrals and ions move independently and
friction is strong, and waves do not propagate. $\lambda_{1}$
corresponds to waves with angular frequencies comparable to the rate
at which neutrals transfer momentum to ions, while $\lambda_{2}$ is
the corresponding rate for momentum transfer from ions to neutrals.
Assuming that the magnetic field strength within our neutral cloud is
$\sim 3\;\mu$G, we find that $\lambda_{1} \approx 4 \times
10^{15}\;$cm and $\lambda_{2} \approx 1.3 \times 10^{15}\;$cm. 
In this case fully developed turbulence is prevented, since 
$\lambda_{1}/r_{\rm c} > \eta_{\rm vis}$ when $Re \sim 10^{5}$,
although instabilities continue to be driven from the bottom up.
Fully developed turbulence may be obtained when the magnetic field
is weaker, since $\lambda_{1}$ and $\lambda_{2}$ are both proportional
to $B$, or when the cloud is larger.

%This is
%within a factor of a few of the minimum wavelength of RT and KH
%instabilities for fully developed turbulence (which in a cloud of
%radius $r_{\rm c}$ is $\sim 10^{-4}\;r_{\rm c}$, i.e.  $\sim 1.5
%\times 10^{15}\;{\rm cm}$ for $r_{\rm c}=4.6\;{\rm pc}$), so
%turbulence is driven from the bottom up in this case too.

In numerical simulations of shock-cloud interactions, the growth of KH
and RT instabilities is closely related to the development of the slip
surface around the cloud, and perturbations with wavelengths smaller
than the thickness of the shear layer are stabilized
\citep{Nakamura:2006}.  Thus, small scale instabilities have been
artificially suppressed in previous work on hydrodynamic shock-cloud
interactions. Our use of a $k-\epsilon$ turbulence model in this paper
attempts to address this shortcoming.

\subsection{The turbulent boundary layer}
\citet{Hartquist:1988} argued that the turbulent boundary layer which
forms around a cloud has a thickness of the order of $r_{\rm
c}/\sqrt{\rm Re_{t}}$, where ${\rm Re_{t}}$ is an effective
``turbulent'' Reynolds number arising from the fact that the
turbulence itself gives rise to an effective viscosity.  Since ${\rm
Re_{t}} \sim 10^{3}$ \citep{Hartquist:1988}, the thickness of the
turbulent boundary layer is a few percent of the cloud radius. More
detailed calculations and laboratory experiments show that the opening
angle of a turbulent mixing layer in a mildly supersonic flow (such as
occurs behind a high Mach number shock) is of order $10^{\circ}$
\citep{Canto:1991} - we find good agreement with this (see
Section~\ref{sec:morph}). Convergence tests reveal that a minimum
numerical resolution of about $120$ cells per cloud radius is needed
for convergence of various global quantities \citep[see, e.g.,][although 
the necessary resolution may to some extent also depend on
the numerical scheme]{Nakamura:2006}.  This is consistent with
simulations of this resolution and higher beginning to resolve the
turbulent boundary layer.

\subsection{Cloud destruction and mixing}
\label{sec:destandmix}
The main stages in the destruction of a cloud by a shock and the
subsequent mixing of its material into the surrounding flow are
reviewed in Section~\ref{sec:stages}.  The most disruptive KH and RT
instabilities are those with wavelengths of the order of the cloud 
radius. However, there are a number of ways in
which the growth of KH and RT instabilities may be hindered or
amplified. For instance, clouds with diffuse boundaries are less
susceptible to KH instabilities and survive longer
\citep{Nakamura:2006}.  Two dimensional MHD simulations found that the
growth of KH and RT instabilities is strongly inhibited when there is
a dynamically important magnetic field, due to the field providing an
additional tension at the interface between the cloud and the
surrounding flow \citep{MacLow:1994}. Somewhat surprisingly, fully
three dimensional magneto-hydrodynamic (MHD) simulations of a
wind-cloud interaction revealed that an ordered magnetic field can
actually {\em enhance} hydrodynamic instabilities, as background field
lines become trapped in deformations in the surface of the cloud
\citep[][though these were low resolution calculations]{Gregori:1999}.
Higher resolution shock-cloud simulations recently presented by
\citet{Shin:2008} show that irrespective of the field geometry, and
the morphology of the cloud fragments which are produced in the
interaction, the rate of mixing is reduced compared to the
non-magnetic case. However, sufficiently weak magnetic fields have no
dynamical influence, and only offer a potential reduction in the
collision mean-free-path. Strong
thermal conduction suppresses hydrodynamic instabilities \citep[see
Section~\ref{sec:reynoldsnumber} and
also][]{Orlando:2005,Marcolini:2005}, but the degree of this effect is
sensitive to the orientation of any magnetic field \citep{Orlando:2008}.

KH and RT instabilities are always stronger when radiative cooling is
important, and the cloud breaks up into numerous dense, cold fragments
\citep{Mellema:2002,Fragile:2004,Fragile:2005,vanLoo:2007}. In the 
corresponding simulations the
fragments appear to survive for an appreciable time, but are poorly
resolved, so the timescales corresponding to their further evolution
are somewhat uncertain. The way in which clouds are destroyed and
mixed into their surroundings is sensitive also to the density
contrast between the cloud and the surrounding medium, in that clouds
with higher values of $\chi$ suffer direct stripping of material from
their surfaces by hydrodynamic ablation \citep[][also compare
Figs.~\ref{fig:nokeps},~\ref{fig:nokeps_chi1e1}
and~\ref{fig:nokeps_chi1e2} in this work]{Klein:1994}.

Cloud destruction and the mixing of the cloud material into the
surrounding flow are two distinct processes which were not always
distinguished in previous work.  In \citet{Nakamura:2006}, the cloud
destruction timescale is taken to be the time when the largest
fragment drops below a certain fraction of the initial cloud mass,
while the timescale for the mixing of former cloud material into the
surrounding medium is estimated by comparing the integrated mass above
a particular threshold density with the initial cloud mass.  However,
even weak magnetic fields can prevent the actual mixing of stripped
material with the surrounding medium, and reconnection on small-scales
is necessary if the plasmas are to mix fully.  This may, of course,
take some considerable time to achieve.

\section{The Numerical Setup}
\label{sec:setup}
\subsection{The numerical scheme}
\label{sec:scheme}
The shock-cloud interaction is modelled by solving numerically the
Euler equations of inviscid fluid flow, supplemented by a sub-grid
turbulent viscosity model as appropriate. When the sub-grid model is
included, the continuity, scalar, momentum, energy, turbulent energy
and turbulent dissipation equations are respectively:
%(Eqs.~\ref{eq:turb_energy} 
%and~\ref{eq:turb_diss}, and the source terms on the right-hand-side of
%Eqs.~\ref{eq:mtm} and~\ref{eq:energy}): 
\begin{equation}
\label{eq:mass}
\frac{\partial \rho}{\partial t} + \nabla \cdot (\rho {\bf u}) = 0,\\  %continuity
\end{equation} 
\begin{equation}
\label{eq:scalar}
\frac{\partial \rho \kappa}{\partial t} + \nabla \cdot (\rho \kappa {\bf u}) -
\nabla \cdot (\mu_{\rm T} \nabla \kappa) = 0,\\ %advected scalar
\end{equation} 
\begin{equation}
\label{eq:mtm}
\frac{\partial \rho {\bf u}}{\partial t} + \nabla \cdot (\rho {\bf uu}) + \nabla P - \nabla \cdot {\bf \tau} = {\bf S}_p,\\ %mtm
\end{equation} 
\begin{equation}
\label{eq:energy}
\frac{\partial E}{\partial t} + \nabla \cdot [(E + P){\bf u} - {\bf u}
\cdot \tau] -  {\gamma \over {\gamma - 1}}  \nabla \cdot (\mu_{\rm T} \nabla
T) = S_E,\\ %energy
\end{equation} 
\begin{equation}
\label{eq:turb_energy}
\frac{\partial \rho k}{\partial t} + \nabla \cdot (\rho k {\bf u}) - \nabla \cdot (\mu_{\rm T} \nabla k) = S_{k},\\ %turbulent energy
\end{equation} 
\begin{equation}
\label{eq:turb_diss}
\frac{\partial \rho \epsilon}{\partial t} + \nabla \cdot (\rho \epsilon {\bf u})  - \nabla  \cdot (\mu_{\epsilon} \nabla  \epsilon) = S_{\epsilon}.%turbulent dissipation 
\end{equation} 
%\end{eqnarray}
Here $\rho$ is the mass density, ${\bf u}$ is the velocity, $E$ is the total
energy density
\begin{equation}
E = \frac{P}{\gamma - 1} + \frac{1}{2}\rho{\bf u}^2,
\end{equation} 
$P$ is the thermal pressure, $k$ is the turbulent energy per unit
mass, $\epsilon$ is the turbulent dissipation rate per unit mass, and
the turbulent diffusion coefficients are
\begin{equation}
\mu_{\rm T} = \rho C_{\rm \mu} \frac{k^{2}}{\epsilon},~~~\mu_\epsilon = {\mu_{\rm T}
\over 1.3},
\end{equation}
where $C_\mu = 0.09$. $\kappa$ is an advected scalar used to distinguish
between cloud and ambient material.

The  momentum   equation  source  term,   ${\bf  S}_p$,  is   zero  in
Cartesians. In cylindrical symmetry it is
\begin{equation}
{\bf S}_p = 
\left( {
\begin{array}{c}
\displaystyle{\mu_{\rm T} \left[ { {2 \over {3 r}} \nabla \cdot {\bf u} - 2
{u_r \over r^2} } \right] + {1 \over r} \left[ {P + {2 \over {3
r}} \rho k} \right]} \\
\\
0\\
\end{array}
} \right).
\end{equation}
The $k$ and $\epsilon$ source terms are respectively
\begin{equation}
S_k = P_t - \rho \epsilon,
\end{equation}
and
\begin{equation}
S_\epsilon = {\epsilon \over k} (C_1 P_t - C_2 \rho \epsilon),
\end{equation}
where $C_1 = 1.4$ and $C_2 = 1.94$.

The turbulent production term
\begin{equation}
P_{\rm t} = \mu_{\rm T}  \left[ {  {{\partial u_i}  \over {\partial  x_j}} \left(
{{{\partial  u_i}  \over  {\partial  x_j}}  +  {{\partial  u_j}  \over
{\partial x_i}} } \right) } \right] - {2 \over 3} \nabla \cdot {\bf u}
(\rho       k       +        \mu_{\rm T}       \nabla       \cdot       {\bf
u}),
\end{equation}
where the summation convention is assumed. The terms involving
$\mu_{\rm T}$ are due to the rate of working of the turbulent
stresses, and the terms involving $\nabla \cdot {\bf u}$ take into
account volume changes on the turbulence.  In cylindrical symmetry,
the production term has to be complemented by an extra geometric term
\begin{equation}
2 \mu_{\rm T} {u_r^2 \over r^2}.
\end{equation}

The turbulent stress tensor, $\tau$, is
\begin{equation}
\tau_{ij} =  \mu_{\rm T} \left(  { {{\partial u_i}  \over {\partial  x_j}} +
{{\partial  u_j}  \over  {\partial  x_i}}  } \right)  -  {2  \over  3}
\delta_{ij} (\mu_{\rm T} \nabla \cdot {\bf u} + \rho k).
\end{equation}

All computations were performed in 2D cylindrical symmetry using an Eulerian
adapative mesh refinement (AMR) hydrodynamic code, with a linear
Godunov solver and piece-wise linear cell interpolation
\citep[see][]{Falle:1991}. Although it is of lower order than the piecewise
parabolic method (PPM), it performs well in multi-dimensional
problems, as it is only partially operator split, and it performs
better than PPM for problems where there is rapid advection across the
computational grid \citep{Runacres:2005} 
%\citep[a recent comparison of grid and particle based codes can 
%be found in][]{Agertz:2007}.  

The entire computational domain is covered by the two coarsest grids,
$G^0$ and $G^1$. The solution at each position is calculated on all
grids that exist there, and the difference between these solutions is
used to control refinement \citep[note that refinement criteria based
on the local gradients of only selected variables (e.g., density) do
not properly resolve turbulent flow - see][]{Iapichino:2008}.  Finer grids
are dynamically added where they are needed, and removed where they
are not. Each refinement level increases the resolution in each of the
spatial directions by a factor of
2, and the refinement is done on a cell-by-cell rather than on a patch
basis.  The time-step on grid $G^{\rm n}$ is $\Delta t_{0}/2n$, where
$\Delta t_0$ is the time-step on $G^0$, in order to ensure Courant
number matching at the boundaries between coarse and fine grids.

The $k$-$\epsilon$ model is designed to model the mean flow in fully 
developed, high Reynolds number, turbulence. It has been calibrated by 
comparing the computed growth of shear layers with experiments of 
high Reynolds number flows \citep{Dash:1983}.  Although the flow in our 
problem is somewhat more complicated than in these experiments, it is 
described by exactly the same equations. Of course, any turbulence model 
can only be approximately correct, but it should give more reliable 
results than an inviscid calculation.
% Note that the nature of
%turbulence is such that the characteristics of a turbulent flow at a
%given snap-shot in time can be very different from the average
%properties obtained from a large ensemble of snapshots. Therefore, the
%$k$-$\epsilon$ model actually approximates the time-averaged
%flow. This property is not a problem in our simulations, since the
%destruction timescale of the cloud is long compared to the lifetime of
%eddies.

In the sub-grid model, turbulent energy is generated by the action of
the turbulent viscosity on the mean flow and is converted to heat at
the dissipation rate, $\epsilon$.  Since the turbulent energy resides
mainly in large eddies, while the dissipation occurs in the small
ones, one can think of $k$ and $\epsilon$ as describing the
large-scale and small-scale turbulence respectively. Since the
aim of the sub-grid model is to mimic a three dimensional turbulent
flow, the effects of the turbulence (such as enhanced transport
coefficients) are treated correctly, even though the grid is
cylindrically symmetric. However, the turbulent motions that are
resolved on the grid are actually vortex rings, and not eddies.
Further details of the model implementation can be found in
\citet{Falle:1994}.

\subsection{Initial and boundary conditions}
We consider a Mach 10 shock interacting with a cloud with a density
contrast $\chi$ of either $10$, $10^{2}$ or $10^{3}$, and compute
simulations with different spatial resolutions using either an
inviscid code or one which includes a $k$-$\epsilon$ turbulence model.
The effect of different levels of turbulence in the
post-shock gas is also explored. Table~\ref{tab:models} summarizes the
calculations performed. Most computations are for clouds with steep
density profiles, but we have also examined the effect of a shallower 
density profile on the resulting evolution (see Section~\ref{sec:shallow}).
A calculation using an inviscid code with grid-scale post-shock
turbulence is also presented (see Section~\ref{sec:grid-scale-turb}).

\begin{table}
\begin{center}
\caption[]{Summary of the shock-cloud simulations performed. The resolution 
is the number of cells per cloud radius on the finest grid. Models with ``sh''
in their name were computed for clouds with a shallow density gradient. Model 
c1rtb32 was computed using an inviscid code with grid-scale turbulence.}
\label{tab:models}
\begin{tabular}{lllll}
\hline
\hline
Model & $\chi$ & $p_{1}$ & resolution & turbulence \\
\hline
c1no & $10^{1}$ & 10 & 32,64,128 & no \\
c1lo & $10^{1}$ & 10 & 32,64,128 & low \\
c1hi & $10^{1}$ & 10 & 64,128 & high \\
c2no & $10^{2}$ & 10 & 32,64,128,256 & no \\
c2lo & $10^{2}$ & 10 & 32,64,128,256 & low \\
c2hi & $10^{2}$ & 10 & 16,32,64,128,256 & high \\
c3no & $10^{3}$ & 10 & 16,32,64,128 & no \\
c3lo & $10^{3}$ & 10 & 16,32,64,128 & low \\
c3hi & $10^{3}$ & 10 & 32,64,128 & high \\
c2nosh & $10^{2}$ & 1 & 64 & no \\
c2losh & $10^{2}$ & 1 & 64 & low \\
c2hish & $10^{2}$ & 1 & 64 & high \\
c3nosh & $10^{3}$ & 1 & 64 & no \\
c3losh & $10^{3}$ & 1 & 64 & low \\
c3hish & $10^{3}$ & 1 & 64 & high \\
c1rtb & $10^{1}$ & 10 & 32 & grid-scale \\
\hline
\end{tabular}
\end{center}
\end{table}

The calculations are computed on an $r-z$ cylindrically symmetric
grid, with a domain of $0 \leq r \leq 24$, $-94 \leq z \leq 6$ when
$\chi=10$, $0 \leq r \leq 24$, $-120 \leq z \leq 6$ when
$\chi=10^{2}$, and $0 \leq r \leq 48$, $-594 \leq z \leq 6$ when
$\chi=10^{3}$.  This ensures that the cloud is well dispersed and
mixed into the post-shock flow before the shock reaches the edge of
the numerical grid.  In this way the global quantities detailed in
Section~\ref{sec:global} are accurately computed.  All calculations
are for an ideal gas with $\gamma=5/3$, and are scaled so that the
fluid variables have values reasonably close to unity.

Several additional parameters must be specified when using a
turbulence model. One of these is the maximum eddy size, which here is
set equal to the cloud radius. Another choice concerns the initial
level of turbulence in the gas, and two extreme cases are considered.
In the first case (hereafter identified by ``low $k$-$\epsilon$'', or
``lo'' in the model name) the postshock gas initially has an extremely
low level of turbulence, with a ratio of turbulent energy density to
thermal energy density $e_{\rm tb}/e_{\rm th} \sim 10^{-6}$.  In the
second case (identified by ``high $k$-$\epsilon$'', or ``hi'' in the
model name) this ratio is 0.13 (values much higher than this cause the
shock to accelerate too much and the interaction does not occur at the
intended Mach number).  High levels of post-shock turbulence may arise
if the shock is propagating through an inhomogeneous medium (e.g.,
when there are density variations further upstream). As we shall see,
this ratio is similar to the turbulent energy fraction attained by the
cloud material after the shock encounter (see
Section~\ref{sec:energy_frac}), so there is a degree of
self-consistency in these models. In both cases, initially $e_{\rm
tb}/e_{\rm th} = 0.04$ within the cloud (this is set low enough to not
affect the dynamics - cf. the initial development of the interaction
in models c3no128 and c3lo128), and the cloud and intercloud medium
are set in pressure equilibrium.

\subsection{Cloud density profile}
\begin{figure}
\psfig{figure=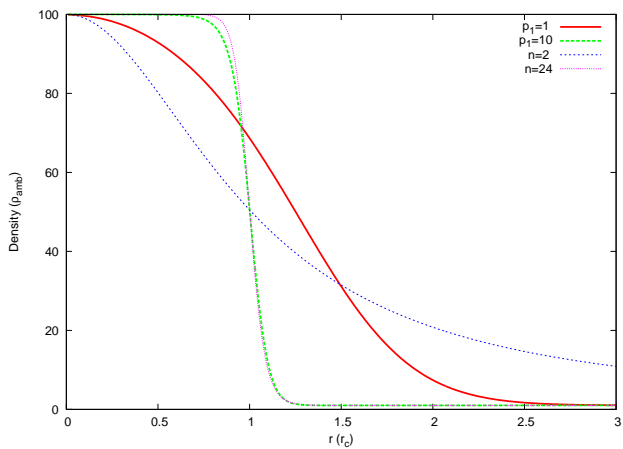,width=8.5cm}
\caption[]{Comparison of cloud density profiles obtained using
Eq.~\ref{eq:cloudprofile} ($p_{1} = 1$ and 10) and Eq.~1 in 
\citet{Nakamura:2006} ($n = 2$ and 24) with $\chi=10^{2}$.}
\label{fig:cloudprofs}
\end{figure}

Clouds in the ISM do not have infinitely sharp edges \citep[see, e.g.,
the discussion in][]{Nakamura:2006}. The density profile adopted in
this work is
\begin{equation}
\label{eq:cloudprofile}
\rho(r) = \rho_{\rm amb}[\psi + (1 - \psi)\eta],
\end{equation}
where
\begin{equation}
\eta = \frac{1}{2}\left[1 + \frac{\alpha - 1}{\alpha + 1}\right],\nonumber
\end{equation}
$\alpha = {\rm exp}\;\{{\rm min}[20.0,p_{1}((r/r_{\rm c})^{2}-1)]\}$,
and $r$ is the distance from the centre of the cloud of radius $r_{\rm
c}$. $\psi$ is adjusted to obtain a specific density contrast for the
centre of the cloud with respect to the ambient medium, $\chi =
\rho_{\rm max}/\rho_{\rm amb}$. The parameter $p_{1}$ controls the
steepness of the profile at the edge of the
cloud. Eq.~\ref{eq:cloudprofile} tends to give a flatter density
profile within the centre of the cloud, and a steeper profile as the
cloud merges into the ambient medium, than profiles obtained using
Eq.~1 in \citet{Nakamura:2006}. For $p_{1} \gtsimm 5$, $\psi \approx
\chi$. Clouds with reasonably sharp edges are obtained with $p_{1}=10$
(similar to those from Eq.~1 in \citet{Nakamura:2006} with $n=24$),
while $p_{1}=1$ produces a much more extended cloud which is closer to
the \citet{Nakamura:2006} profile with $n=2$ (see
Fig.~\ref{fig:cloudprofs}).

\begin{table*}
\begin{center}
\caption[]{Dependence of the global cloud and core properties on the
level of turbulence and the density contrast of the cloud (see
Table~\ref{tab:models}). In each case the cloud was hit by a Mach 10
shock. The time-dependent quantities are evaluated at $t=t_{\rm mix}$
rather than $t=t_{\rm m}$ \citep[c.f.][]{Nakamura:2006}, since $a_{\rm
cloud}$ continues to rise in some simulations. Values in parentheses
are obtained from integrations over the ``core'' mass rather than the
``cloud'' mass. Model names containing ``sh'' were computed using a
shallow density profile ($p_{1} = 1$). Model c1rt32 was computed using
an inviscid code and grid-scale post-shock turbulence.}
\label{tab:results}
\begin{tabular}{lcccccccc}
\hline
\hline
Model & $t_{\rm drag}/t_{\rm cc}$ & $t_{\rm mix}/t_{\rm cc}$ & $t_{\rm m}/t_{\rm cc}$ & $a/r_{\rm c}$ & $c/r_{\rm c}$ & $c/a$ & $\langle\rho\rangle/\rho_{max}$ & $\langle v_{\rm z}\rangle/v_{\rm b}$ \\
\hline
c1no128 & 1.03 (1.10) & (6.82) & 3.82 & 1.74 (1.72) & 2.82 (2.92) & 1.62 (1.70) & 0.65 (0.87) & 0.634 (0.685) \\
c1lo128 & 1.04 (1.11) & (5.96) & 3.57 & 1.66 (1.74) & 2.20 (2.40) & 1.33 (1.38) & 0.68 (1.03) & 0.579 (0.629) \\
c1hi128 & 0.72 (0.86) & (4.37) & 3.52 & 1.88 (1.90) & 1.36 (1.30) & 0.72 (0.68) & 0.56 (0.66) & 0.621 (0.651) \\
c2no128 & 3.06 (3.10) & (5.05) & 4.53 & 4.36 (4.21) & 5.07 (3.58) & 1.16 (0.85) & 0.073 (0.130) & 0.635 (0.606) \\
c2lo128 & 3.08 (3.09) & (4.86) & 4.82 & 3.92 (3.35) & 4.12 (3.05) & 1.05 (0.91) & 0.075 (0.142) & 0.593 (0.554) \\
c2hi128 & 2.47 (2.65) & (5.73) & - & 4.05 (4.27) & 6.24 (1.66) & 1.54 (0.39) & 0.058 (0.078) & 0.606 (0.590) \\
c3no128 & 6.58 (9.48) & (8.59) & 10.23 & 2.60 (1.22) & 75.9 (12.6) & 29.2 (10.3) & 0.0081 (0.0401) & 0.394 (0.189) \\
c3lo128 & 6.84 (7.15) & (7.83) & - & 4.01 (2.82) & 49.1 (6.88) & 12.2 (2.44) & 0.0078 (0.0248) & 0.430 (0.271) \\
c3hi128 & 4.58 (4.93) & (5.95) & 7.25 & 6.84 (7.79) & 38.2 (5.57) & 5.58 (0.71) & 0.0064 (0.0122) & 0.584 (0.578) \\
c2nosh64 & 4.05 (5.51) & (7.92) & 8.60 & 3.09 (2.75) & 11.7 (2.33) & 3.80 (0.85) & 0.104 (0.392) & 0.455 (0.402) \\
c2losh64 & 4.00 (4.33) & (7.37) & 9.50 & 2.89 (2.21) & 10.8 (2.02) & 3.74 (0.91) & 0.103 (0.446) & 0.441 (0.377) \\
c2hish64 & 3.08 (3.85) & (4.87) & - & 2.49 (1.46) & 6.50 (0.98) & 2.61 (0.67) & 0.114 (0.476) & 0.379 (0.313) \\
c3nosh64 & 6.72 (9.84) & (7.86) & - & 5.16 (2.42) & 48.6 (8.64) & 9.41 (3.57) & 0.0124 (0.145) & 0.370 (0.216) \\
c3losh64 & 7.56 (14.44) & (8.99) & 5.16 & 3.89 (1.05) & 64.2 (6.12) & 16.5 (5.86) & 0.0121 (0.085) & 0.302 (0.150) \\
c3hish64 & 6.55 (8.36) & (4.95) & 11.82 & 2.63 (2.13) & 22.8 (2.94) & 8.67 (1.38) & 0.0241 (0.144) & 0.204 (0.155) \\
c1rt32 & 0.66 (0.72) & (5.84) & - & 2.21 (2.58) & 1.29 (0.90) & 0.58 (0.35) & 0.61 (0.80) & 0.614 (0.666) \\
\hline
\end{tabular}
\end{center}
%\medskip
\end{table*}

\subsection{Global quantities}
\label{sec:global}
The cloud evolution is studied through various integrated quantities
\citep[see][]{Klein:1994,Nakamura:2006}. Averaged quantities 
$\langle f \rangle$, are constructed by
\begin{equation}
\langle f\rangle = \frac{1}{m_{\beta}}\int_{\kappa \geq \beta} \kappa \rho f \;dV,
\end{equation}
where the mass identified as being part of the cloud is
\begin{equation}
m_{\beta} = \int_{\kappa \geq \beta} \kappa \rho \;dV.
\end{equation}
$\kappa$ is an advected scalar, which has an initial value of 
$\rho/(\chi \rho_{\rm amb})$ for cells within a distance of 
$2.25 r_{\rm c}$ from the centre of the cloud, and a value of
zero at greater distances. Hence, $\kappa=1$ in the centre of the cloud,
and declines outwards. The above integrations are performed only over
cells in which $\kappa$ is at least as great as the threshold value,
$\beta$. Setting $\beta = 0.5$ probes only the densest parts of the
cloud and its fragments, while setting $\beta = 2/\chi$
probes the whole cloud including its low density envelope, and 
regions where only a small percentage of cloud material is mixed into the
ambient medium.

To measure the shape of the cloud,
effective radii normal to and along the axis of symmetry
are defined respectively as
\begin{equation}
a = \left(\frac{5}{2}\langle r^{2}\rangle \right)^{1/2}, \;\;\;\;\;\; 
c = \left[5\left(\langle z^{2}\rangle - \langle z\rangle^{2}\right)\right]^{1/2}.
\end{equation}
In inviscid calculations a measure of the turbulence of the cloud is obtained from
the velocity dispersions in the radial and axial directions, defined 
respectively as
\begin{equation}
\delta v_{\rm r} = \left<v_{\rm r}^{2}\right>^{1/2},\;\;\;\;\;
\delta v_{\rm z} = \left(\left<v_{\rm z}^{2}\right> - \langle v_{\rm z} \rangle^{2}\right)^{1/2}.
\end{equation}
The mean density is defined as
\begin{equation}
\langle\rho\rangle = \frac{m_{\beta}}{V_{\beta}},
\end{equation}
where $V_{\beta}$ is the volume of the region with $\kappa \geq \beta$.

All quantities computed with $\beta = 0.5$ are identified with the
subscript ``core'' (e.g., $a_{\rm core}$), while those computed with
$\beta = 2/\chi$ are given the subscript ``cloud'' 
(e.g., $a_{\rm cloud}$). 

\subsection{Timescales}
Several timescales are obtained from the simulations.
The characteristic radial expansion timescale, $t_{\rm m}$, is defined
as the time at which the cloud radius normal to the axis of symmetry, $a$,
has increased to 90 per cent of its maximum value. 
The time for the cloud velocity relative to that of the postshock
ambient flow to decrease by a factor of 1/$e$ is defined as the
``drag time'', $t_{\rm drag}$. The ``mixing time'', $t_{\rm mix}$, is
defined as the time when the mass of the core of the cloud,
$m_{\rm core}$, reaches half of its initial value. The zero-point of all 
the time measurements quoted in this work occurs when the intercloud 
shock is level with the centre of the cloud.

\begin{figure*}
\psfig{figure=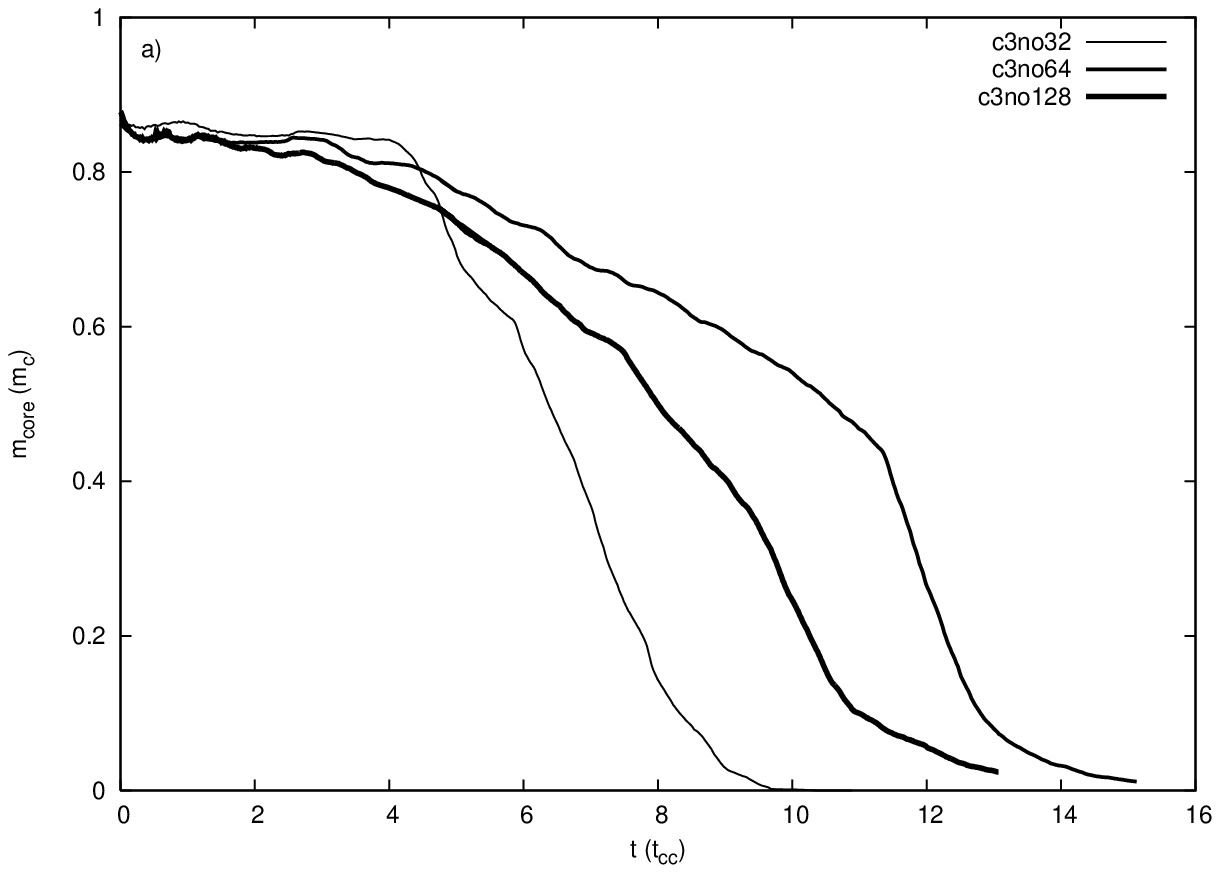,width=8.5cm}
\psfig{figure=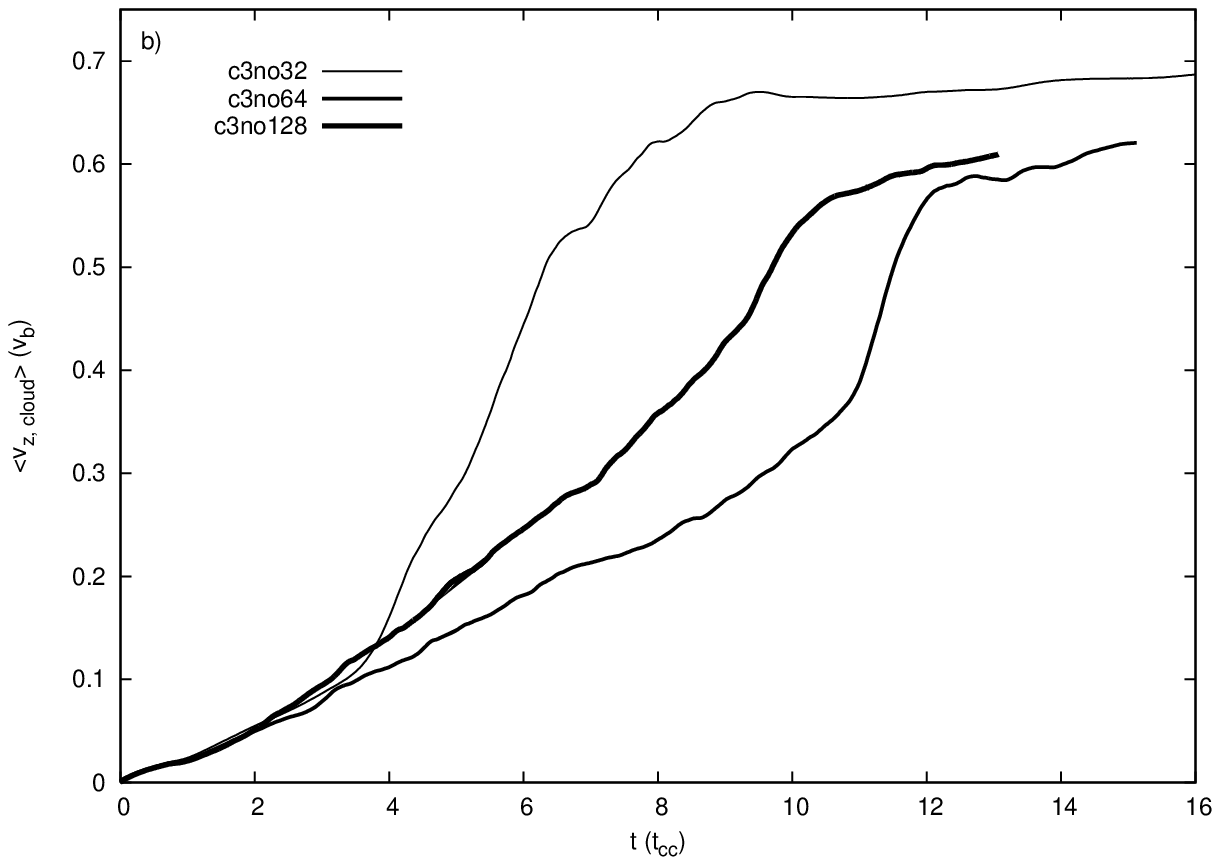,width=8.5cm}
\psfig{figure=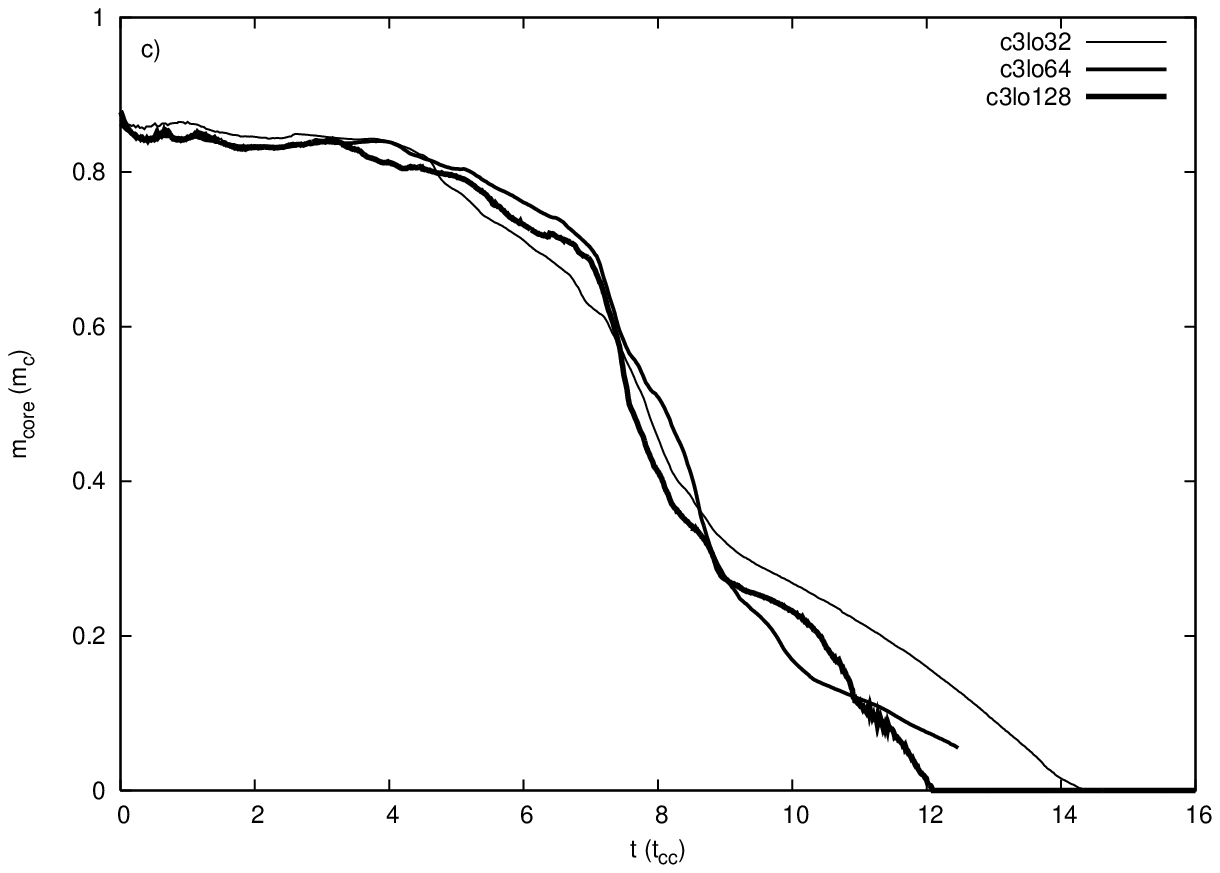,width=8.5cm}
\psfig{figure=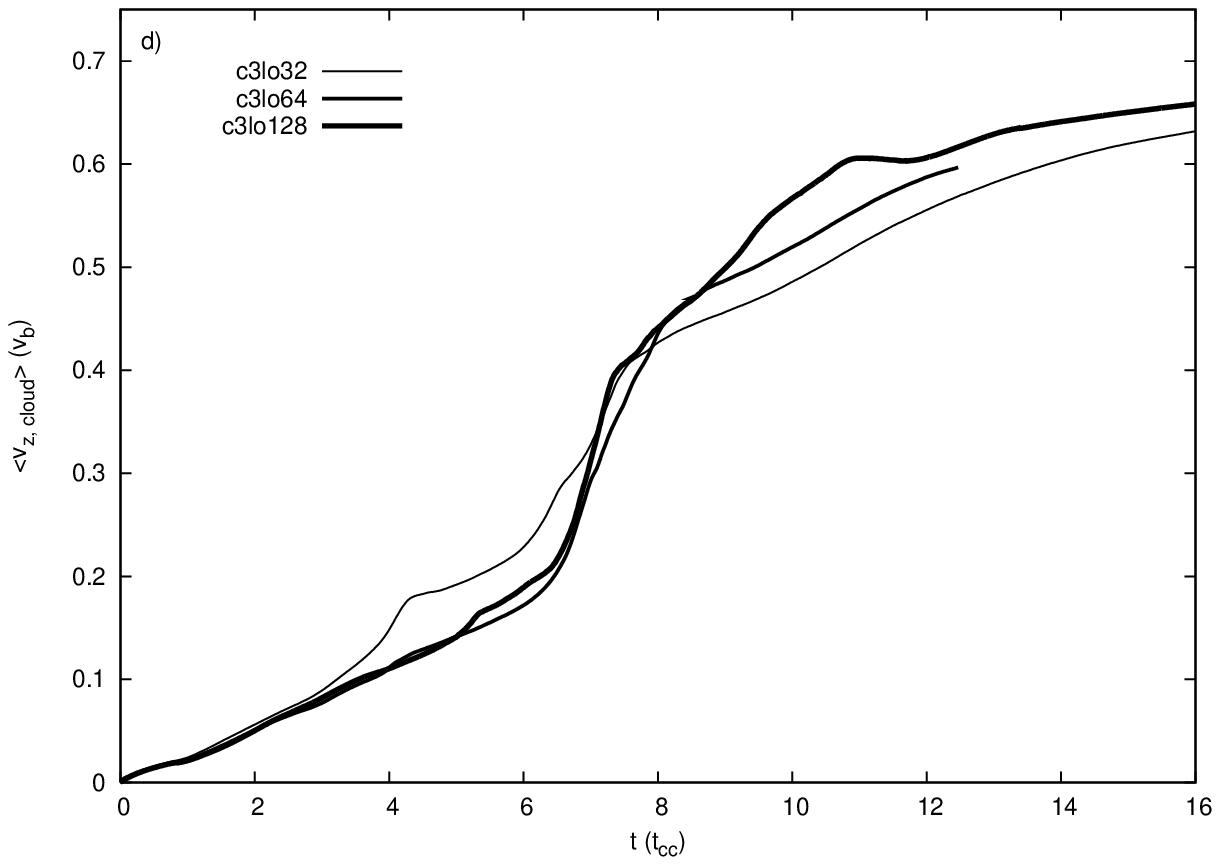,width=8.5cm}
\caption[]{Convergence tests for a shock-cloud interaction with
$\chi=10^{3}$ and $p_{1}=10$ using an inviscid (panels a and b) and
$k$-$\epsilon$ (panels c and d) code. The time evolution of the core
mass (panels a and c) and the mean velocity of the cloud (panels b and
d) are shown.  Note the much tighter correlation of $m_{\rm core}$ and
$\langle v_{\rm z, cloud}\rangle$ from the $k$-$\epsilon$ code.}
\label{fig:rescomp}
\end{figure*}

\begin{figure*}
\psfig{figure=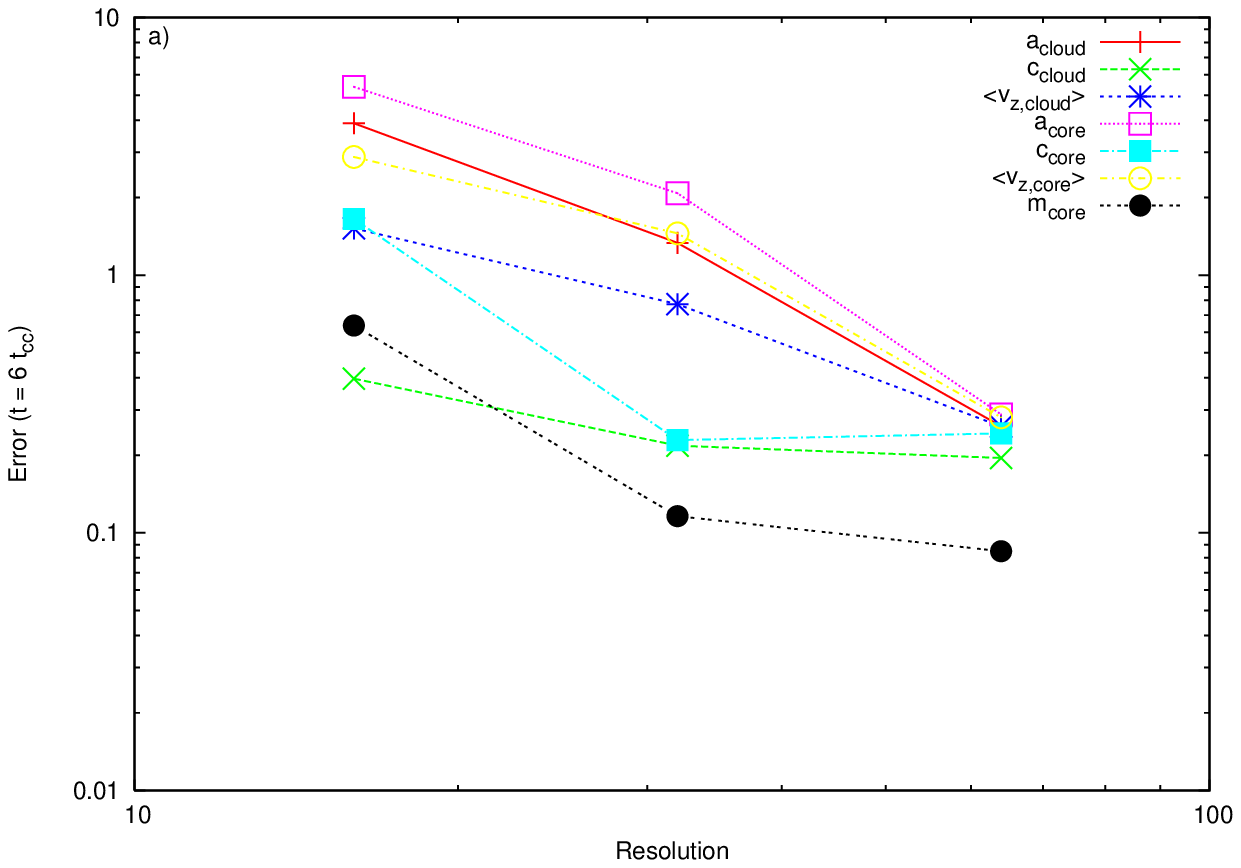,width=8.5cm}
\psfig{figure=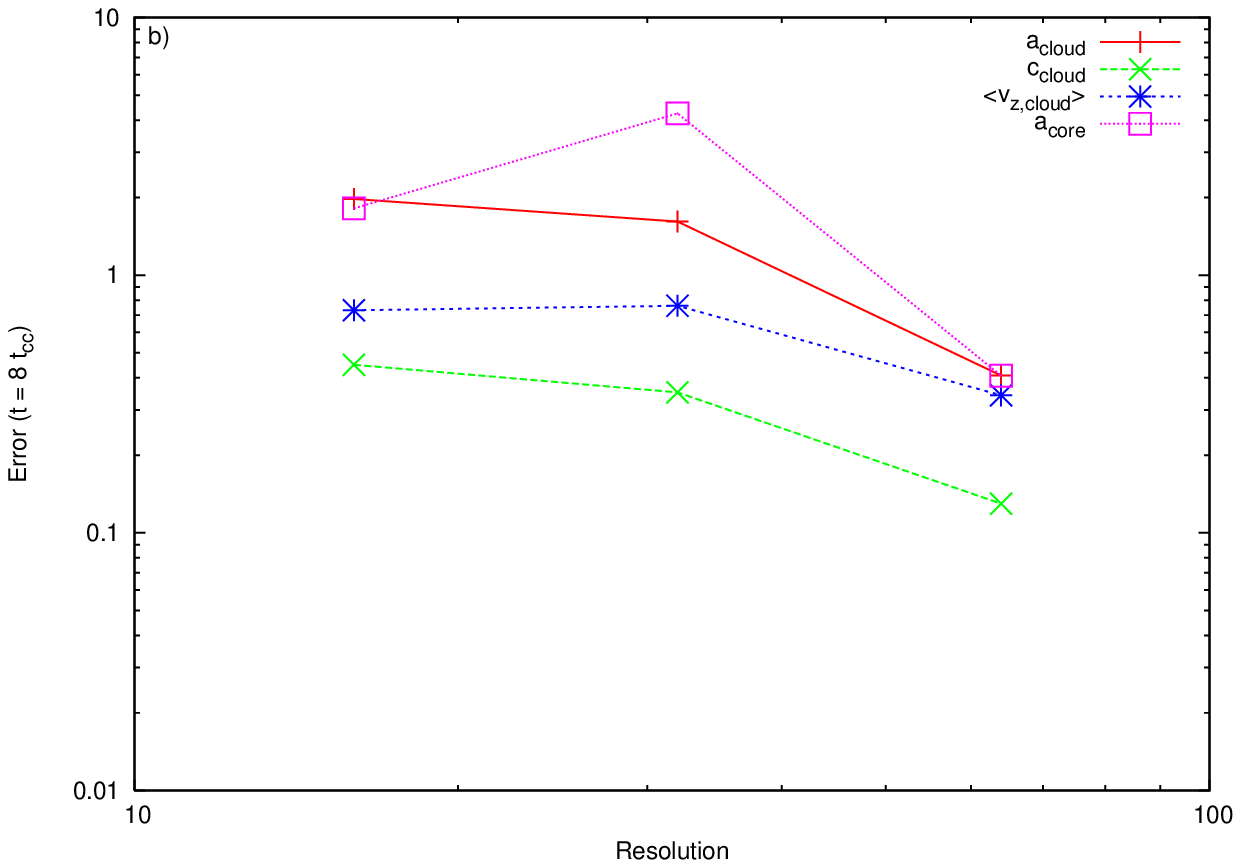,width=8.5cm}
\psfig{figure=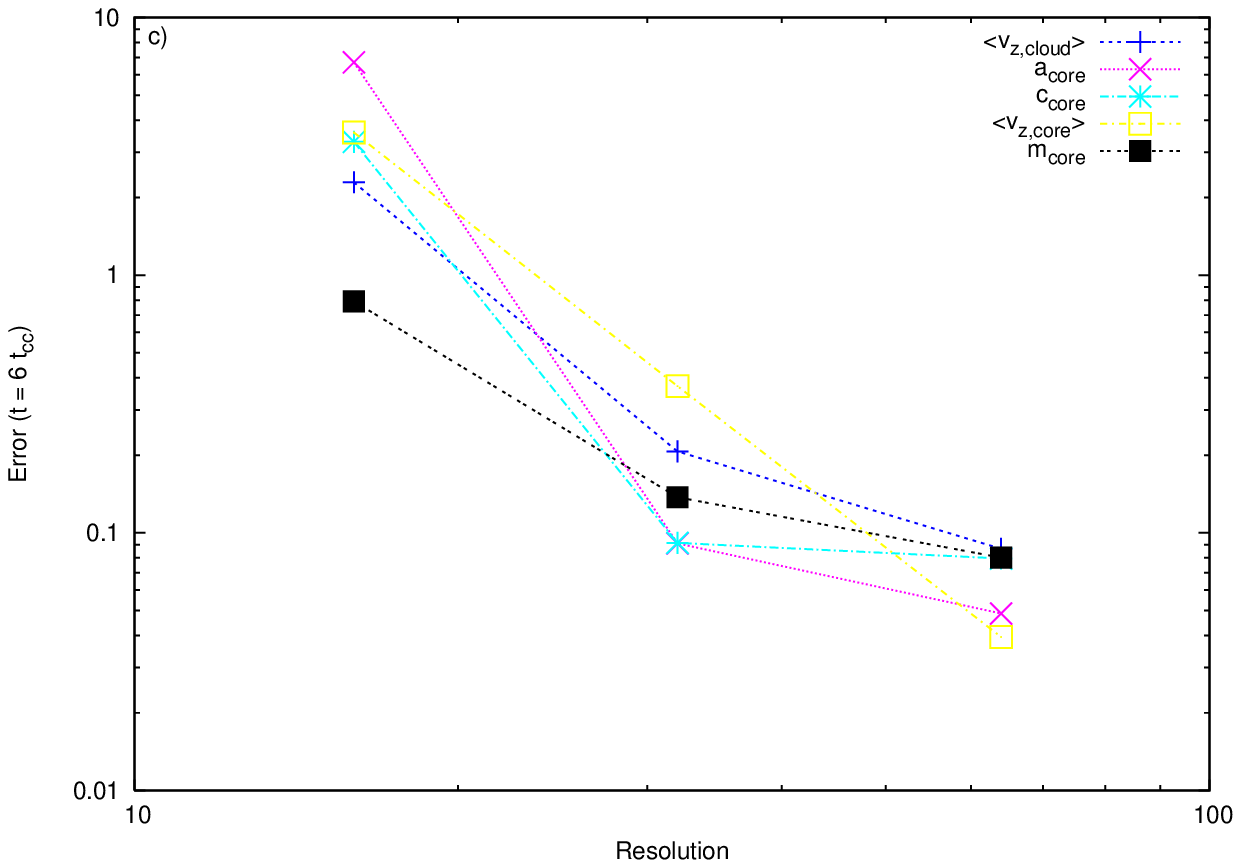,width=8.5cm}
\psfig{figure=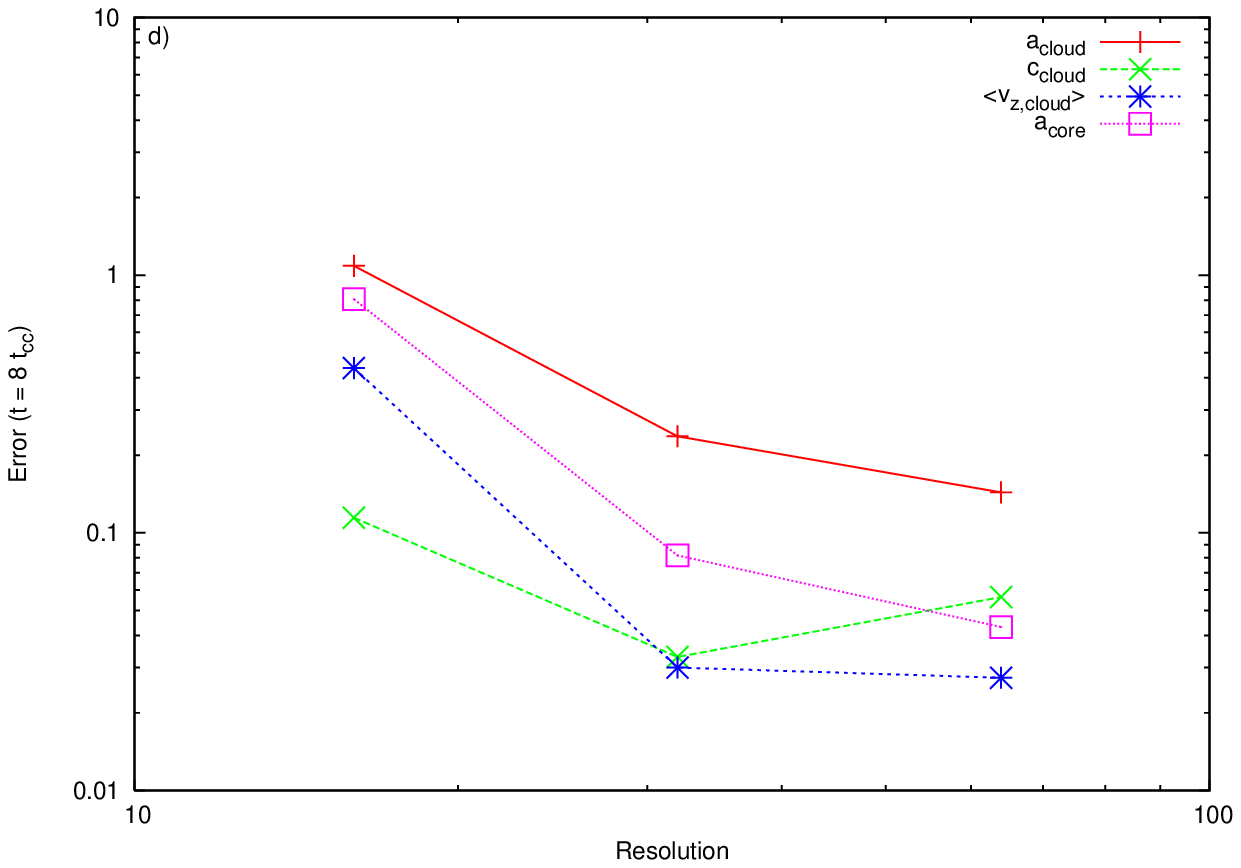,width=8.5cm}
\caption[]{Relative error versus spatial resolution (number of cells
per cloud radius on the finest grid) for a number of global quantities
measured from a shock-cloud interaction with $\chi=10^{3}$ and
$p_{1}=10$ at $t=6\;t_{\rm cc}$ (left panels) and $8 t_{\rm cc}$
(right panels). The top panels show results from the inviscid code,
while the bottom panels are the results from the $k$-$\epsilon$ code
with ``low $k$-$\epsilon$'' conditions (models c3lo).}
\label{fig:rescomp2}
\end{figure*}

\subsection{Convergence tests}
It is important to demonstrate that the calculations are performed at
spatial resolutions that are high enough to resolve key features of
the interaction. Increasing the resolution in inviscid calculations
leads to smaller scales of instabilities. Quantities which are
sensitive to these small scales (such as the mixing rate between cloud
and ambient gas) may not be converged, while quantities which are
insensitive to gas motions at small scales (e.g., the shape of the
cloud) are more likely to show convergence. Previous studies
\citep[e.g.,][]{Klein:1994, Nakamura:2006} have indicated that about
100 cells per cloud radius are needed for convergence of the
simulations.  Here we carry out a similar study for calculations which
use a subgrid turbulence model.

Fig.~\ref{fig:rescomp} shows the evolution of the core mass and mean
cloud velocity as a function of spatial resolution for both inviscid
and $k$-$\epsilon$ calculations. These parameters are a good test of
the convergence, since both the rate of mixing and the momentum
transfer between the cloud and the ambient flow are sensitive to small
scale instabilities. It is therefore not too surprising to see that
these quantities are poorly converged in the low resolution inviscid
calculations \citep[see also][]{Shin:2008}. In contrast, the subgrid
turbulence model leads to results which are much less dependent on the
spatial resolution.  This is also demonstrated in
Fig.~\ref{fig:rescomp2} which shows, for a number of parameters, the
relative error defined as the fractional difference between the value
measured at resolution $N$ and the value at the finest resolution,
$f$:
\begin{equation}
\Delta Q_{N} = \frac{|Q_{N} - Q_{f}|}{|Q_{f}|}.
\end{equation}
The convergence is much better for the $k$-$\epsilon$ calculations
(see Fig.~\ref{fig:rescomp2}), leading to much less variation with
resolution as seen in Fig.~\ref{fig:rescomp}. Both inviscid and
$k$-$\epsilon$ calculations demonstrate better convergence at
$\chi=10^{2}$ (not shown).

Figs.~\ref{fig:rescomp} and~\ref{fig:rescomp2} support claims from
previous studies that of order 100 cells per cloud radius are needed
for convergence. However, the neglect of the effect of turbulent
eddies on the mean flow means that it is not clear that inviscid
simulations, particularly at high values of $\chi$, are actually
converging to a physically realistic solution.

\begin{figure*}
%paper scales in x.dat: without scales = 3.3x12.0, with scales = 4.5x12.0
\psfig{figure=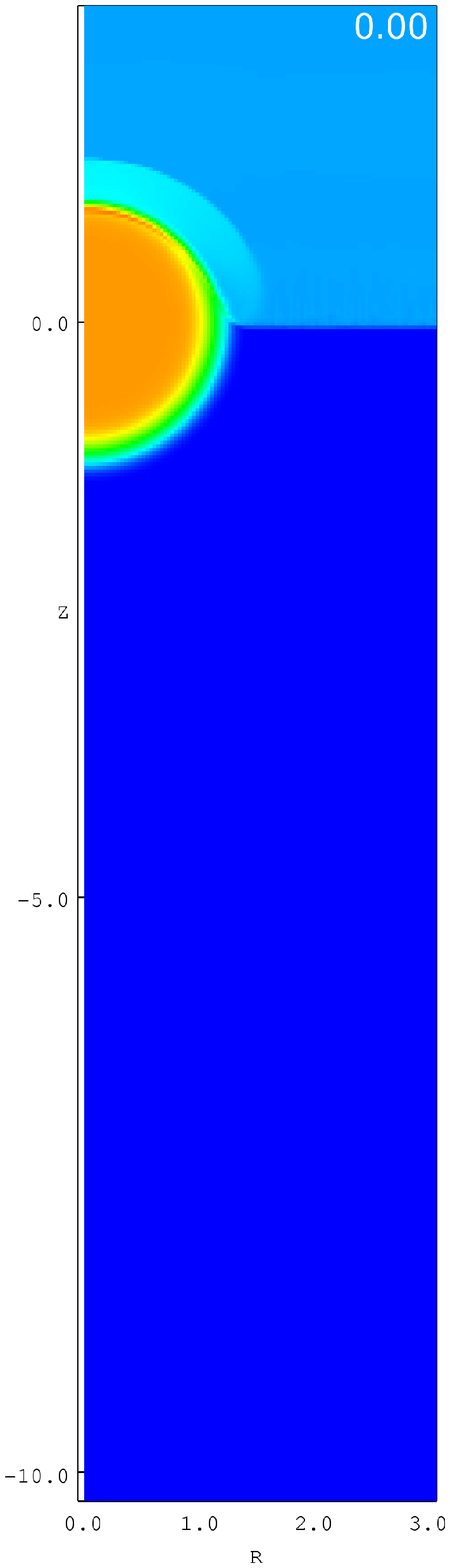,width=3.0cm}
\psfig{figure=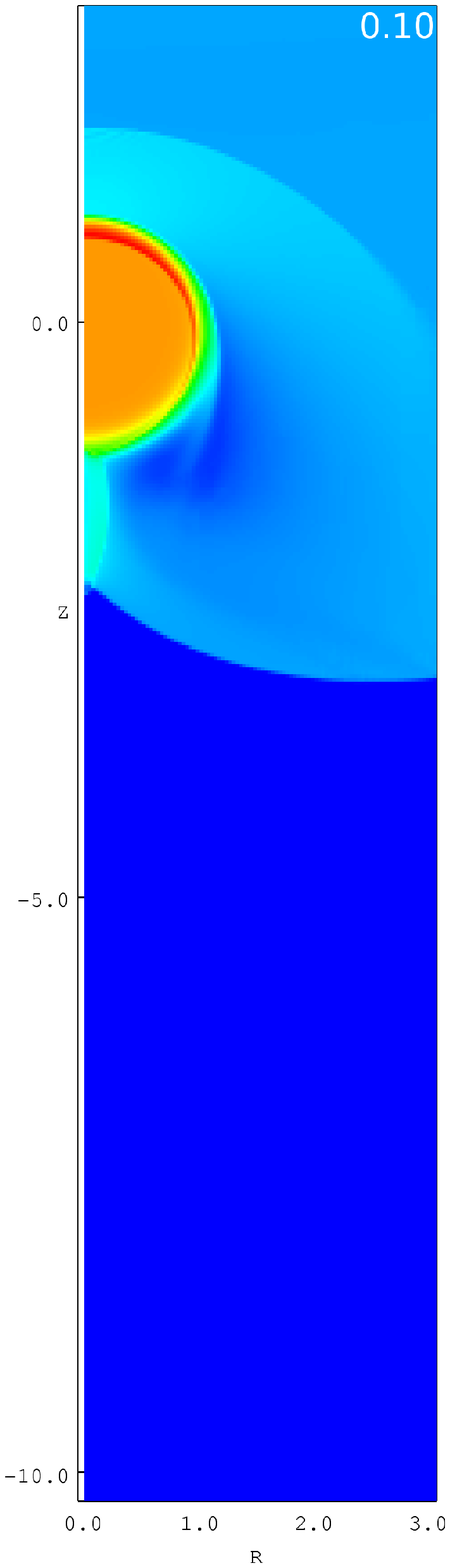,width=3.0cm}
\psfig{figure=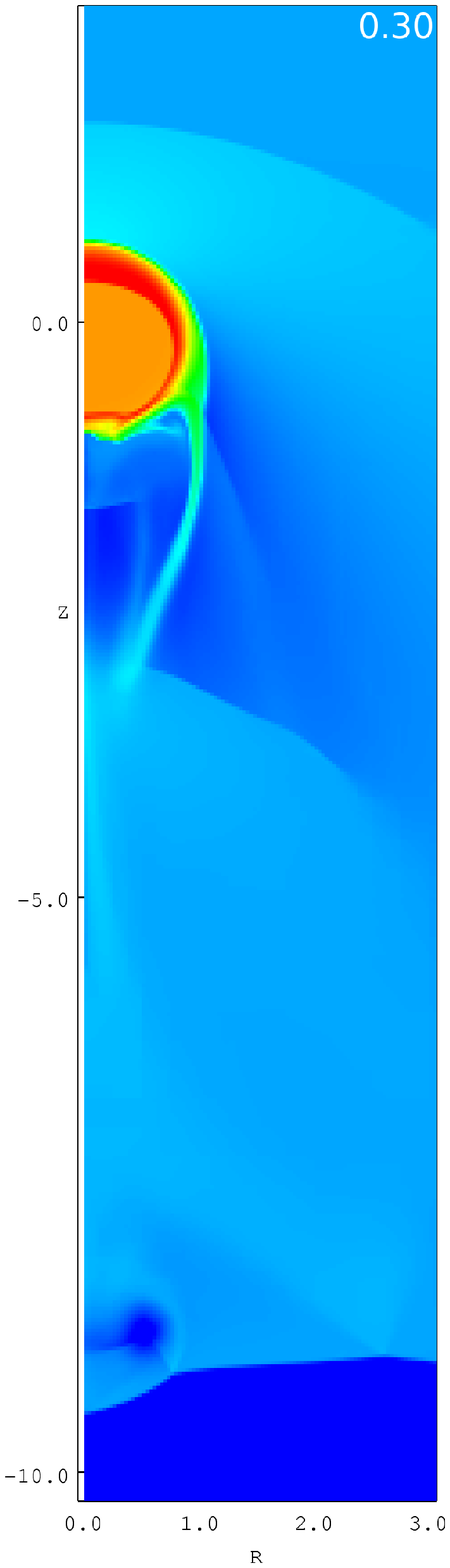,width=3.0cm}
\psfig{figure=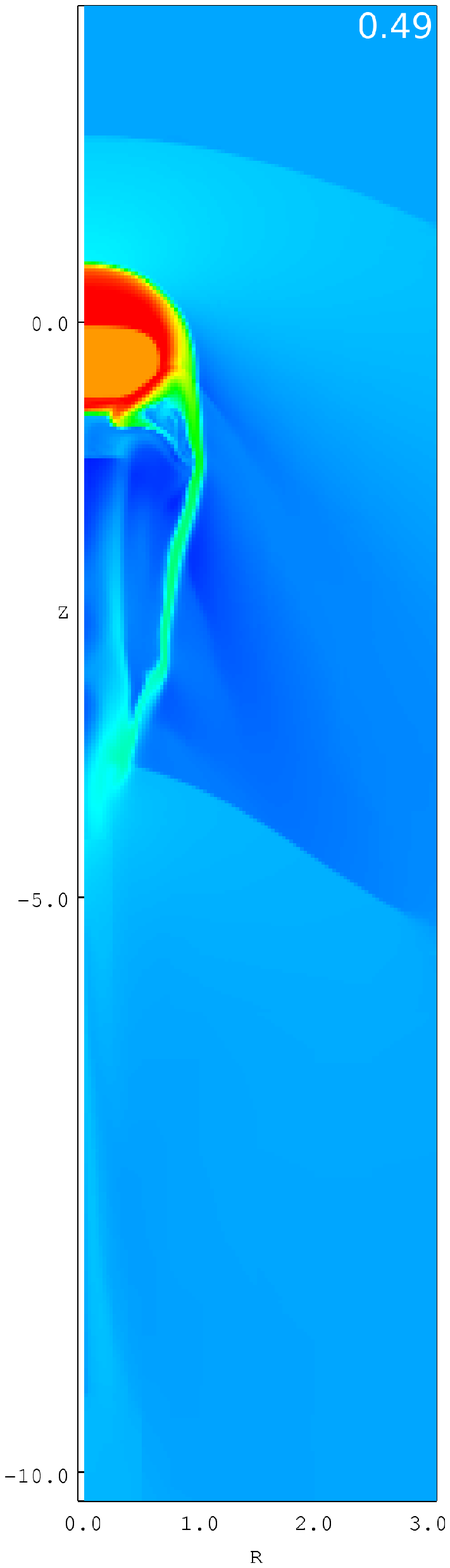,width=3.0cm}
\psfig{figure=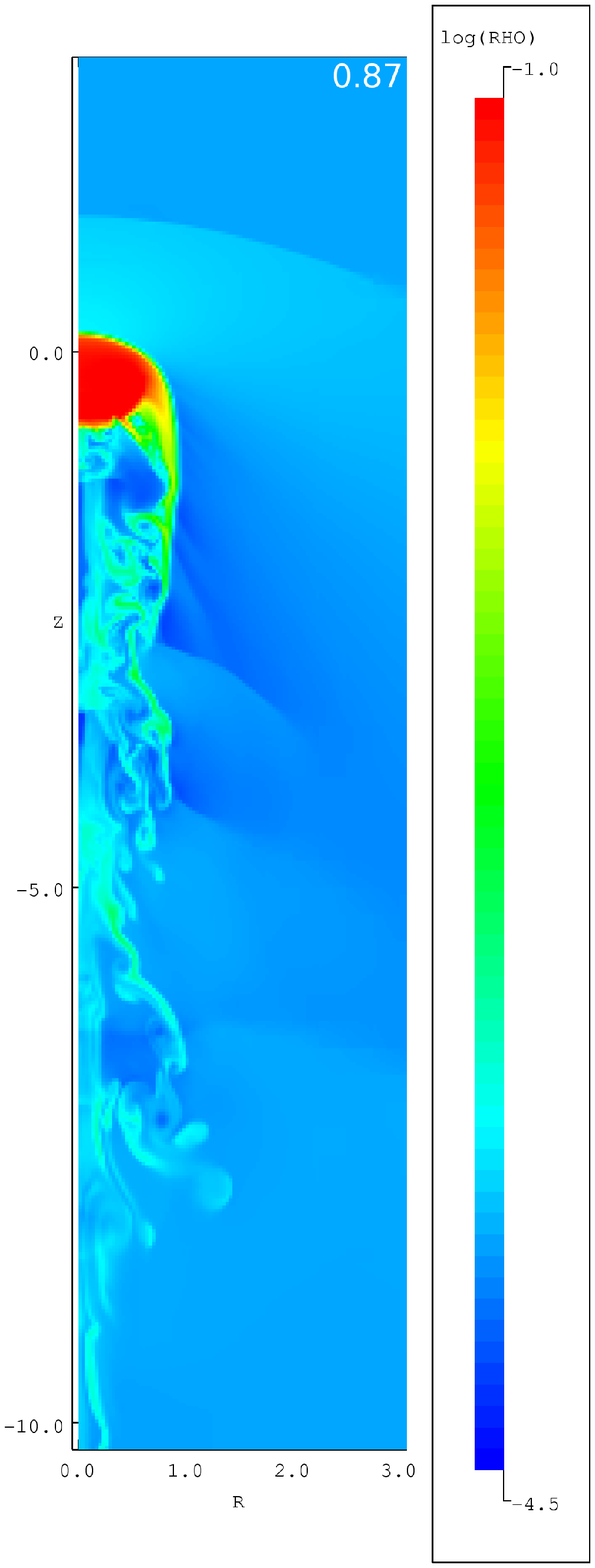,width=4.09cm}
\psfig{figure=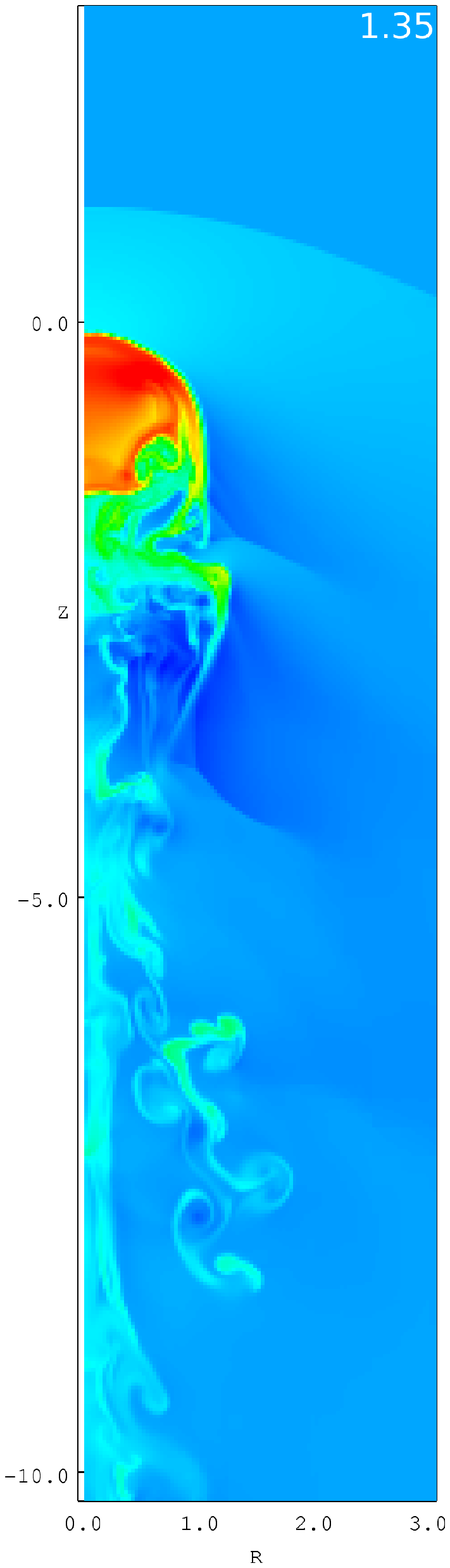,width=3.0cm}
\psfig{figure=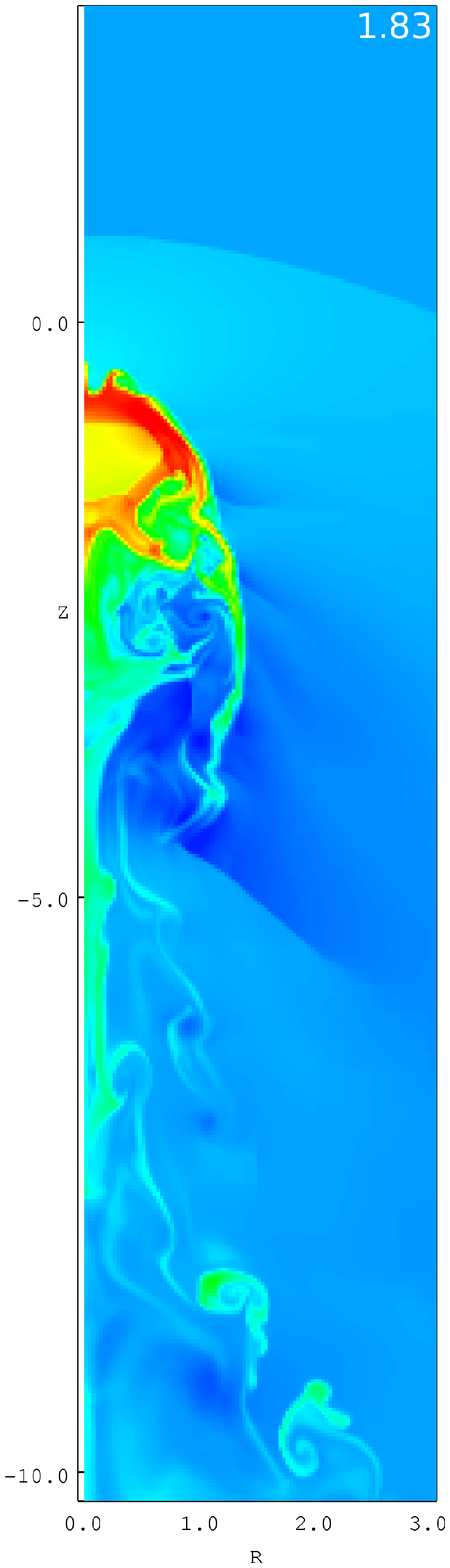,width=3.0cm}
\psfig{figure=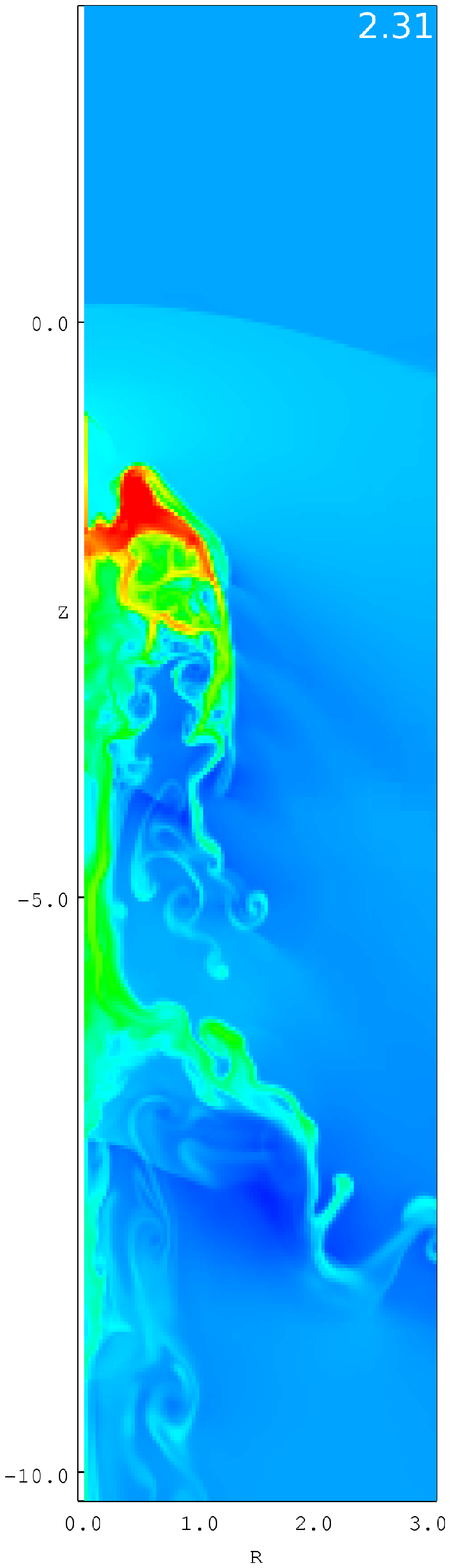,width=3.0cm}
\psfig{figure=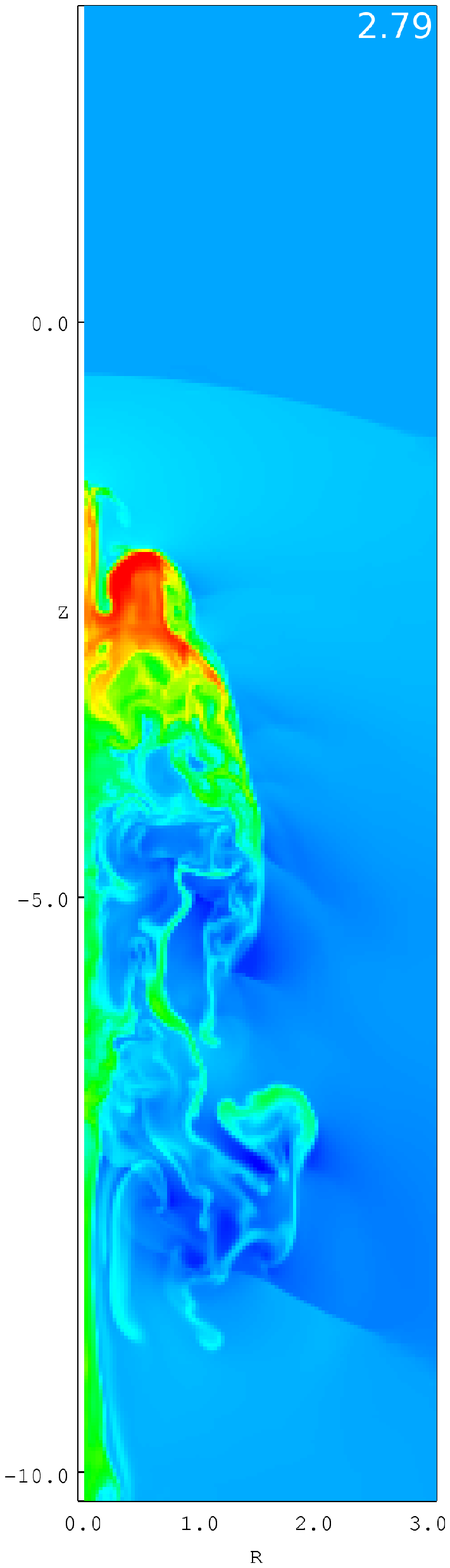,width=3.0cm}
\psfig{figure=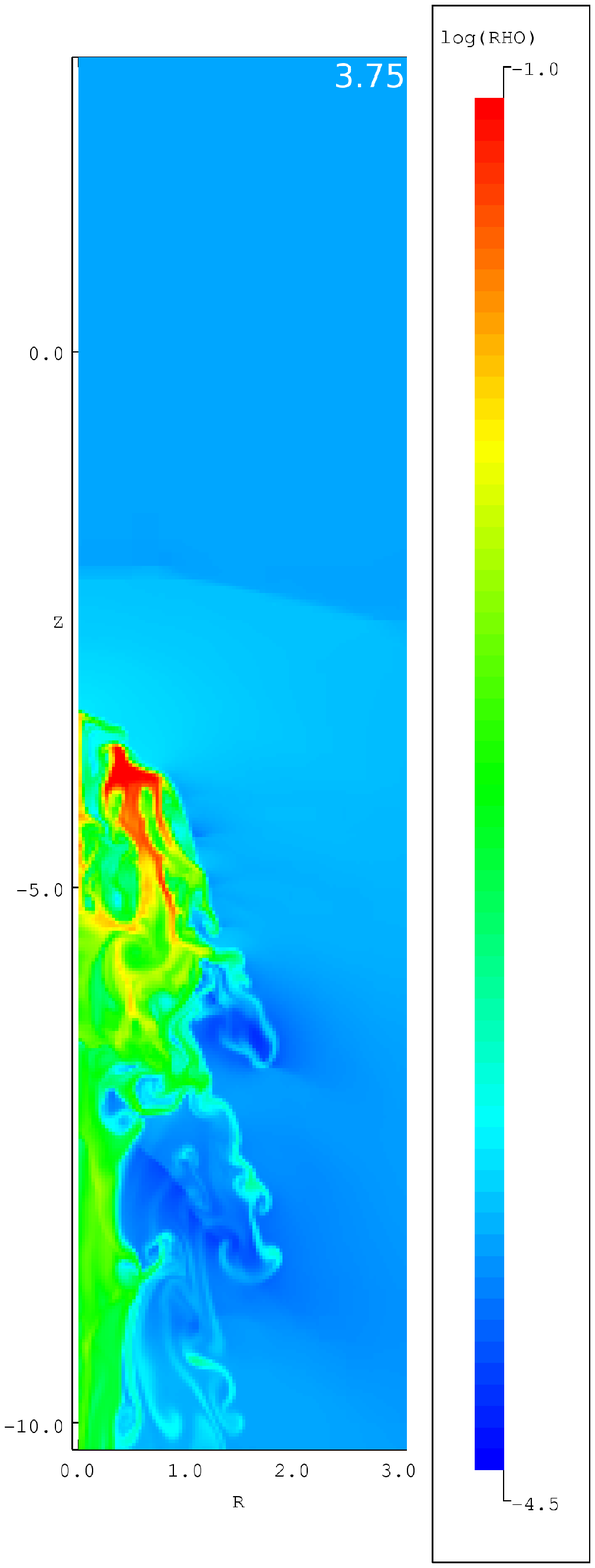,width=4.09cm}
\caption[]{Snapshots of the density distribution from an inviscid
calculation of a Mach 10 adiabatic shock hitting a cloud with a
density contrast of $10^{3}$ with respect to the ambient medium and a
density profile with $p_{1}=10$ (model c3no128). The resolution is 128
cells per cloud radius. The evolution proceeds left to right and top
to bottom with $t = 0.0, 0.1, 0.3, 0.49, 0.87, 1.35, 1.83, 2.31,
2.79$, and $3.75\;t_{\rm cc}$.}
\label{fig:nokeps}
\end{figure*}

\begin{figure*}
%paper scales in x.dat: without scales = 3.3x12.0, with scales = 4.5x12.0
\psfig{figure=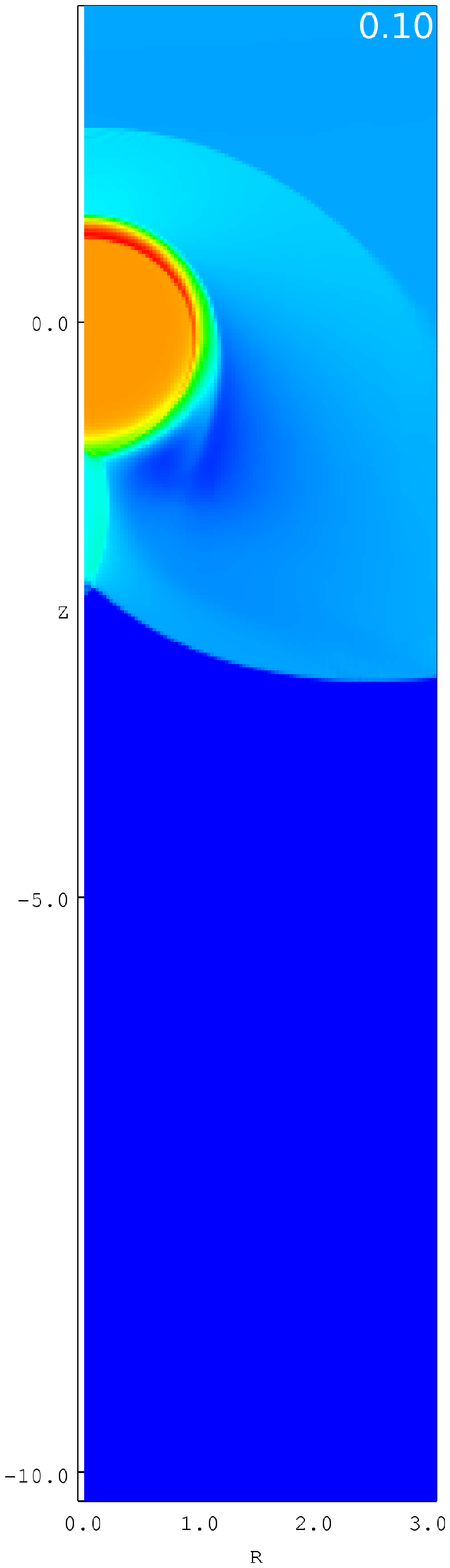,width=3.0cm}
\psfig{figure=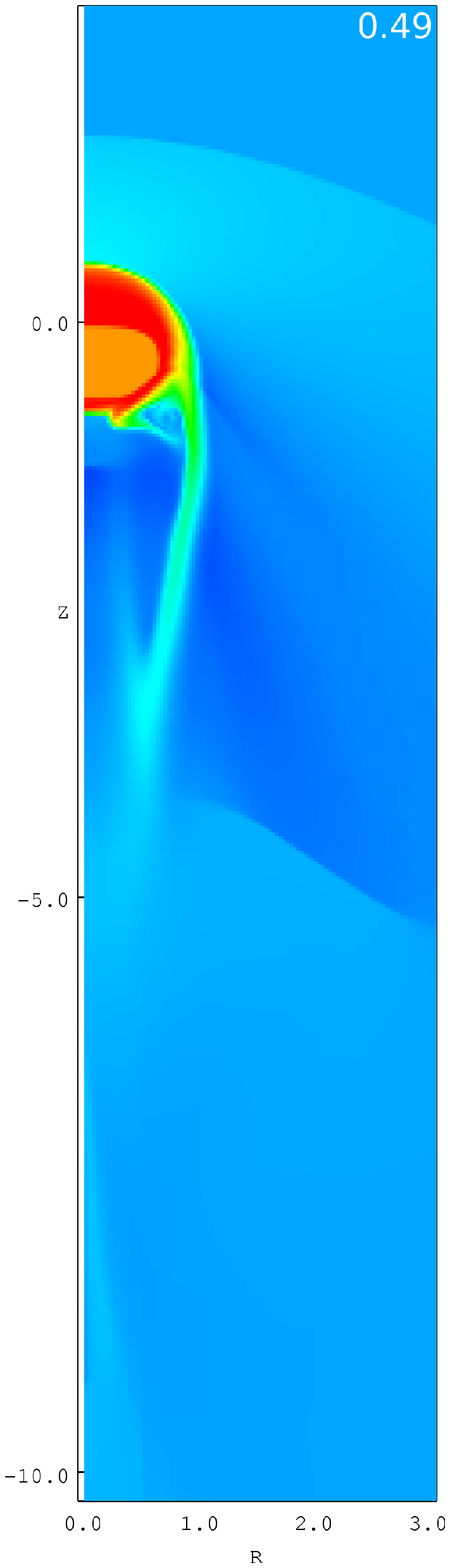,width=3.0cm}
\psfig{figure=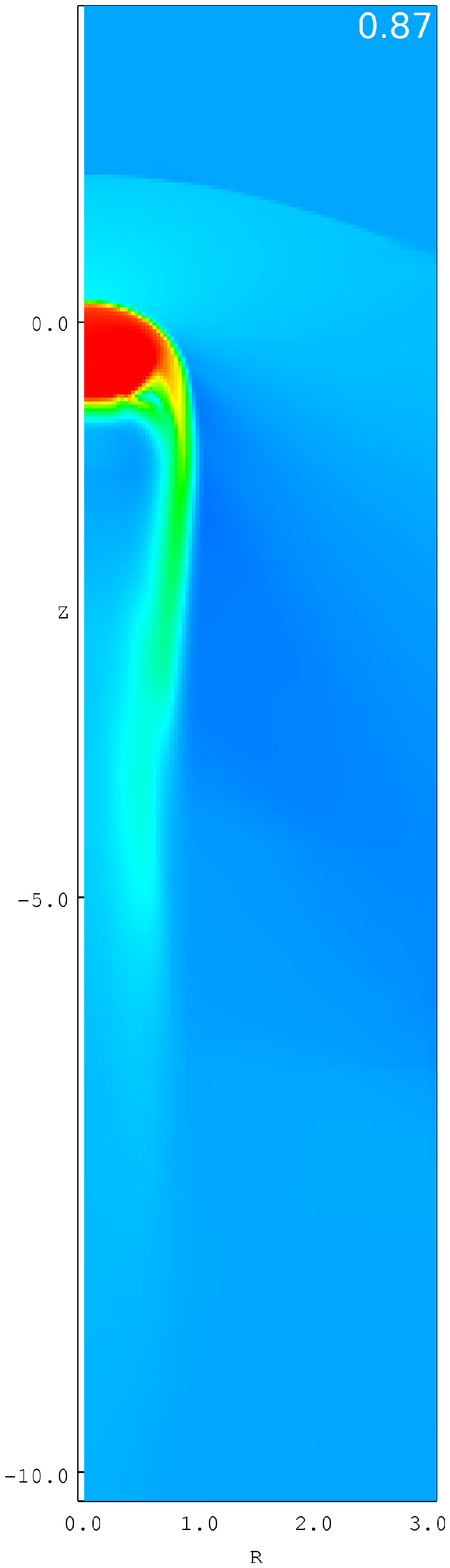,width=3.0cm}
\psfig{figure=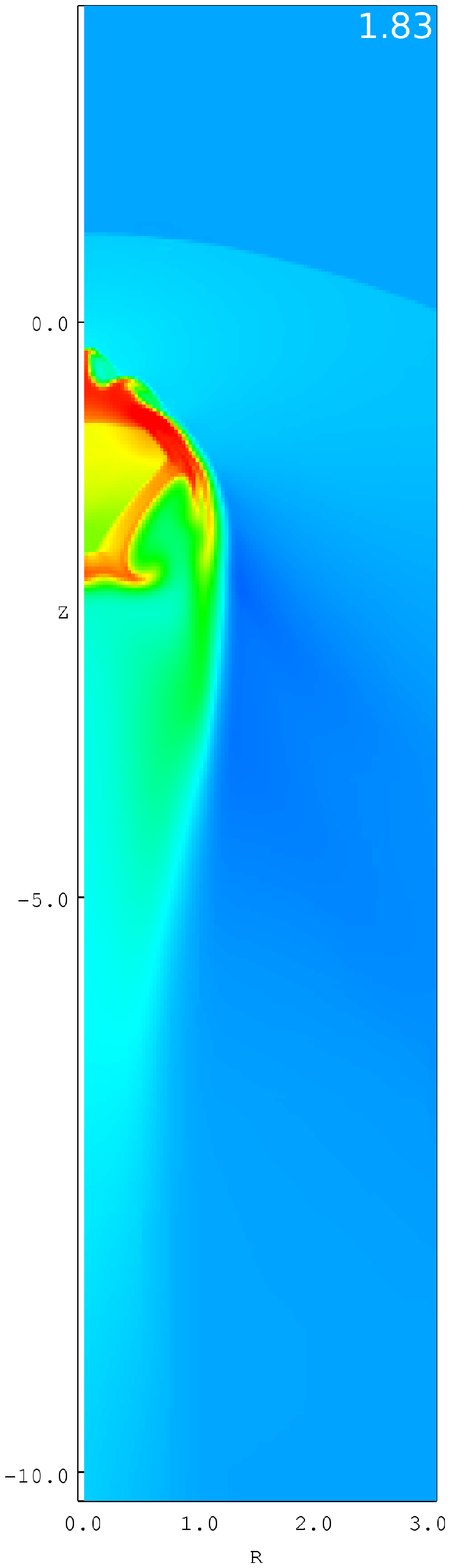,width=3.0cm}
\psfig{figure=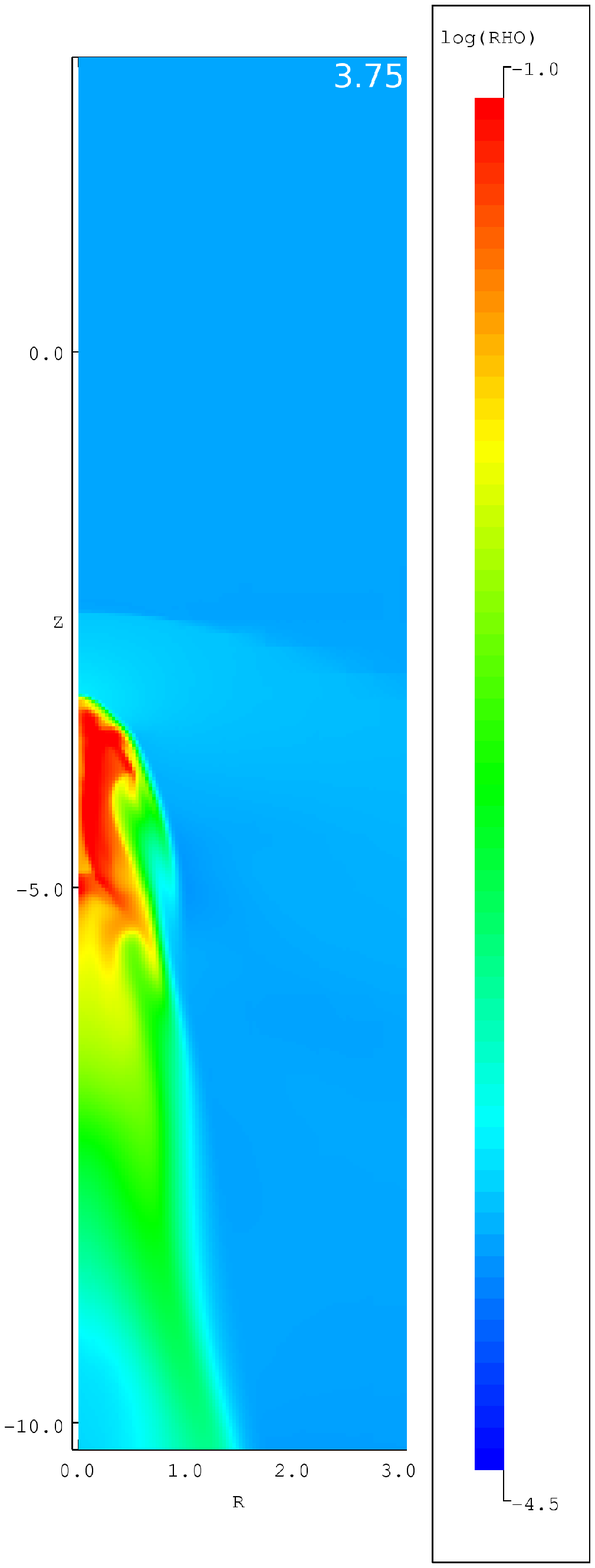,width=4.09cm}
\caption[]{As Fig.~\ref{fig:nokeps} but for a $k$-$\epsilon$
calculation with low initial postshock turbulence (model c3lo128). The
times of the snapshots are $t = 0.1, 0.49, 0.87, 1.83$, and
$3.75\;t_{\rm cc}$.}
\label{fig:lowkeps}
\end{figure*}

\begin{figure*}
%paper scales in x.dat: without scales = 3.3x12.0, with scales = 4.5x12.0
\psfig{figure=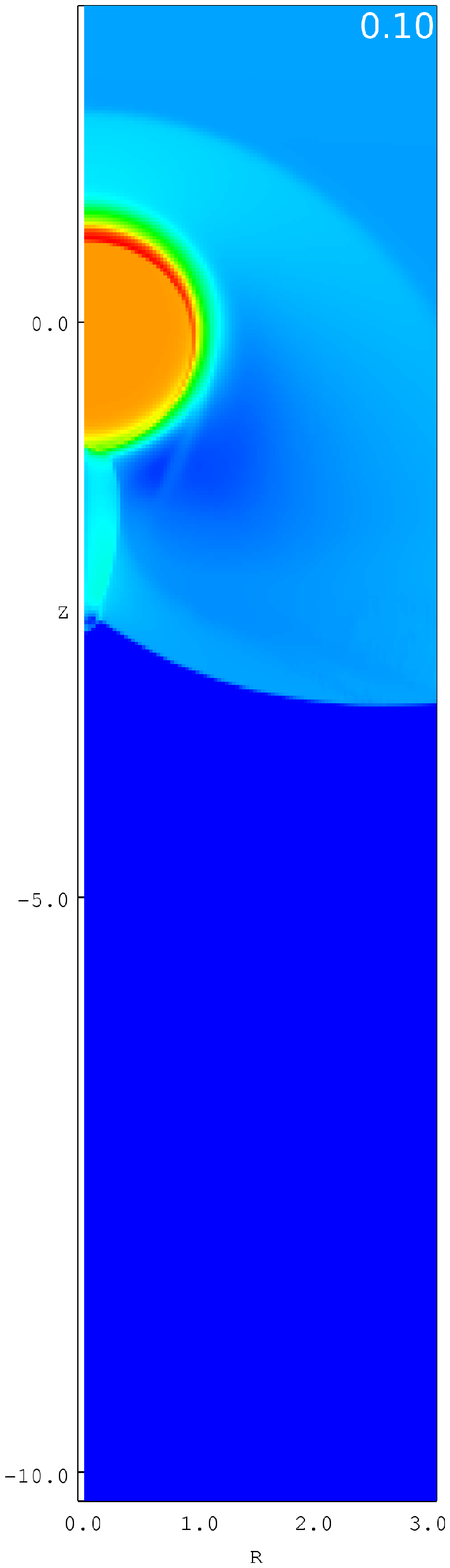,width=3.0cm}
\psfig{figure=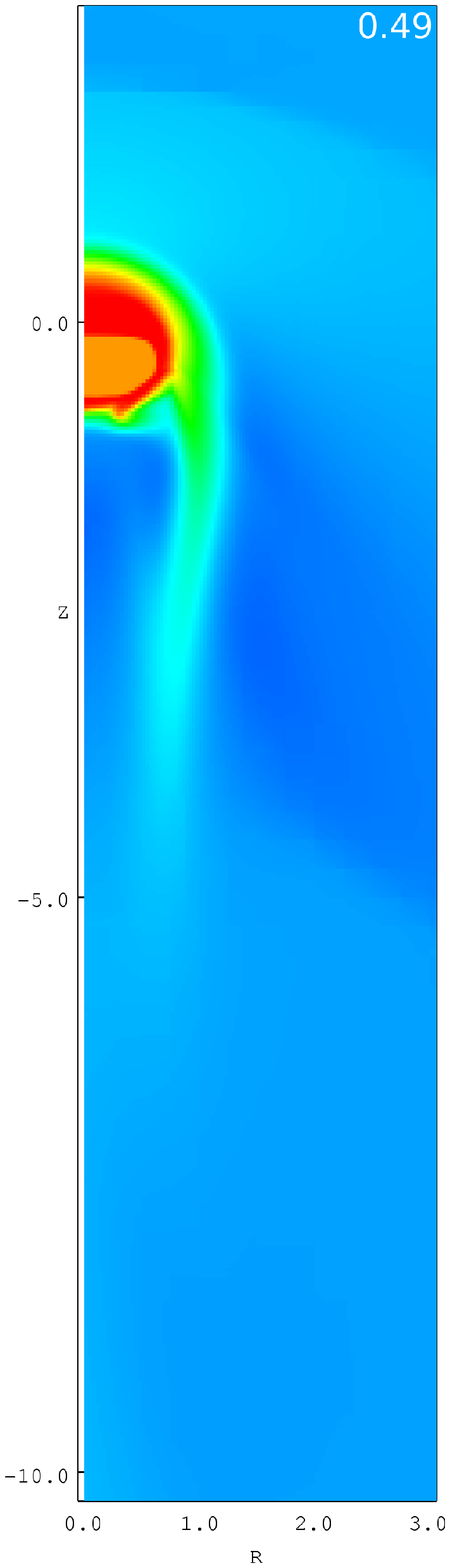,width=3.0cm}
\psfig{figure=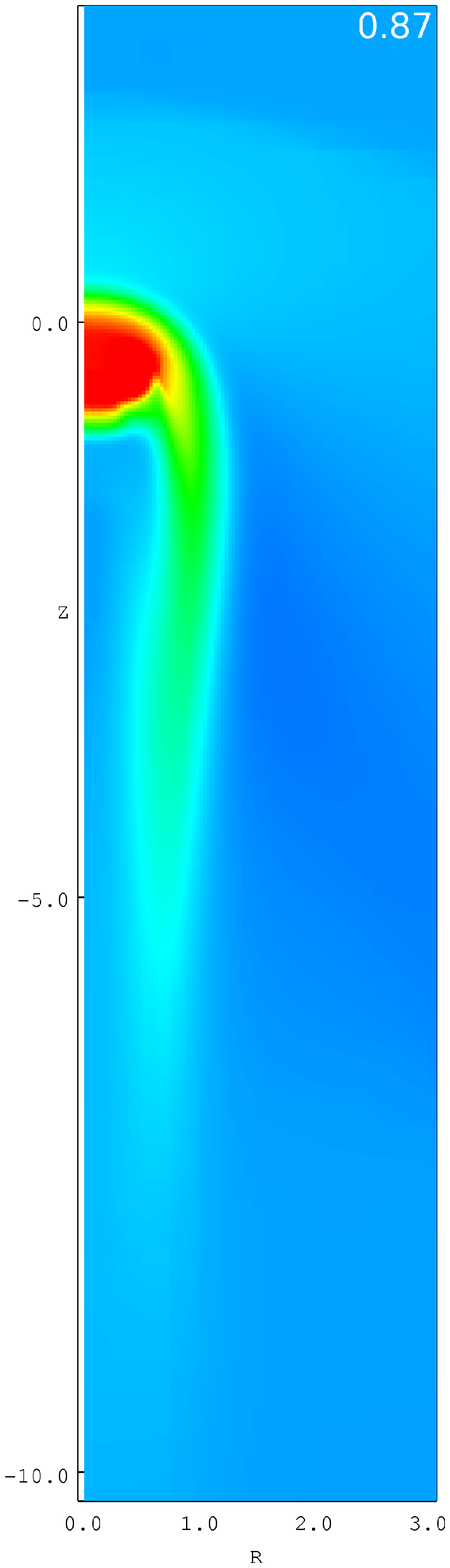,width=3.0cm}
\psfig{figure=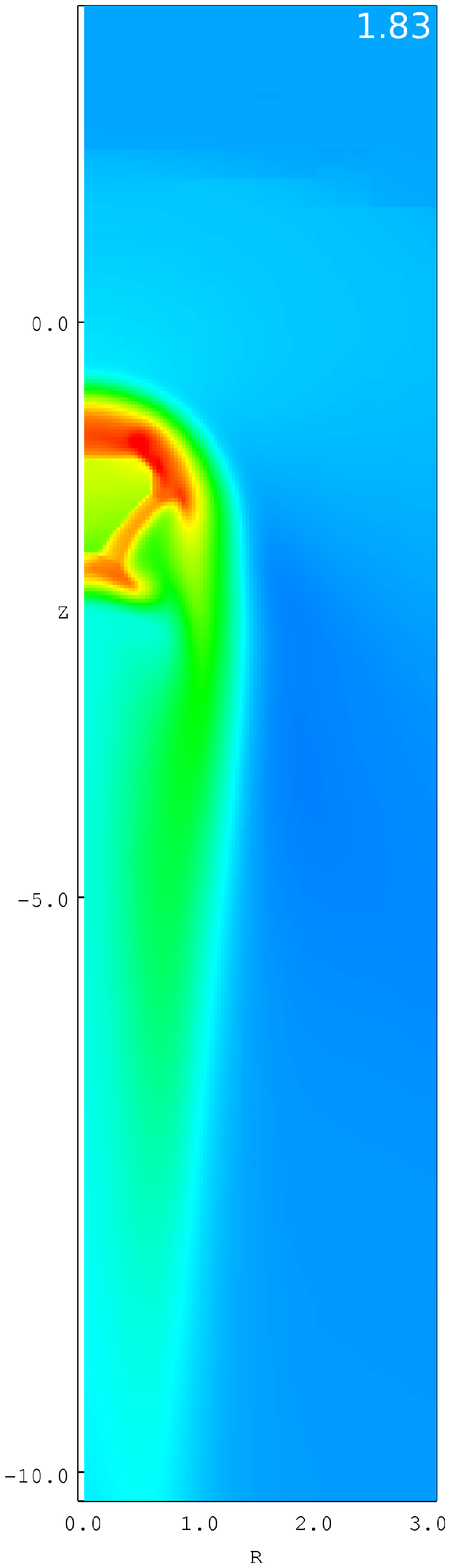,width=3.0cm}
\psfig{figure=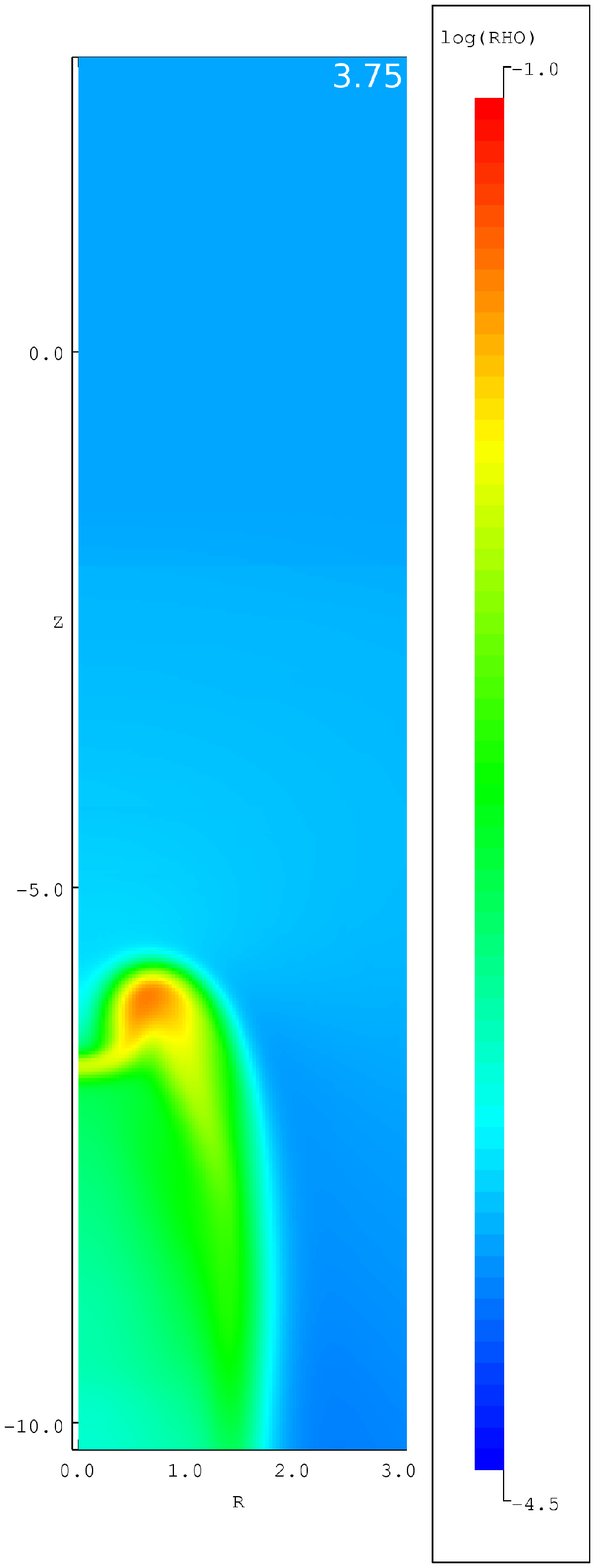,width=4.09cm}
\caption[]{As Fig.~\ref{fig:nokeps} but for a $k$-$\epsilon$
calculation with high initial postshock turbulence (model
c3hi128). The times of the snapshots are $t = 0.1, 0.49,
0.87, 1.83$, and $3.75\;t_{\rm cc}$.}
\label{fig:hikeps}
\end{figure*}

\section{results}
\label{sec:results}
\subsection{Stages}
\label{sec:stages}
The main stages of the interaction of a shock of velocity $v_{\rm b}$
with a uniform cloud of density contrast $\chi$, in the adiabatic,
un-magnetized, non-conducting case are: (1) an ``initial transient
stage'', where the incident shock propagates into the cloud with
velocity $v_{\rm s} = v_{\rm b}/\chi^{1/2}$, and a bow shock or bow
wave propagates upstream into the ambient medium; (2) the
``compression stage'', where the cloud is compressed mainly in the
$z$-direction by the transmitted shock and by a shock driven into the
back of the cloud; (3) the ``expansion stage'', where the
highly-pressured cloud expands downstream and laterally; and (4) the
``destruction stage'', where the cloud is destroyed and its material
mixed into the surrounding flow. 
%The flow structure in the initial
%compression phase is illustrated in Fig.~6 of
%\citet*{Poludnenko:2002}.  
In other situations, for example when there is efficient cooling, the
evolution can be significantly different (see
Section.~\ref{sec:destandmix}).  In all work to date the cloud is
destroyed by the shock. The addition of gravitational forces is likely
to be needed if the cloud is to survive.

\begin{figure*}
%paper scales in x.dat: without scales = 3.3x12.0, with scales = 4.5x12.0
\psfig{figure=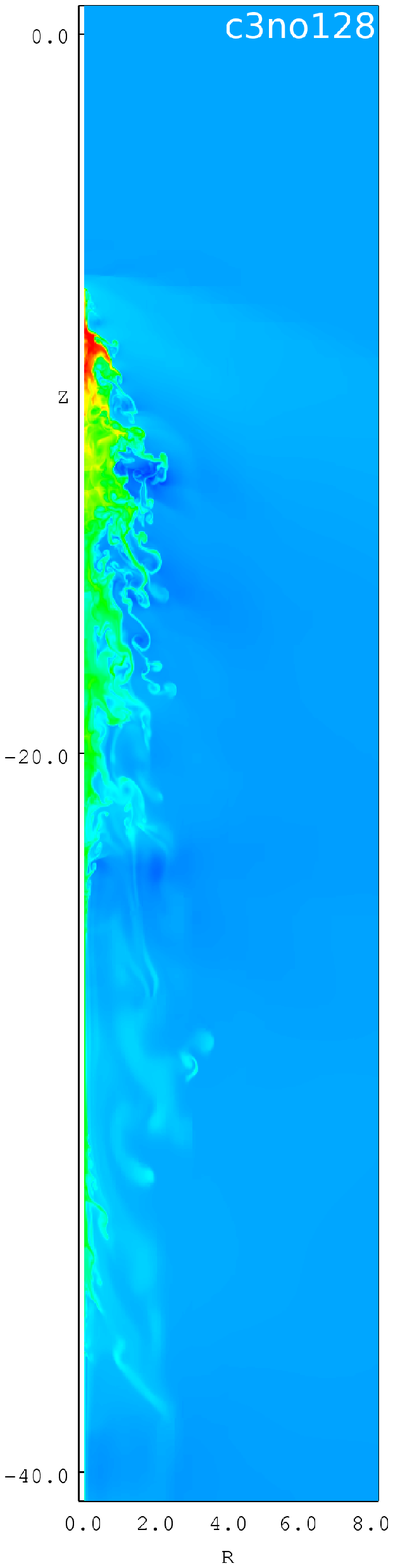,width=4.8cm}
\psfig{figure=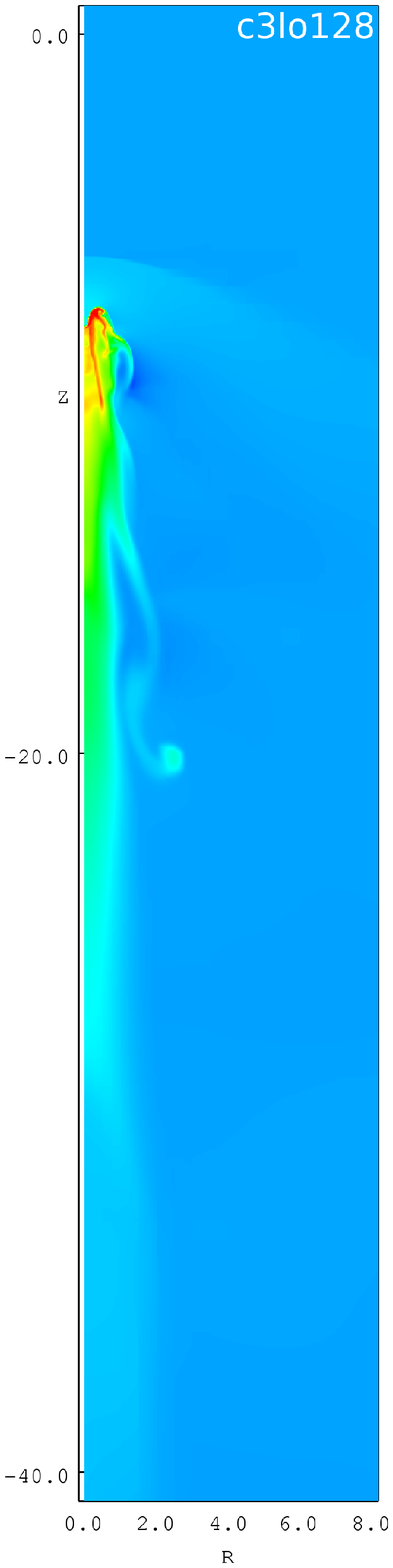,width=4.8cm}
\psfig{figure=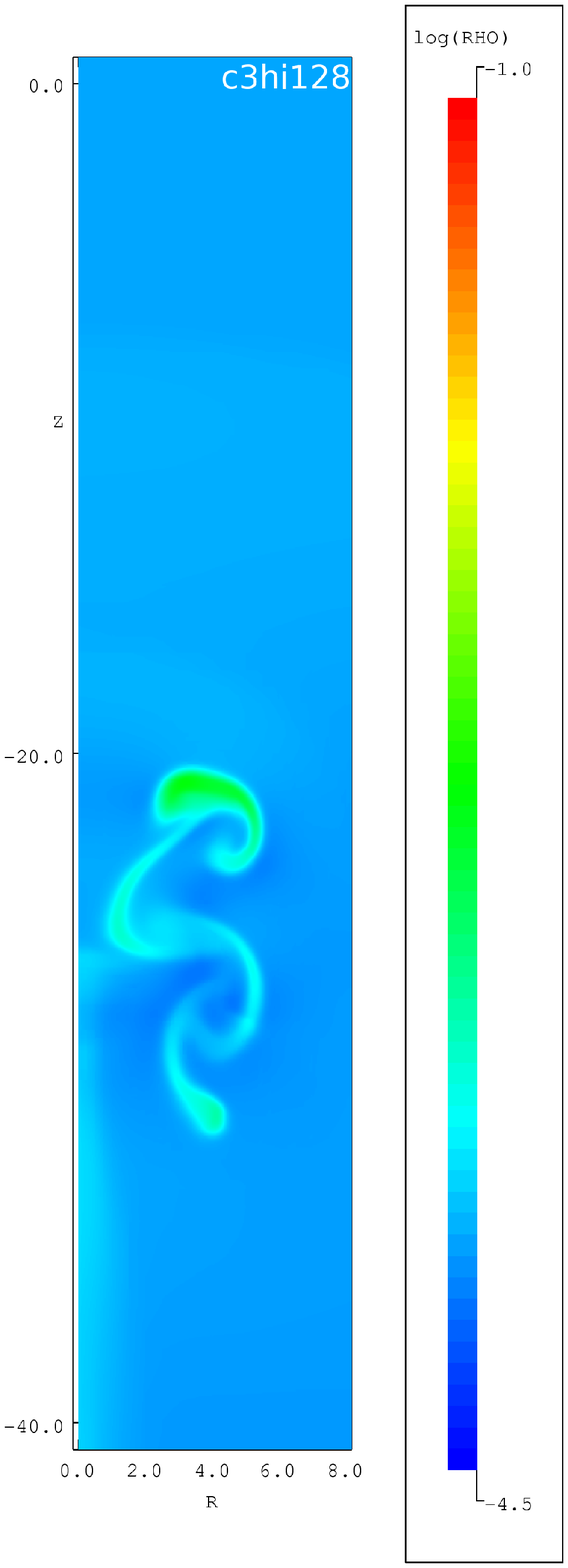,width=6.5436cm}
\caption[]{Comparison between the inviscid (model c3no128, left panel) and
$k$-$\epsilon$ calculations with low and high initial postshock
turbulence (models c3lo128 and c3hi128, middle and right panels, 
respectively) at $t = 5.66\;t_{\rm cc}$.}
\label{fig:comp_latet}
\end{figure*}

\begin{figure*}
\psfig{figure=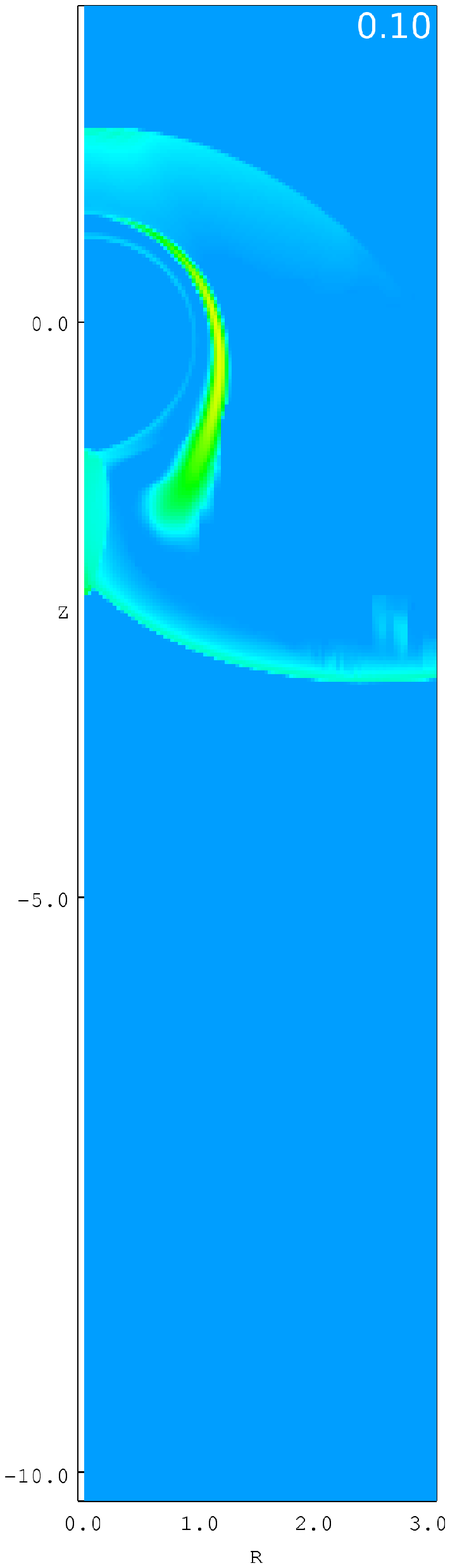,width=3.0cm}
\psfig{figure=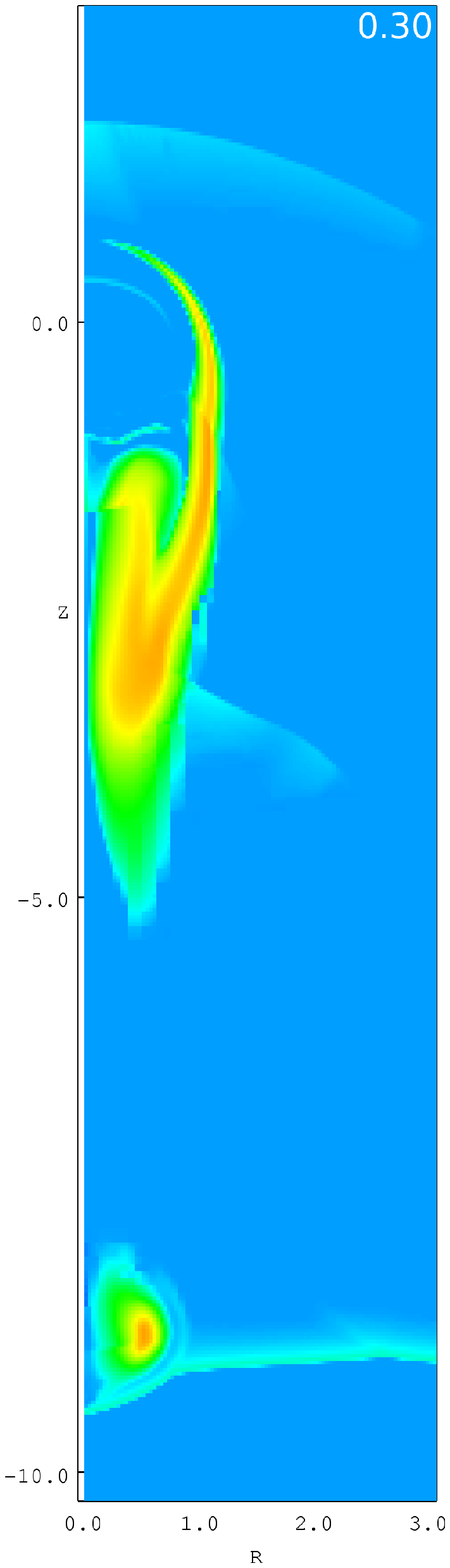,width=3.0cm}
\psfig{figure=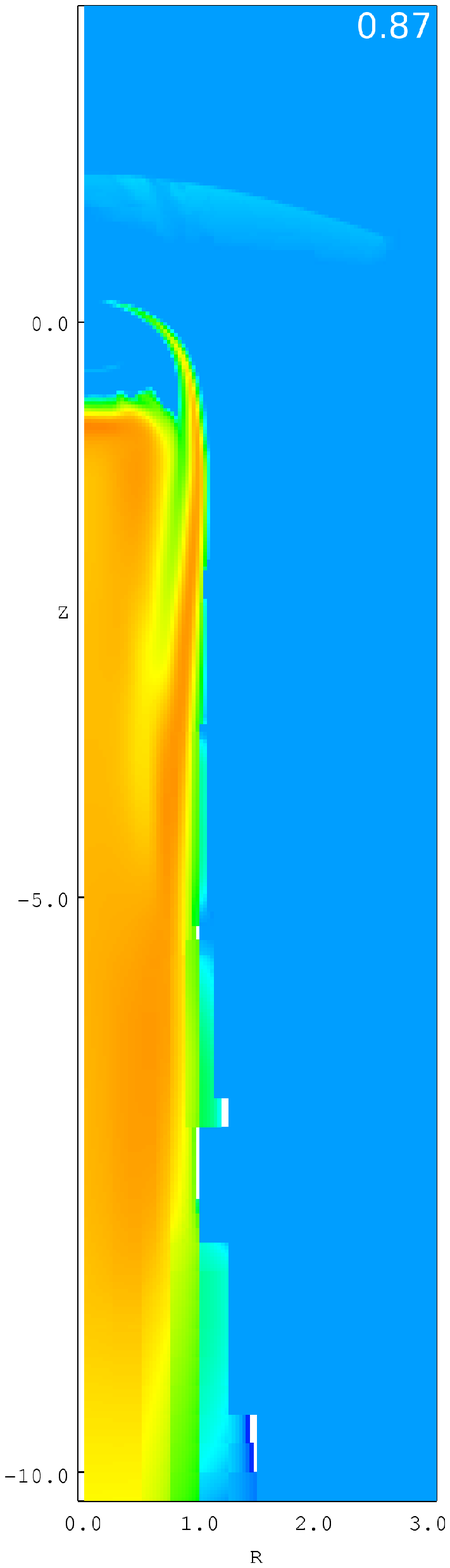,width=3.0cm}
\psfig{figure=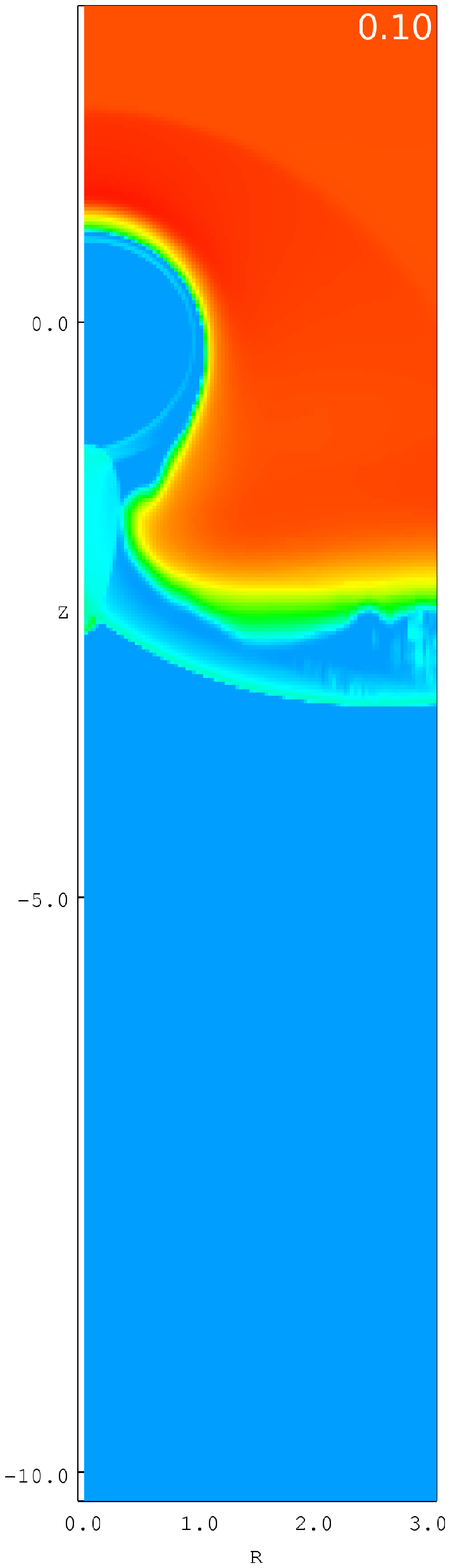,width=3.0cm}
\psfig{figure=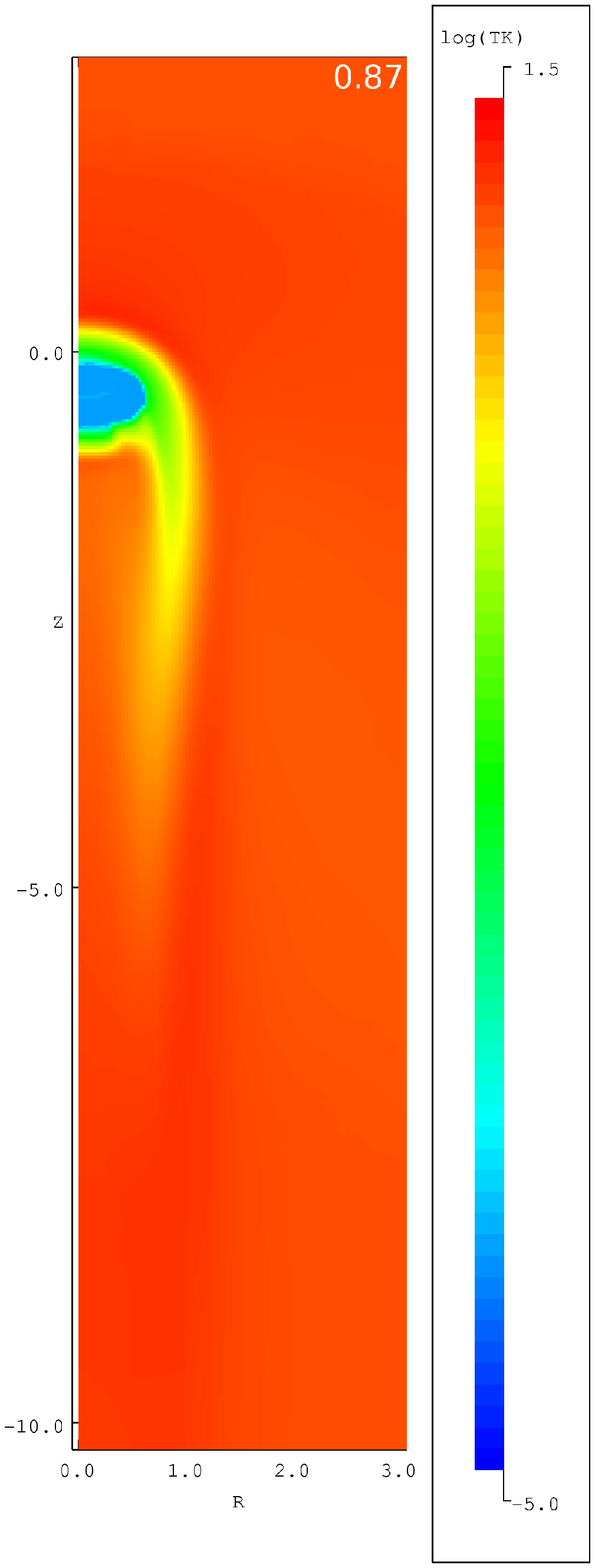,width=4.09cm}
\caption[]{Snapshots of the turbulent energy per unit mass, $k$, from
the $k$-$\epsilon$ calculations with $\chi=10^{3}$ and $p_{1}=10$. The
three left-most panels show the evolution with low initial postshock
turbulence (model c3lo128) proceeding left to right with $t = 0.1,
0.3$, and $0.87\;t_{\rm cc}$. The two right-most panels show the
evolution with high initial postshock turbulence (model c3hi128)
proceeding left to right with $t = 0.1$ and $0.87\;t_{\rm cc}$. The
white regions in the middle panel are artifacts of the plotting
routine.}
\label{fig:lowkeps_tk}
\end{figure*}

\subsection{Cloud morphology and turbulence}
\label{sec:morph}
In Fig.~\ref{fig:nokeps}, snapshots of the density distribution at
different times are shown for an inviscid calculation with 128 grid
cells per cloud radius (model c3no128). The evolution of the cloud
broadly follows the stages outlined above. The first two stages last
until $t \approx t_{\rm cc}$ (i.e. the top 5 panels in
Fig.~\ref{fig:nokeps}).  The expansion of the cloud in stage 3 is
supersonic with respect to the sound speed within the cloud, 
and a low density interior surrounded by a higher density shell
forms (see the snapshot at $t=1.83\;t_{\rm cc}$). The high density
shell then collapses in on itself (see the snapshot at $t=2.31\;t_{\rm
cc}$).  Material is continuously ablated off the surface of the cloud
by the fast-flowing surroundings, and a turbulent wake with prominent
RT and KH instabilities forms. Fig.~\ref{fig:comp_latet} shows that
at later times the  mass-loss from the cloud resembles a single
tail-like structure (this is in contrast to models with lower values
of $\chi$ - see Section~\ref{sec:chi}).

Fig.~\ref{fig:lowkeps} shows that the initial interaction of the shock
with the cloud in the low $k$-$\epsilon$ case (model c3lo128) is
similar to the inviscid case, since the initial post-shock turbulence
is low and $k$ and $\epsilon$ are small.  However, the viscosity
introduced by the sub-grid turbulence model prevents the subsequent
development of the resolution-dependent RT and KH instabilities seen
in Fig.~\ref{fig:nokeps}. Instead, the loss of material from the cloud
occurs much more smoothly. This is exactly as expected given that the
purpose of the $k$-$\epsilon$ model is to approximate the
time-averaged flow.

Simulations with a high initial level of post-shock turbulence are
different again, as seen in Fig.~\ref{fig:hikeps} where the results of
model c3hi128 are displayed. The transport/diffusion coefficients are
considerably higher in this simulation, and this leads to a faster
rate of ablation from the cloud, and ultimately its more rapid
destruction. The high level of upstream turbulence also
smooths/broadens the bowshock ahead of the cloud and the tailshock
which forms downstream, so that their time-averaged
positions are represented.

Fig.~\ref{fig:comp_latet} compares the morphology of the clouds in
these 3 simulations at $t = 5.66\;t_{\rm cc}$. The global features of
models c3no128 and c3lo128 are reasonably comparable at this time, but
the cloud in model c3hi128 is clearly at a more advanced stage of
destruction (cf. Table~\ref{tab:results} and
Fig.~\ref{fig:kepscomp3}). Interestingly, material stripped from the
cloud lies off-axis in model c3hi128. This develops from the off-axis
density peak seen at $t=1.83$ and $3.75\;t_{\rm cc}$ shown in
Fig.~\ref{fig:hikeps}, though the initial disturbance occurs at even
earlier times. Clearly the turbulence in this model affects the
properties of the shocks driven into the cloud and the slip surface
that forms around it, with small differences at early stages being
amplified during the subsequent evolution.

The development of the sub-grid turbulence in model c3lo128 is shown in the
three left-most panels of Fig.~\ref{fig:lowkeps_tk}, where the
turbulent energy per unit mass, $k$, is displayed. $k$ is created in
regions of high shear, particularly in a thin turbulent boundary layer
along the slip surface around the cloud (see the left-most panel of
Fig.~\ref{fig:lowkeps_tk}). The turbulent intensity quickly saturates
at a level that is almost independent of its initial value.

The $\nabla \cdot {\bf u}$ terms in the production term for $k$ (see
Sec.~\ref{sec:scheme}) means that $k$ is also generated behind shocks,
as can be seen in 4 specific regions in the snapshot at $t =
0.1\;t_{\rm cc}$: behind the incident shock sweeping through the
ambient medium; behind the reflected shock formed as the incident
shock converges on the axis behind the cloud; behind the bow shock
formed upstream of the cloud; and behind the slow shock driven into
the cloud.

The reflected shock on the axis behind the cloud becomes increasingly
oblique as the point of convergence moves away from the rear of the
cloud, and interacts with the incident shock to create a double Mach
reflected shock that propagates along the axis \citep{Klein:1994}. A
powerful supersonic vortex ring forms just behind the Mach reflected
shock, in which a region of high turbulence is generated (see the
second from left panel of Fig.~\ref{fig:lowkeps_tk}).  While the
turbulence generated behind shocks decays very rapidly, the turbulence
associated with the vortex ring is much more persistent, as is the
turbulence generated at the slip surface around the cloud.

Fig.~\ref{fig:lowkeps_tk} shows that at later times the turbulence
generated at the slip surface proceeds to develop into a highly
turbulent wake with a radius comparable to the initial cloud
radius. The setup time for the wake is $\sim t_{\rm cc}$, and the
sub-grid turbulent energy of the cloud material grows and then
dissipates as the cloud is mixed into its surroundings (see
Figs.~\ref{fig:kepscomp3}g and~\ref{fig:kepsenergy}). Note that the
core of the cloud has very little turbulence associated with it.
%Due to the cascade of energy to smaller scales, small-scale eddies
%($\epsilon$) are present in all regions containing larger-scale eddies
%($k$).

The finite timescale for the development of significant turbulence
means that simulations with the $k$-$\epsilon$ subgrid turbulence
model with a low initial level of postshock turbulence produce similar
morphologies to those obtained from inviscid calculations at early
times ($t \ltsimm 0.5\;t_{\rm cc}$). However, the increase in the
transport coefficients in regions of high turbulence leads to
increasing divergence from inviscid calculations at later times, and
ultimately to a faster destruction of the cloud.

In contrast, a high level of environmental turbulence immediately
affects the evolution of the cloud, since the transport coefficients
around the cloud are also high. The two right-most panels of
Fig.~\ref{fig:lowkeps_tk} show the highly turbulent post-shock flow
engulfing the cloud in model c3hi128. Both the limb of the cloud and
the bowshock upstream of the cloud become broader and less distinct as
the high level of turbulence leads to strong diffusion across these
boundaries (see Fig.~\ref{fig:hikeps}).
Another major difference compared to model c3lo128 is that the
turbulence downstream of the cloud at $t = 0.87\;t_{\rm cc}$ is
roughly as strong as that in the post-shock flow. Hence the turbulent
wake which is seen so clearly in the centre panel of
Fig.~\ref{fig:lowkeps_tk} is indistinguishable from the surroundings
in the right panel of Fig.~\ref{fig:lowkeps_tk}.  Note also that the
value of $k$ created downstream of the incident shock (see the 2nd
from right panel of Fig.~\ref{fig:lowkeps_tk}) is much smaller 
than the initial post-shock value. This reflects the fact that
such high levels of post-shock turbulence are not naturally generated
by a shock sweeping through a perfectly homogeneous medium.

The two rightmost panels in Fig.~\ref{fig:lokepsp21} show the
turbulent energy per unit mass, $k$, on a linear scale at
$t=1.81\;t_{\rm cc}$ for low $k$-$\epsilon$ calculations with
$\chi=10^{3}$ and different density profiles (models c3lo128 and
c3losh64).  This highlights the fact that the strongest turbulence is
generated where the shear velocity is high, that the central region of
the wake immediately behind the cloud has a somewhat lower level of
turbulence, and that further downstream the turbulence has penetrated
throughout the wake. The opening angle of the turbulent layer in model
c3lo128 is estimated as $\approx 13^{\circ}$, which is in good
agreement with experimental results \citep[see][and references
therein]{Canto:1991}, where an opening angle of $\approx 11^{\circ}$
is obtained for a Mach 1.3 flow past a stationary medium (the gas
behind a Mach 10 shock has a Mach number of 1.31). Future work will
examine whether this agreement with experiment persists as the Mach
number is varied.

\subsection{Dependence on cloud density contrast}
\label{sec:chi}
A range of density contrasts between the cloud and the ambient medium
is expected. For instance, $\chi \sim 10^{2}$ for cold atomic clouds
embedded in the warm neutral or photoionzied medium where $T\sim10^{4}\;$K,
or for warm clouds embedded in the coronal gas where $T\sim10^{6}\;$K.
For molecular clouds embedded in warm gas, $\chi\sim10^{3}$, while
cold atomic clouds embedded in coronal gas have $\chi\sim10^{4}$.

\begin{figure*}
%paper scales in x.dat: without scales = 3.3x12.0, with scales = 4.5x12.0
\psfig{figure=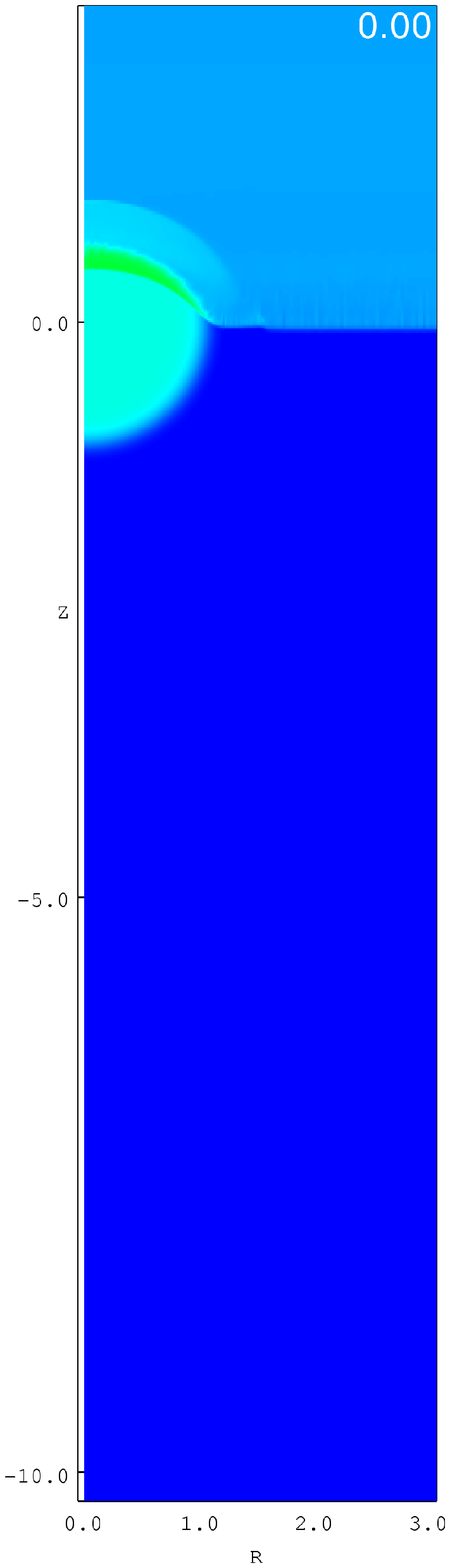,width=3.0cm}
\psfig{figure=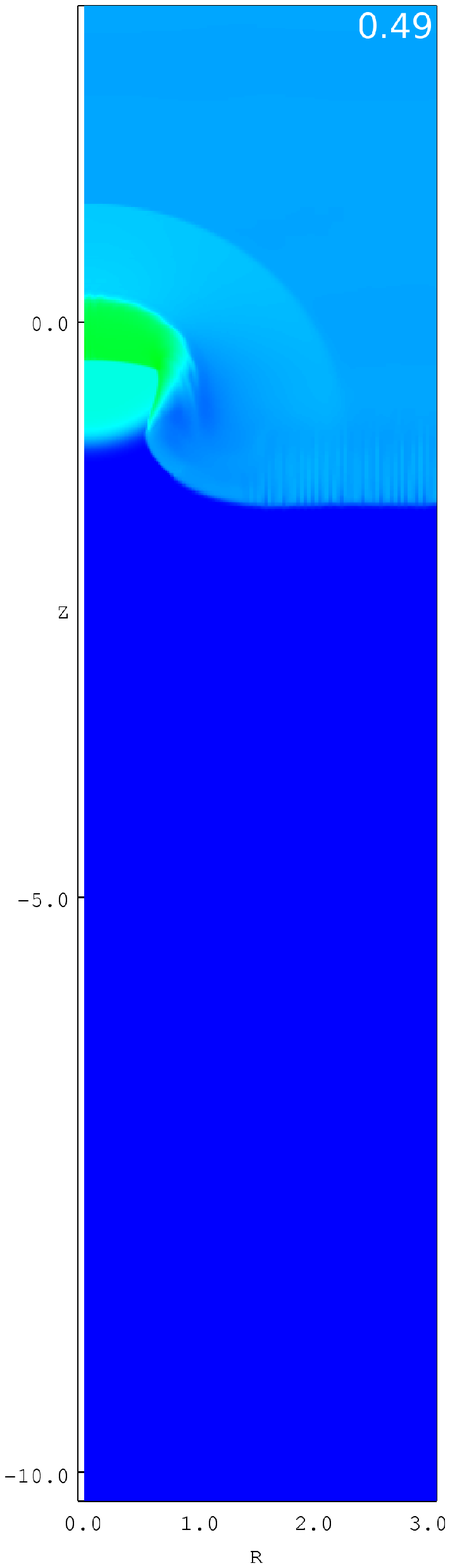,width=3.0cm}
\psfig{figure=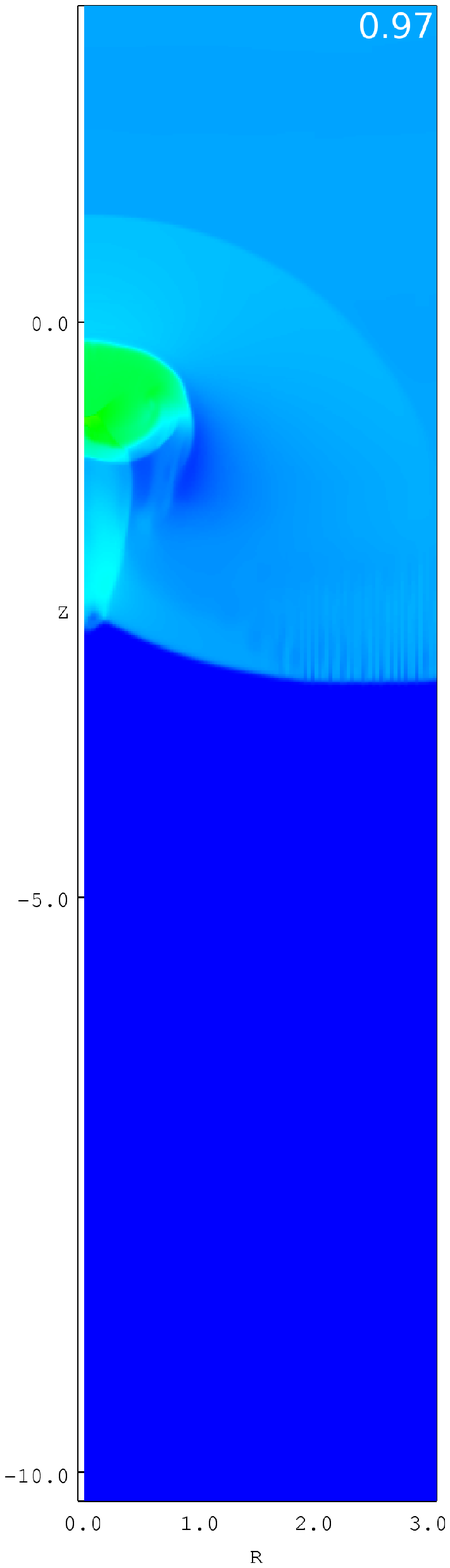,width=3.0cm}
\psfig{figure=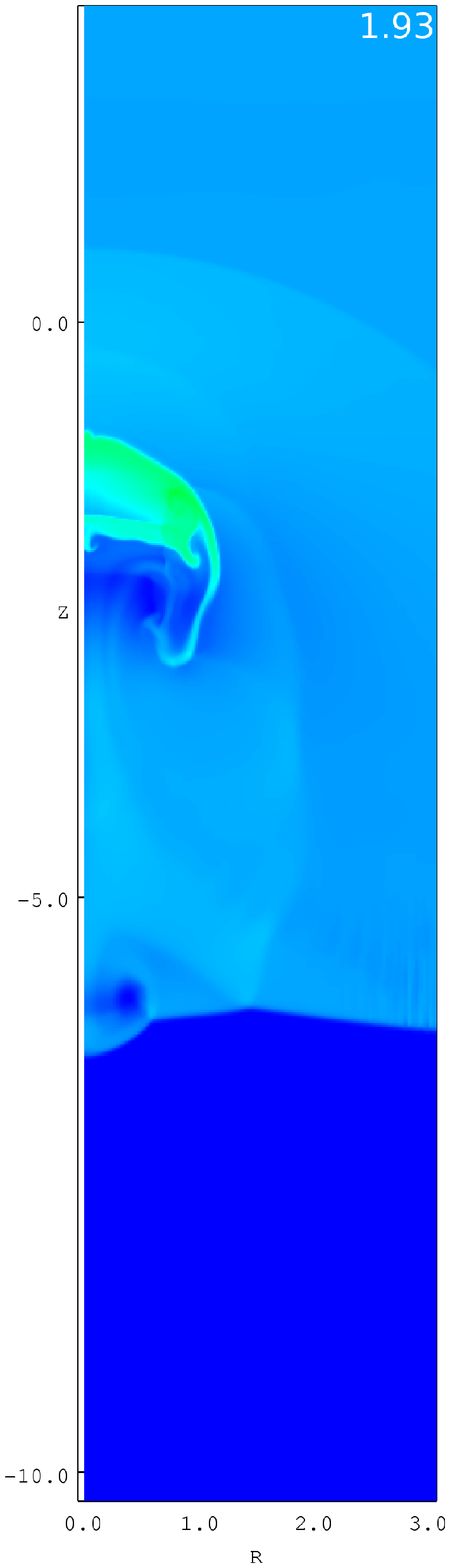,width=3.0cm}
\psfig{figure=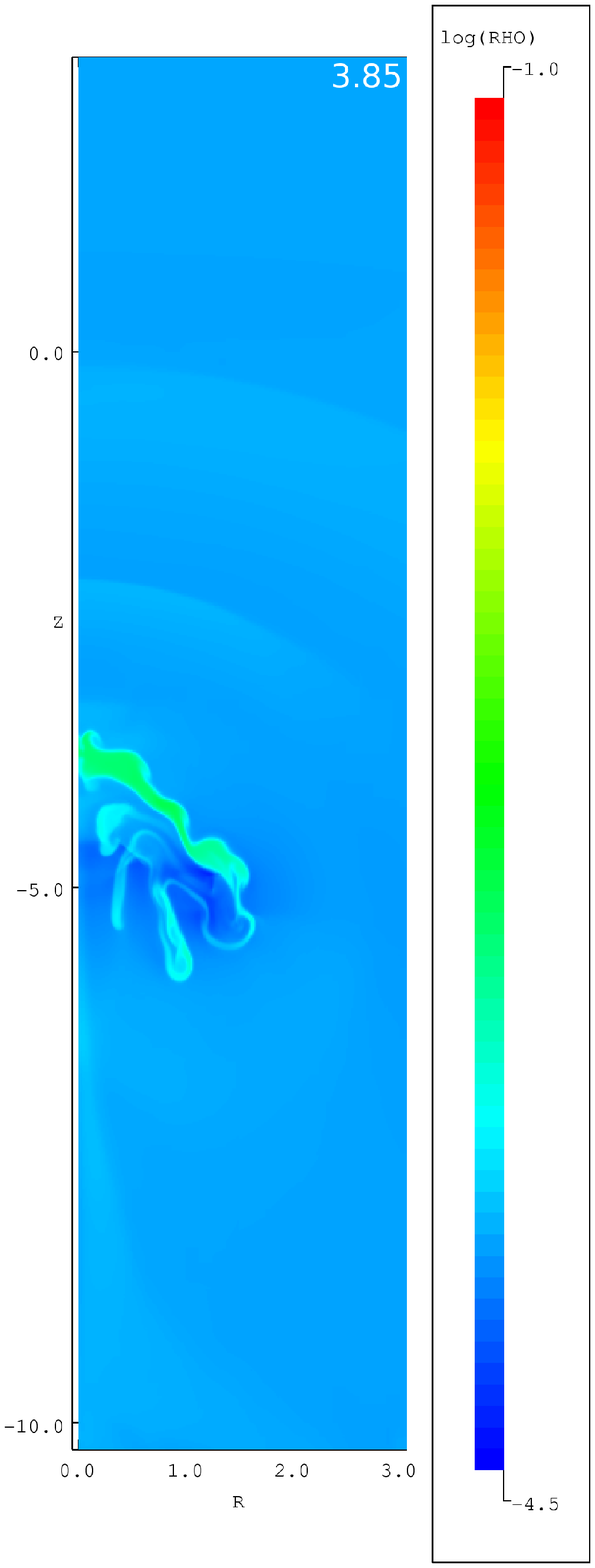,width=4.09cm}
\caption[]{Snapshots of the density distribution from an inviscid
calculation of a Mach 10 adiabatic shock hitting a cloud with a
density contrast of $10$ with respect to the ambient medium and
with a density profile specified by $p_{1}=10$ (model c1no128). The
resolution is 128 cells per cloud radius. The evolution proceeds left
to right with $t = 0.0, 0.49, 0.97, 1.93$, and $3.85\;t_{\rm cc}$.}
\label{fig:nokeps_chi1e1}
\end{figure*}

\begin{figure*}
%paper scales in x.dat: without scales = 3.3x12.0, with scales = 4.5x12.0
\psfig{figure=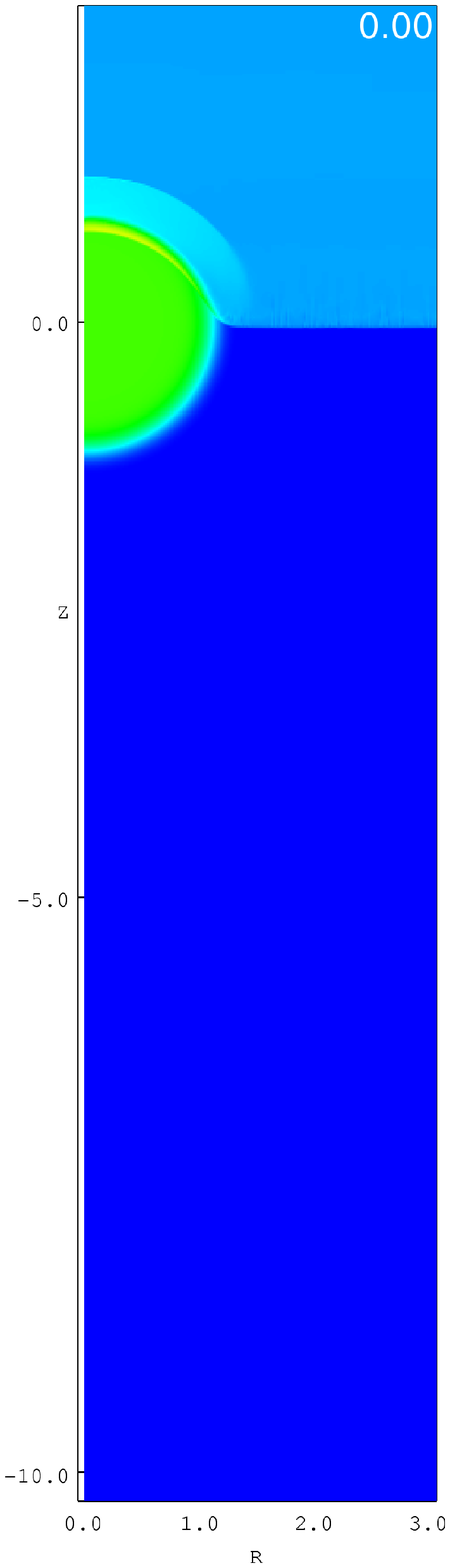,width=3.0cm}
\psfig{figure=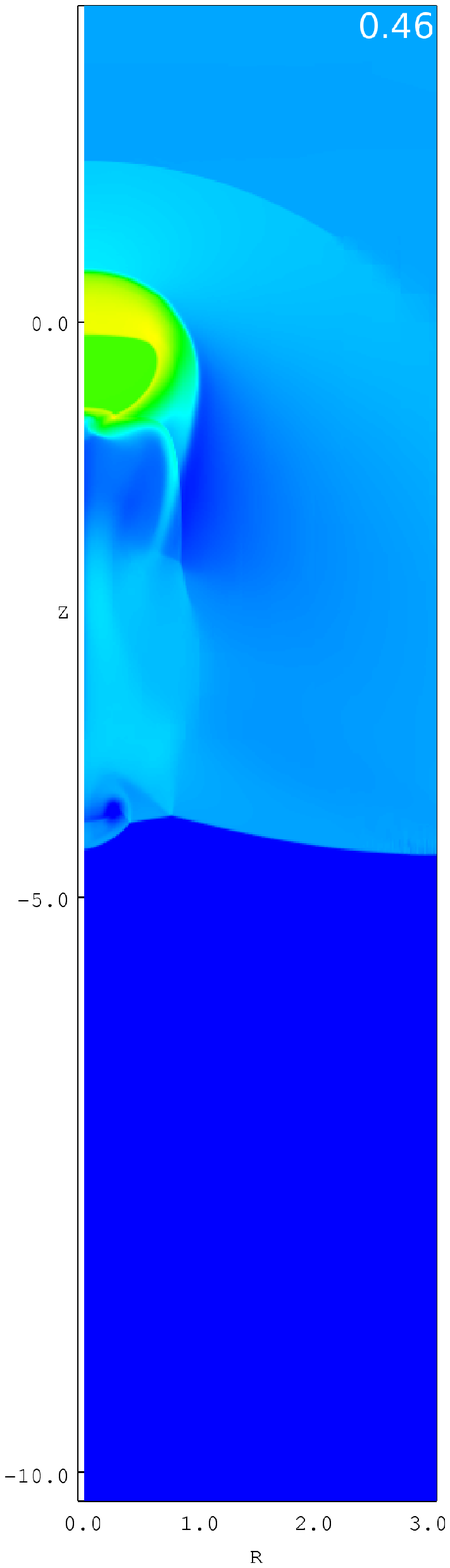,width=3.0cm}
\psfig{figure=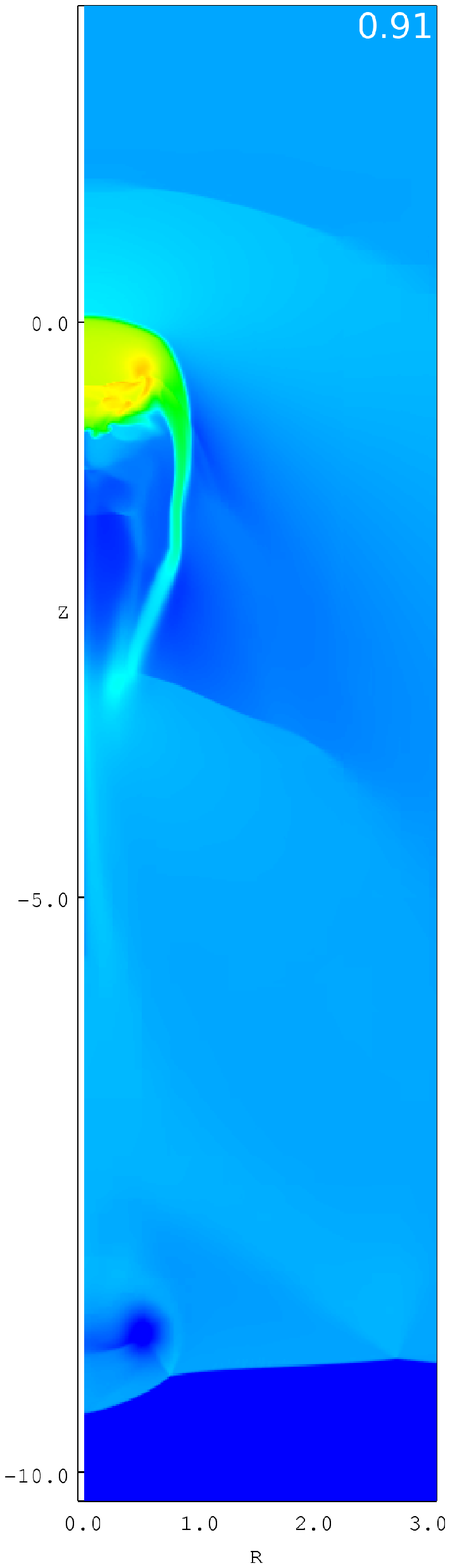,width=3.0cm}
\psfig{figure=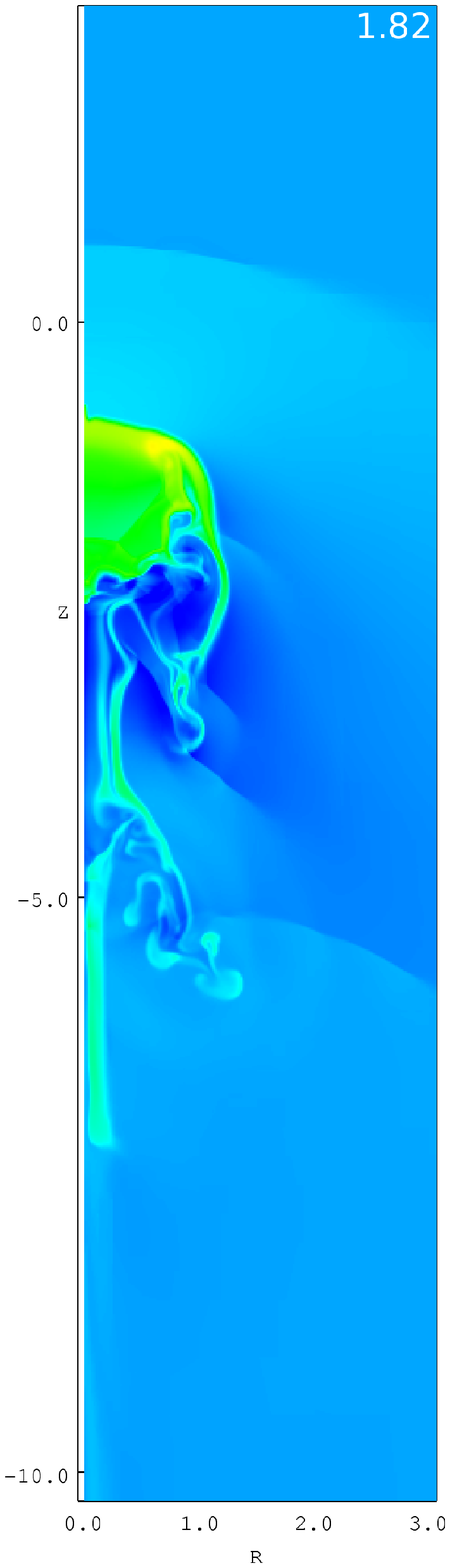,width=3.0cm}
\psfig{figure=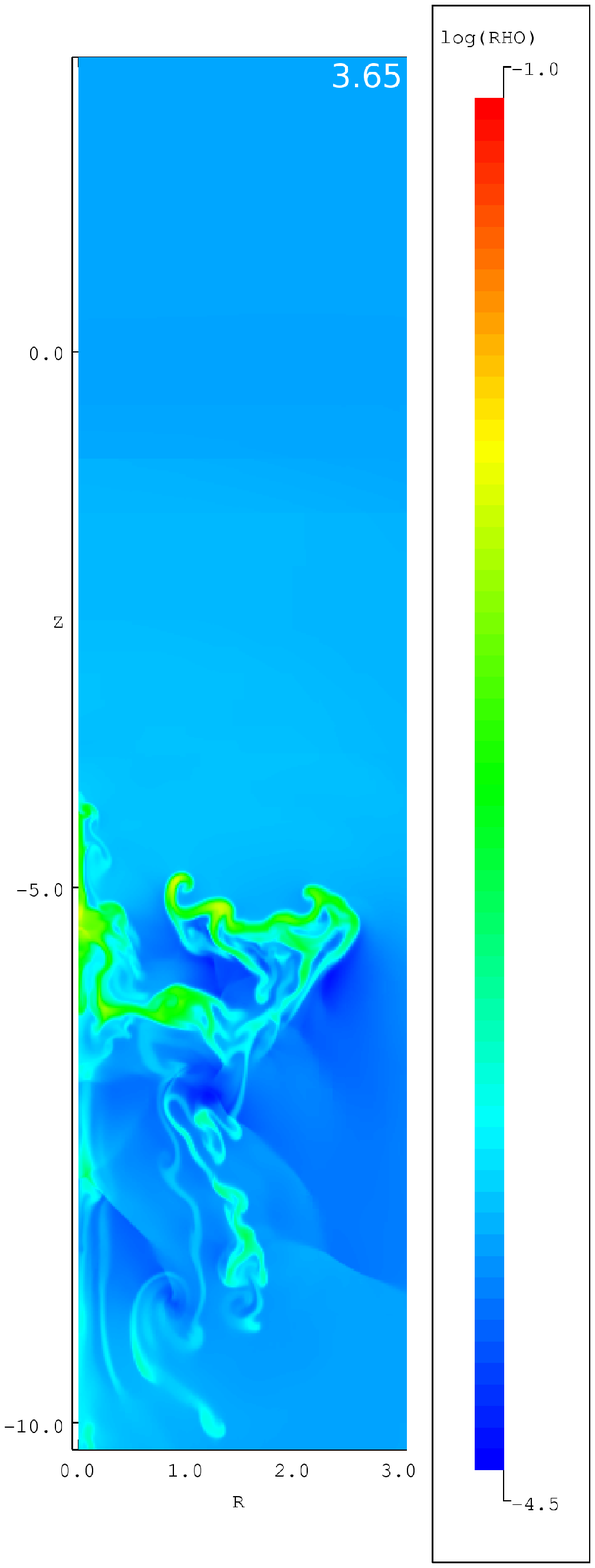,width=4.09cm}
\caption[]{Same as Fig.~\ref{fig:nokeps_chi1e1} but for a density
contrast of $\chi = 10^{2}$ (model c2no128). The evolution proceeds left
to right with \linebreak $t = 0.0, 0.46, 0.91, 1.82$, and $3.65\;t_{\rm cc}$.}
\label{fig:nokeps_chi1e2}
\end{figure*}

Fig.~\ref{fig:nokeps_chi1e1} shows the destruction of a cloud with a
density contrast $\chi=10$, while Fig.~\ref{fig:nokeps_chi1e2}
shows the corresponding case for a cloud with $\chi=10^{2}$, both
computed with an inviscid code. The colour scaling in both figures is
identical to that in previous figures for easier comparison.  The
timescale for the cloud material to mix into the ambient flow scales
roughly with the cloud crushing timescale, $t_{\rm cc}$, in agreement
with previous works \citep{Klein:1994,Nakamura:2006}. However, the
normalized growth timescale for RT and KH instabilities decreases with
increasing $\chi$, as is apparent from a comparison of
Figs.~\ref{fig:nokeps},~\ref{fig:nokeps_chi1e1}
and~\ref{fig:nokeps_chi1e2}.  The normalized drag time, $t_{\rm
drag}/t_{\rm cc}$, increases with $\chi$, as does the axial
stretching of the cloud, $c/a$ (see Table~\ref{tab:results}).
Figs.~\ref{fig:kepscomp1}-\ref{fig:kepscomp3} also reveal that the
ratio of the velocity dispersion in the axial to normal directions,
$\delta v_{\rm z}/\delta v_{\rm r}$, increases with $\chi$. 
These results are all in agreement with earlier works.

The effect of a highly turbulent post-shock flow is greatest at high
$\chi$ (e.g., compare the values of the mean cloud and core velocities
in Table~\ref{tab:results} for the ``high $k$-$\epsilon$'' models
against the ``low $k$-$\epsilon$'' and inviscid models as a function
of $\chi$). This is because clouds with a high density contrast
survive for a considerable time after the initial passage of the
shock, and thus are subject to considerable ``buffeting'' by the
highly turbulent postshock environment, whereas at lower values of
$\chi$, the cloud is destroyed relatively quickly after the initial
passage of the shock. This is also manifest in the increasing
disparity with $\chi$ in the evolution of various global quantities
from the ``hi'' models on the one hand, and the ``no'' and ``lo''
models on the other hand, as shown in
Figs.~\ref{fig:kepscomp1}-\ref{fig:kepscomp3}.

\begin{figure*}
%paper scales in x.dat: without scales = 3.3x12.0, with scales = 4.5x12.0
\psfig{figure=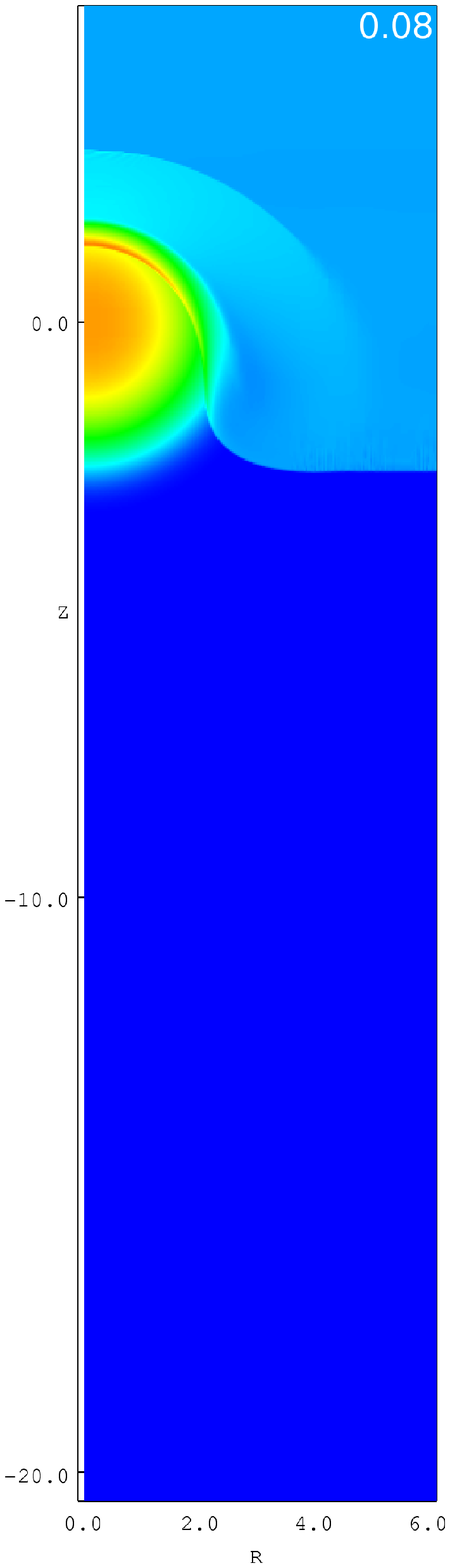,width=3.0cm}
\psfig{figure=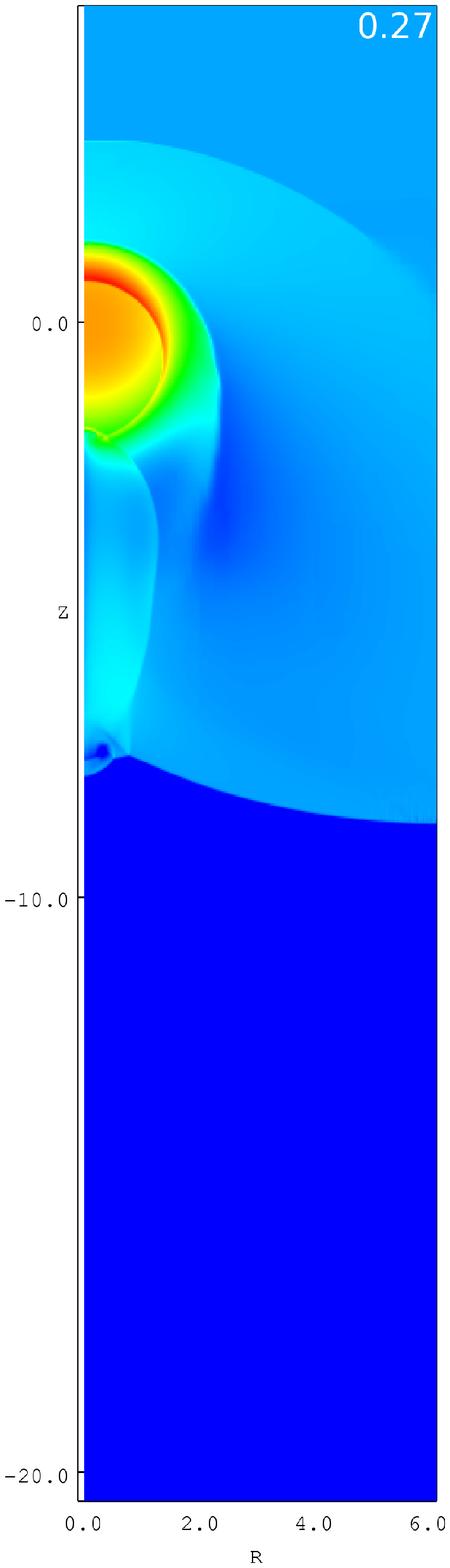,width=3.0cm}
\psfig{figure=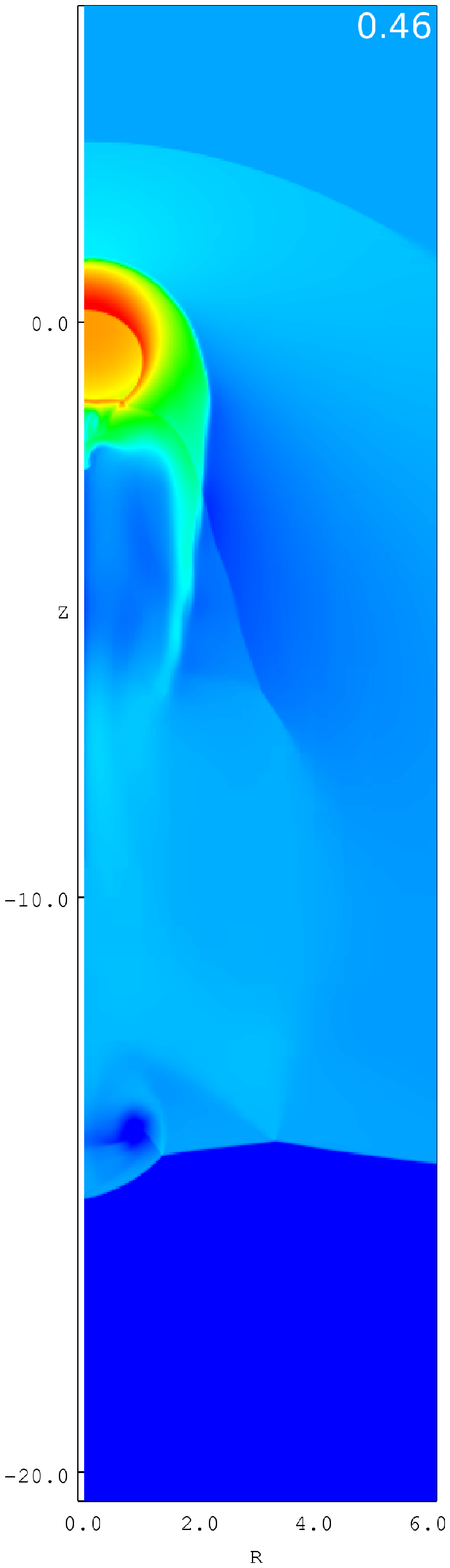,width=3.0cm}
\psfig{figure=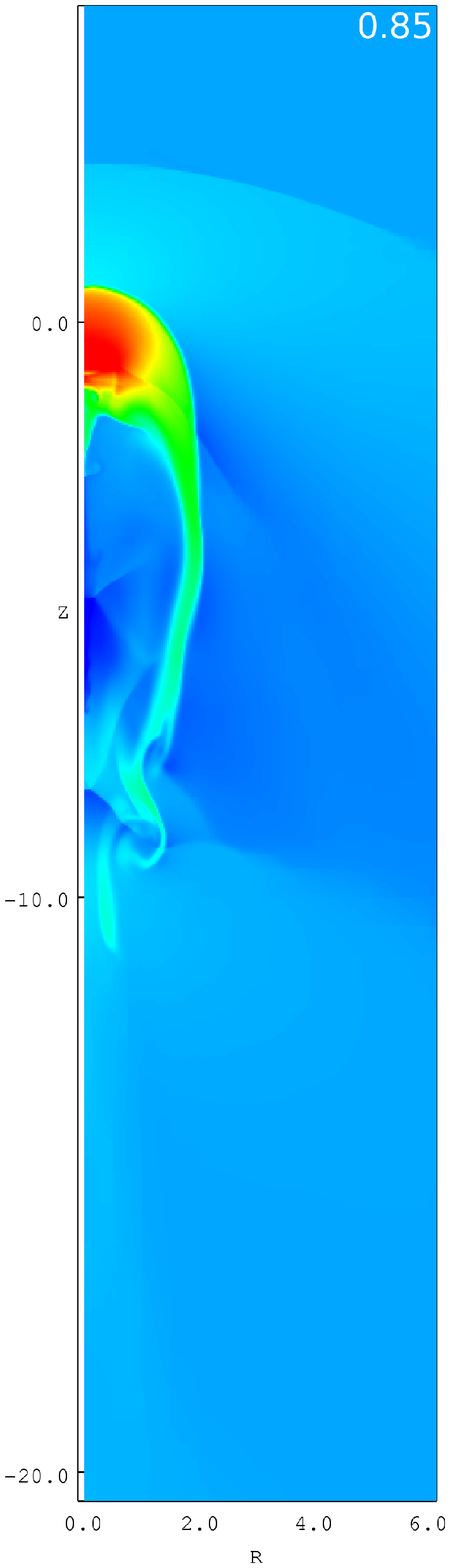,width=3.0cm}
\psfig{figure=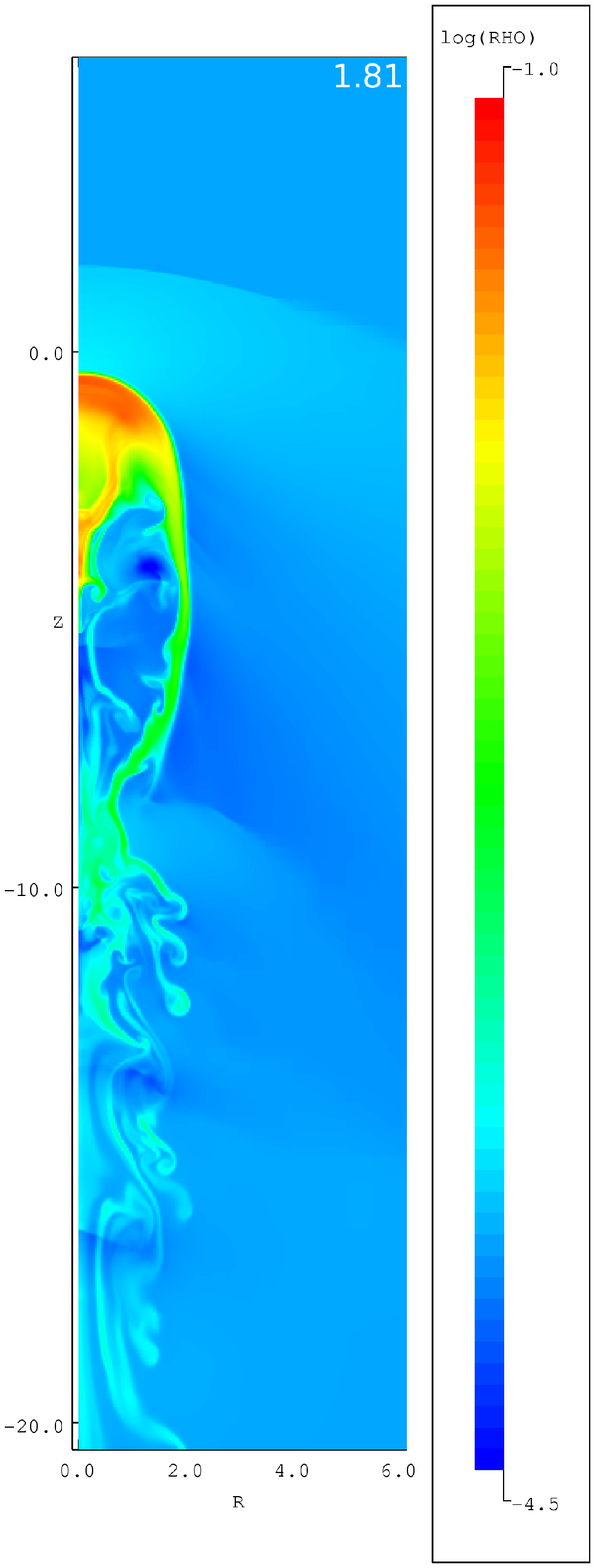,width=4.09cm}
\caption[]{Snapshots of the density distribution from an inviscid
calculation of a Mach 10 adiabatic shock hitting a cloud with a
density contrast of $10^{3}$ with respect to the ambient medium and
with a shallow density gradient, $p_{1}=1$ (model c3nosh64). The
formal resolution is 64 cells per $r_{\rm c}$, but due to the shallow profile
the effective resolution is roughly 128 cells per cloud radius.  The
evolution proceeds left to right with $t = 0.08, 0.27, 0.46, 0.85$,
and $1.81\;t_{\rm cc}$. Note the different spatial scale of the plots compared
to Fig.~\ref{fig:nokeps}.}
\label{fig:nokepsp21}
\end{figure*}

\subsection{Dependence on cloud profile}
\label{sec:shallow}
In Fig.~\ref{fig:nokepsp21}, snapshots of the density distribution at
different times are shown for an inviscid calculation of a Mach 10
shock hitting a cloud with a shallow density gradient (model c3nosh64;
see Table~\ref{tab:models}).  While the resolution is such that
$r_{\rm c}$ is equal to the width of 64 cells on the finest grid, the
cloud in fact extends to $r \gtsimm 2\;r_{\rm c}$, so the effective
resolution is similar to the previous models
(see Fig.~\ref{fig:cloudprofs}).

\begin{figure}
%paper scales in x.dat: without scales = 3.3x12.0, with scales = 4.5x12.0
%For spanning 1 column
\psfig{figure=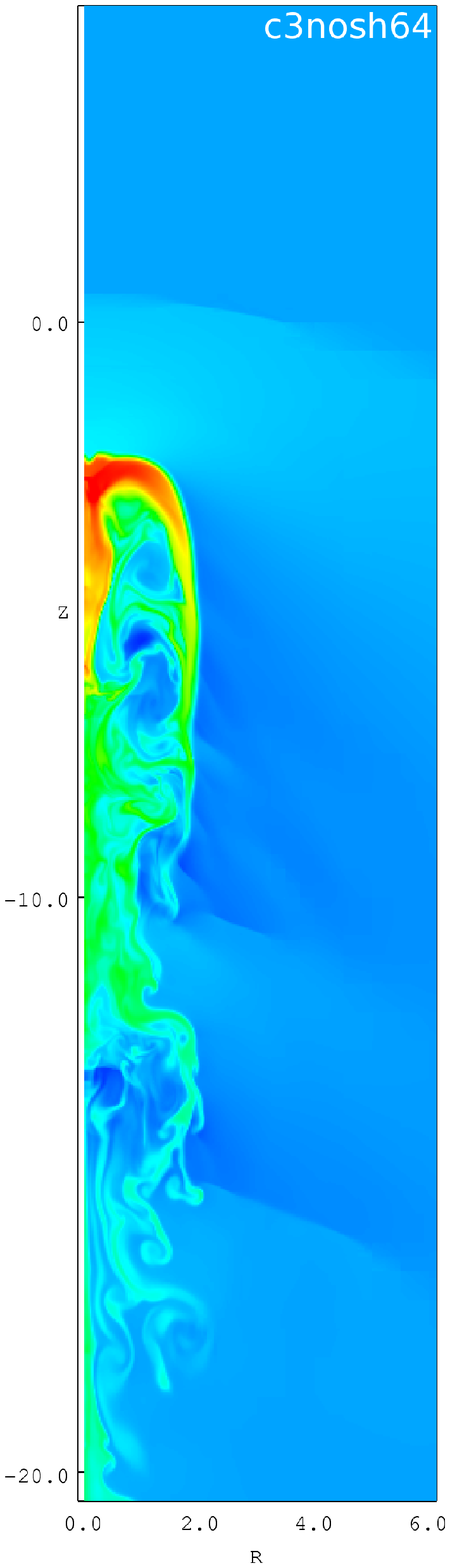,width=3.5cm}
\psfig{figure=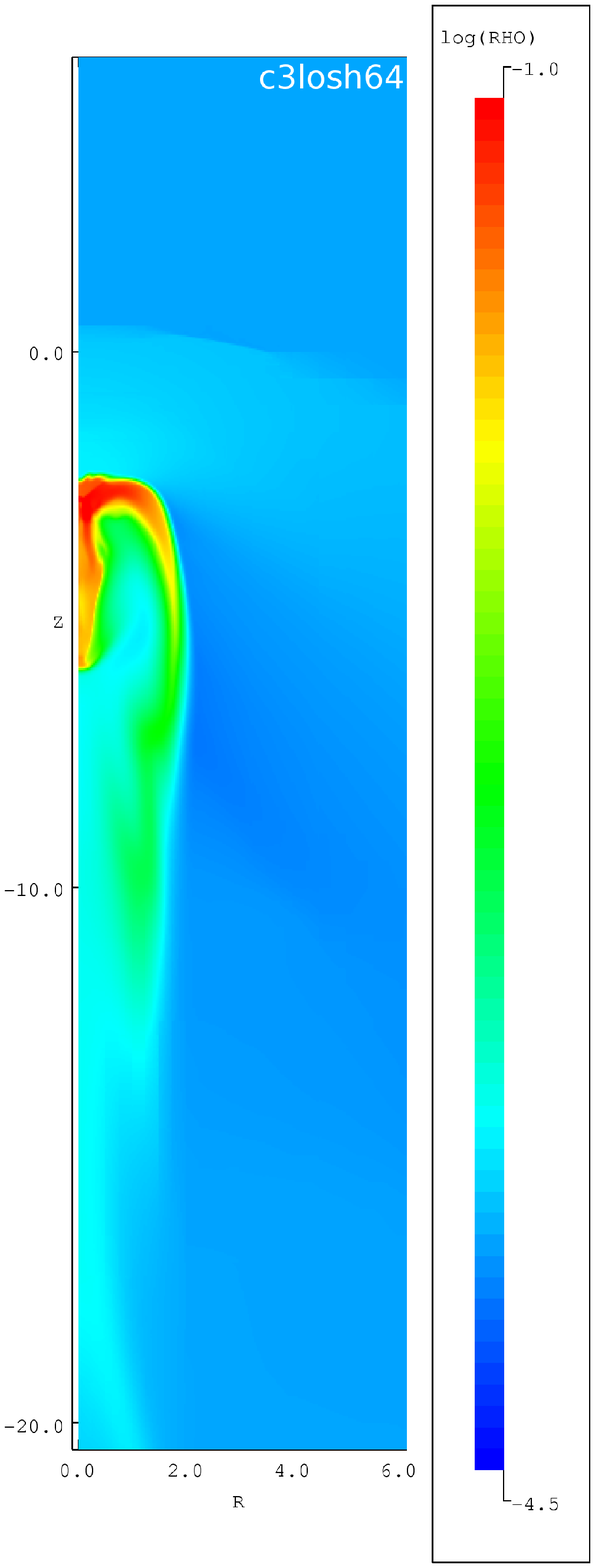,width=4.7714cm}
\caption[]{Comparison between the inviscid (left panel, model
c3nosh64) and $k$-$\epsilon$ calculation with low initial postshock
turbulence (right panel, model c3losh64) for a Mach 10 adiabatic shock
hitting a cloud with $\chi=10^{3}$ and a shallow density gradient 
($p_{1}=1$) at $t = 2.77\;t_{\rm cc}$.}
\label{fig:comp_latet_p21}
\end{figure}

The interaction of a shock with a smooth cloud was previously studied
by \cite{Nakamura:2006} for the case where $\chi=10$. In this case the
cloud offered little impediment to the oncoming shock, with the result
that the transmitted shock and the intercloud shock had similar mean
speeds. As a result, the intercloud shock did not converge on the
$z$-axis behind the cloud, and the shock compression from the
downstream side was weak, leading to a slow lateral expansion of the
cloud. In contrast, the cloud is a much more robust obstacle when
$\chi=10^{3}$, and we find that it maintains many aspects of the
evolution seen in sharper-edged clouds (c.f. Figs.~\ref{fig:nokeps}
and~\ref{fig:nokepsp21}). Fig.~\ref{fig:comp_latet_p21} compares the
density structure at $t=2.77\;t_{\rm cc}$ for inviscid and low
$k$-$\epsilon$ calculations (models c3nosh64 and c3losh64).  At later
times the material stripped from the cloud resembles a single
tail-like structure, as was also the case for a cloud with sharper edges.

\begin{figure*}
%paper scales in x.dat: without scales = 3.3x12.0, with scales = 4.5x12.0
\psfig{figure=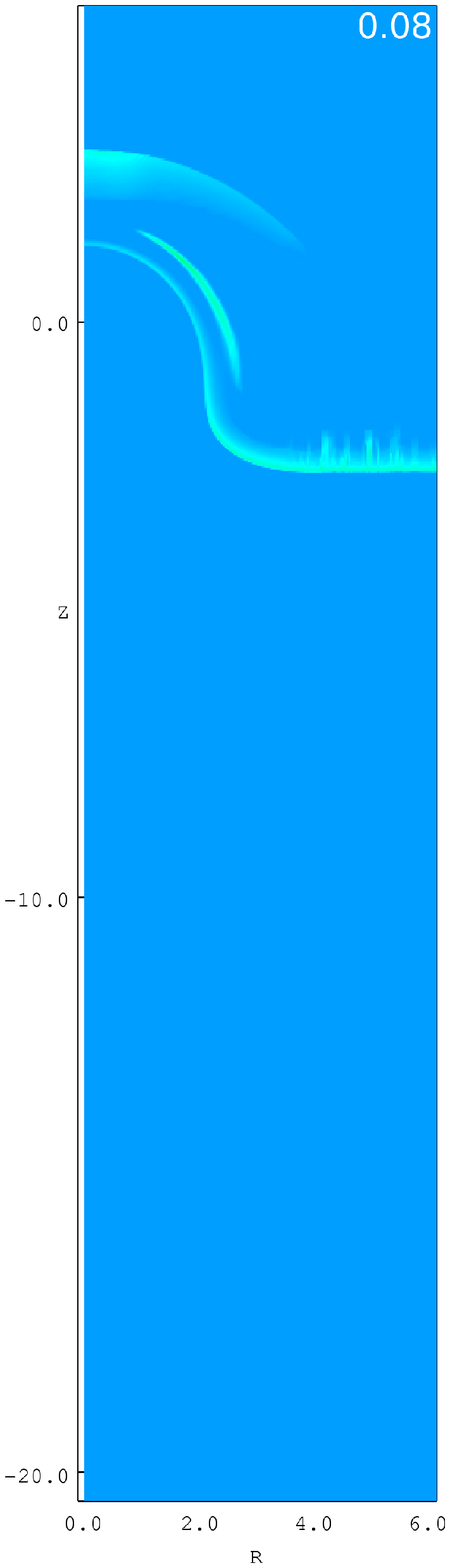,width=3.0cm}
\psfig{figure=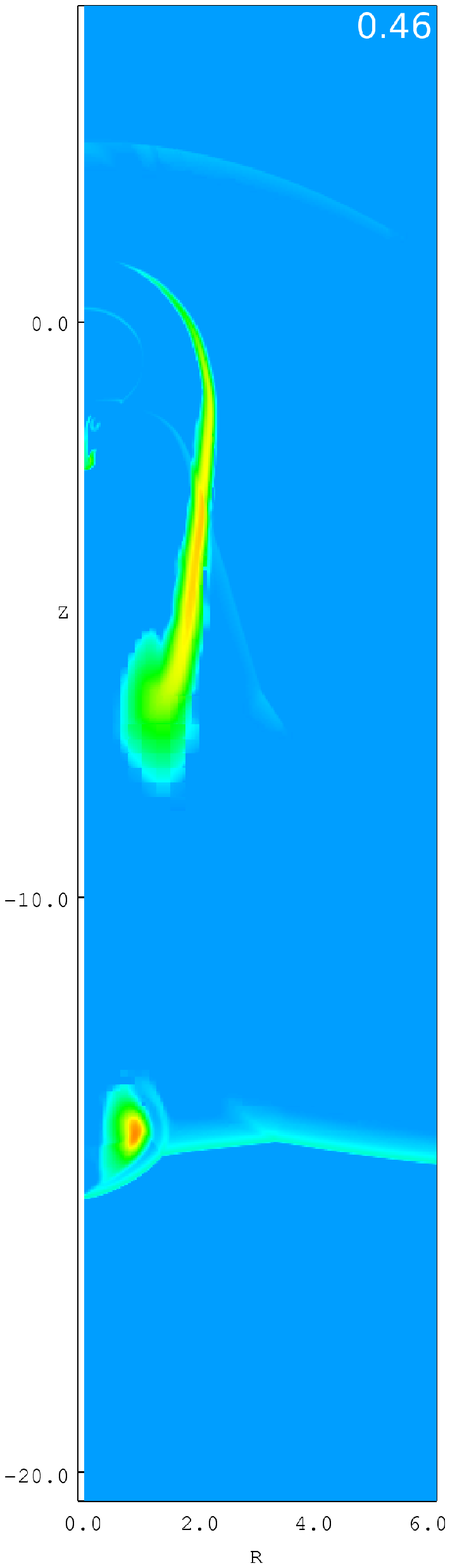,width=3.0cm}
\psfig{figure=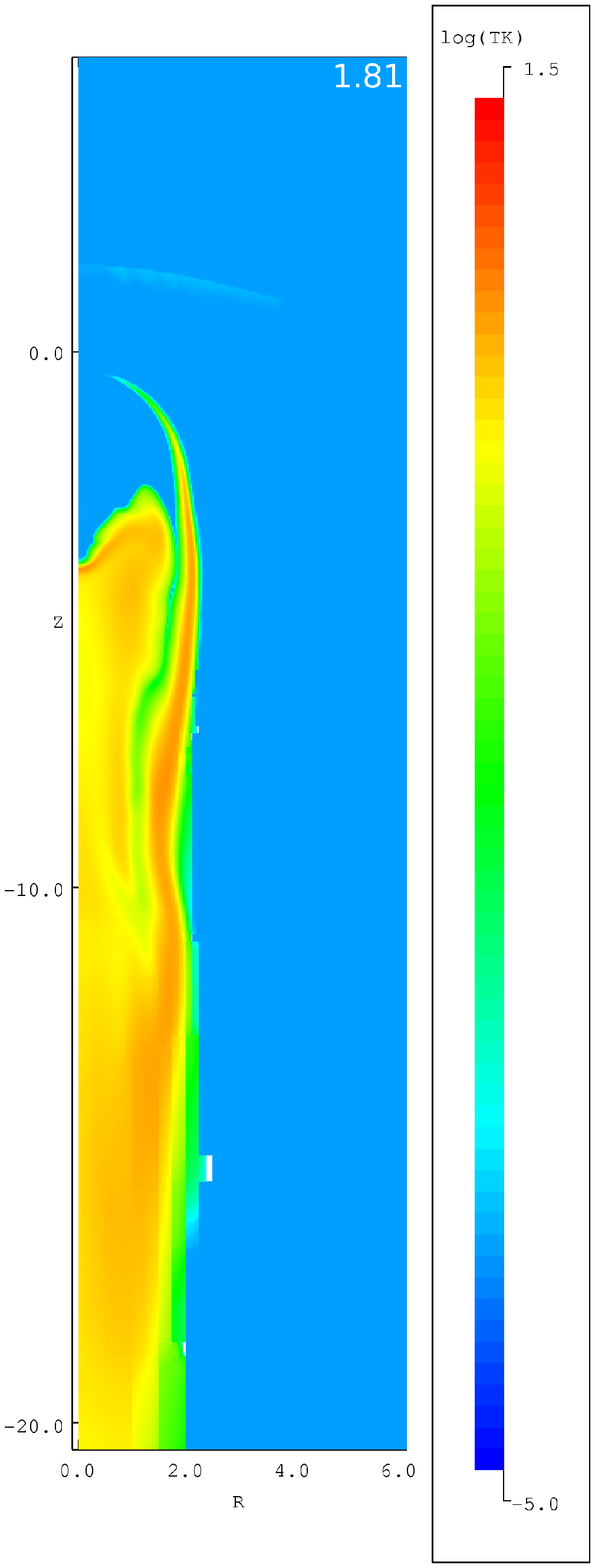,width=4.09cm}
\psfig{figure=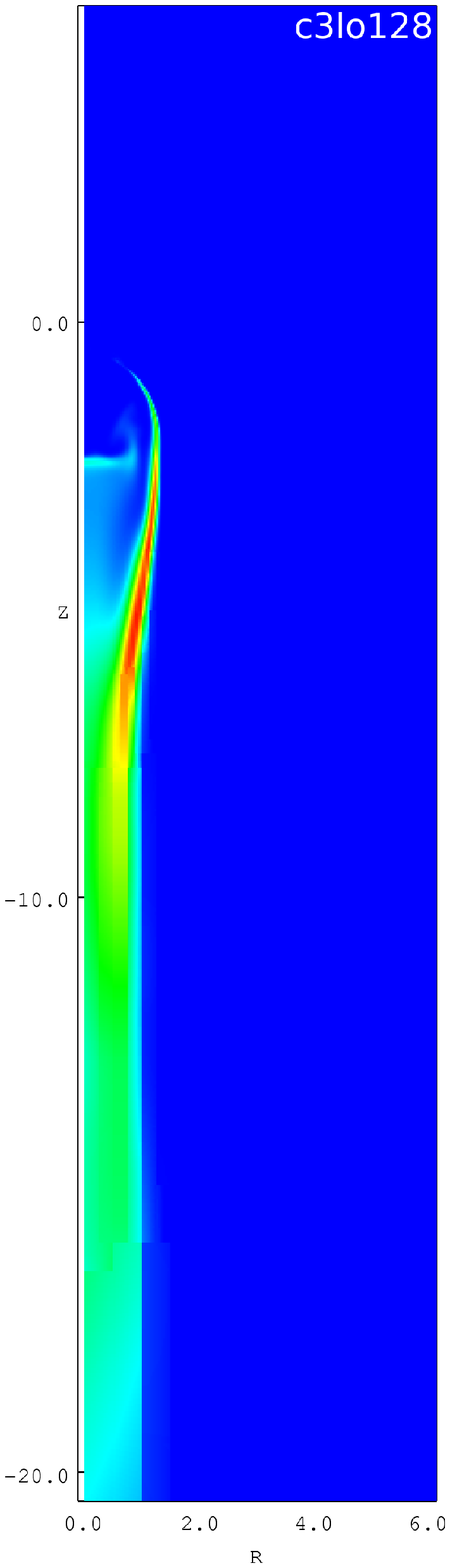,width=3.0cm}
\psfig{figure=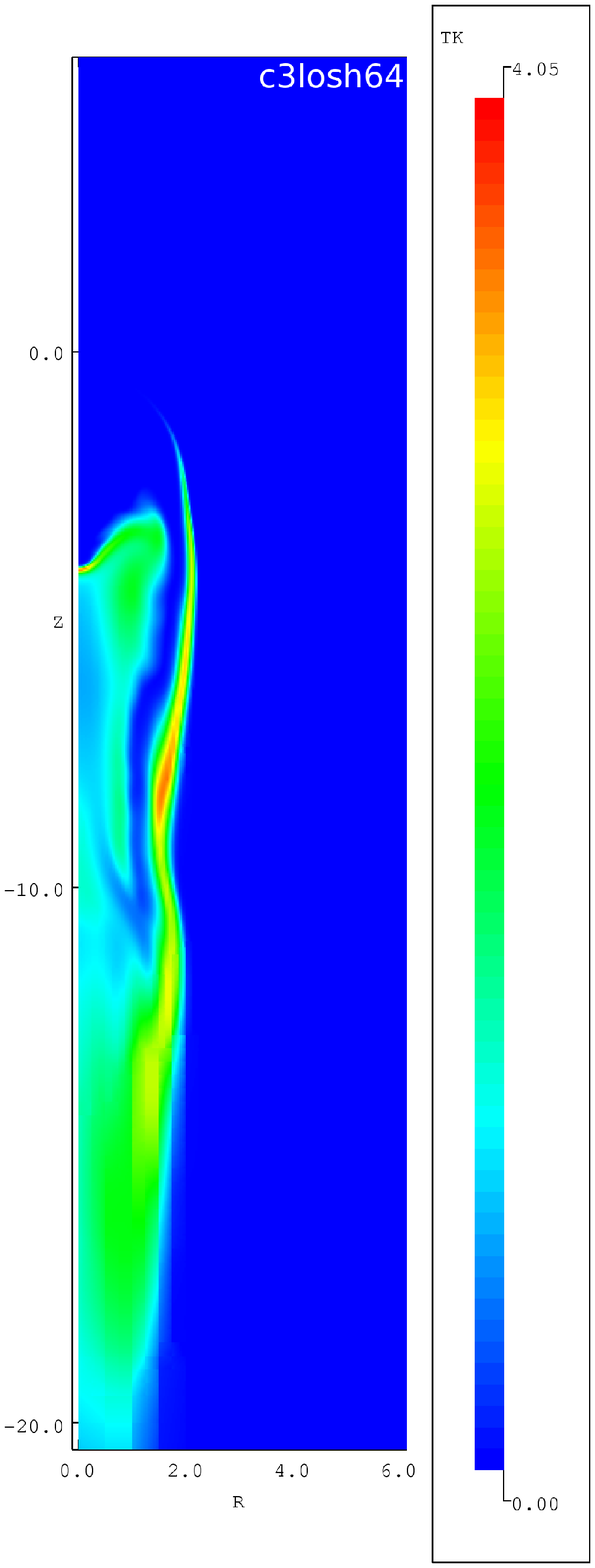,width=4.09cm}
\caption[]{Snapshots of the turbulent energy per unit mass, $k$. The
three left-most panels show the evolution of $k$ in model c3losh64
proceeding left to right with $t = 0.08, 0.46$ and $1.81\;t_{\rm cc}$.
In the two rightmost panels a linear scale is used for $k$ at
$t=1.81\;t_{\rm cc}$ - model c3lo128 is shown in the second from right
panel, while the rightmost panel shows again model c3losh64.  Note
also the different spatial scale of the plots compared to
Fig.~\ref{fig:lowkeps_tk}.}
\label{fig:lokepsp21}
\end{figure*}

Nevertheless, the shallower density gradient does lead to a milder
interaction. This is also manifest as a slower growth of turbulence
around the cloud (c.f. Figs~\ref{fig:lowkeps_tk}
and~\ref{fig:lokepsp21} for $k$-$\epsilon$ models with low initial
turbulence), with the turbulent wake not completely forming until
$t\approx2\;t_{\rm cc}$. An exact comparison of the respective
timescales is complicated by the fact that the cloud with the shallow
density gradient is also larger and more massive, and so the true
value of $t_{\rm cc}$ will be different between the models. In fact,
we find that the forward and rear shocks driven into the cloud
converge at $t \approx 0.8\;t_{\rm cc}$ in model c3no128, and at $t
\approx 1.0\;t_{\rm cc}$ in model c3nosh64 (we continue to calculate
$t_{\rm cc}$ using Eq.~\ref{eq:tcc} with $\chi=10^{3}$ and $r_{\rm
c}=1$, despite this equation being applicable only to clouds with
sharp edges). Since these times are not too discrepant, it
becomes clear that the growth of turbulence around the smoother cloud
is indeed slower, in agreement with the statement by \cite{Nakamura:2006}
that it takes more time to form the slip surface around the
cloud. The maximum turbulent energy per unit mass (i.e. $k_{\rm max}$) is
also higher in model c3lo128.

\subsection{A non-uniform post-shock flow}
\label{sec:grid-scale-turb}
To obtain further confirmation of the previous results, we have
simulated the destruction of a cloud hit by a shock with non-uniform
post-shock flow.  To generate the necessary random motions of the
post-shock flow, perturbations in the density and velocity are mapped
onto the flow and allowed to evolve. These initial pertubations
produce normalized standard deviations of 0.18, 0.26, and 0.13 in the
density, pressure and velocity of the post-shock flow. The
perturbations subsequently decay as the shock sweeps up the smooth
inter-cloud ambient medium and because of the viscosity inherent in
any numerical code, and the standard deviations of the fluctuations
noted above decrease. We use a non-AMR setup (i.e. a fine, rather than
a coarse, $G^0$ grid), since the perturbations downstream of the shock
would cause a large amount of grid refinement, thus erasing the
benefits that are usually possible with AMR. These two issues (decay
of the post-shock turbulence, and the inability to effectively use
AMR) further highlight the benefits of using a sub-grid turbulence
model.

In Fig.~\ref{fig:rt} we show the initial interaction of a shock with
grid-scale post-shock turbulence and a cloud with a density contrast,
$\chi=10$. The bowshock around the cloud is much less distinct, and
the symmetry of the interaction present in the previous figures is
broken. The shock shows small-scale curvature along its surface and is
marginally faster than its counterpart with smooth post-shock flow
(the effective Mach number of the shock is about 5 per cent higher,
and reflects the extra ``turbulent'' energy mapped into the initial
post-shock flow). However, the most important difference is that the
highly turbulent environment destroys the cloud much more rapidly than
when the post-shock flow is smooth (c.f. Figs.~\ref{fig:nokeps_chi1e1}
and~\ref{fig:rt}, and also examine Fig.~\ref{fig:kepscomp1}),
confirming the results obtained with the $k$-$\epsilon$ sub-grid
turbulence model.

\subsection{Cloud statistics}
Figs.~\ref{fig:kepscomp1}-\ref{fig:kepscomp3} show the evolution of
various global quantities of the cloud as a function of numerical
resolution for the inviscid and $k$-$\epsilon$ models with
$\chi=10,\;10^{2}$ and $10^{3}$. In addition, the results from a
simulation with $\chi=10^{3}$ and a shallow density profile are shown
in Fig.~\ref{fig:kepscomp3sh}. Various numerical quantities from these
simulations are noted in Table~\ref{tab:results}.  As already shown,
the solutions that the $k$-$\epsilon$ simulations attain depend on the
initial level of turbulence in the post-shock flow which overruns the
cloud. For models with low initial turbulence, the evolution of the
cloud is similar to that attained from the inviscid code, particularly
at lower values of $\chi$. However, it is clear from the plots in
Figs.~\ref{fig:kepscomp1}-\ref{fig:kepscomp3sh} that when the
environment surrounding the cloud is turbulent itself, the mixing of
cloud material into the surrounding flow proceeds at a much faster
pace (as shown, for instance, by the more rapid growth in the cloud
size (see panels a and c) and reduction in the core mass (see panel
g)) due to the enhanced transport and diffusion coefficients. It is
also clear that the cloud with the shallow density gradient is
typically less susceptible to high levels of environmental turbulence
than clouds with sharper boundaries.

A detailed discussion of the statistics is presented in the following
subsections. The results obtained from models with $\chi=10^{2}$ are
discussed initially in each subsection, and then a comparison is made
to results obtained with lower and higher density contrasts and with a
shallower density profile. Results from the inviscid model with
grid-scale post-shock turbulence (model c1rt32) are also discussed
where appropriate.

\subsubsection{Cloud shape}
Fig.~\ref{fig:kepscomp2}a-f) shows the time evolution of the rms cloud
and core radii, $a$ and $c$, and their ratios, for simulations with
$\chi=10^{2}$. The transverse dimensions, $a_{\rm cloud}$ and $a_{\rm
core}$, decrease during the shock compression stage, and then increase
during the expansion stage.  $a_{\rm cloud}$ reaches a maximum of
$\approx 4\;r_{\rm c}$ (at $t\approx6\;t_{\rm cc}$) in all 3 models,
and stays at roughly this level until at least $t = 12\;t_{\rm
cc}$. Although $a_{\rm core}$ reaches a similar maximum, its value
subsquently drops precipitously once the mass within the core becomes
a small fraction ($\ltsimm 15\%$) of its initial value.

After the initial compression stage, which lasts until $t\approx
t_{\rm cc}$, the cloud becomes increasingly elongated in the direction
of the propagation of the shock. The values of $c_{\rm cloud}$ and
$c_{\rm cloud}/a_{\rm cloud}$ are still growing at $t=12\;t_{\rm cc}$,
when the simulations were stopped.  At this point the cloud material
is dispersed over a distance of $\sim 15\;r_{\rm c}$ in the axial
direction in all 3 simulations. However, Fig.~\ref{fig:nokeps_chi1e2}
shows that even at earlier times the material stripped from the cloud
is highly fragmented and does not resemble a single tail-like
structure (in contrast, the stripped material better resembles a tail
when $\chi=10^{3}$ - see Fig.~\ref{fig:comp_latet}).  

The axial radius of the core, $c_{\rm core}$, displays the same
initial behaviour as $c_{\rm cloud}$, though its rate of expansion is
less rapid.  Thereafter $c_{\rm core}$ behaves in a similar way to
$a_{\rm core}$: it reaches a maximum in all three simulations by
$t\ltsimm8\;t_{\rm cc}$, and then declines as material which was
formerly within the core is mixed into the surrounding flow to the
extent that it can no longer be identified as ``core'' material.
The values of $a_{\rm core}$ and $c_{\rm core}$ can
show abrupt changes as the cloud fragments, and are generally (though
not always) smaller than the corresponding values of $a_{\rm cloud}$
and $c_{\rm cloud}$.

The core remains compressed in the axial direction (i.e.  $c_{\rm
core}/a_{\rm core} < 1$) in simulations c2no128 and c2lo128 until $t
\approx 6\;t_{\rm cc}$, after which $c_{\rm core}/a_{\rm core}$
rapidly increases to values much larger than unity. In contrast, in
model c2hi128, $c_{\rm core}/a_{\rm core} < 1$ at all times, and
declines to a minimum value of 0.32. The core is always less elongated
than the shocked cloud (i.e. has a smaller value of $c/a$) in the
c2hi128 simulation, but this is true only for times prior to $\approx
6\;t_{\rm cc}$ and $\approx 8\;t_{\rm cc}$ in the c2lo128 and c2no128
simulations, respectively.

Clouds with $\chi=10$ do not expand as much in the lateral direction,
with the values of $a_{\rm cloud}$ and $a_{\rm core}$ remaining below
$2\;r_{\rm c}$ as the cloud is destroyed (Fig.~\ref{fig:kepscomp1}a
and~b). In contrast, $a_{\rm cloud}$ continues to increase in the
latter stages of the interaction in model c1rt32, due to the strong
diffusion in this simulation. The enhanced diffusion
from the model with grid-scale turbulence probably reflects the
larger eddy sizes in this simulation, and reveals that the process
of cloud destruction is sensitive to the properties of any turbulence
in the surrounding medium. 

$c_{\rm cloud}$ increases to a value of 10 in model c1no128
at the time that the simulation is terminated ($t = 28\;t_{\rm cc}$), 
while $c_{\rm cloud}/a_{\rm cloud}$ asymptotes at $\approx 5.5$
at $t\approx 25\;t_{\rm cc}$. In model c1lo128,
$c_{\rm cloud}/a_{\rm cloud}$ follows fairly closely the behaviour of
model c1no128 until $t\approx17\;t_{\rm cc}$, after which there is a
dramatic plunge as mixing of the material reduces $\kappa$ below the
threshold for material to be identified as ``cloud material''. In
model c1hi128,  $c_{\rm cloud}/a_{\rm cloud}$ reaches a much lower 
maximum of 1.5 at $t\approx12\;t_{\rm cc}$.

\begin{figure*}
%paper scales in x.dat: without scales = 3.3x12.0, with scales = 4.5x12.0
\psfig{figure=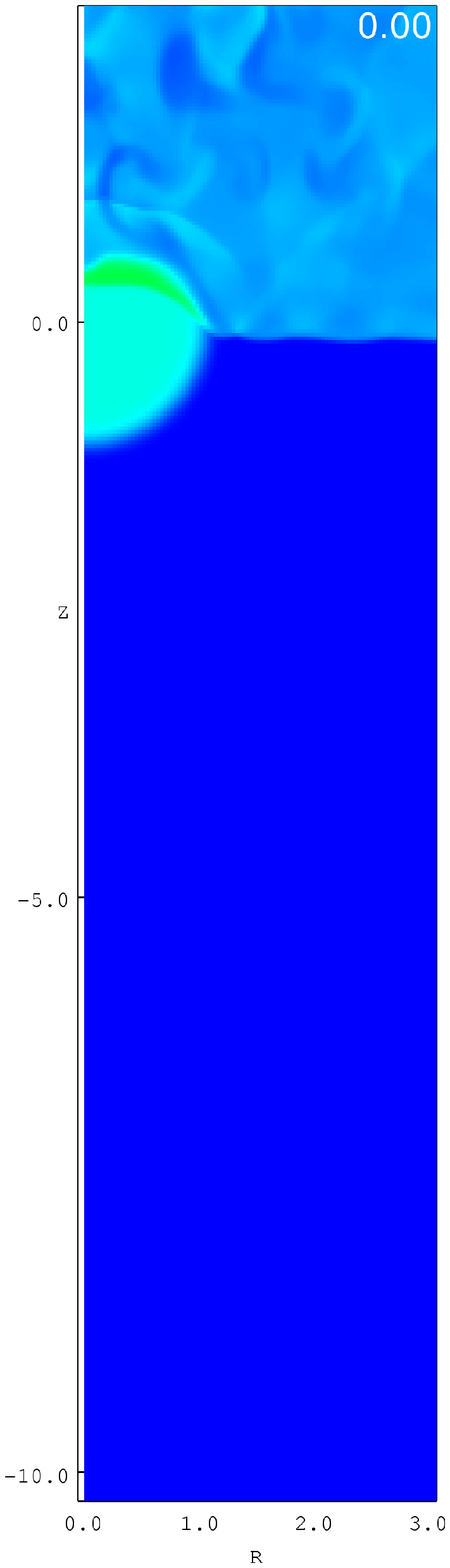,width=3.0cm}
\psfig{figure=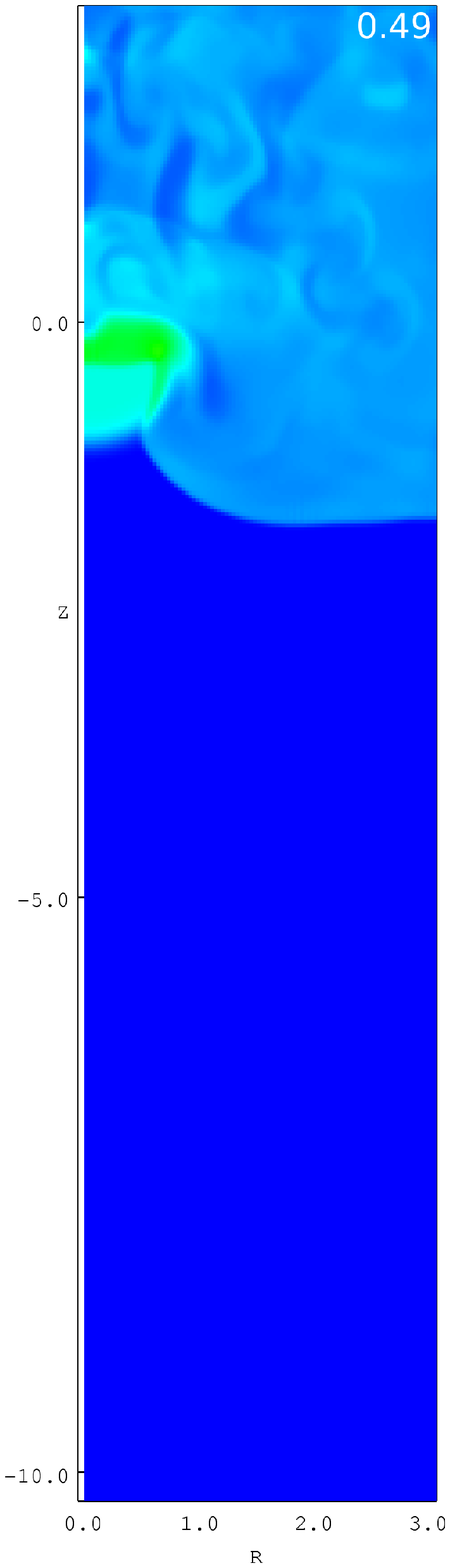,width=3.0cm}
\psfig{figure=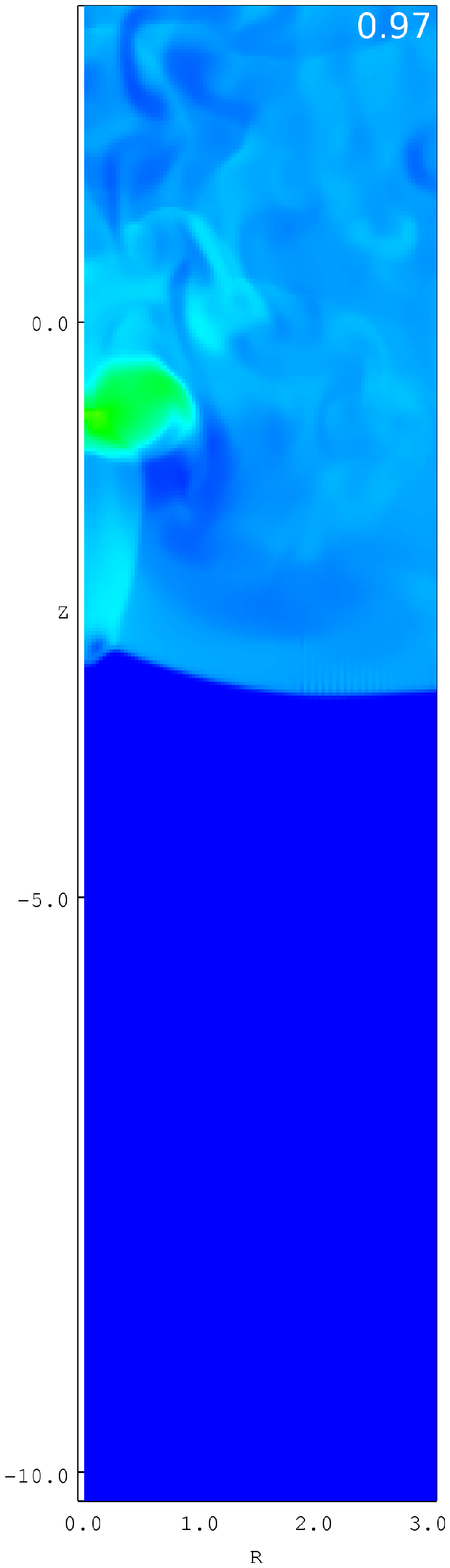,width=3.0cm}
\psfig{figure=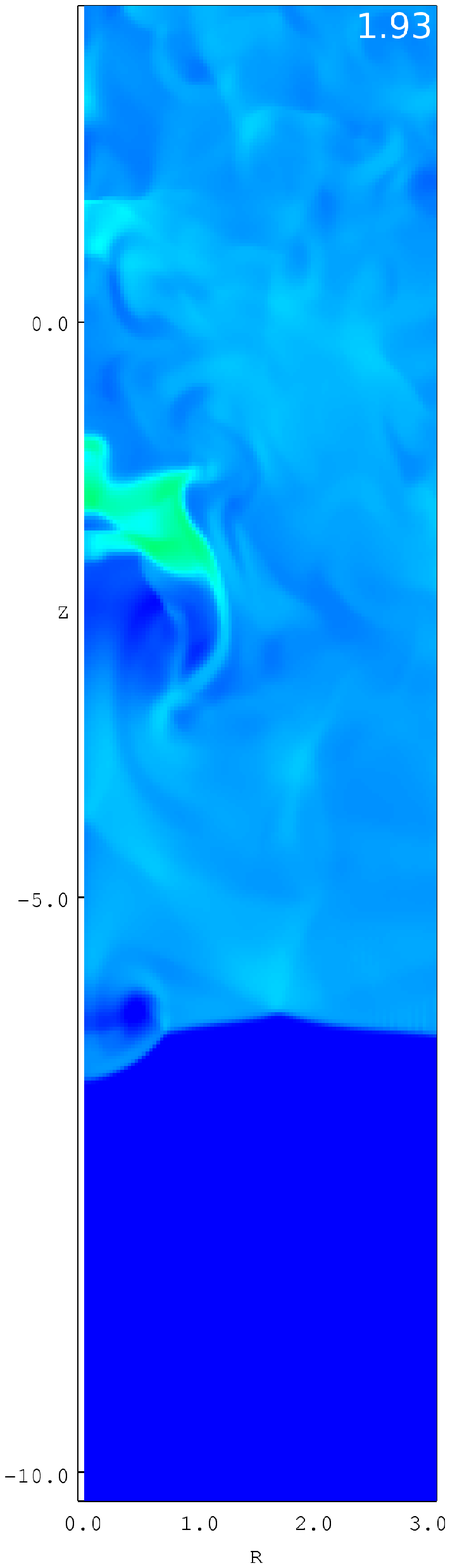,width=3.0cm}
\psfig{figure=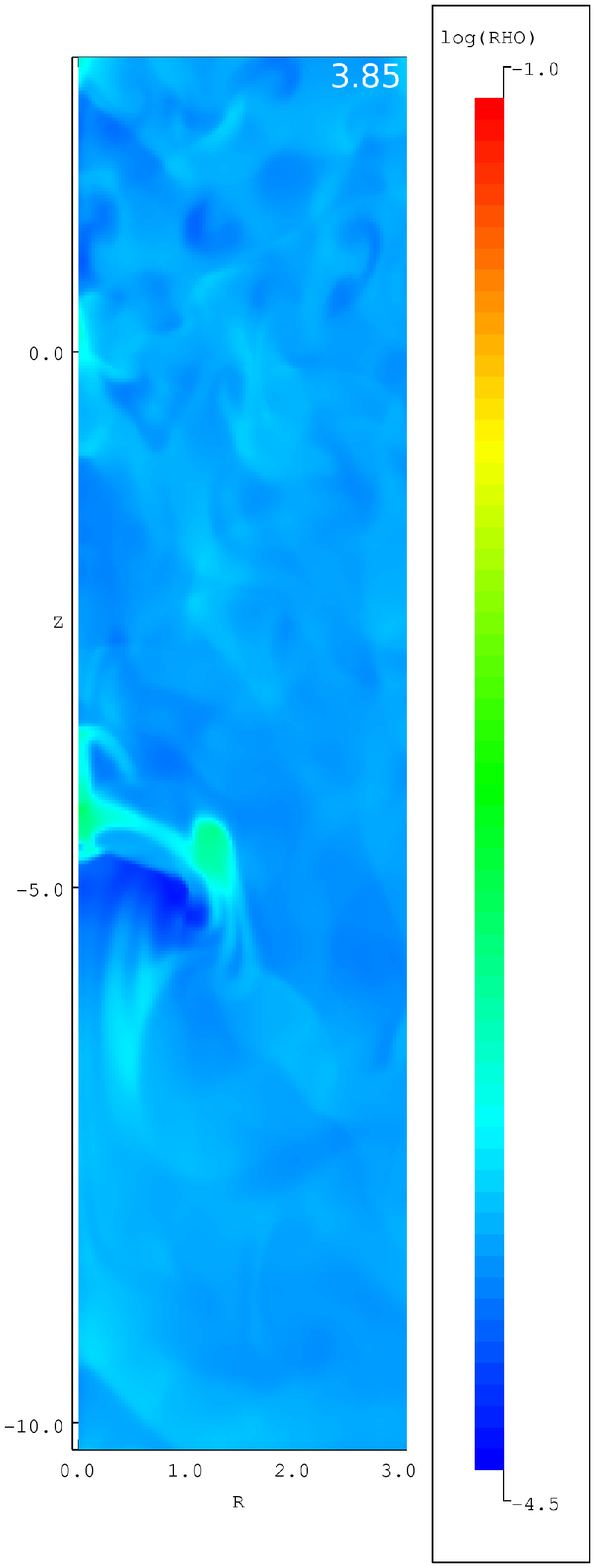,width=4.09cm}
\caption[]{Snapshots of the density distribution from an inviscid
calculation of a Mach 10 adiabatic shock hitting a cloud with a
density contrast of $10$ with respect to the ambient medium and
with grid-scale post-shock turbulence (model c1rt32)  The
evolution proceeds left to right with $t = 0.0, 0.49, 0.97, 1.93$,
and $3.85\;t_{\rm cc}$.}
\label{fig:rt}
\end{figure*}

Clouds with $\chi=10^{3}$ show different behaviour again
(Fig.~\ref{fig:kepscomp3}a-e).  In models c3no128 and c3lo128,
$a_{\rm cloud}$ and $a_{\rm core}$ are comparable to the initial cloud
radius until a considerable time into the evolution ($t \ltsimm
6\;t_{\rm cc}$). $a_{\rm cloud}$ eventually reaches values of 
$\approx 4$ and $6\;r_{\rm c}$ in models c3no128 and c3lo128, respectively.
In contrast, the core and the more extended cloud grow much more
rapidly with time in model c3hi128, due to the enhanced transport
coefficients in this case, and the lateral extent of the cloud
eventually exceeds 10 times the initial cloud radius. However, the
faster dispersal of the ablated material into the cloud's surroundings
does not last, and eventually the growth rate of $c_{\rm cloud}$ and
$c_{\rm core}$ slows below that from models c3no128 and c3no128 as
efficient mixing with the surrounding flow prevents the most dispersed
material from being identified as material from the original cloud.
For this reason, the value of $c_{\rm cloud}$ eventually falls below
the values obtained in the inviscid and ``low $k$-$\epsilon$'' models
for each of the cloud density contrasts considered (this change occurs
at $4.5 \ltsimm t/t_{\rm cc} \ltsimm 7$).  This process of mixing also
limits the maximum value of $c_{\rm core}$ obtained in the
simulations.

Fig.~\ref{fig:kepscomp3sh}a-e) shows the evolution of the cloud shape
for a much shallower initial density profile of the cloud but with the
same $\chi=10^{3}$ as for Fig.~\ref{fig:kepscomp3}.  The behaviour of
the cloud shape is generally fairly similar to the models of the
sharper-edged cloud (see Fig.~\ref{fig:kepscomp3}).

Figs.~\ref{fig:kepscomp1}-\ref{fig:kepscomp3sh} and
Table~\ref{tab:results} show that the values of $c_{\rm cloud}/a_{\rm
cloud}$ and $c_{\rm core}/a_{\rm core}$ generally increase with $\chi$
and decrease with increasing levels of turbulence in the model. Since
clouds with a high density contrast survive a long time, $c_{\rm
cloud}/a_{\rm cloud}$ can exceed 20 when $\chi=10^{3}$.

Figs.~\ref{fig:kepscomp1}a),~\ref{fig:kepscomp2}a)
and~\ref{fig:kepscomp3}a) also show the predicted lateral expansion of
the cloud determined from Eq.~6.1 of \citet{Klein:1994}. This equation
is only appropriate for $t \ltsimm t_{\rm m}$ (i.e. until the lateral
expansion of the cloud reaches a maximum). A mismatch at early times
is also expected, since this expression does not properly account for
the initial compression stage. With these caveats in mind, the
predicted rate of expansion appears to be slightly too fast compared
to the model results with $\chi=10$ and $10^{3}$, but slower than is
obtained from the models with $\chi=10^{2}$. 
%However, the value of
%$a_{\rm cloud}$ from model c3lo128 is almost in agreement with the
%theoretical prediction over the period $7 \ltsimm t/t_{\rm cc} \ltsimm 11$. 
The most striking observation is that the lateral expansion of
the cloud in model c3hi128 is very much faster than the theoretical
prediction over the period $4 \ltsimm t/t_{\rm cc} \ltsimm 8$, which
reveals that the highly turbulent surroundings diffuse the cloud
material at a speed faster than the shocked cloud's internal sound
speed.

\begin{figure*}
%8.5cm for 2-col layout, 5.7cm for 3-col layout
\psfig{figure=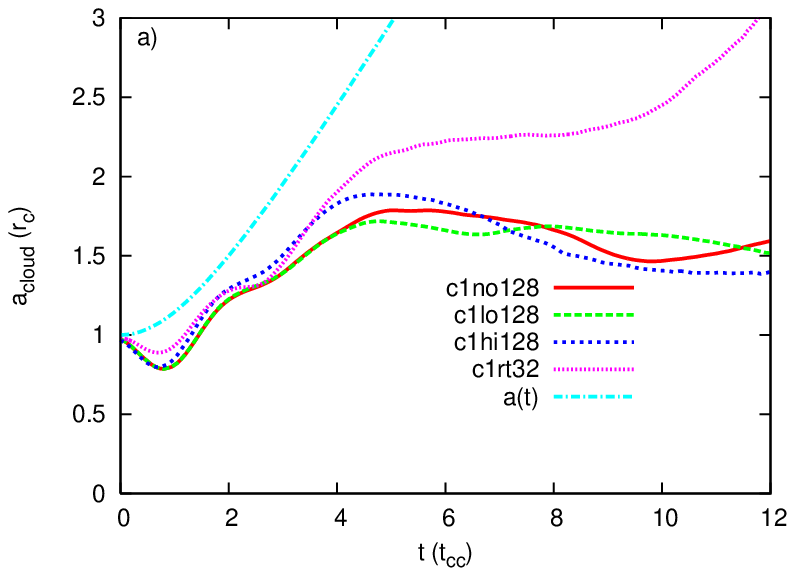,width=5.7cm}
\psfig{figure=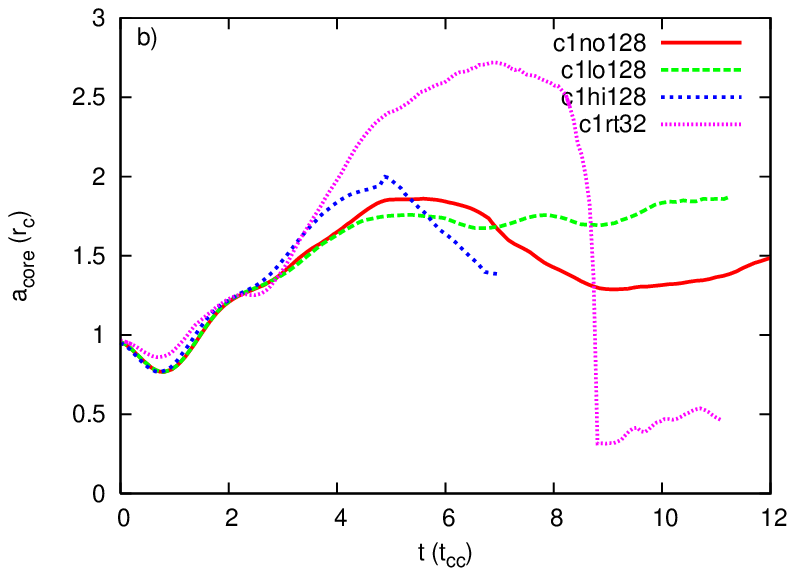,width=5.7cm}
\psfig{figure=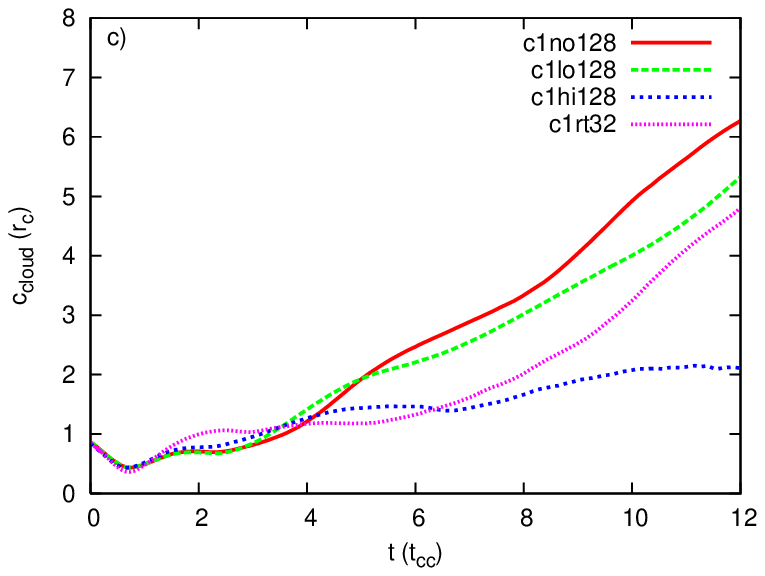,width=5.7cm}
\psfig{figure=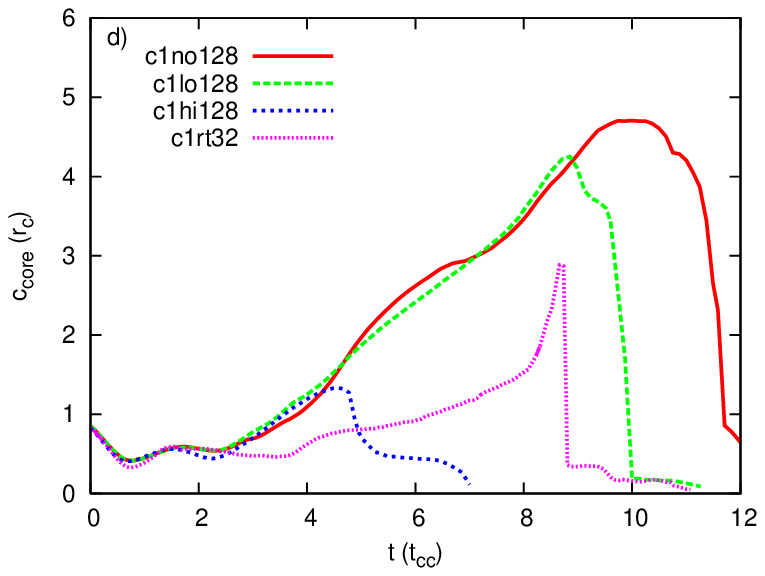,width=5.7cm}
\psfig{figure=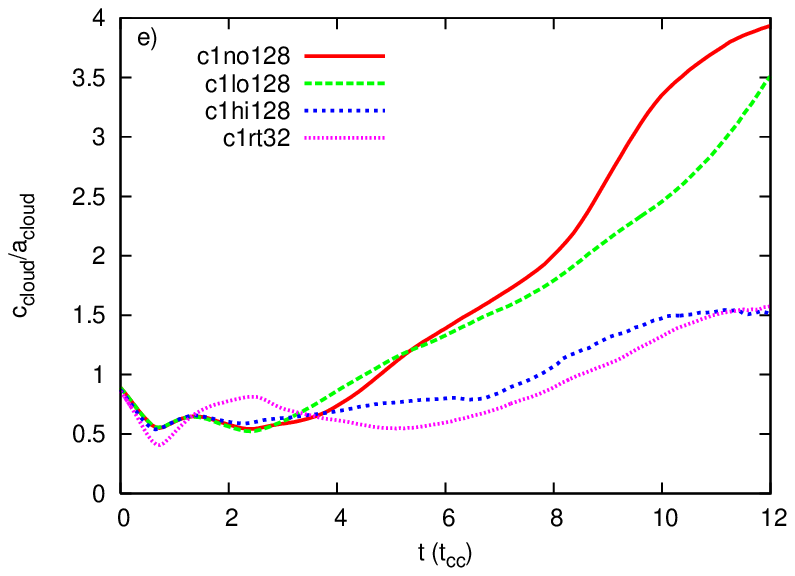,width=5.7cm}
\psfig{figure=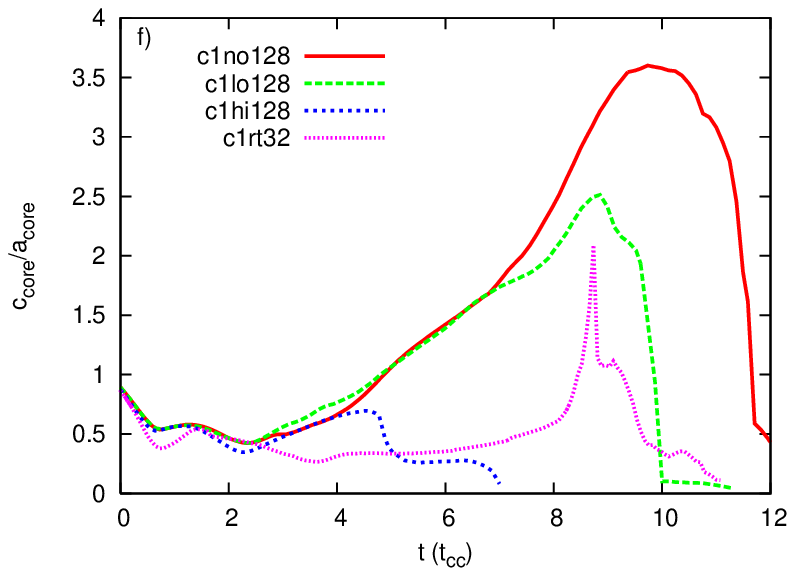,width=5.7cm}
\psfig{figure=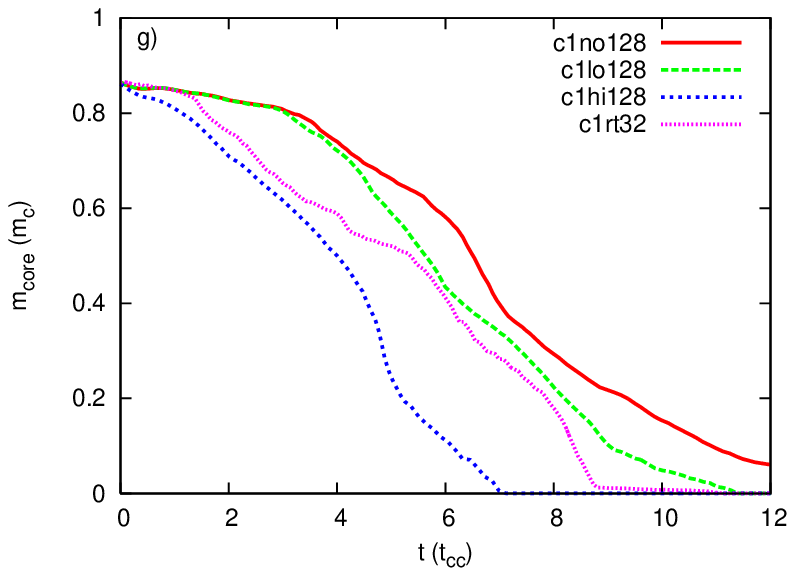,width=5.7cm}
\psfig{figure=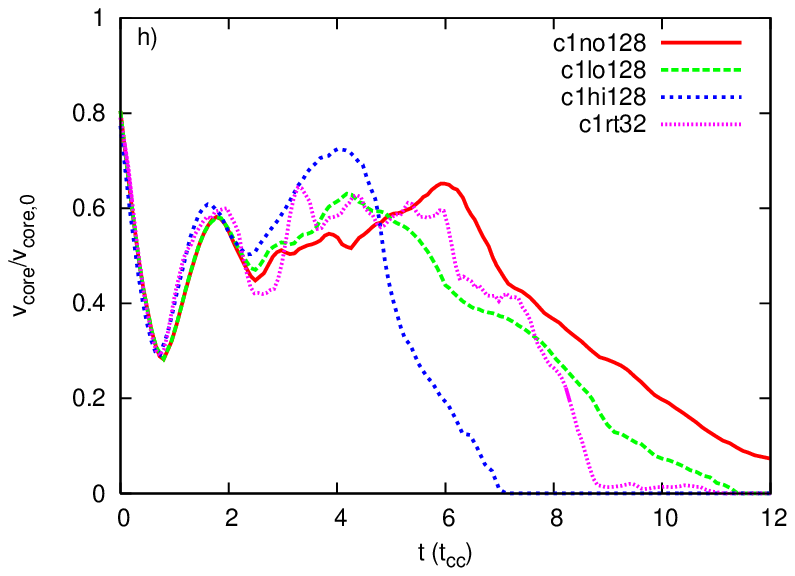,width=5.7cm}
\psfig{figure=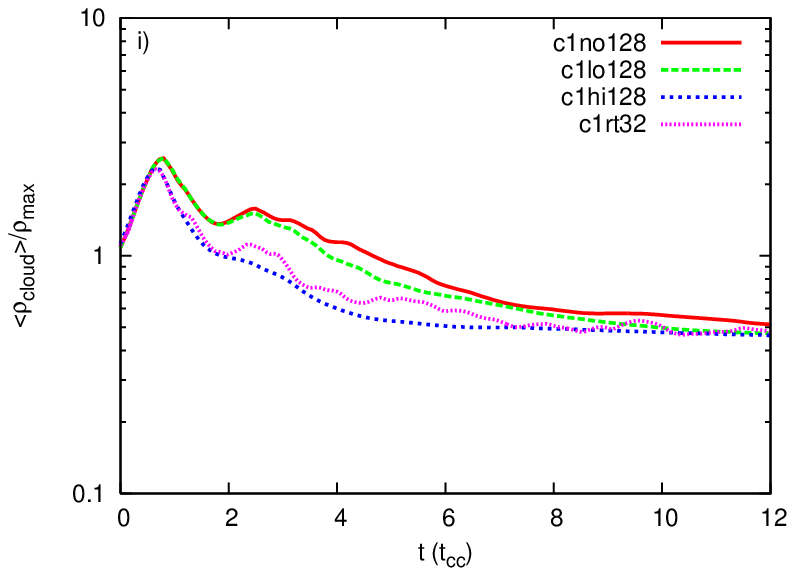,width=5.7cm}
\psfig{figure=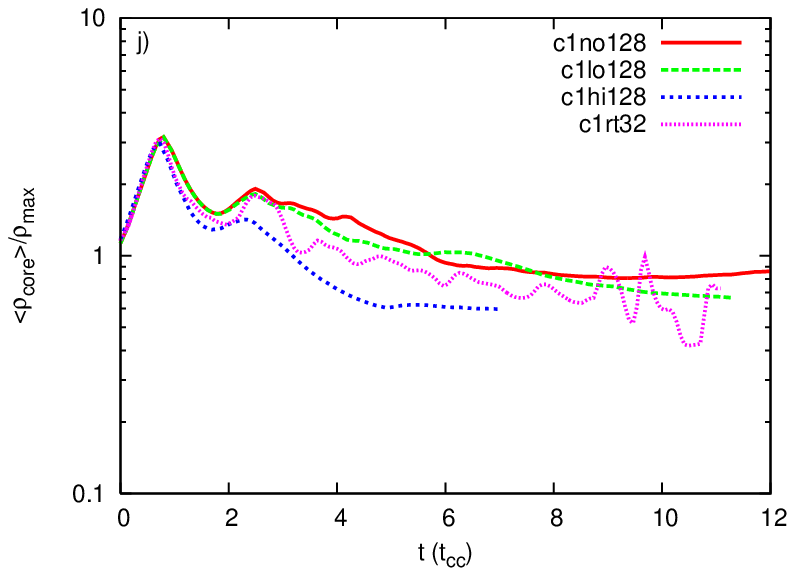,width=5.7cm}
\psfig{figure=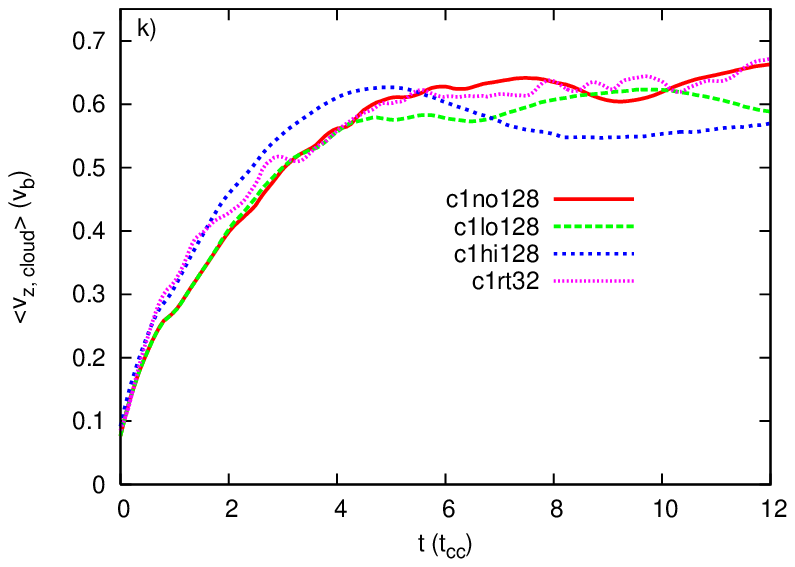,width=5.7cm}
\psfig{figure=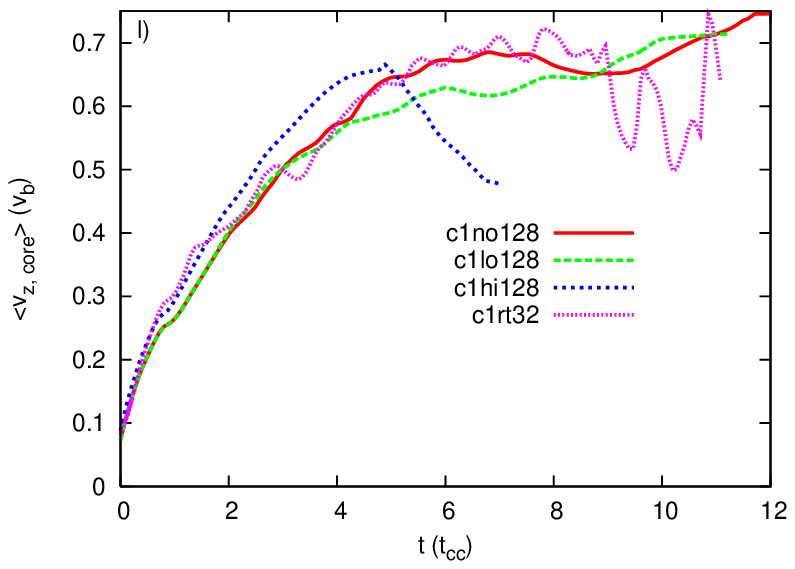,width=5.7cm}
\psfig{figure=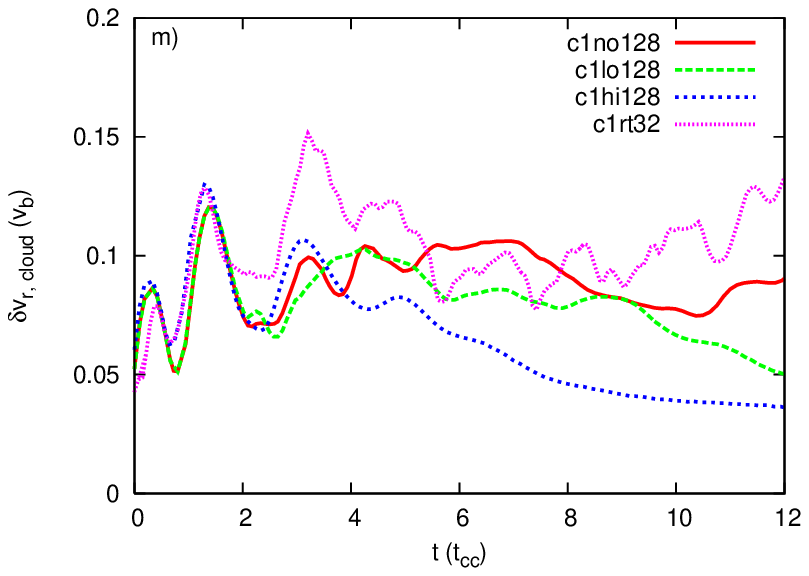,width=5.7cm}
\psfig{figure=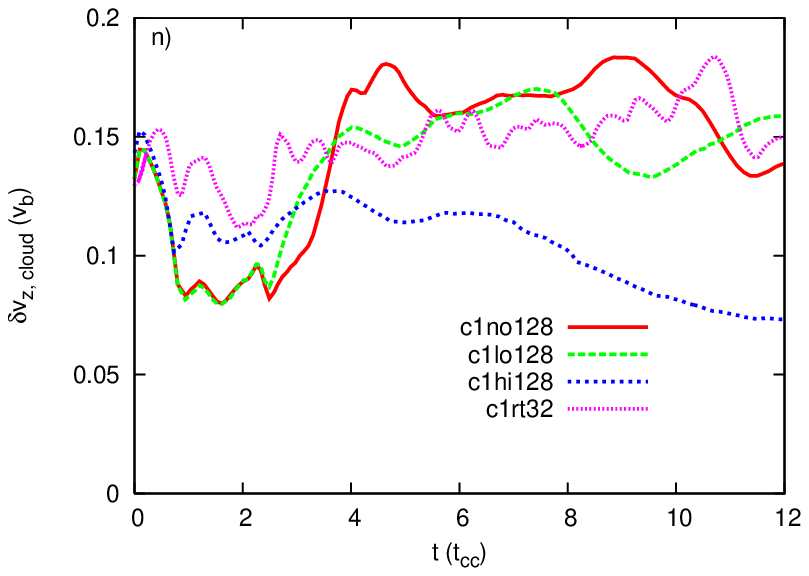,width=5.7cm}
\psfig{figure=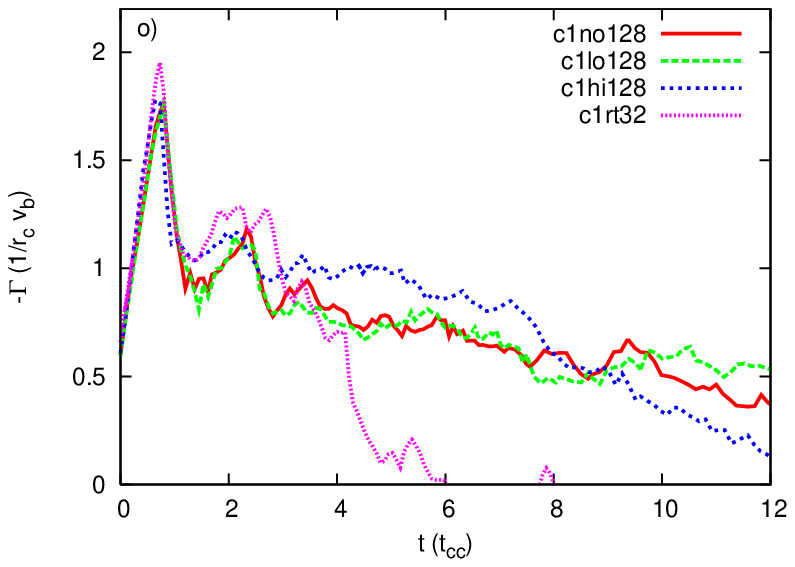,width=5.7cm}
\caption[]{Time evolution of the cloud shape, mass, volume, mean
density, mean axial velocity, velocity dispersion and circulation 
for invsicid and $k$-$\epsilon$ models with a density contrast $\chi=10$, a 
steep density gradient ($p_{1}=10$), and 128 cells per cloud
radius.}
\label{fig:kepscomp1}
\end{figure*}

\begin{figure*}
%8.5cm for 2-col layout, 5.7cm for 3-col layout
\psfig{figure=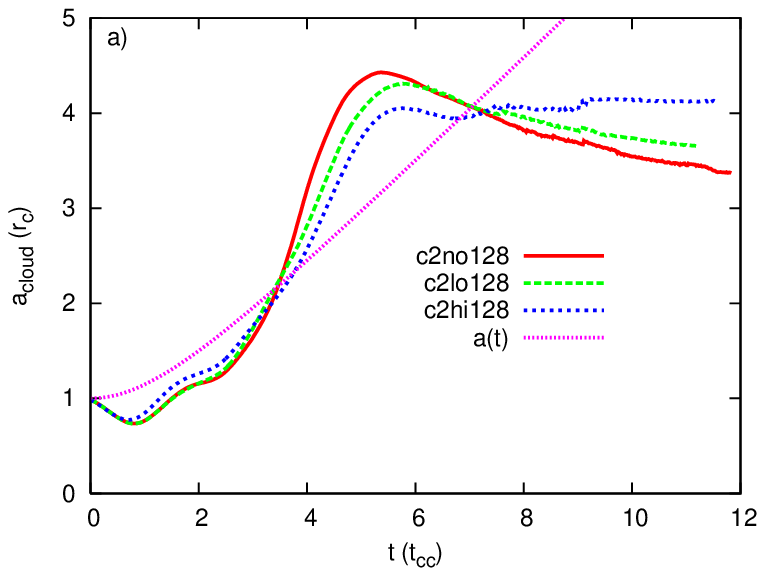,width=5.7cm}
\psfig{figure=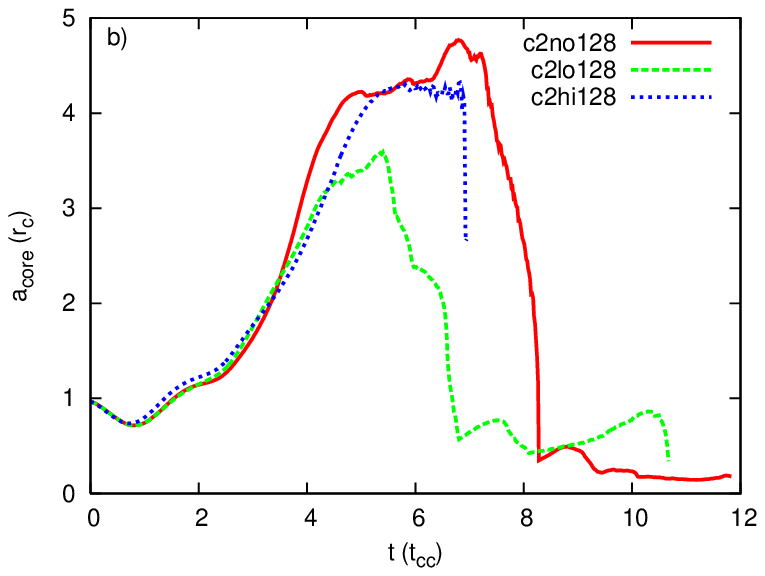,width=5.7cm}
\psfig{figure=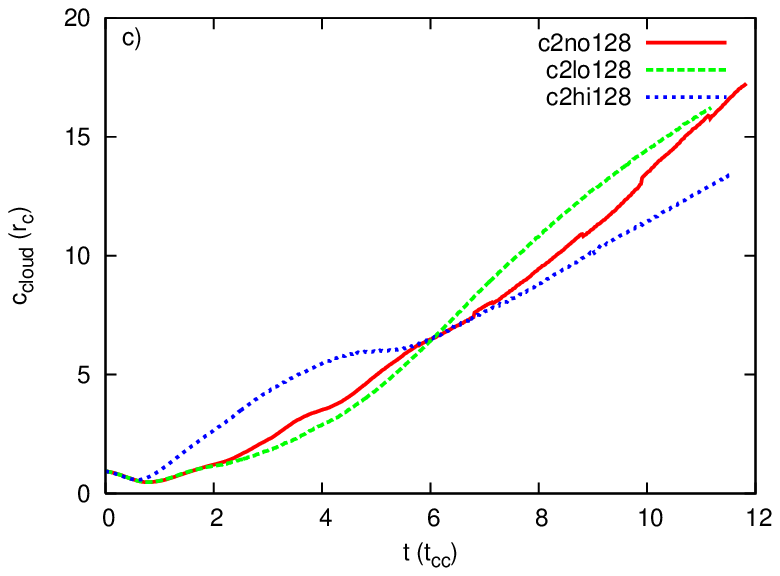,width=5.7cm}
\psfig{figure=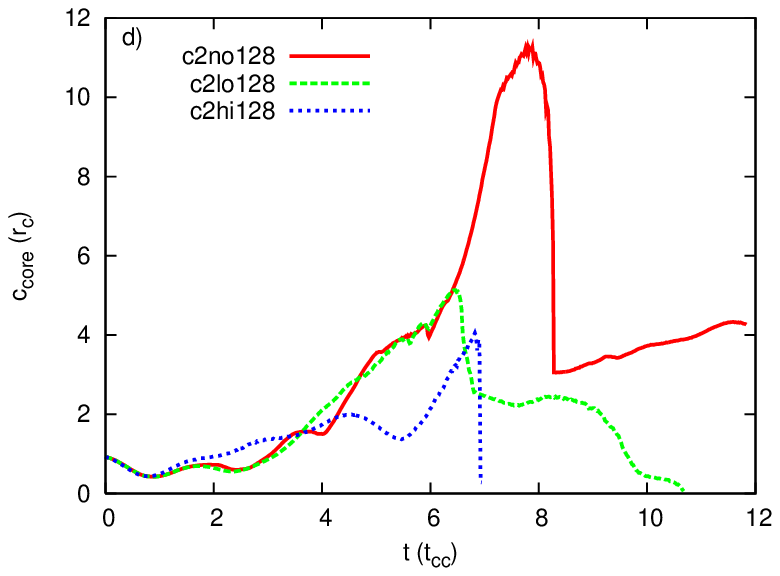,width=5.7cm}
\psfig{figure=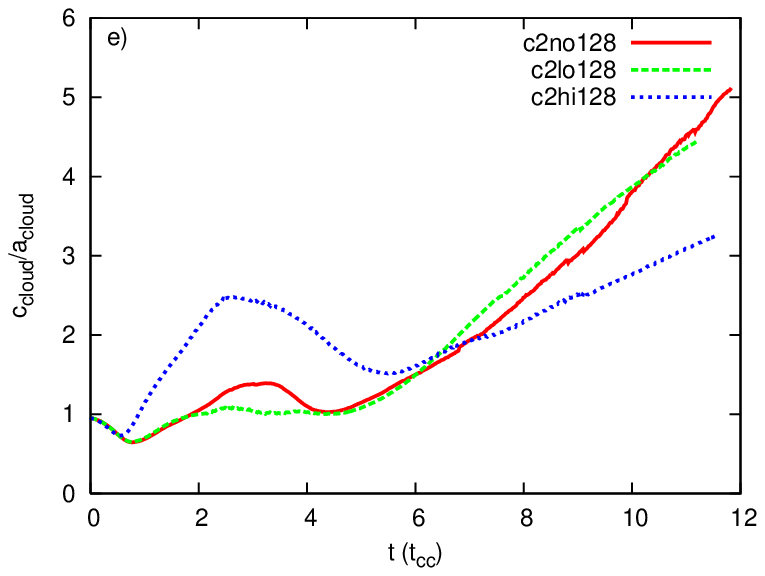,width=5.7cm}
\psfig{figure=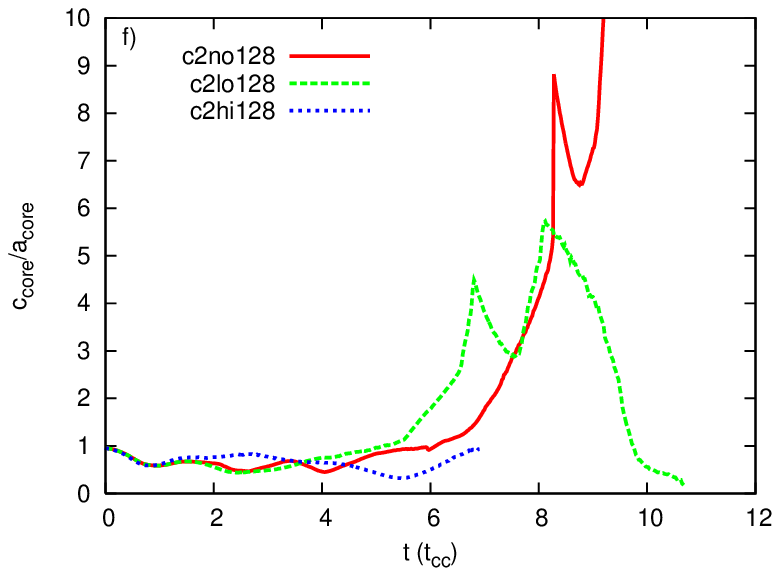,width=5.7cm}
\psfig{figure=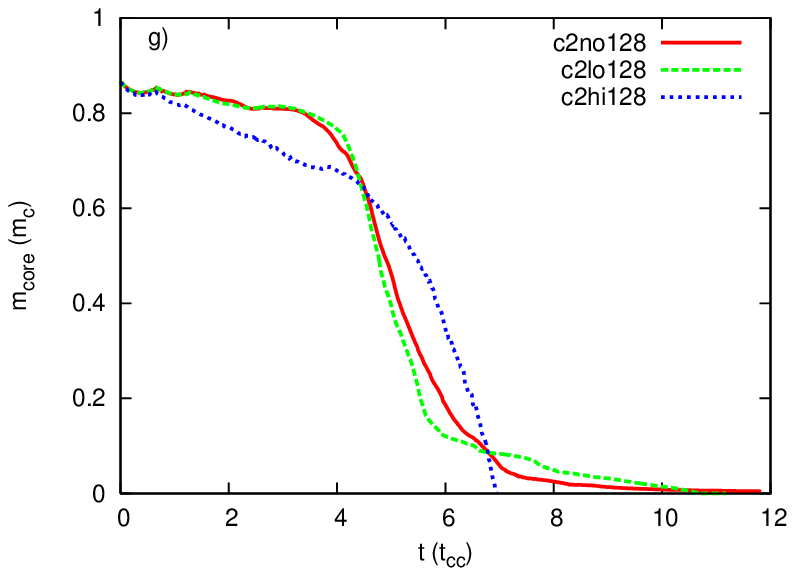,width=5.7cm}
\psfig{figure=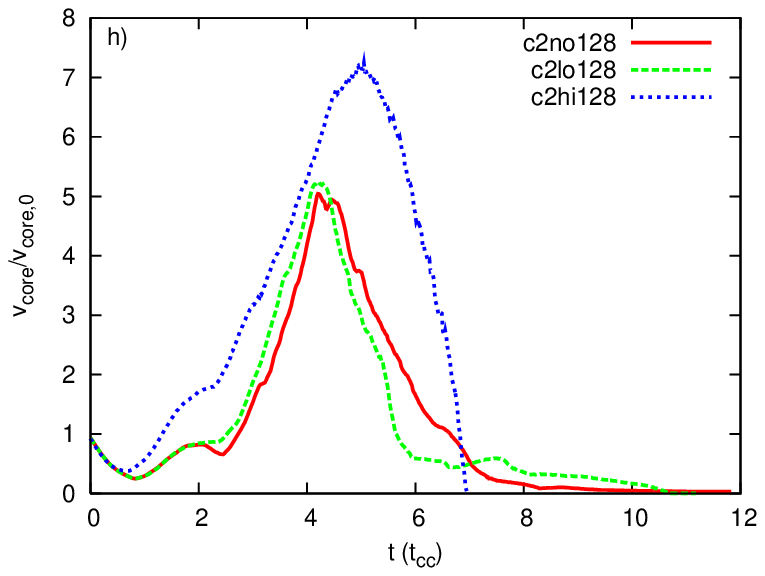,width=5.7cm}
\psfig{figure=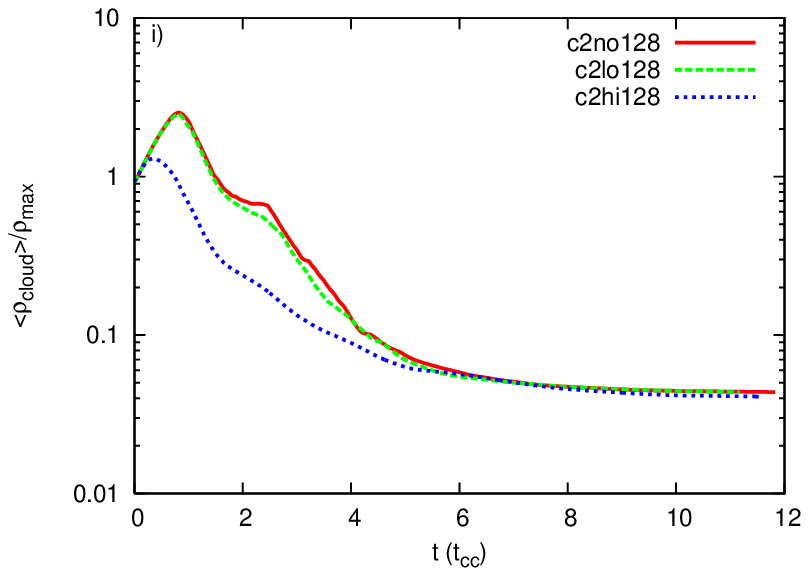,width=5.7cm}
\psfig{figure=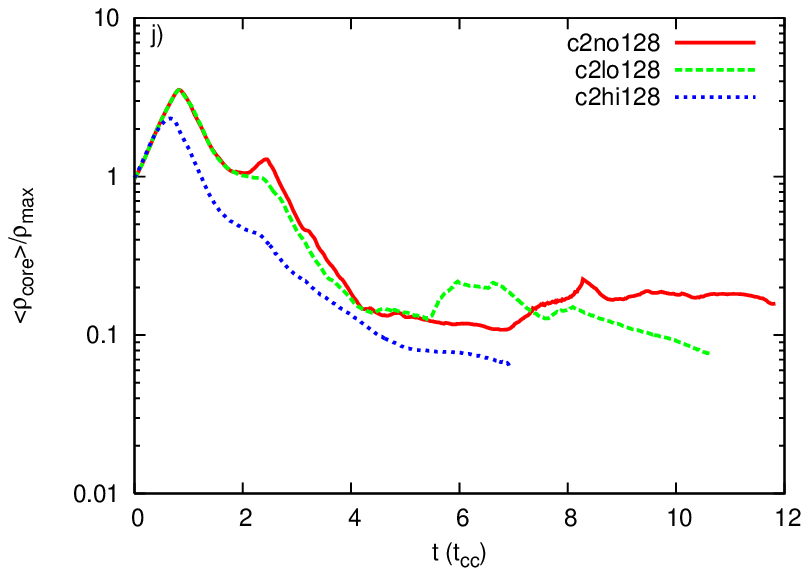,width=5.7cm}
\psfig{figure=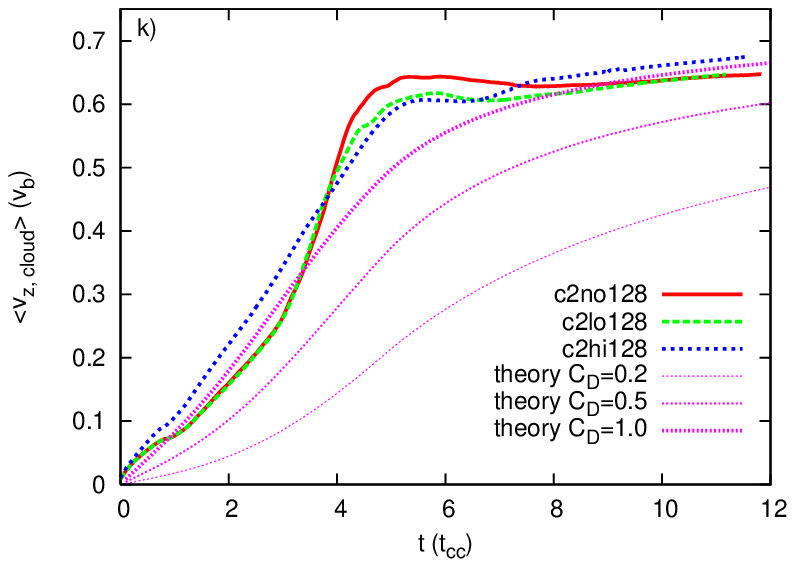,width=5.7cm}
\psfig{figure=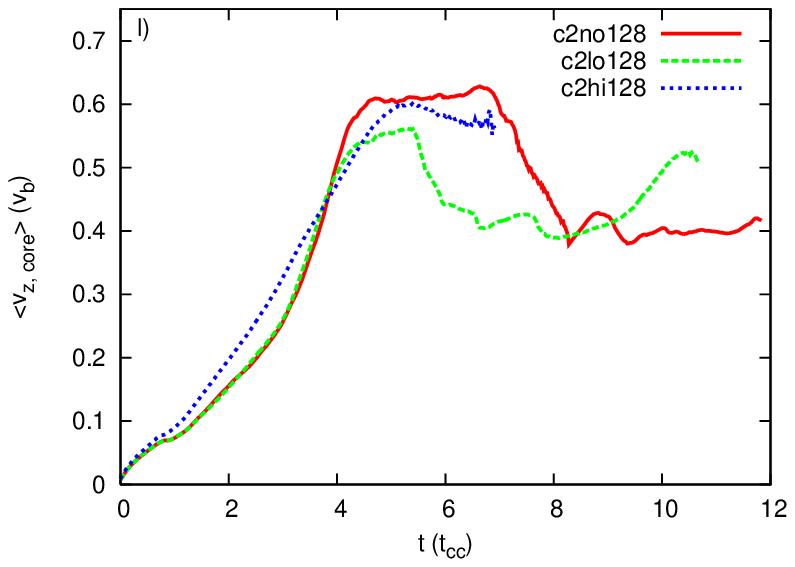,width=5.7cm}
\psfig{figure=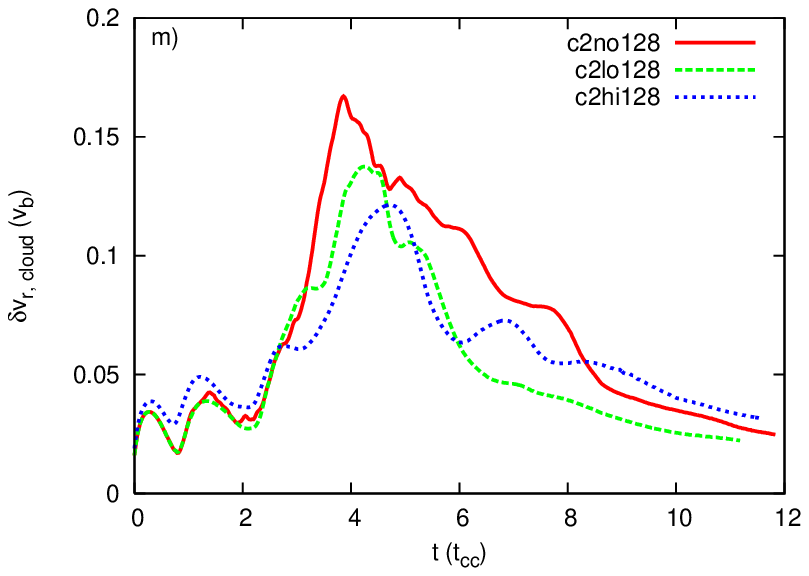,width=5.7cm}
\psfig{figure=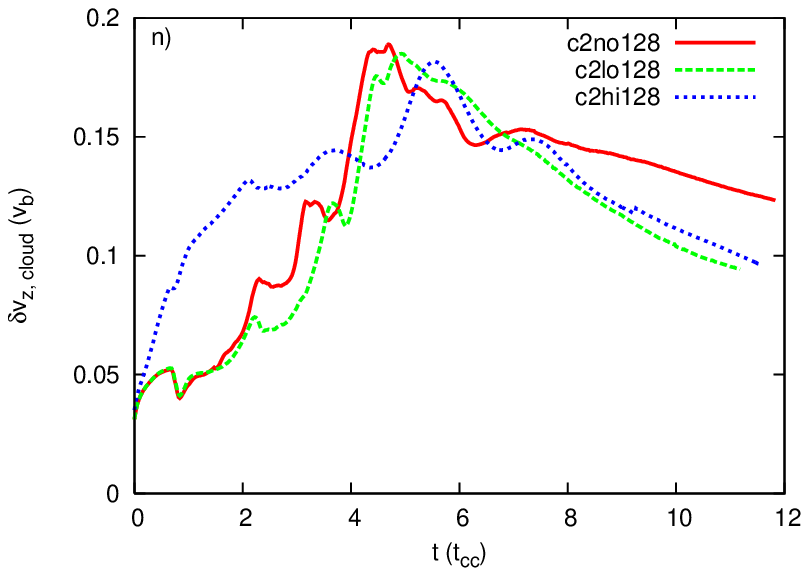,width=5.7cm}
\psfig{figure=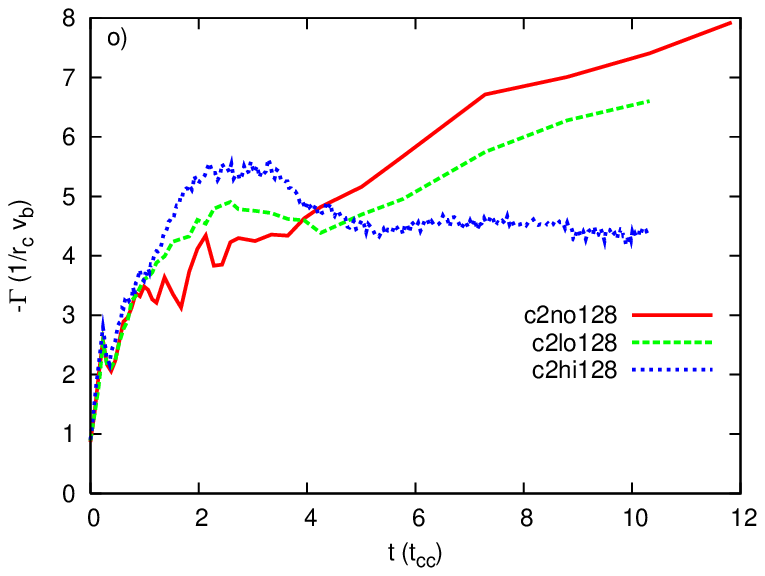,width=5.7cm}
\caption[]{As Fig.~\ref{fig:kepscomp1} but for $\chi=10^{2}$.}
\label{fig:kepscomp2}
\end{figure*}

\begin{figure*}
\psfig{figure=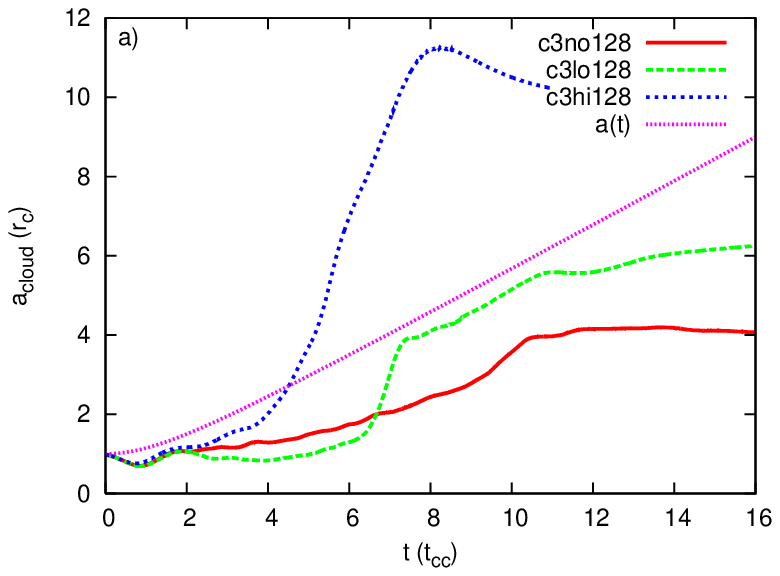,width=5.7cm}
\psfig{figure=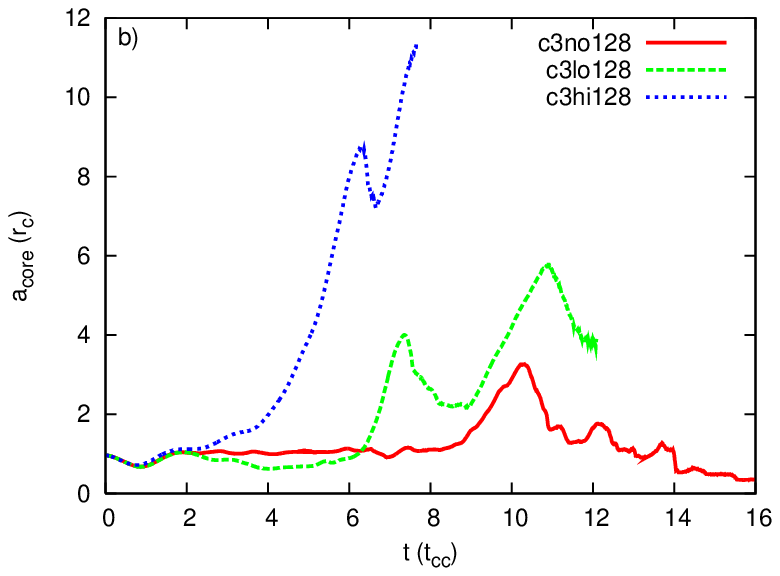,width=5.7cm}
\psfig{figure=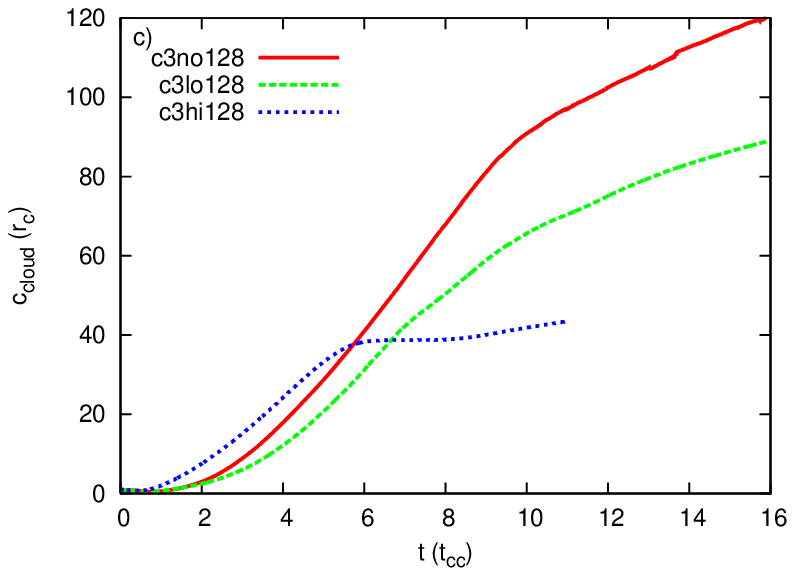,width=5.7cm}
\psfig{figure=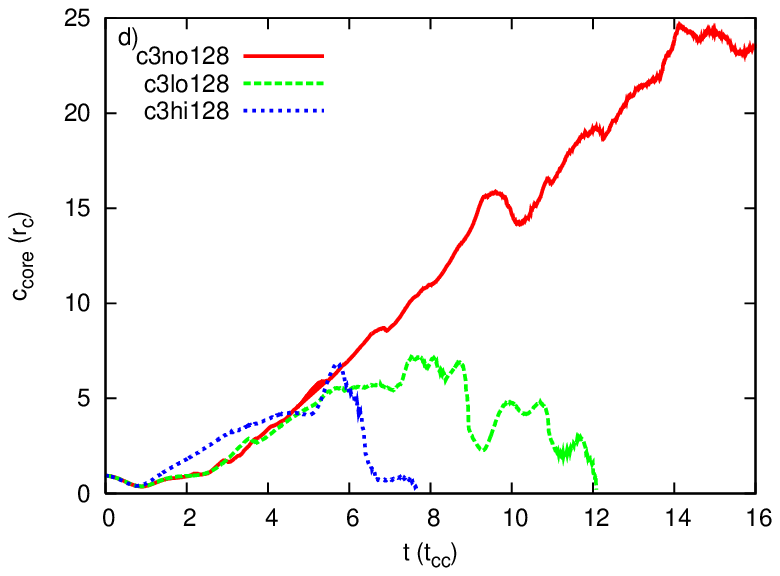,width=5.7cm}
\psfig{figure=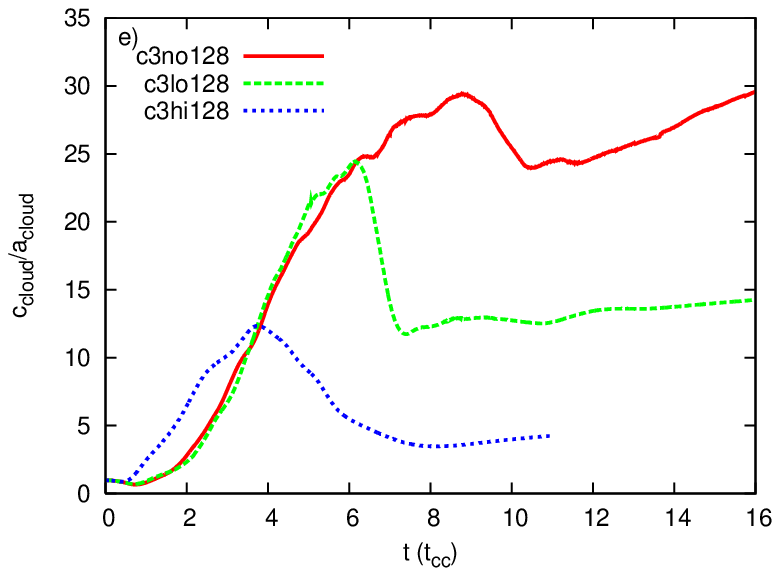,width=5.7cm}
\psfig{figure=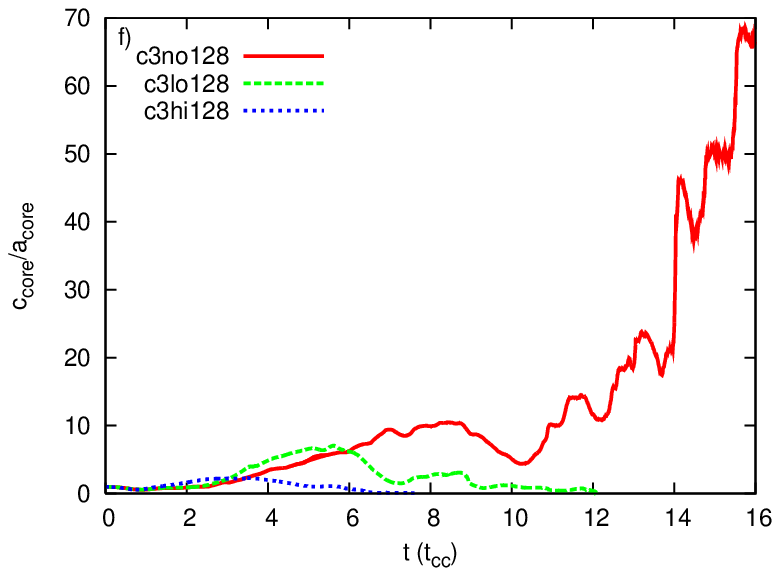,width=5.7cm}
\psfig{figure=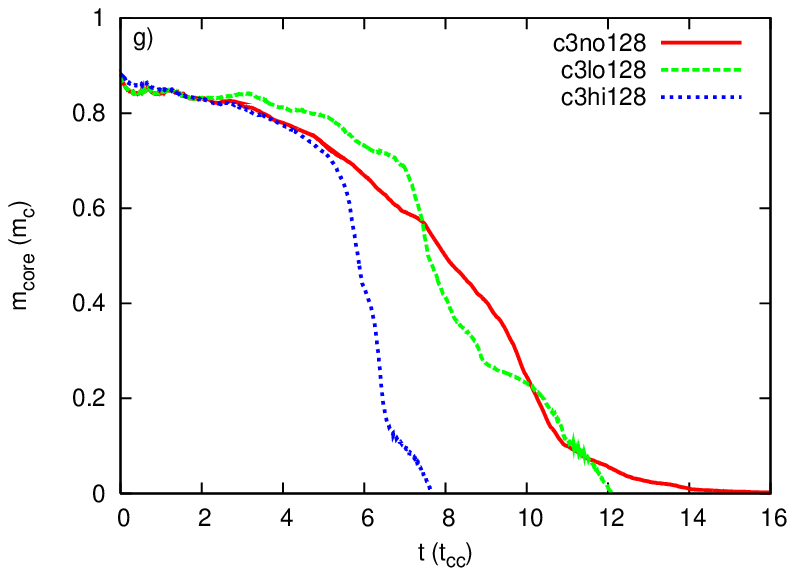,width=5.7cm}
\psfig{figure=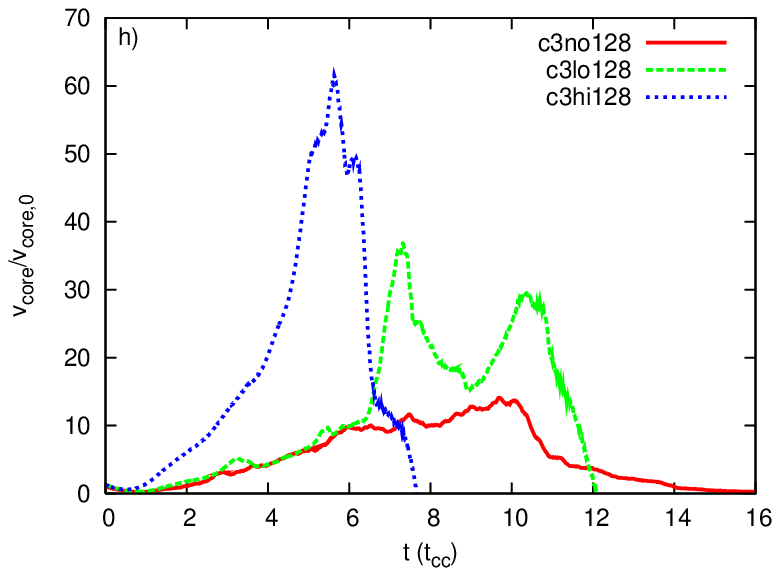,width=5.7cm}
\psfig{figure=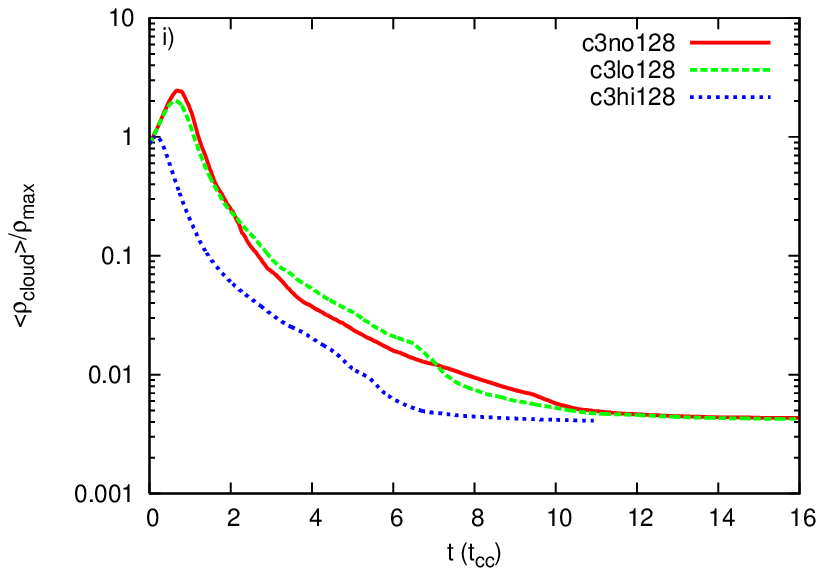,width=5.7cm}
\psfig{figure=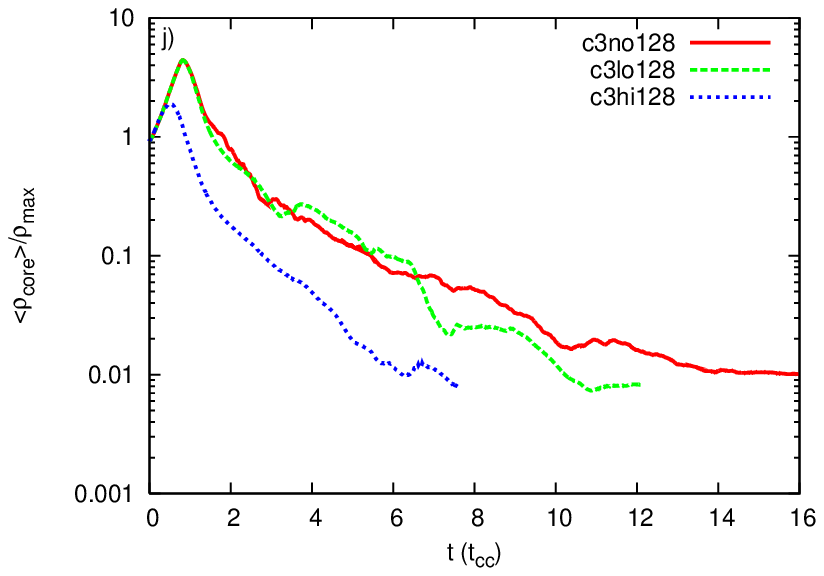,width=5.7cm}
\psfig{figure=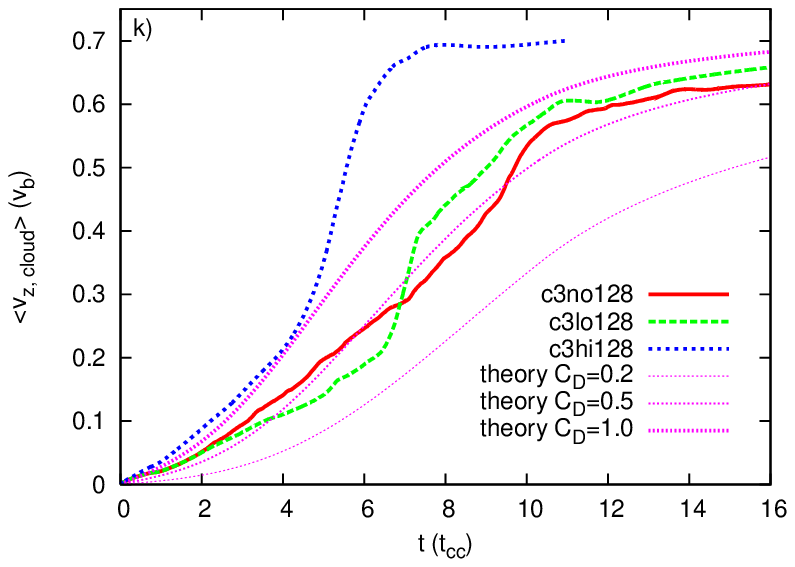,width=5.7cm}
\psfig{figure=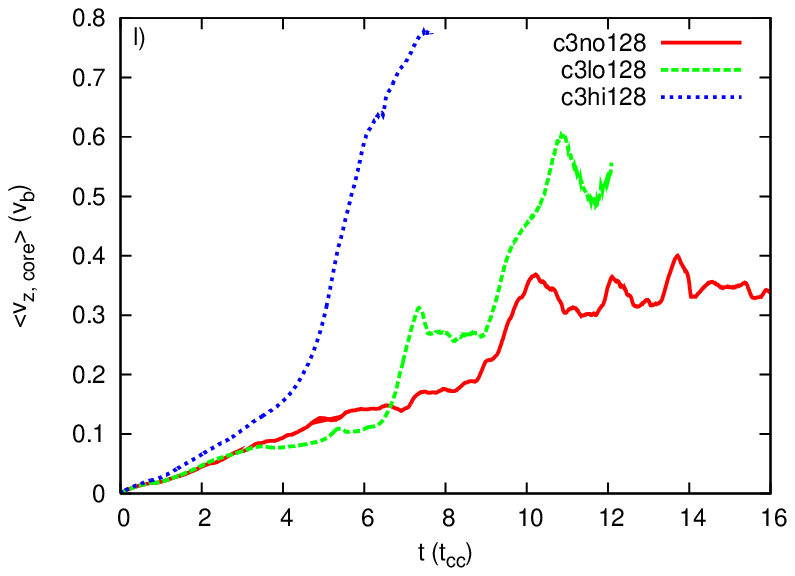,width=5.7cm}
\psfig{figure=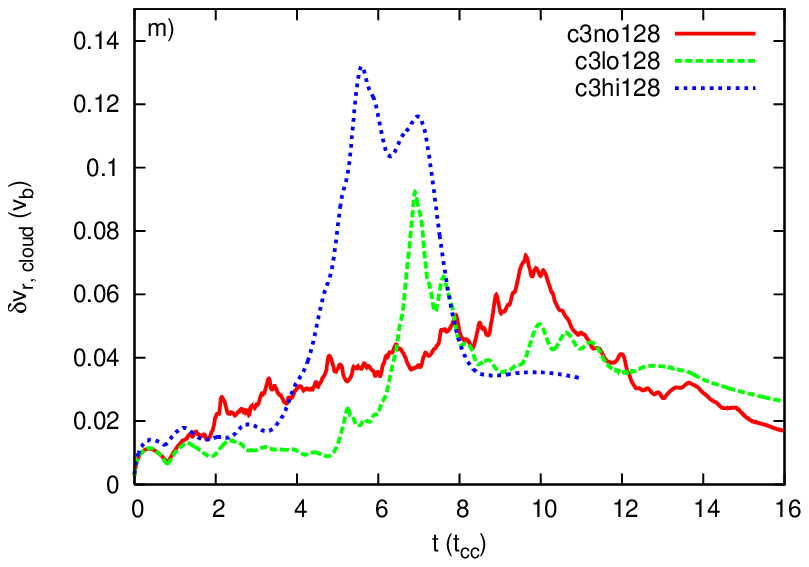,width=5.7cm}
\psfig{figure=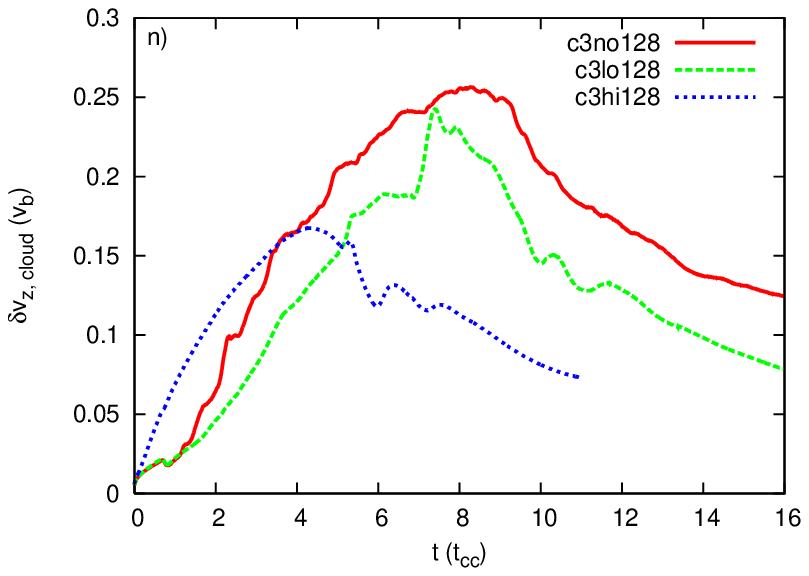,width=5.7cm}
\psfig{figure=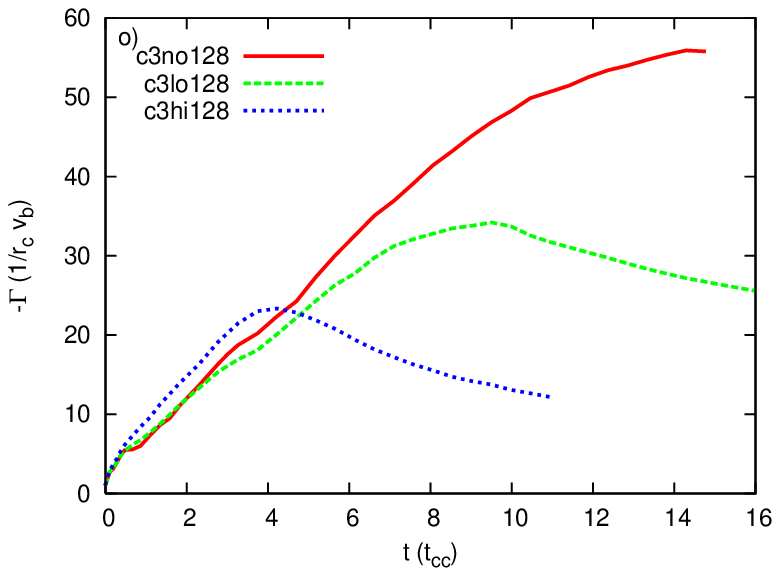,width=5.7cm}
\caption[]{As Fig.~\ref{fig:kepscomp1} but for $\chi=10^{3}$.}
\label{fig:kepscomp3}
\end{figure*}

\begin{figure*}
\psfig{figure=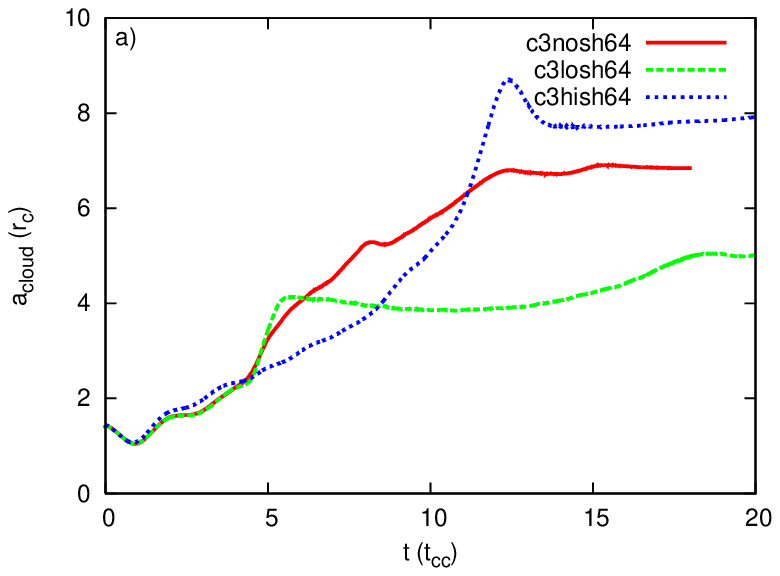,width=5.7cm}
\psfig{figure=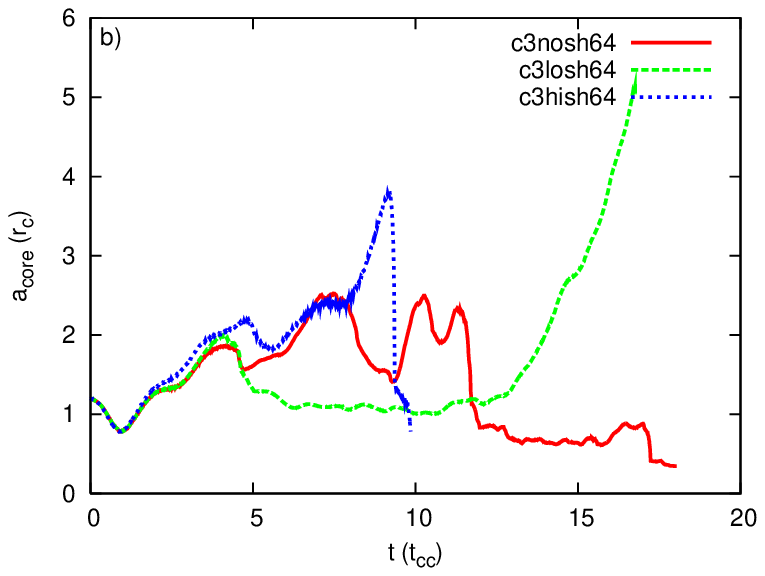,width=5.7cm}
\psfig{figure=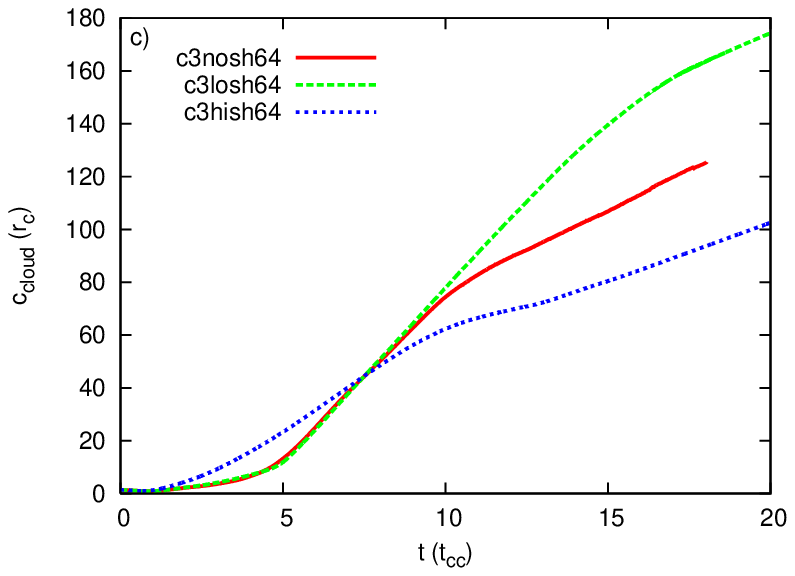,width=5.7cm}
\psfig{figure=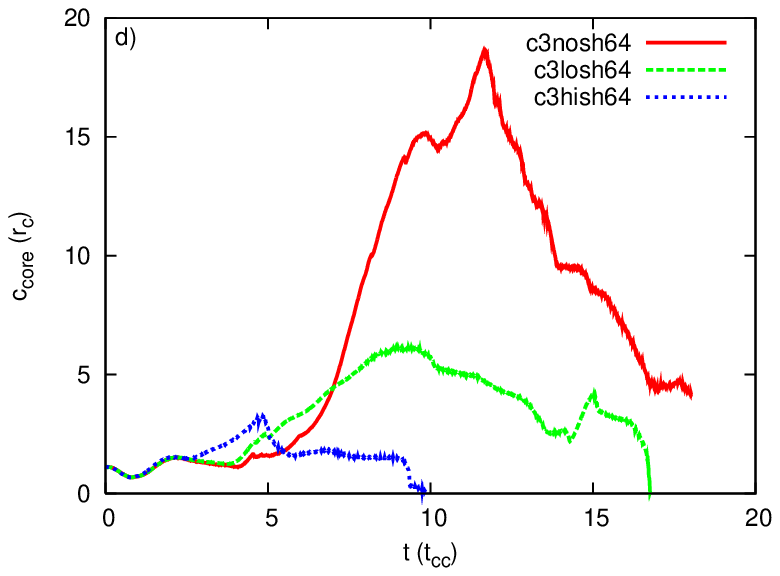,width=5.7cm}
\psfig{figure=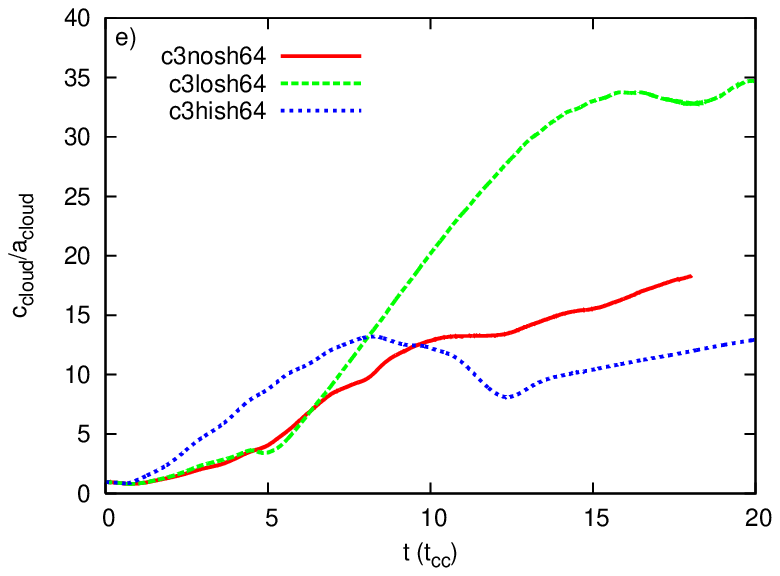,width=5.7cm}
\psfig{figure=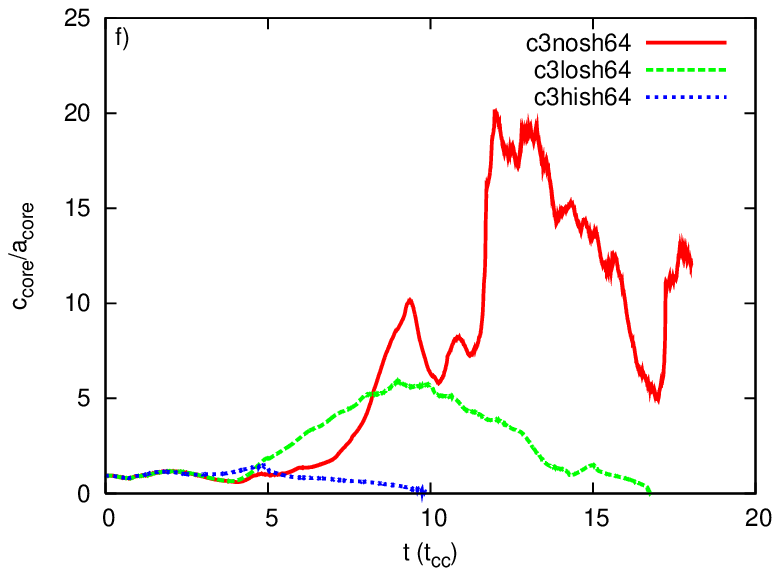,width=5.7cm}
\psfig{figure=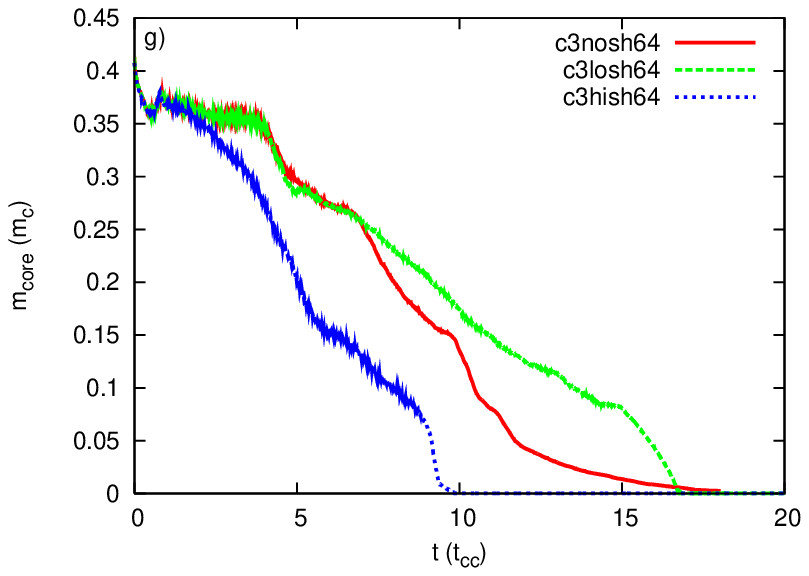,width=5.7cm}
\psfig{figure=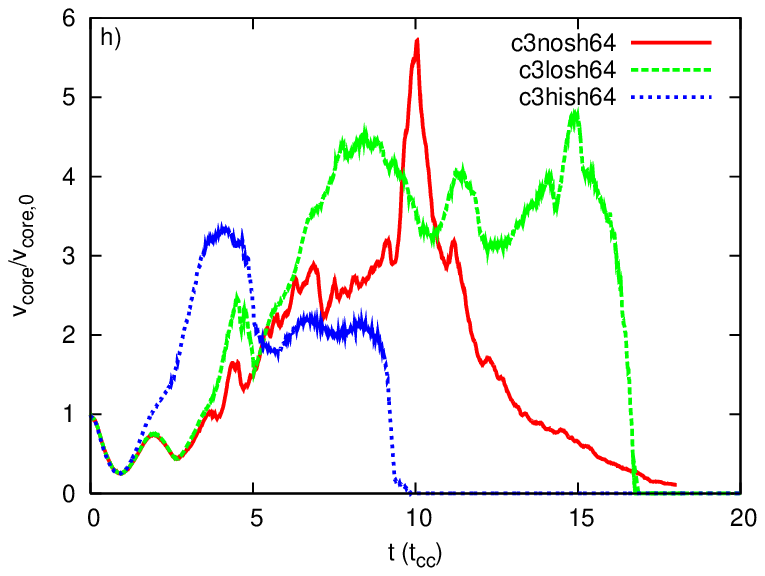,width=5.7cm}
\psfig{figure=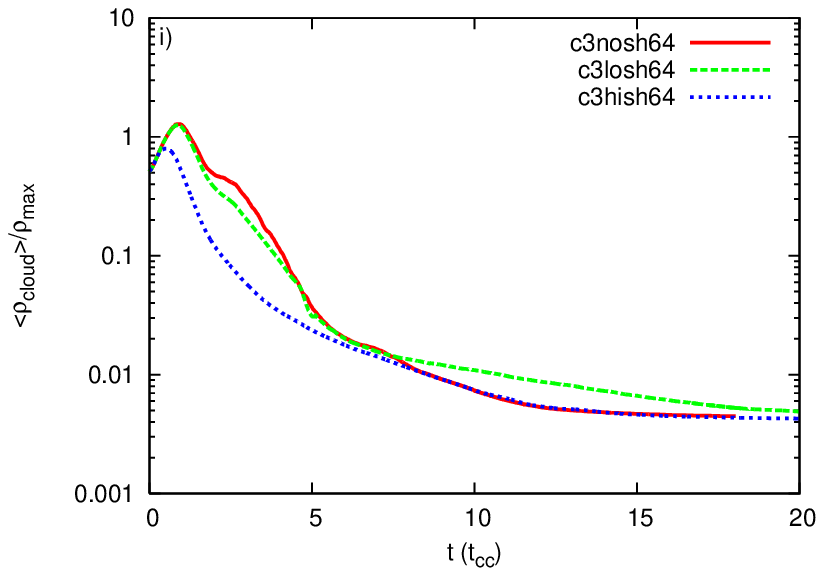,width=5.7cm}
\psfig{figure=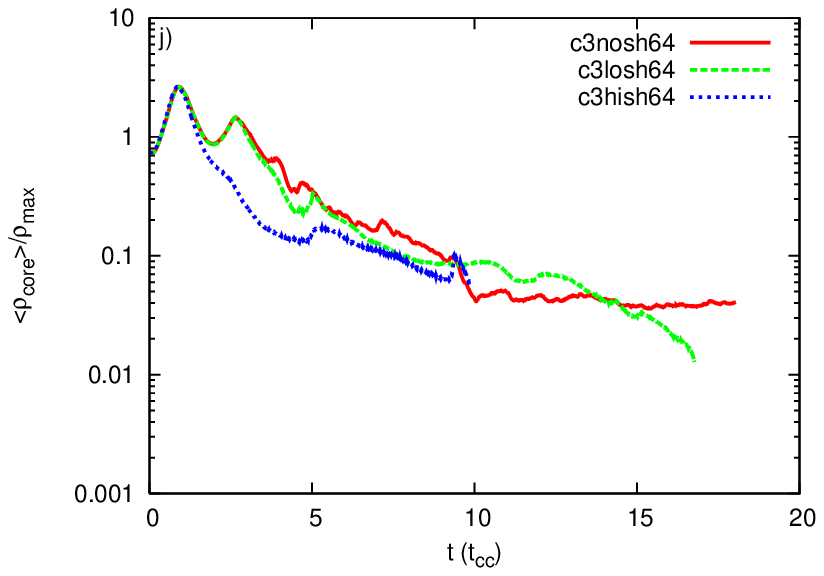,width=5.7cm}
\psfig{figure=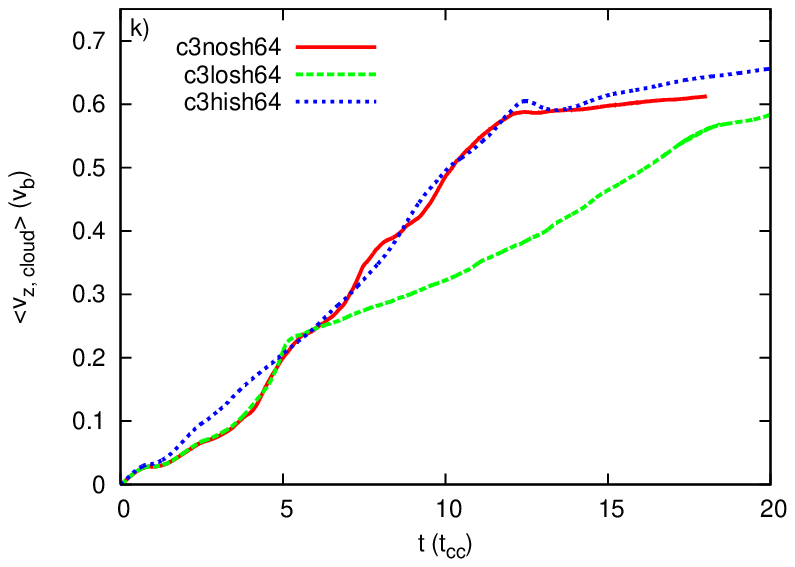,width=5.7cm}
\psfig{figure=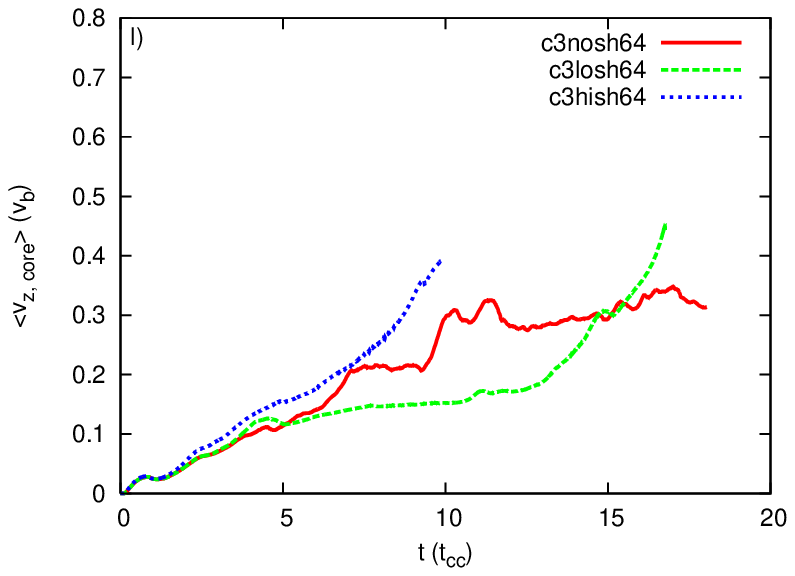,width=5.7cm}
\psfig{figure=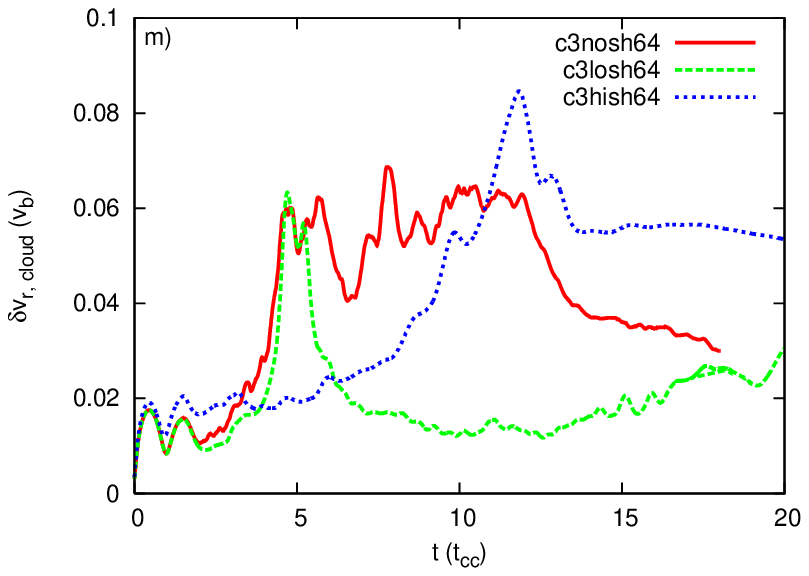,width=5.7cm}
\psfig{figure=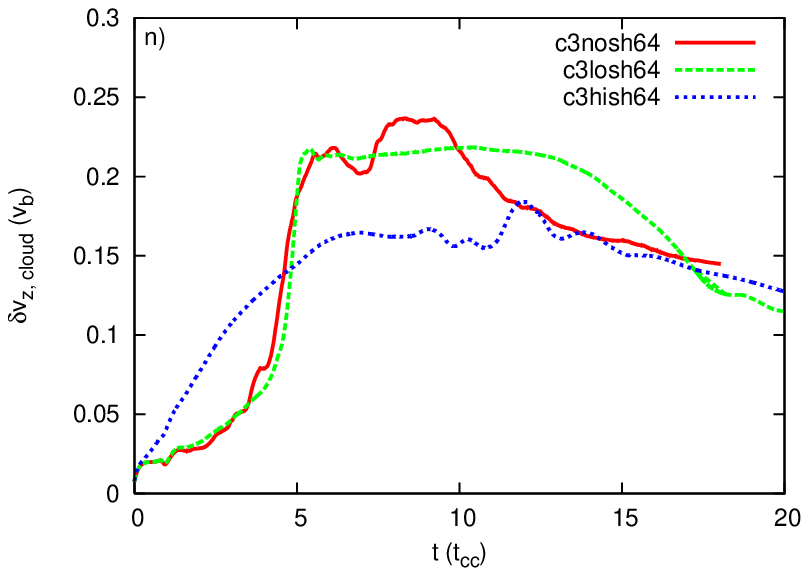,width=5.7cm}
\psfig{figure=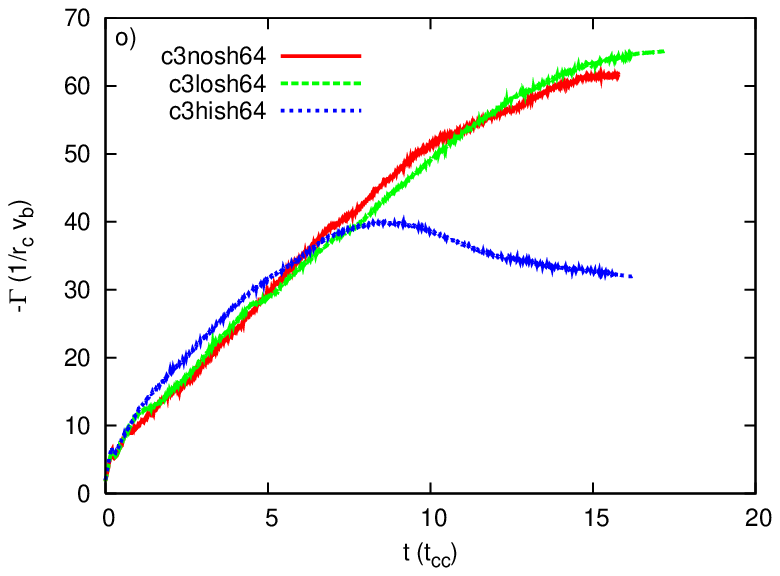,width=5.7cm}
\caption[]{As Fig.~\ref{fig:kepscomp1} but for $\chi=10^{3}$ and a shallow
density gradient ($p_{1}=1$).}
\label{fig:kepscomp3sh}
\end{figure*}

\subsubsection{Cloud mass}
Panel g) in Figs.~\ref{fig:kepscomp1}-\ref{fig:kepscomp3sh} shows the
time evolution of the core mass, which reaches half its initial value
at $t=t_{\rm mix}$. The $k$-$\epsilon$ models with low initial
turbulence show similar evolution to the inviscid models. In contrast,
the models with high initial turbulence show a much faster decline in
the core mass. 
%(although the period of most rapid decline occurs later
%in model c2hi128 than in models c2no128 and c2lo128). 
We again note that the results from the sub-grid turbulence models are
much less sensitive to spatial resolution than those from inviscid
models (Fig.~\ref{fig:rescomp}). This is also revealed in the mixing
time, $t_{\rm mix}$, which, for instance, is $8.19, 8.50$ and
$7.93\;t_{\rm cc}$ for models c3lo32, c3lo64, and c3lo128, but is
$6.81, 11.47$ and $8.69\;t_{\rm cc}$ for models c3no32, c3no64, and
c3no128 (i.e. the latter 3 models show much greater spread in
$t_{\rm mix}$).

\begin{figure*}
\psfig{figure=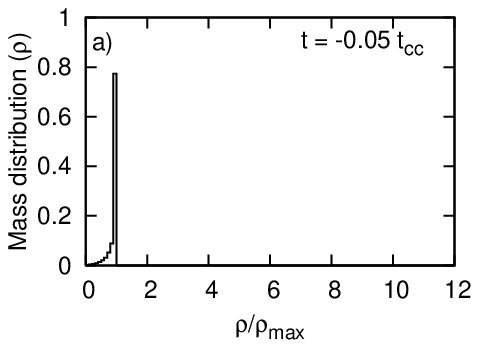,width=3.4cm}
\psfig{figure=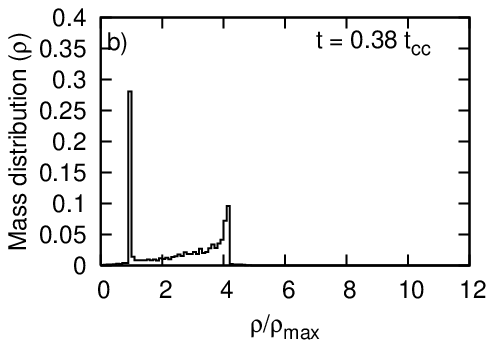,width=3.4cm}
\psfig{figure=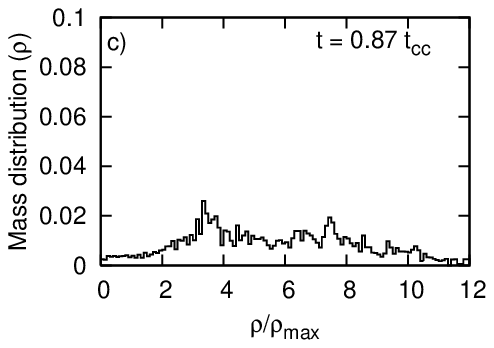,width=3.4cm}
\psfig{figure=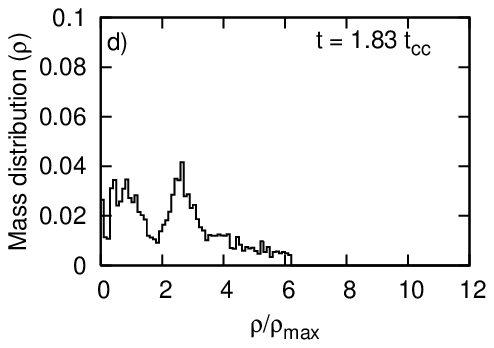,width=3.4cm}
\psfig{figure=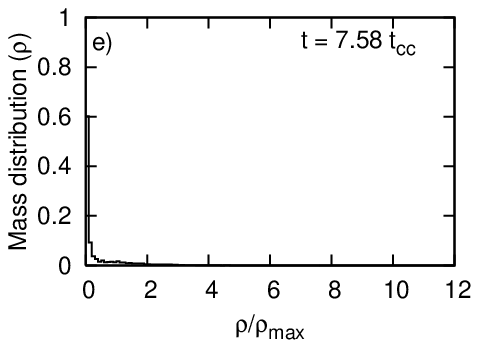,width=3.4cm}
\psfig{figure=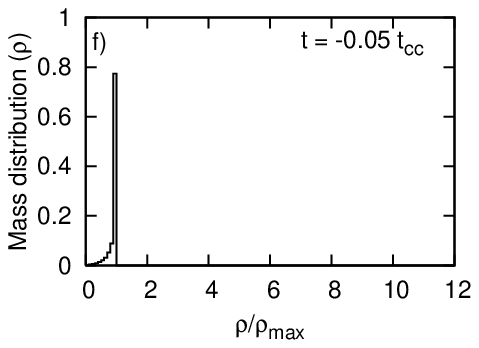,width=3.4cm}
\psfig{figure=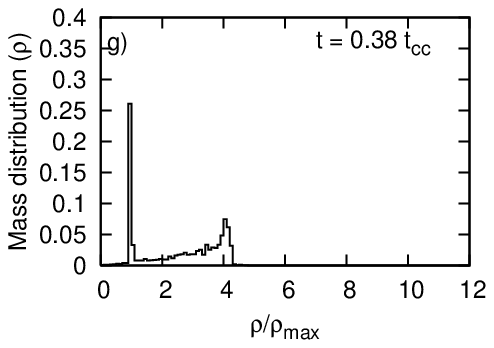,width=3.4cm}
\psfig{figure=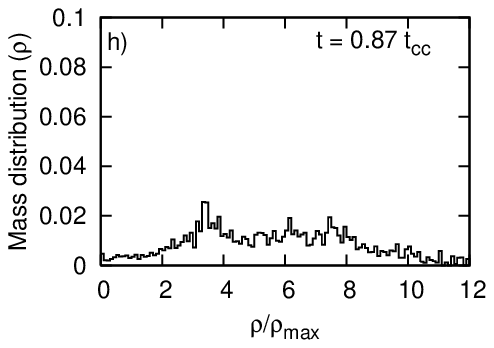,width=3.4cm}
\psfig{figure=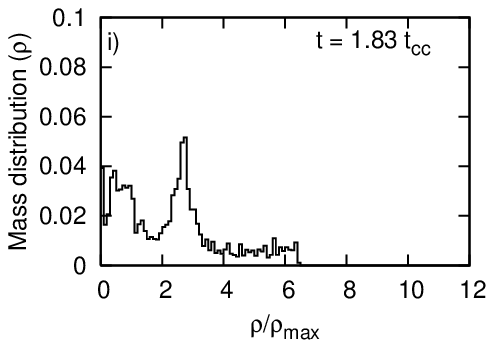,width=3.4cm}
\psfig{figure=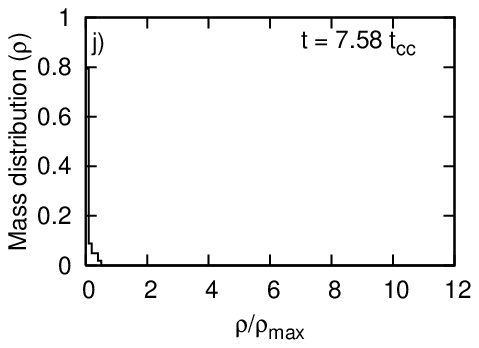,width=3.4cm}
\psfig{figure=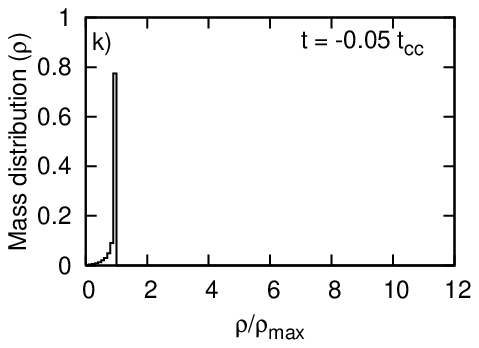,width=3.4cm}
\psfig{figure=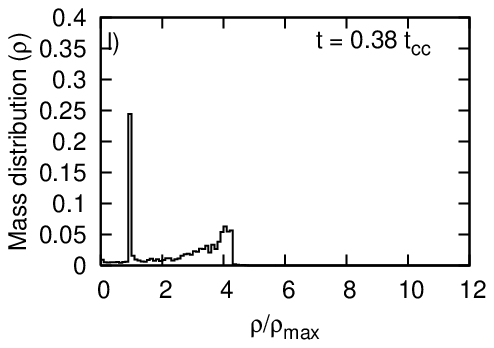,width=3.4cm}
\psfig{figure=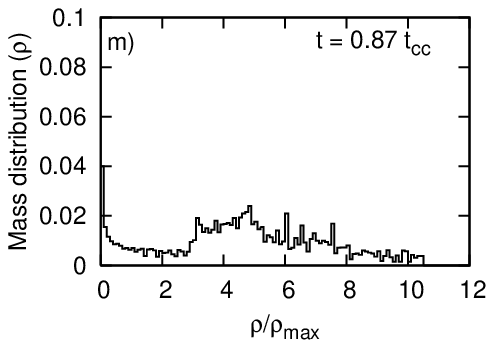,width=3.4cm}
\psfig{figure=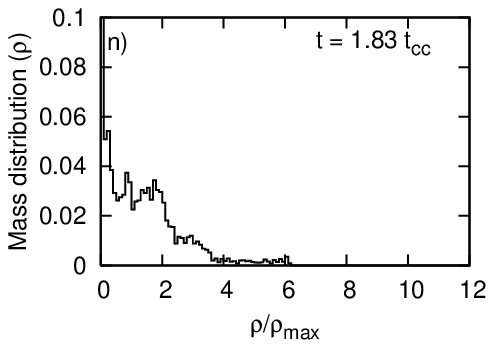,width=3.4cm}
\psfig{figure=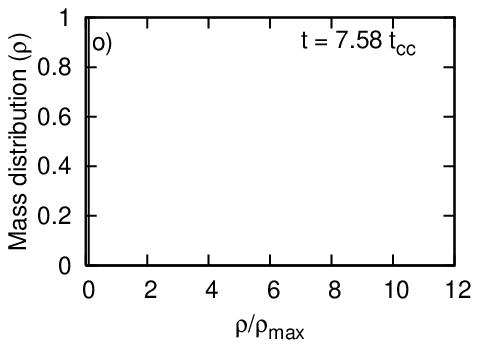,width=3.4cm}
\caption[]{Time evolution of the mass distribution function for models c3no128
(panels a-e), c3lo128 (panels f-j), and c3hi128 (panels k-o). The
histograms indicate the fraction of mass contained within each density
bin of width $0.1 \rho_{\rm max}$, normalized by the total cloud mass,
$m_{\rm c}$. From left to right the mass fractions were computed
at $t=-0.05, 0.38, 0.87, 1.83$ and $7.58\;t_{\rm cc}$.}
\label{fig:massfrac}
\end{figure*}

Fig.~\ref{fig:massfrac} shows how the cloud material mixes with the
surrounding gas in models c3no128, c3lo128, and c3hi128. The
histograms indicate the fraction of cloud mass over a range of density
bins of width $0.1 \rho_{\rm max}$. For $p_{1}=10$, about 80 per cent
of the cloud mass has a density $\approx \rho_{\rm max}$ at the start
of each simulation.  At $t=0.38\;t_{\rm cc}$, the shock driven into
the cloud has increased the density of the gas it has encountered by a
factor of $\sim 4$, while a substantial part of the cloud remains unaffected
and at its initial density. Eventually the transmitted shock sweeps
through the entire cloud, and densities more than 10 times
higher than the initial core density occur as the transmitted shock
interacts with the shock driven into the back of the cloud. At $t
\approx t_{\rm cc}$ the mass distribution function is approximately flat over a
wide range in density.  As the shocked cloud expands the density
decreases. The turbulent mixing of cloud material into the surrounding
flow causes the density of the cloud material to eventually approach
the shocked intercloud density ($\rho \approx 4\times10^{-3}\rho_{\rm
max}$). This evolution proceeds slightly faster in model c3hi128, as
previously noted.

The double peaked structure of the mass spectrum at $t = 1.83\;t_{\rm
cc}$ in models c3no128 and c3lo128 is consistent with previous
findings \citep{Nakamura:2006}, with the higher density peak at $\rho
\approx 2.5\rho_{\rm max}$ representing the main core, and the lower
density peak at $\rho \approx 0.7\rho_{\rm max}$ representing material
stripped from it and mixed with surrounding material. While there is
little difference between the evolution of the mass spectrum in models
c3no128 and c3lo128, in contrast there is no sign of a double-peaked
mass spectrum in model c3hi128. This is further evidence that
the destruction and mixing of interstellar clouds depends on the
level of turbulence present.

The rate of mass-loss from the core in models c2no128, c2lo128, and
c2hi128 can be compared to the analytical formula for hydrodynamic
ablation given by \citet{Hartquist:1986}. To make such a comparison
our simulations must be appropriately scaled. We therefore assume that
the cloud is ionized, has a radius $r_{\rm c} = 2\;$pc, a core density
$\rho_{\rm c}=4\times10^{-25}\;{\rm g\;cm^{-3}}$, and a temperature of
$8000\;$K, and is in pressure equilibrium with surroundings of density
$4\times10^{-27}\;{\rm g\;cm^{-3}}$ and temperature $8 \times
10^{5}\;$K. Both the cloud and its surroundings are assumed to have
solar abundances. The Mach 10 shock then travels at a speed of
$1360\;\kmps$ through the ambient medium, heating the medium to $T =
2.6 \times 10^{7}\;$K, and giving it a velocity and Mach number of
$\approx 1000\;\kmps$ and 1.3, respectively. The rate of mass-loss
from a cloud of mass $M_{\rm c}$ through hydrodynamic ablation is
$\Mdot_{\rm ab} = M_{\rm c}/t \approx l^{2}\rho_{\rm l}v_{\rm exp}$,
where $t$ is a characteristic destruction/mixing timescale, $\rho_{\rm
l}$ is the characteristic density of ablated material within a mixing
region of size $l^{2}$ around the cloud, and $v_{\rm exp}$ is the
expansion speed of material off the surface of the cloud. Momentum
conservation requires that $\rho_{\rm l} v_{\rm exp} \approx \rho v$
where $\rho$ and $v$ are the density and velocity of the surrounding
flow. If the surrounding flow is supersonic, $v_{\rm exp} \approx
c_{\rm c}$, since material cannot leave the cloud much faster than the
sound speed of the cloud, $c_{\rm c}$.  Hence $\Mdot_{\rm ab} \approx
(M_{\rm c} c_{\rm c})^{2/3} (\rho v)^{1/3}$.  With the above
parameters, $\Mdot_{\rm ab} \approx 1.2 \times 10^{-6}\;\Msolpyr$, and
the cloud survives for approximately $1.6 \times 10^{5}\;$yrs. In
comparison the cloud crushing timescale, $t_{\rm cc} = 1.4 \times
10^{4}\;$yrs, so that the cloud survives for about 10 cloud crushing
timescales before being destroyed.

\begin{figure*}
\psfig{figure=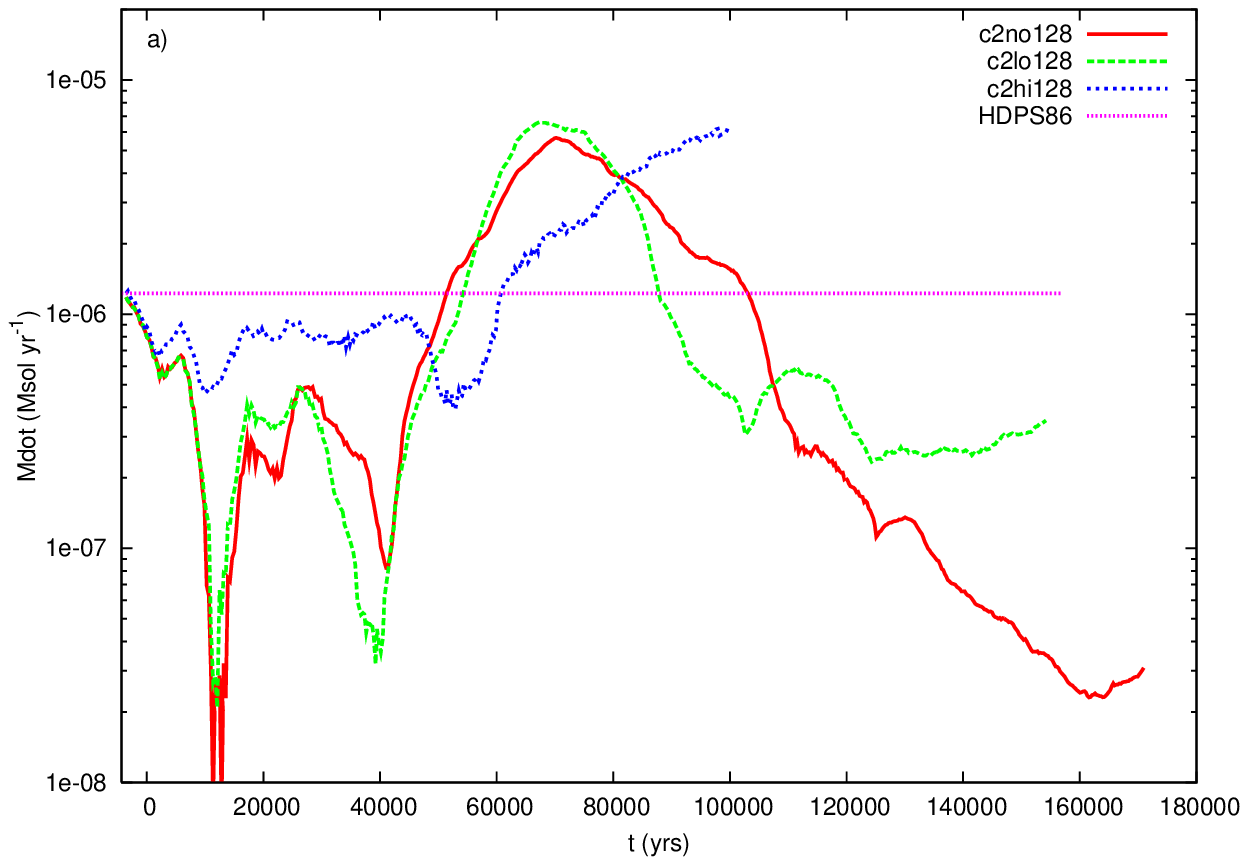,width=8.5cm}
\psfig{figure=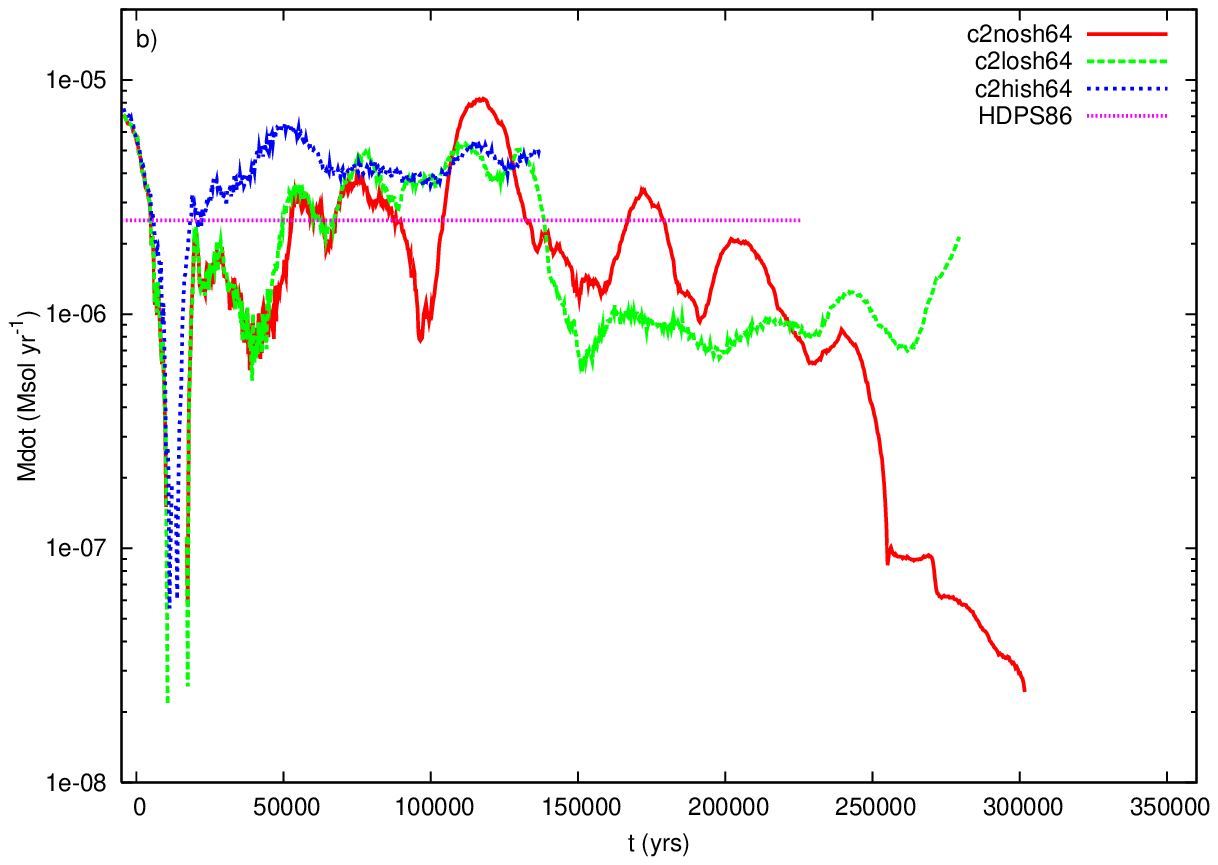,width=8.5cm}
\caption[]{a) Mass loss rate for models c2no128, c2lo128, and c2hi128
compared to the mass-loss formula of \citet{Hartquist:1986}. b) 
Mass loss rate for models c2nosh64, c2losh64, and c2hish64
compared to the mass-loss formula of \citet{Hartquist:1986}. In both
panels $t=0$ corresponds to the time at which the intercloud shock is
level with the centre of the cloud.}
\label{fig:massloss}
\end{figure*}

Fig.~\ref{fig:massloss}a) compares $\Mdot_{\rm ab}$ with the
numerically determined mass-loss rates, where it
is apparent that there is good basic agreement between the analytical
and numerical mass-loss rates. Not surprisingly, the mass-loss rates
from the numerical models are time-dependent: they decrease as the
cloud is compressed by the shocks driven into it, and then increase
significantly as the cloud re-expands. The higher mass-loss rates
compared to $\Mdot_{\rm ab}$ during this latter period reflect the
increased surface area of the cloud which is not taken into account in
the analytical theory. The mass-loss rate in model c2hi128 is higher
than in models c2no128 and c2lo128 for the first $t \approx
55,000\;$yrs of the interaction - this is due to the increased
transport coefficients which cause the stripping of material at a
faster rate. During this time, the mass-loss rate from model c2hi128
is also remarkably constant. 
%During the interval $50,000 \ltsimm
%t/{\rm yr} \ltsimm 80,000$ ($t=0$ corresponds to the time at which the
%intercloud shock is level with the centre of the cloud), the mass-loss
%rate in model c2hi128 is lower than in models c2no128 and c2lo128.
The maximum mass-loss rate in all 3 models is $\approx 6\times
10^{-6}\;\Msolpyr$, or 5 times the rate predicted from the formula in
\citet{Hartquist:1986}. In model c2hi128 this occurs at the end of the
cloud's life. In contrast, the peak rate of mass-loss occurs at about
$70,000\;$yr ($t\approx5\;t_{\rm cc}$), in models c2no128 and c2lo128,
which is similar to the time at which half of the core mass has been
mixed (see Table~\ref{tab:results}).  In models c2lo128 and
(especially) c2no128, the cloud enjoys a more sedate ending of its life,
with the mass-loss rate declining as the core mass decreases. The time
at which half of the core is mixed ($t_{\rm mix}$) and the destruction
time of the core (when $m_{\rm core}=0$) are more widely separated in
these simulations.

Fig.~\ref{fig:massloss}b) shows an identical analysis for a cloud with
a smoother density profile (models c2nosh64, c2losh64, and c2hish64).
The mass of the cloud is 2.9 times greater, and the predicted
mass-loss rate is twice as high. The cloud is therefore expected to
survive for 40 per cent longer (i.e. $230,000\;$yrs). The initial rate
of mass-loss from the simulations is approximately 3 times as high as
the predicted value, which reflects the ease at which the tenuous
outer layers of the cloud are removed. The variations in the rate of
mass-loss also appear to be reduced, again consistent
with a milder interaction. Otherwise, similar behaviour to the
previous case of a sharp-edged cloud is seen, though
the cloud survives appreciably longer than predicted in models
c2nosh64 and c2losh64.

To conclude this section, we note that the cloud in model c2hi128 is
destroyed in about 65 per cent of the predicted time, while in models
c2no128 and c2lo128 the destruction time is in good agreement with the
analytical prediction. Clouds with a smoother density profile survive
for about 30 per cent longer than the predicted time, though if they
are overrun by a highly turbulent environment they may survive only 60
per cent of the predicted time.  A similar analysis for the
simulations with $\chi=10^{3}$ reveals that in model c3hi128 the cloud
is destroyed more than twice as fast as predicted by the analytical
theory appropriate for a smooth post-shock flow. In model c3lo128 it
is destroyed in about 75 per cent of the predicted time, while in
model c3no128 the destruction time is in good agreement with the
analytical prediction. The Mach number dependence of the mass-loss rate
is examined in a subsequent paper (Pittard, in preparation).

\subsubsection{Cloud volume and mean density}
In each simulation the volume of the cloud core increases after the
initial compression stage to a maximum value, then decreases as the
core material is gradually ablated and mixed into the surrounding
flow. The time of this reversal is typically just prior to $t_{\rm
mix}$.  The strong turbulence present in models c2hi128 and c3hi128
vigorously rips material off the surface of the cloud and advects it
downstream, so that the $\kappa = 0.5$ isosurface is greatly
extended. This leads to a rapid increase in the volume of the
``core'', $V_{\rm core}$, and a commensurate decrease in the core
density, $\rho_{\rm core}$ (see Figs.~\ref{fig:kepscomp2}h and j
and~\ref{fig:kepscomp3}h and j). In contrast, the destruction of the
cloud when $\chi=10$ is so rapid that there is not enough time for the
highly turbulent surroundings to produce much of an effect (see
Fig.~\ref{fig:kepscomp1}h and j), whereas a smooth density profile
tends to delay the time at which the effects from highly turbulent
surroundings become obvious (see Fig.~\ref{fig:kepscomp3sh}h and j).

\subsubsection{Cloud velocity}
The acceleration of the cloud occurs in two stages. The cloud is first
accelerated to a velocity $v_{\rm s}$ by the shock driven into it. The
flow of shocked intercloud gas then further accelerates the cloud
until they have the same velocity (i.e. $0.725v_{\rm b}$ when $M=10$).
Since $v_{\rm s}/v_{\rm b} = 1/\chi^{1/2}$, the second stage dominates
when $\chi\gtsimm7.5$. If the cross-section of the cloud were to
remain constant, the characteristic drag time for sharp-edged clouds
would be $t_{\rm drag,0} \sim \chi^{1/2}\;t_{\rm cc}$
\citep{Klein:1994}. However, since the cloud expands laterally, the
actual drag time is considerably shorter \citep[][see also
Table~\ref{tab:results}]{Klein:1994}.
%The effect of a smooth cloud
%boundary on the drag has been investigated by \citet{Nakamura:2006}.

Fig.~\ref{fig:kepscomp2}k) shows that the acceleration of the cloud is
virtually identical in models c2no128 and c2lo128 for $t\ltsimm
4\;t_{\rm cc}$. This is also the case for the core acceleration
(Fig.~\ref{fig:kepscomp2}l).  In contrast, the stronger turbulent
mixing in model c2hi128 increases the momentum coupling of the surface
of the cloud with the ambient flow so that the cloud accelerates more
quickly to the postshock intercloud speed. This effect is even more
pronounced in model c3hi128 (see Fig.~\ref{fig:kepscomp3}k). The
acceleration of clouds with smooth density profiles is more gentle
(cf. Figs.~\ref{fig:kepscomp3}k and~\ref{fig:kepscomp3sh}k), as
expected. In all cases the cloud accelerates faster than the core
(i.e. $\langle v_{\rm z,cloud}\rangle > \langle v_{\rm
z,core}\rangle$), as expected.

\citet{Klein:1994} and \citet{Nakamura:2006} presented a semi-analytic
theory for the acceleration of sharp-edged and smooth-edged clouds,
respectively.  The formulation from \citet{Klein:1994} converted to
the rest frame of the initial cloud is shown in
Figs.~\ref{fig:kepscomp2}k) and~\ref{fig:kepscomp3}k), where the value
of $t_{\rm m}$ obtained from models c2no128 and c3no128 is used. In
each case, the results obtained with three different values for the
drag coefficient are shown: $C_{\rm D} = 0.2,0.5$, and
1.0. \citet{Klein:1994} and \citet{Nakamura:2006} adopt $C_{\rm D} =
1.0$, but it is not clear that this is the best choice. 
For solid spheres, experiments have determined that $C_{\rm D} =
1.0$ occurs when ${\rm Re} \approx 100$ \citep{Landau:1959}, which is
a far lower Reynolds number than occurs in the astrophysical settings
that we are interested in (see Section~\ref{sec:reynoldsnumber}).
However, interstellar clouds are compressible and are ablated in
shock-cloud interactions, so one would not necessarily expect the 
drag coefficient to be similar to the solid sphere case. Nevertheless, 
clouds with a high density contrast should, at least initially, 
behave somewhat like solid spheres. If this is the case, more
appropriate values for $C_{\rm D}$ would be $\approx 0.5$ when
$2\times10^{4} < {\rm Re} < 2\times10^{5}$, and $\approx 0.2$ for
${\rm Re} \gtsimm 2\times10^{5}$.

In fact, we find that there is poor agreement between the theoretical
acceleration and the results from model c2no128 for values of
$C_{\rm D} \ltsimm 1$ (Fig.~\ref{fig:kepscomp2}k).  The closest match is
obtained with $C_{\rm D} \approx 1.0$, though the theoretically
estimated acceleration is too high for $t \ltsimm 4\;t_{\rm cc}$, and
too low at later times. This discrepancy is the opposite of that noted
in \citet{Klein:1994}. The poor match is at least in part due to the
poor agreement between the predicted and model evolution of $a_{\rm
cloud}$ (Fig.~\ref{fig:kepscomp2}a). Since $C_{\rm D} \approx 0.2-0.5$ 
for high Reynolds number flow over a solid sphere, we conclude that
clouds with $\chi=10^{2}$ are a poor approximation to a hard sphere.

In contrast, Fig.~\ref{fig:kepscomp3}k) shows that the theoretical
acceleration of the cloud agrees much better with the numerical
results from models c3no128 and c3lo128 where $\chi =10^{3}$. The best
agreement is obtained with $C_{\rm D} \approx 0.5$, which is
consistent with the assumption of fully developed turbulence in the
$k$-$\epsilon$ turbulence model. We therefore conclude that clouds
with $\chi=10^{3}$ are a better approximation to a hard sphere.

\subsubsection{Cloud velocity dispersion}
The interaction of shocks with clouds produces substantial vorticity
and velocity dispersion, which may be a key mechanism for generating
turbulent motions in the ISM
\citep[e.g.,][]{Kornreich:2000,MacLow:2004}. Although the sub-grid
turbulence model deals with turbulent motions and mixing on scales
smaller than the cell size of the numerical grid, larger scale
turbulent motions can be directly measured from the velocity
dispersions in the axial and radial directions, $\delta v_{\rm z}$ and
$\delta v_{\rm r}$ respectively. The time evolution of these
quantities is shown in panels m) and n) of
Figs.~\ref{fig:kepscomp1}-\ref{fig:kepscomp3sh}.

Our results for model c2no128 can be compared to Figs.~9 and~10 in
\citet{Klein:1994}. The level of qualitative and quantitative
agreement is basically good. In particular, we confirm that the radial
velocity dispersion is generally somewhat less than the axial velocity
dispersion. However, we find that the maximum values of $\delta v_{\rm
z}$ and $\delta v_{\rm r}$ occur somewhat later in our
simulations. Comparing models c2no128, c2lo128, and c2hi128, we see
that $\delta v_{\rm z}$ is more sensitive to a higher initial level of
turbulence in the environment around the cloud than $\delta v_{\rm
r}$.

A comparison with the $\chi=10$ and $10^{3}$ simulations reveals that
$\delta v_{\rm z}/\delta v_{\rm r}$ increases with $\chi$, as is also
apparent in the work of \citet{Klein:1994}. $\delta v_{\rm z}$ peaks
at about the same value in c2hi128 and c3hi128, indicating that the
random instability-induced grid-scale motions which develop are 
limited by the high viscosity introduced by the subgrid turbulence model 
in these simulations. $\delta v_{\rm z}$  peaks at higher
values in models c3no128 and c3lo128, compared to models c2no128 and
c2lo128, due to greater growth of RT and KH instabilities resulting from
the longer drag and mixing time in the former models. 
The velocity dispersion appears to be slightly reduced for
clouds with smooth boundaries, with some sign that the time evolution of
$\delta v_{\rm z}$ attains a broader maximum.  In all simulations,
$\delta v_{\rm r}$ peaks slightly prior to the time of the maximum
radial extent of the cloud, $t_{\rm m}$.

\begin{figure*}
\psfig{figure=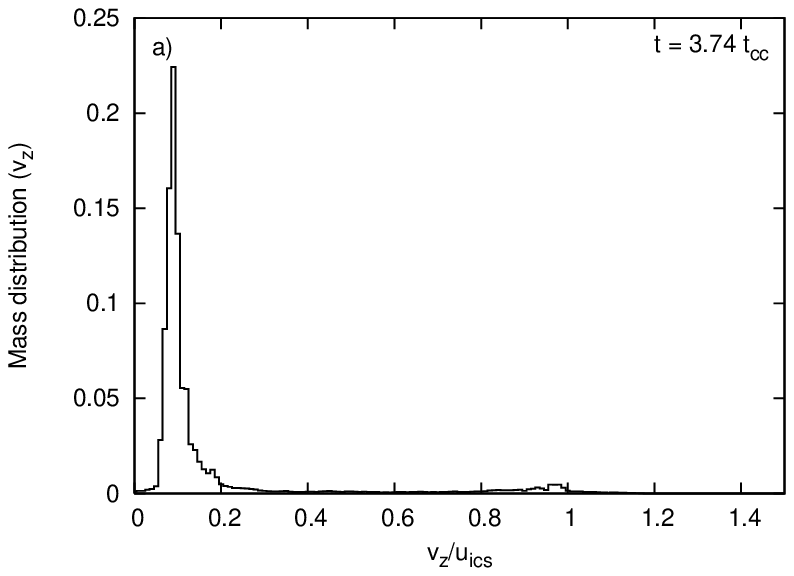,width=5.7cm}
\psfig{figure=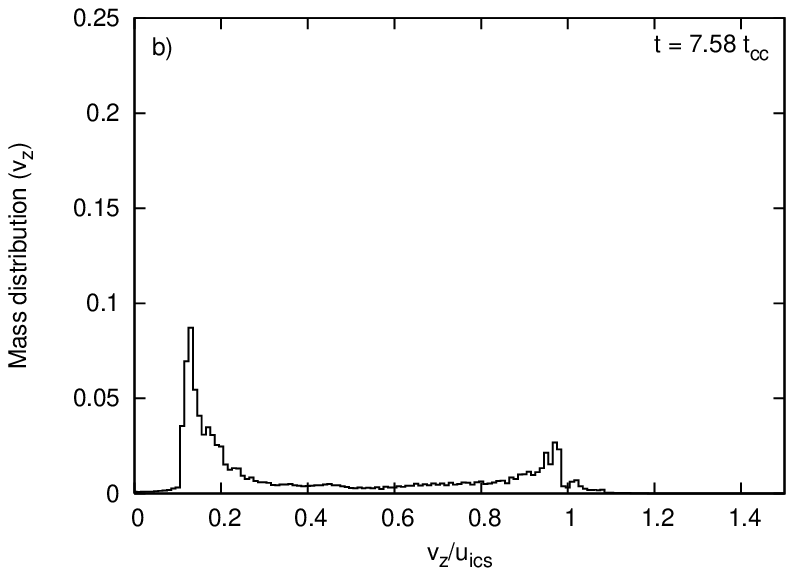,width=5.7cm}
\psfig{figure=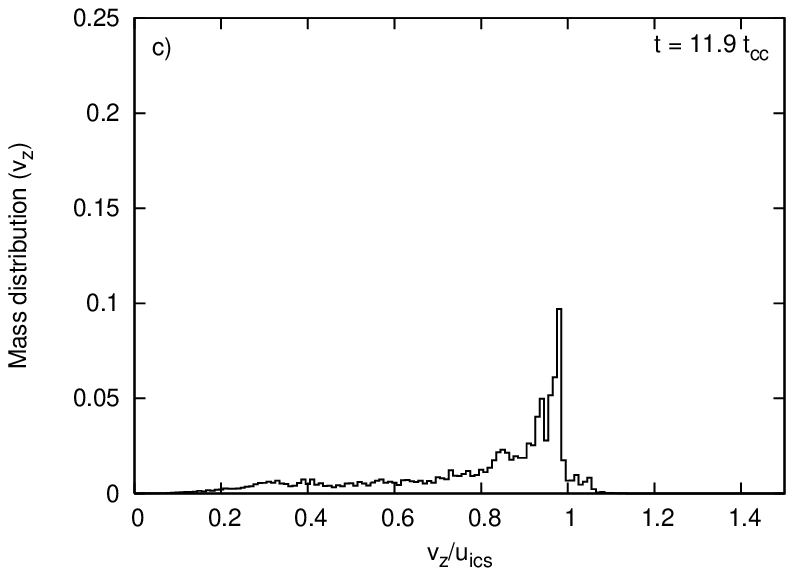,width=5.7cm}
\psfig{figure=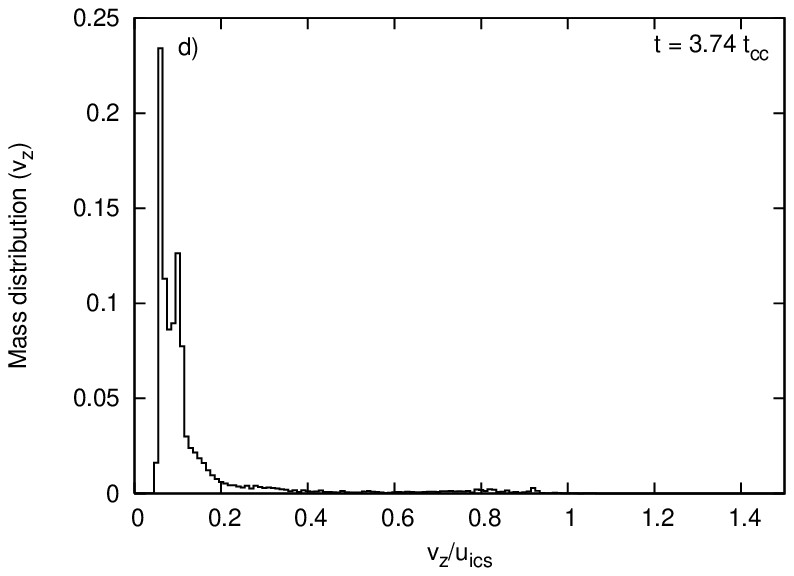,width=5.7cm}
\psfig{figure=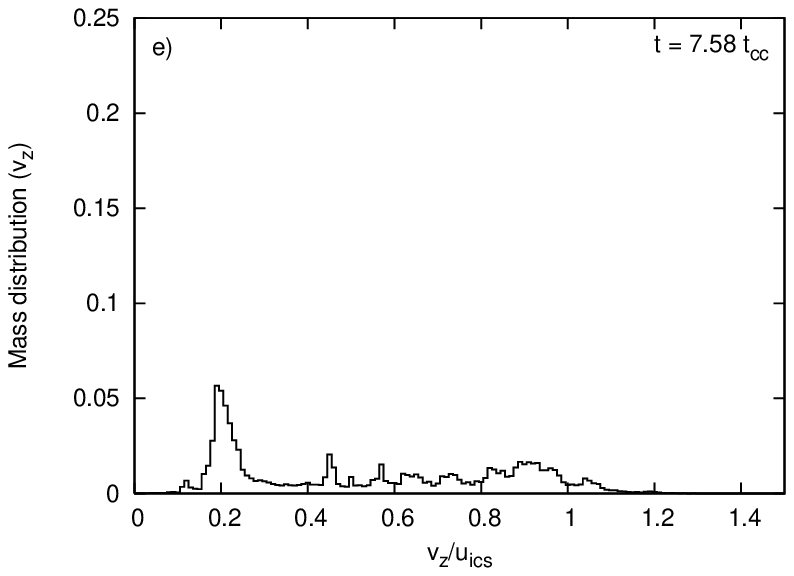,width=5.7cm}
\psfig{figure=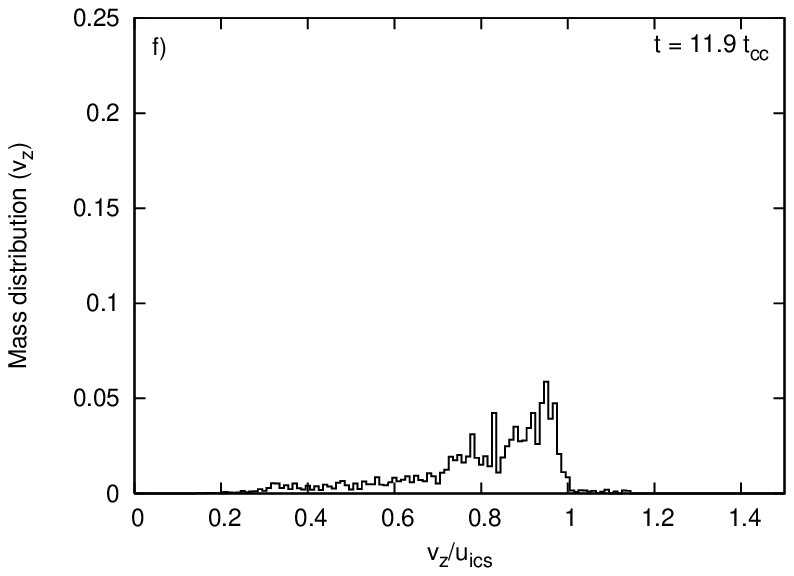,width=5.7cm}
\psfig{figure=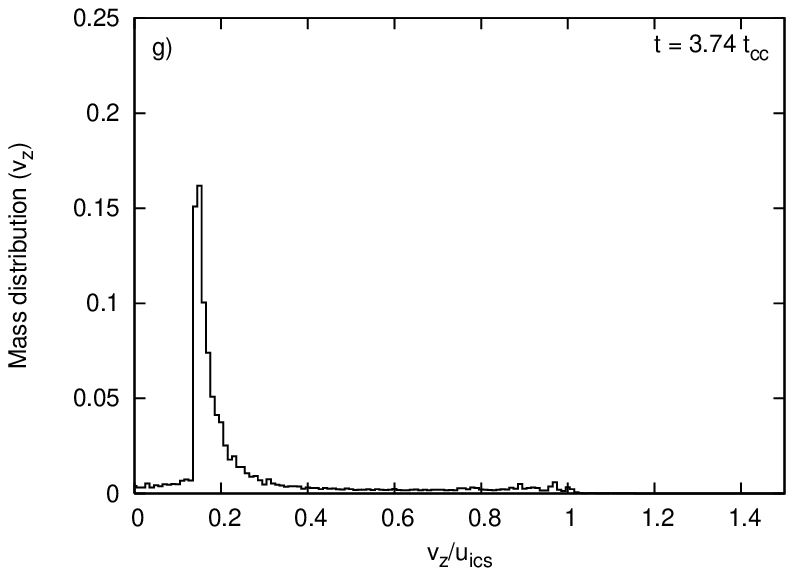,width=5.7cm}
\psfig{figure=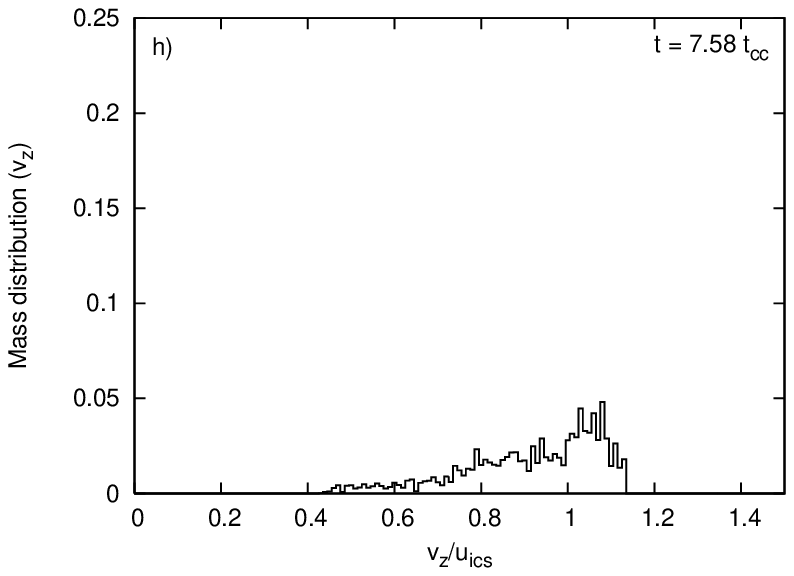,width=5.7cm}
\psfig{figure=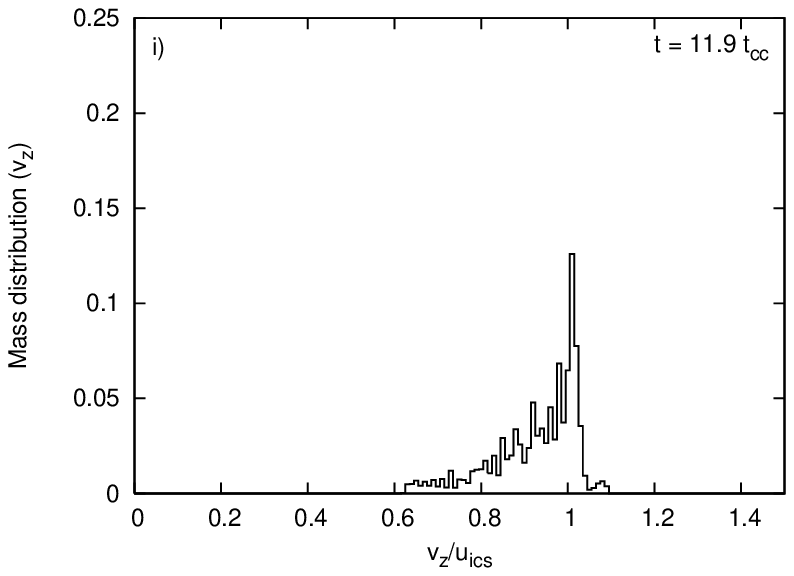,width=5.7cm}
\caption[]{Mass distribution as a function of $v_{\rm z}$ for models
c3no128 (panels a-c), c3lo128 (panels d-f), and c3hi128 (panels g-i)
at $t = 3.74, 7.58$ and $11.9\;t_{\rm cc}$.  The histograms denote the
mass contained within a velocity bin of width $0.01\;u_{\rm ics}$ for
a line of sight parallel to the z-axis. The integrated mass is the
mass of the cloud, $m_{\rm c}$.}
\label{fig:mass_spec}
\end{figure*}

Fig.~\ref{fig:mass_spec} shows the mass distribution function in
velocity space within the entire cloud integrated along the $z$-axis for
models with $\chi=10^{3}$. The histograms indicate the mass contained
within a corresponding velocity bin with a width of $0.01\;u_{\rm
ics}$, where $u_{\rm ics}$ is the postshock ambient velocity.  The
shock initially driven into the cloud has a speed $v_{\rm
b}/\sqrt{\chi} = 0.042\;u_{\rm ics}$. By $t=3.74\;t_{\rm cc}$ the
majority of the mass within the cloud in models c3no128 and c3lo128
has been further accelerated by additional shocks and momentum
transfer from the surrounding flow to roughly twice this speed. In
contrast, the material stripped off the surface of the cloud moves at
speeds up to $u_{\rm ics}$, with a small fraction exceeding $u_{\rm
ics}$ as a result of turbulent motions. Further stripping and momentum
transfer results in a gradual acceleration of the majority of the
cloud material to speeds near $u_{\rm ics}$ as time progresses.  Not
unexpectedly, the acceleration of cloud material in model c3lo128 is
more rapid than in model c3no128, and even more so in model c3hi128,
where even the slowest moving cloud material has a speed in excess of
$0.6\;u_{\rm ics}$ at $t\approx 12\;t_{\rm cc}$.

\subsubsection{Cloud vorticity}
A key feature of the interaction of a shock with a cloud is the
development of powerful vortex rings. Vorticity (${\mathbf \omega} \equiv
{\mathbf \nabla \times v}$) can be produced (or destroyed) when pressure
and density gradients are not aligned (i.e. by a curl in the
acceleration), and by viscosity.  The corresponding circulation, ${\bf
\Gamma} \equiv \int {\bf \omega \cdot dA}$.  \citet{Klein:1994} showed
that the vorticity production can be classified into 4
components. Vorticity at the interface between the cloud and the
surrounding flow is produced by the initial passage of the shock
($\Gamma_{\rm shock}$) and the subsequent postshock flow ($\Gamma_{\rm
post}$).  The third component is connected with the triple points
associated with the Mach-reflected shocks behind the cloud
($\Gamma_{\rm ring}$), while the fourth component is the vorticity
produced in the cloud ($\Gamma_{\rm cloud}$) and is smaller than the
other components by a factor $\sim \chi^{-1/2}$. 
\citet{Nakamura:2006} showed that the vorticity produced by a shock
overunning a cloud with smooth boundaries is qualitatively similar to
that produced from clouds with sharp boundaries. \citet{Klein:1994} 
constructed analytical expressions
for 3 of the above components which we reproduce below:
\begin{eqnarray}
%\begin{eqnarray*}
\Gamma_{\rm shock} & \approx & - \frac{9}{4}(1 - \chi^{-1/2})v_{\rm b}r_{\rm c},\\
\Gamma_{\rm post} & \approx & - \frac{9}{64}\left(\frac{\chi^{1/2} t_{\rm drag}}{t_{\rm cc}}\right)v_{\rm b}r_{\rm c},\\
\Gamma_{\rm ring} & = & \frac{3}{4}v_{\rm b}r_{\rm c}.
\end{eqnarray}
$\Gamma_{\rm post}$ is the dominant contribution to the circulation
when $\chi$ is large.  \citet{Klein:1994} did not attempt to model
$\Gamma_{\rm cloud}$, since this component is small. For clouds with
smooth boundaries, $\Gamma_{\rm post}$ must be further multiplied by
$(1 - \chi^{-1/2})^{2}$ \citep{Nakamura:2006}. 
%Although not directly
%relevant to the single-cloud calculations presented in this work,
%\citet{Collins:2005} noted that the vorticity production of a 
%collection of shocked clouds is dependent on the ratio of the cloud
%separation to the critical distance presented by \citet*{Poludnenko:2002}.

The time evolution of the circulation from our numerical models is
shown in panel o) of Figs.~\ref{fig:kepscomp1}-\ref{fig:kepscomp3sh}.
Table~\ref{tab:circ} lists values estimated for $\Gamma_{\rm post}$
and the total circulation, $\Gamma_{\rm tot} =\Gamma_{\rm shock} +
\Gamma_{\rm post} + \Gamma_{\rm ring}$. In each case we use the values
for $t_{\rm drag}$ in Table~\ref{tab:results}. Obviously, the
circulation computed using the cloud and core drag times is similar
when $t_{\rm drag}$ for the cloud and the core are also similar (see,
e.g., models of sharp-edged clouds with $\chi=10$ and 100). However,
when $t_{\rm drag}$ for the cloud and the core differ (as occurs for
models with a shallow cloud density profile), we find that
the numerically determined circulation is in much better agreement
with theoretical estimates if the drag time for the core is
used. Figs.~\ref{fig:kepscomp1}-\ref{fig:kepscomp3sh} indeed show that
the mean velocity of the cloud and core as a function of time become
increasingly disparate with increasing $\chi$. Clearly it is the
motion of the core with respect to the postshock ambient medium which
dominates the generation of circulation - after all, this is where the
highest velocity shear occurs.

\begin{table}
\begin{center}
\caption[]{Theoretical estimates for the total circulation and the
component produced by the postshock flow.  In each case the calculated
value uses the drag-time for the cloud (core), as given in
Table~\ref{tab:results}.  $-\Gamma_{\rm shock} = 1.54, 2.03,$ and 2.18
when $\chi=10, 10^{2}$ and $10^{3}$, respectively, while the value of
$\Gamma_{\rm ring}$ is always $0.75$.}
\label{tab:circ}
\begin{tabular}{lll}
\hline
\hline
Model & $-\Gamma_{\rm post}$ & $-\Gamma_{\rm tot}$ \\
\hline
c1no128 & 0.45 (0.49) & 1.24 (1.28) \\
c1lo128 & 0.47 (0.49) & 1.26 (1.28) \\
c1hi128 & 0.32 (0.38) & 1.11 (1.17) \\
c2no128 & 4.30 (4.35) & 5.58 (5.63) \\
c2lo128 & 4.33 (4.35) & 5.61 (5.63) \\
c2hi128 & 3.47 (3.73) & 4.75 (5.01) \\
c3no128 & 29.2 (42.1) & 30.6 (43.5) \\
c3lo128 & 30.4 (31.8) & 31.8 (33.2) \\
c3hi128 & 20.4 (21.8) & 21.8 (23.2) \\
c2nosh64 & 4.61 (6.28) & 5.89 (7.56) \\
c2losh64 & 4.56 (4.93) & 5.84 (6.21) \\
c2hish64 & 3.51 (4.39) & 3.08 (3.85) \\
c3nosh64 & 28.0 (41.0) & 29.4 (42.4) \\
c3losh64 & 31.5 (60.2) & 32.9 (61.6) \\
c3hish64 & 27.3 (34.9) & 28.7 (36.3) \\
\hline
\end{tabular}
\end{center}
\end{table}

Table~\ref{tab:circ} reveals that the model results and the
theoretical estimates are generally in good agreement. The
contribution of the 3 main components is best seen in
Fig.~\ref{fig:kepscomp1}o). The initial rise to maximum is caused by
the initial passage of the shock ($\Gamma_{\rm shock}$), while the
subsequent drop is caused by the formation of the supersonic vortex
ring behind the cloud, $\Gamma_{\rm ring}$.  The increase in
circulation after this minimum (until $t \approx 2\;t_{\rm cc}$) 
reveals the vorticity production due to the post-shock flow over the cloud.

The total circulation in the models is often in excellent agreement
with the predictions (for instance, the peak circulation in models
c3lo128, c3hi128, c3losh64, and c3hish64 are $33.8, 23.3, 65.2,$ and
$40.0$, while the theoretical predictions are $33.2, 23.2, 61.6,$ and
$36.3$, respectively).  In some models (e.g., c2no128, c2lo128,
c3no128 and c3nosh64) the circulation is somewhat higher than
predicted. Spurious vorticity can be generated at the boundary of
grids of different refinement levels \citep{Plewa:2001}, but this does
not seem to be the case here (since, for instance, one would
expect a larger discrepancy for model c3lo128 than for model c2lo128).
  
The total circulation in models with $\chi=10^{3}$ tends to reach a
lower maximum when there is high postshock turbulence, due to the
more rapid destruction of the cloud in such circumstances.  Models
where the cloud has a shallow density profile generate more
circulation, principally because of the larger cloud mass and the
longer drag times in these models.  The decay in $\Gamma_{\rm tot}$ at
later times in some models may reflect the inherent viscosity
diffusing vorticity as turbulence is dissipated into heat.

\subsubsection{Energy fractions}
\label{sec:energy_frac}
As the cloud material accelerates it gains kinetic energy, while its
thermal energy increases due to shock heating, adiabatic compression,
and heat transport from turbulent mixing with the hotter surrounding
flow. There is also likely to be some numerical transport of heat. The
turbulent energy increases as shear motions generate a turbulent
cascade. Eventually, the cloud material should acquire the same
kinetic and thermal energy density as the ambient medium, and the
turbulent energy should dissipate as heat.  For a strong adiabatic
shock, the ratio of kinetic to thermal energy in the postshock flow,
$E_{\rm k}/E_{\rm th} = \gamma M_{\rm ps}^{2}/3$, where $M_{\rm ps}$
is the postshock Mach number.  In our simulations $M_{\rm ps}=1.32$,
so at late times we expect that $E_{\rm k}/E_{\rm th} \simeq 1$.

\begin{figure*}
\psfig{figure=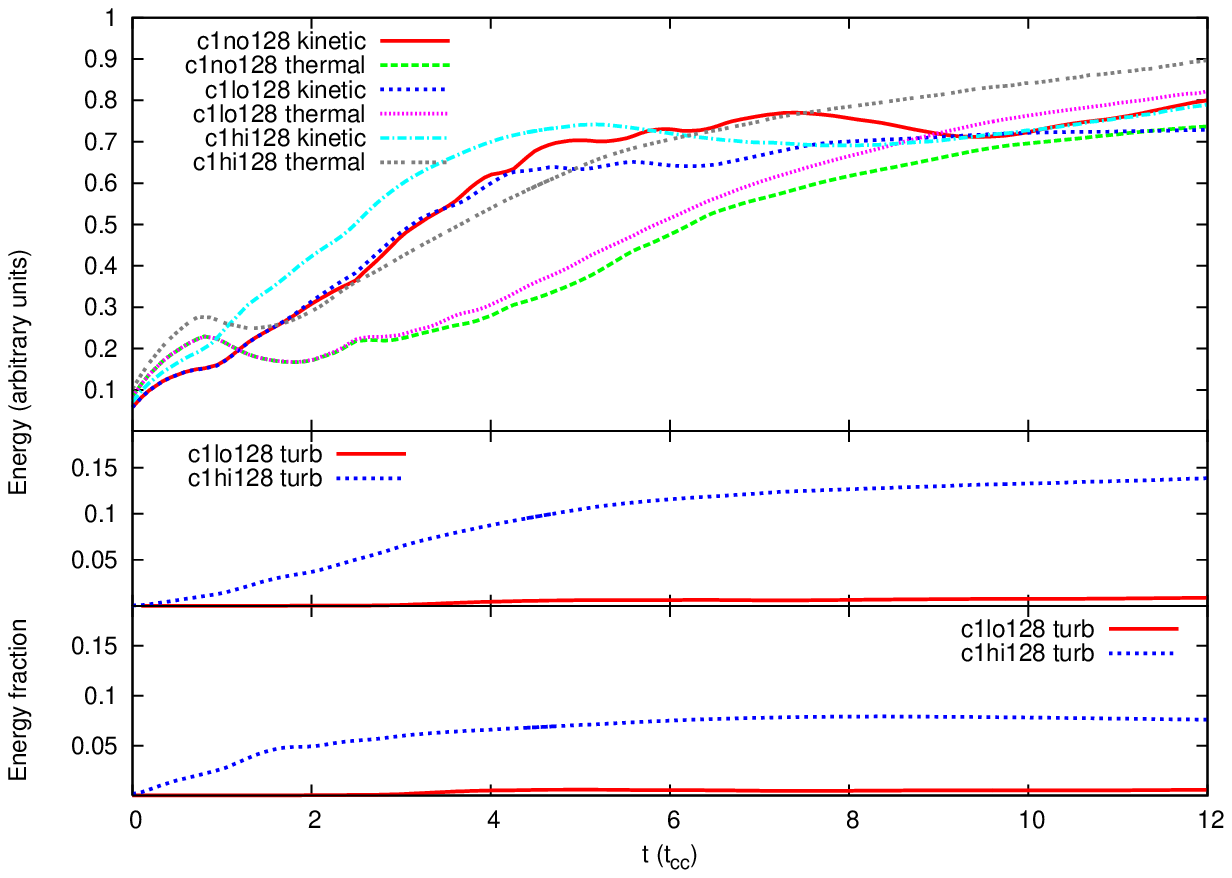,width=8.5cm}
\psfig{figure=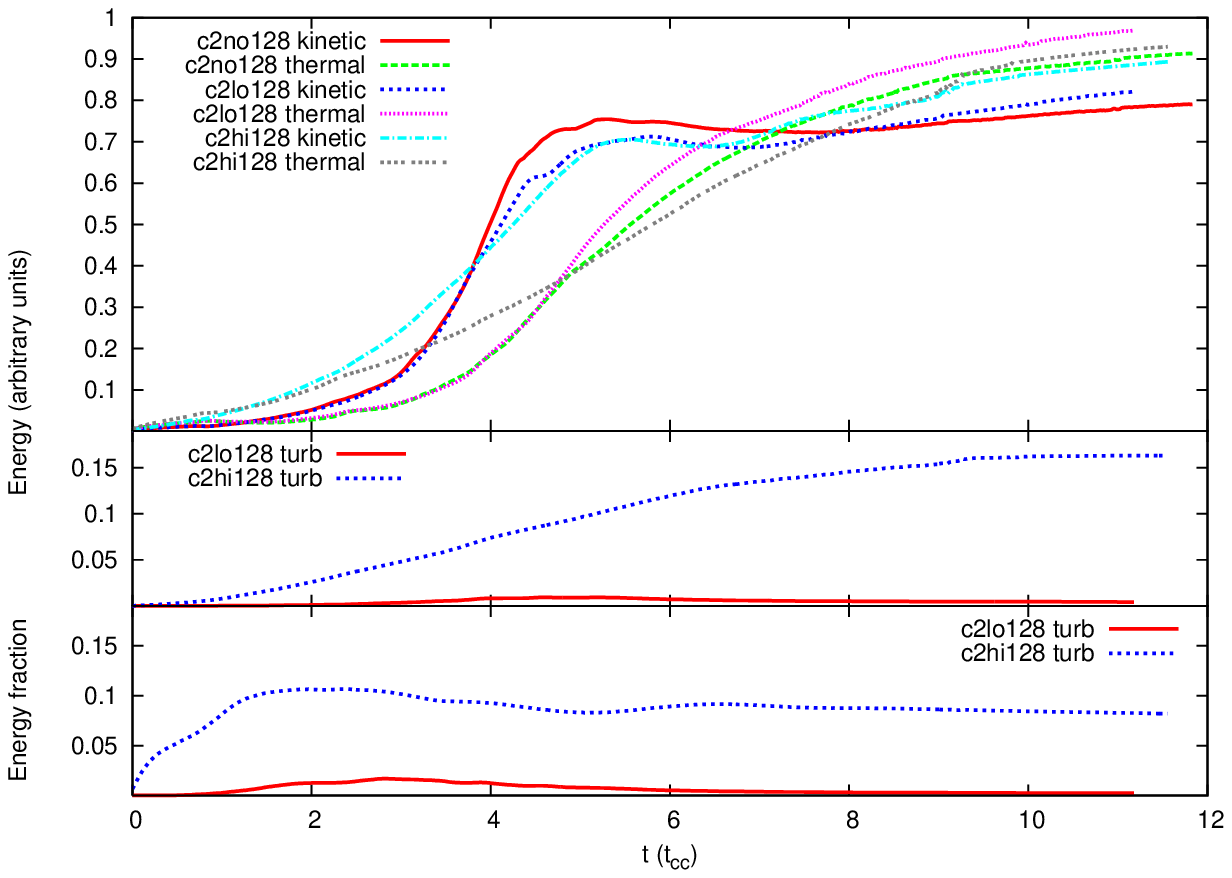,width=8.5cm}
\psfig{figure=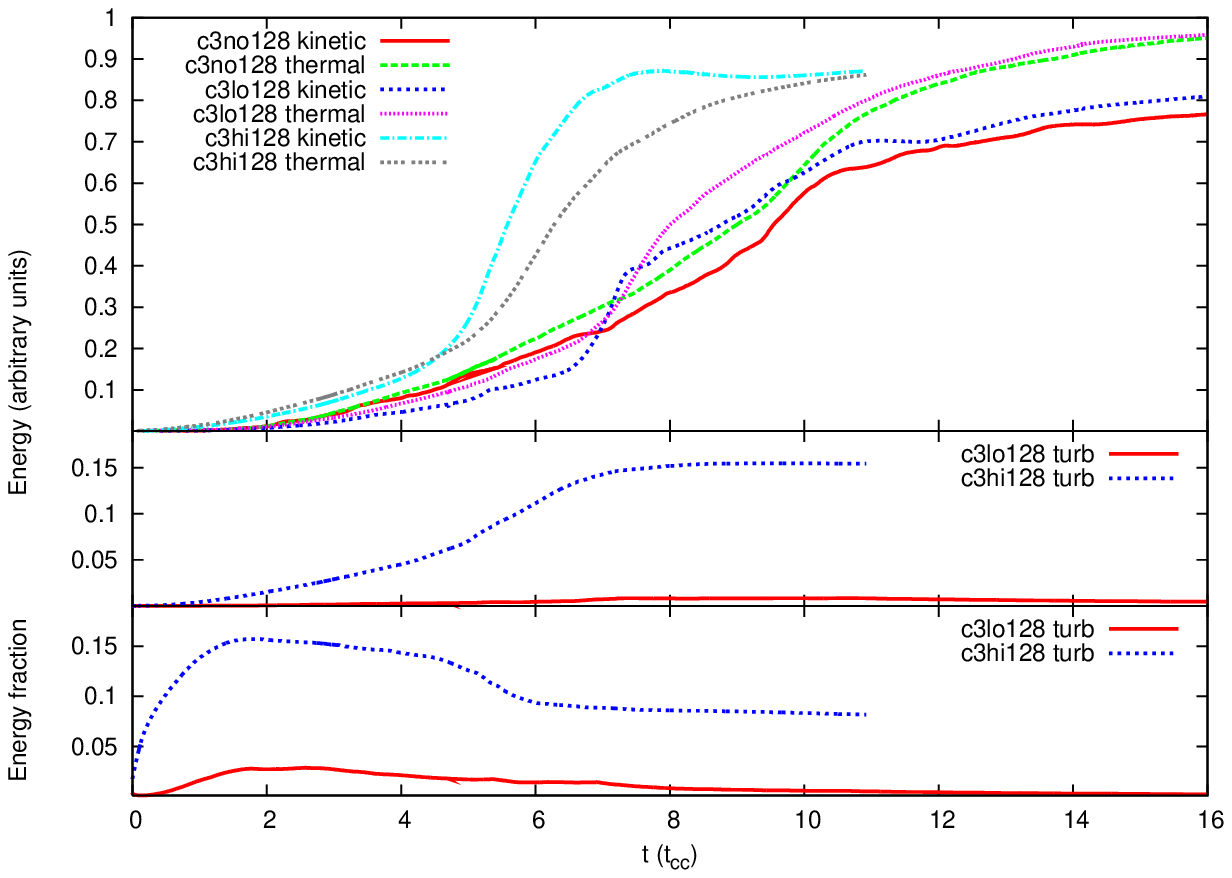,width=8.5cm}
\psfig{figure=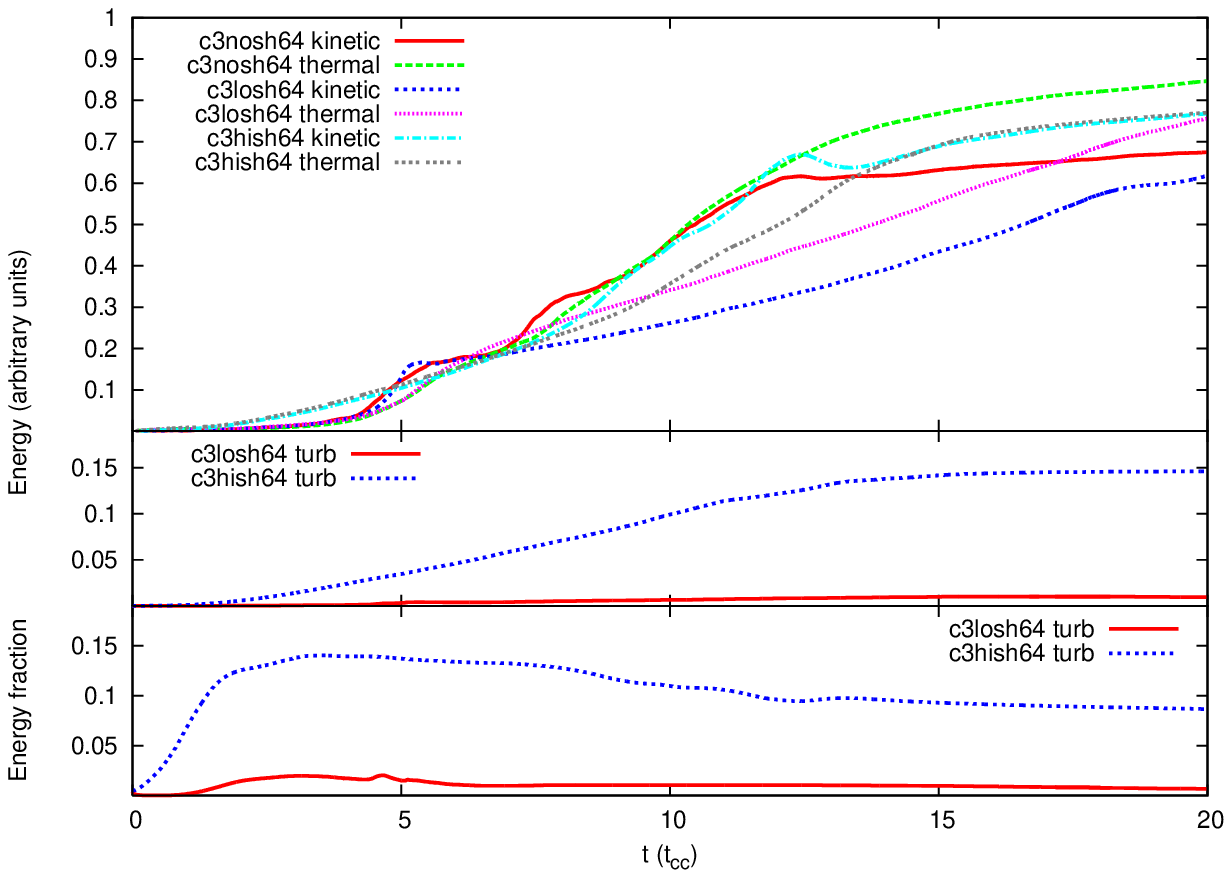,width=8.5cm}
\caption[]{Time evolution of the cloud kinetic and thermal energies
for invsicid and $k$-$\epsilon$ models.  The cloud turbulent energy at
subgrid scales and its fraction of the total cloud energy is also
shown for the $k$-$\epsilon$ models.}
\label{fig:kepsenergy}
\end{figure*}

Fig.~\ref{fig:kepsenergy} shows the time evolution of these energies
and the fraction of the total energy in small-scale (subgrid)
turbulent motions for simulations with $\chi=10,10^{2}$ and $10^{3}$.
In models c3no128 and c3lo128, the thermal energy of the cloud
generally exceeds its kinetic energy at any particular instant in
time, while in model c3hi128 the kinetic energy exceeds the thermal
energy during the period $4 \ltsimm t/t_{\rm cc} \ltsimm 11$. In model
c3hish64, the kinetic energy always exceeds the thermal energy. It is
clear that the cloud material has still not fully mixed into the
surrounding flow at the end of most of the simulations, since the
thermal energy often significantly exceeds the kinetic energy at this
point (obvious exceptions are models c2hi128, c3hi128, and c3hish64).
 
The fraction of the cloud energy in small-scale subgrid turbulent motions
increases with $\chi$ for the ``low $k$-$\epsilon$'' models, with
maxima of 0.0062, 0.017, and 0.029 for models c1lo128, c2lo128, and
c3lo128, respectively. When the environment is highly turbulent,
the turbulent energy fraction is 0.079, 0.107 and 0.157 for models
c1hi128, c2hi128, and c3hi128, respectively. 
The latter value is similar to the fractional energy
in turbulence in the post-shock flow prior to its impact with the cloud,
and demonstrates how high levels of upstream turbulence may be
maintained through sequential shock-cloud interactions. Clouds with
smoother density profiles produce a slightly smaller peak fractional
energy in turbulent motions (0.020 and 0.140 for models c3losh64 and
c3hish64, respectively).

\section{Discussion}
\label{sec:discussion}
The interaction of shocks, winds and jets with clouds, and the
collapse and/or destruction of the cloud, is of fundamental importance
to studies on star formation, the ISM, feedback and galaxy evolution,
and the evolution of diffuse astrophysical sources, such as planetary
nebulae, wind-blown-bubbles, supernova remnants, H{\sc ii} regions,
galactic winds, and the intracluster medium.

This work has examined the interaction of a shock with a cloud.
However, clouds with a high density contrast easily survive the
initial passage of the shock, and then find themselves immersed in a
subsonic or mildly supersonic post-shock flow which to all intents and
purposes resembles a wind of the same Mach number.  Hence the
simulations presented in this work are also relevant to scenarios
where clouds are ablated by a wind. In a future paper we will examine
in detail the differences between shock-cloud and wind-cloud
interactions.

In the following subsections we discuss shock-cloud interactions in
supernova remnants (SNRs), and the tails formed behind clouds in various
types of wind-cloud interactions. Jet-cloud interactions, though not
discussed in this work, can be seen in Herbig-Haro objects and AGN.
In some sources, clouds are more commonly referred to as bullets,
clumps, globules, or knots.

\subsection{SNR-cloud interactions}
\label{sec:snr-cloud}
The study of shock cloud interactions has typically focussed on SNRs.
In most cases the shocks driven into the clouds are radiative, so such
interactions are not directly comparable to the adiabatic simulations
presented in this paper. Nevertheless, in recent years there has been
much progress in unravelling the nature of these interactions and it
is worth discussing some of their interesting features.

An interesting observation is that some clouds show clear signs of a
bowshock (for instance, the Cynus Loop's southeast cloud
\citep{Graham:1995}, and also clouds on the western limb
\citep*{Levenson:2002}), while others do not (e.g., the XA cloud in
the Cygnus Loop \citep*{Danforth:2001}, the southwestern cloud of the
Cygnus Loop \citep{Patnaude:2002}, and Vela FilD \citep{Miceli:2006}).
Possible explanations for the lack of a bowshock are the earliness of
the interaction \citep{Klein:1994}, a smooth cloud density profile
\citep{Nakamura:2006}, or a high ellipticity of the cloud
\citep{Miceli:2006}. Thermal conduction can also reduce the visibility
of bow shocks \citep{Orlando:2005} (though \citet{Marcolini:2005} show
that in the case of a wind-cloud interaction it can enhance its
visibility), while cosmic rays can smooth shocks out
\citep[e.g.,][]{Wagner:2006}. In addition to these mechanisms, we note
that a highly turbulent postshock flow may also hinder the formation
of a clearly defined bowshock (see, e.g.,
Fig.~\ref{fig:rt}). Variations in pre-shock density due to
interstellar turbulence have been estimated to be $\sim 20$ per cent
\citep{Raymond:2007}. For SNRs, post-shock turbulence could be
generated by pre-shock interstellar turbulence, or, at least in young
SNRs like SN~1006, by clumps of ejecta and/or long RT fingers
affecting the contact discontinuity and blast shock.  Particle
acceleration also produces turbulent motions \citep[see, e.g.,][and references
therein]{Jones:1991}. Finally, we note that
although thermal conduction has been invoked as necessary to explain
the hot corona surounding the Vela FilD cloud \citep{Miceli:2006}, it
is possible that the turbulent transport of heat could instead account
for this.

It is also interesting to note that all of the numerical simulations
of shock-cloud interactions have assumed that the interaction takes
place in the so-called ``small-cloud'' limit, where the size of the
cloud is sufficiently small that the properties of the post-shock gas
do not change significantly in the time that it takes for the cloud to
be crushed or destroyed. This is unlikely to be the case for denser 
clouds. In addition, the interaction of a superbubble with a cloud
is better described as a shell-cloud interaction, and will
be the subject of a forthcoming work.

\subsection{Wind-cloud interactions: tails}
Material stripped off clouds by the passage of a shock or fast wind is
frequently manifest as a long tail behind the cloud 
(see, e.g., Fig.~\ref{fig:comp_latet}). Although this
work does not specifically focus on tails, the tails themselves are
interesting as their properties may allow a diagnosis of the flow past
the cloud \citep*[e.g.,][]{Dyson:1993,Dyson:2006}.  Examples of
resolved tails include those in NGC~7293 \citep*[the Helix nebula;][]
{ODell:2005,Hora:2006,Matsuura:2007}, the complex sub-structure of
knots and wakes seen in the Orion Molecular Cloud OMC1
\citep*[e.g.,][]{Allen:1993,Schultz:1999,Tedds:1999,Lee:2000,Kaifu:2000},
and the radial protrusions from the bipolar nebula surrounding the
supermassive star $\eta$~Carinae
\citep*{Weis:1999,Currie:2000,Redman:2002}.  Comet-like tails are also
seen extending from the Galactic Centre source IRS~7, a red supergiant
\citep*{Yusef-Zadeh:1991,Serabyn:1991}, and from behind Mira, an
asymptotic giant branch star \citep{Martin:2007}.

Tail-like structures are also seen in galactic winds
\citep{Cecil:2001,Ohyama:2002}, and resemble similar features seen in
simulations \citep[e.g.,][]{Strickland:2000,Cooper:2008}. Observations
and simulations of galactic winds suggest that the clumps at the heads
of the filaments are material which is ripped out of the galactic disk
as the galactic wind develops. Beautiful filamentary structures are
also seen in some galaxy clusters, of which the most prominent example
is in the Perseus cluster \citep{Conselice:2001}. Numerical
simulations indicate that a high velocity wind is necessary for these
to form, otherwise material stripped from the cloud sinks under
gravity almost as quickly as the cloud itself \citep{Pope:2008}. Tails
in the intracluster medium may help to regulate the heating of the central
regions of galaxy clusters by dissipating some of the energy injected by
the central active galactic nucleus (AGN), and therefore better 
couple the AGN to the intracluster medium. 

The Reynolds number is likely to be high enough in all of the above
cases that the tail is fully turbulent. Indeed, the emission in Mira's
tail can be explained if molecular hydrogen is excited by the
turbulent mixing of cool molecular gas and shock-heated gas
\citep{Martin:2007}.

\section{Summary and conclusions}
\label{sec:summary}
This is the first in a series of papers investigating the turbulent
destruction of clouds. Here we have investigated the destruction of a
cloud by an adiabatic shock using a hydrodynamical code which
incorporates a subgrid $k$-$\epsilon$ turbulence model, in which an
attempt is made to calculate the properties of the turbulence and the
resulting increase in the transport coefficients. The results are
compared against those from a hydrodynamical code which solves only
the inviscid Euler equations of fluid motion. The motivation for this
study is that fully developed turbulence is
prevented by the artificial viscosity in all numerical codes, though
it is expected in astrophysical environments where the Reynolds
number of the shock-cloud interaction is $\gtsimm 10^{5}$.  The effect 
of a highly turbulent environment sweeping over the cloud is also
investigated - all other works studying the interaction of a shock
with a single cloud have assumed that the post-shock environment is
completely smooth, though this is clearly not the case when there are
upstream inhomogeneties. Our main results are summarized thus:

\begin{enumerate}
\item The evolution and destruction of clouds with inviscid and
$k$-$\epsilon$ models occurs at roughly the same speed when the
post-shock flow is smooth. This is because the turbulence takes some
time to be generated and for its effects to be subsequently felt.
However, it is clear that there are increasing differences between
such models as the density contrast, $\chi$, between the cloud and
inter-cloud medium increases.  We also show that the $k$-$\epsilon$
model results are far less dependent on the spatial resolution than
calculations with an inviscid code. This behaviour is attractive given
the recent interest in simulations with multiple clouds.

\item Turbulence is mainly generated at the slip surface around the
cloud, though a small amount is also generated behind shocks. The
set-up time to form a turbulent wake behind the cloud is $\sim t_{\rm
cc}$. There is slower growth of turbulence around clouds with a smooth
density profile.  The opening angle of the turbulent mixing layer is
consistent with experimental results. The fraction of energy in
small-scale sub-grid turbulent motions initially increases with time
as turbulence continues to be generated by the velocity shear of flow
past the cloud, but then decays as turbulent energy subsequently
dissipates as heat.  The peak value of the turbulent energy fraction
increases with $\chi$ due to the longer drag-time of denser clouds and
the higher velocity shear at the cloud surface, and decreases as the
boundary of the cloud becomes smoother. The interaction of shocks with
smooth clouds is generally milder, as previously reported.

\item Clouds which are subject to a highly turbulent post-shock
environment are destroyed significantly quicker than those within a
smooth flow, due to the enhanced transport and diffusion
coefficients. This effect increases with $\chi$, since high density
clouds are subject to a longer period of ``buffeting''.  Clouds with
small density contrasts (e.g., $\chi=10$) are destroyed so quickly
after the initial shock passage that they experience very little
``buffeting''. Strong environmental turbulence increases the momentum
coupling between the cloud and its surroundings, which increases the
acceleration of the cloud. The degree by which the destruction
of the cloud speeds up should depend on the strength of the turbulence
imposed on the post-shock flow, but here our intention is simply to
demonstrate that it does, in fact, occur more rapidly.

\item The rate of mass-loss from a cloud overrun by a Mach 10 shock
due to hydrodynamic ablation is found to be broadly consistent with
theoretical expectations, but shows large variations over the
destruction period: the peak mass-loss rate is about 5 times higher
than the time-independent value from a simplified theory. Clouds in a highly
turbulent environment can be destroyed approximately twice as
fast. However, significant differences between theoretical and
numerically determined mass-loss rates exist at higher Mach numbers
(Pittard, in preparation).

\item Material stripped off the cloud is fragmented and irregular
in structure when $\chi \ltsimm 10^{2}$. In contrast, the mass-loss
from models with $\chi=10^{3}$ better resembles a single tail-like
feature. The length-to-width ratio of the tail increases with $\chi$ 
(denser clouds live longer, allowing material to be more dispersed in
the axial direction), but decreases with the level of environmental
turbulence.

\item The vorticity generated in the interaction is broadly similar in
models with differing levels of post-shock turbulence when $\chi \ltsimm
10^{2}$. However, models with high environmental turbulence generate
substantially less total circulation when $\chi=10^{3}$ compared to
models with a smooth post-shock environment, due to the more rapid
destruction of the cloud. This may self-limit the total circulation
that can be generated in such interactions.  

\item Confirmation of the general speeding up of cloud destruction in
a highly turbulent post-shock environment is attained by additional
calculations where strong, grid-scale motions and inhomogeneities are
imposed in the post-shock flow of an inviscid calculation. Differences
in the resulting evolution compared to those from the subgrid
turbulence models with high post-shock turbulence reveal that the
interaction is sensitive to the details of the turbulence, such as the
maximum eddy size. The degree by which the destruction speeds up
should again depend on the strength and properties of the turbulence
imposed on the post-shock flow. Future simulations with a non-smooth
inter-cloud medium would also be of interest.
\end{enumerate}

\section{Future work}
\label{sec:future}
Irrespective of the shortcomings of the $k$-$\epsilon$ model
\citep[see, e.g.,][]{Davidson:2004}, it is clear that in shock-cloud
interactions, turbulence (both pre-existing and newly generated) plays
an important role, and adds a new dimension to the parameter space
that has hitherto been studied.  An obvious extension of this work is
to three dimensions: previous work has shown that many features seen
in two-dimensional simulations are unstable in three dimensions
\citep{Stone:1992}. Future work will also examine the Mach number
dependence of shock-cloud interactions and the interaction of winds
and dense shells with clouds.

Synthetic signatures of the interaction should also be compared
against observations. A comparison of the synthetic emission from
models against X-ray observations has, for instance, already been
carried out by \citet{Marcolini:2005} and \citet{Miceli:2006}.  Other
observational signatures also need to be examined in greater
detail: \citet{Westmoquette:2007a,Westmoquette:2007b}
have recently concluded that the broad emission wings to H$\alpha$
line profiles which are observed throughout the central regions of
starburst galaxies arise in a turbulent boundary layer at the
interface between hot gas flowing past cold gas stripped from clouds
\citep{Melnick:1999}, but the range in velocities may instead
reflect the acceleration of material along the tail, rather than the
turbulent motions within the mixing layer.

In many environments flows interact not with a single cloud, but with
many clouds. Hydrodynamic simulations of the complex interactions
between multiple clouds and a tenuous flow have been studied by
\citet{Jun:1996}, \citet{Poludnenko:2002}, \citet{Steffen:2004},
\citet{Pittard:2005}, \citet*{Melioli:2005},
\citet{Tenorio-Tagle:2006}, \citet{Sutherland:2007},
\citet{Yirak:2008} and \citet{Cooper:2008}. Most of these works
focussed on the changes to the global flow, but the interaction between
two or more long-lived clouds in close proximity has been examined in
detail by \citet{Pittard:2005}. In all situations, mass injection into
a flow due to the destruction of clouds enhances the thermal pressure
of the flow at the expense of the flow's ram pressure.  If the flow is
slowed and pressurized enough, it may induce the gravitational
collapse of clouds and trigger new star formation.  This process could
be a central mechanism for feedback in the interstellar medium (e.g.,
in starburst regions), and a self-consistent hydrodynamical model of
it is a long-term goal.

\section*{acknowledgements}
We would like to thank the referee, Ralf Kissmann, for a timely and
useful report. JMP thanks the Royal Society for a University Research
Fellowship.

\label{lastpage}


\begin{thebibliography}{99}
\bibitem[\protect\citeauthoryear{Allen \& Burton}{1993}]{Allen:1993}
Allen D.~A., Burton M.~G., 1993, Nature, 363, 54
\bibitem[\protect\citeauthoryear{Asai, Fukuda \& Matsumoto}{Asai et al.}{2004}]{Asai:2004}
Asai N., Fukuda N., Matsumoto R., 2004, ApJ, 606, L105
\bibitem[\protect\citeauthoryear{Cant\'{o} \& Raga}{1991}]{Canto:1991}
Cant\'{o} J., Raga A.~C., 1991, ApJ, 372, 646 
\bibitem[\protect\citeauthoryear{Cecil et al.}{2001}]{Cecil:2001}
Cecil G., Bland-Hawthorn J., Veilleux S., Filippenko A.~V., 2001, ApJ, 555, 338
\bibitem[\protect\citeauthoryear{Chandran \& Maron}{2004}]{Chandran:2004}
Chandran B.~D.~G., Maron J.~L., 2004, ApJ, 602, 170
\bibitem[\protect\citeauthoryear{Cho et al.}{2003}]{Cho:2003}
Cho J., Lazarian A., Honein A., Knaepen B., Kassinos S., Moin P., 2003, ApJ, 589, L77
\bibitem[\protect\citeauthoryear{Conselice, Gallagher \& Wyse}{Conselice et al.}{2001}]{Conselice:2001}
Conselice C.~J., Gallagher J.~S., Wyse R.~F.~G., 2001, ApJ, 122, 2281
\bibitem[\protect\citeauthoryear{Cooper et al.}{2008}]{Cooper:2008}
Cooper J.~L., Bicknell G.~V., Sutherland R.~S., Bland-Hawthorn J., 2008, ApJ, 674, 157
\bibitem[\protect\citeauthoryear{Cox}{2005}]{Cox:2005}
Cox D.~P.,2005, ARA\&A, 43, 337
\bibitem[\protect\citeauthoryear{Cox \& Smith}{1974}]{Cox:1974}
Cox D.~P., Smith B.~W., 1974, ApJ, 189, L105
\bibitem[\protect\citeauthoryear{Currie et al.}{2000}]{Currie:2000}
Currie D.~G., et al., 2000, ESO Messenger, 100, 12
\bibitem[\protect\citeauthoryear{Davidson}{2004}]{Davidson:2004}
Davidson P.~A., 2004, ``Turbulence. An Introduction For Scientists And Engineers'' (Oxford University Press)
\bibitem[\protect\citeauthoryear{Danforth, Blair \& Raymond}{Danforth et al.}{2001}]{Danforth:2001}
Danforth C.~W., Blair W.~P., Raymond J.~C., 2001, AJ, 122, 938
\bibitem[\protect\citeauthoryear{Dash \& Wolf}{1983}]{Dash:1983}
Dash S.~M., Wolf D.~E., 1983, AIAA paper 83-0704
\bibitem[\protect\citeauthoryear{Dopita \& Sutherland}{2003}]{Dopita:2003}
Dopita M.~A., Sutherland R.~S., 2003, ``Astrophysics of the Diffuse Universe'', (Springer)
\bibitem[\protect\citeauthoryear{Dyson, Hartquist \& Biro}{Dyson et al.}{1993}]{Dyson:1993}
Dyson J.~E., Hartquist T.~W., Biro S., 1993, MNRAS, 261, 430
\bibitem[\protect\citeauthoryear{Dyson et al.}{2006}]{Dyson:2006}
Dyson J.~E., Pittard J.~M., Meaburn J., Falle S.~A.~E.~G., 2006, A\&A, 457, 561
\bibitem[\protect\citeauthoryear{Elmegreen \& Lada}{1977}]{Elmegreen:1977}
Elmegreen B.~G., Lada C.~J., 1977, ApJ, 214, 725
\bibitem[\protect\citeauthoryear{Elmegreen \& Scalo}{2004}]{Elmegreen:2004}
Elmegreen B.~G., Scalo J., 2004, ARA\&A, 42, 211
\bibitem[\protect\citeauthoryear{Ettori \& Fabian}{2000}]{Ettori:2000}
Ettori S., Fabian A.~C., 2000, MNRAS, 317, L57
\bibitem[\protect\citeauthoryear{Falle}{1991}]{Falle:1991}
Falle S.~A.~E.~G., 1991, MNRAS, 250, 581
\bibitem[\protect\citeauthoryear{Falle}{1994}]{Falle:1994}
Falle S.~A.~E.~G., 1994, MNRAS, 269, 607
\bibitem[\protect\citeauthoryear{Fragile et al.}{2005}]{Fragile:2005}
Fragile P.~C., Anninos P., Gustafson K., Murray S.~D., 2005, ApJ, 619, 327
\bibitem[\protect\citeauthoryear{Fragile et al.}{2004}]{Fragile:2004}
Fragile P.~C., Murray S.~D., Anninos P., van Breugel W., 2004, ApJ, 604, 74
\bibitem[\protect\citeauthoryear{Graham et al.}{1995}]{Graham:1995}
Graham J.~R., Levenson N.~A., Hester J.~J., Raymond J.~C., Petre R., 1995, ApJ, 444, 787
\bibitem[\protect\citeauthoryear{Gregori et al.}{1999}]{Gregori:1999}
Gregori G., Miniati F., Ryu D., Jones T.~W., 1999, ApJ, 527, L113
\bibitem[\protect\citeauthoryear{Hartquist \& Dyson}{1988}]{Hartquist:1988}
Hartquist T.~W., Dyson J.~E., 1988, Ap\&SS, 144, 615
\bibitem[\protect\citeauthoryear{Hartquist et al.}{1986}]{Hartquist:1986}
Hartquist T.~W., Dyson J.~E., Pettini M., Smith L.J., MNRAS, 1986, 221, 715
\bibitem[\protect\citeauthoryear{Hartquist et al.}{1998}]{Hartquist:1998}
Hartquist T.~W., Caselli P., Rawlings J.~M.~C., Ruffle D.~P., Williams D.~A., 1998, in The Molecular Astrophysics of Stars and Galaxies, (Clarendon Press, Oxford)
\bibitem[\protect\citeauthoryear{Hora et al.}{2006}]{Hora:2006}
Hora J.~L., Latter W.~B., Smith H.~A., Marengo M., 2006, ApJ, 652, 426
\bibitem[\protect\citeauthoryear{Iapichino et al.}{2008}]{Iapichino:2008}
Iapichino L., Adamek J., Schmidt W., Niemeyer J.~C., 2008, MNRAS, submitted (arxiv:0801.4695)
\bibitem[\protect\citeauthoryear{Jones \& Ellison}{1991}]{Jones:1991}
Jones F.~C., Ellison D.~C., 1991, Space Sci. Rev., 58, 259
\bibitem[\protect\citeauthoryear{Jun, Jones \& Norman}{1996}]{Jun:1996}
Jun B.-I., Jones T.~W., Norman M.~L., 1996, ApJ, 468, L59
\bibitem[\protect\citeauthoryear{Kaifu et al.}{2000}]{Kaifu:2000}
Kaifu N., et al., 2000, PASJ, 52, 1
\bibitem[\protect\citeauthoryear{Klein et al.}{2003}]{Klein:2003}
Klein R.~I., Budil K.~S., Perry T.~S., Bach D.~R., 2003, ApJ, 583, 245 
\bibitem[\protect\citeauthoryear{Klein, McKee \& Colella}{Klein et al.}{1994}]{Klein:1994}
Klein R.~I., McKee C.~F., Colella P., 1994, ApJ, 420, 213
\bibitem[\protect\citeauthoryear{Kornreich \& Scalo}{2000}]{Kornreich:2000}
Kornreich P., Scalo J., 2000, ApJ, 531, 366
\bibitem[\protect\citeauthoryear{Kulsrud \& Pearce}{1969}]{Kulsrud:1969}
Kulsrud R., Pearce W.~P., 1969, ApJ, 156, 445
\bibitem[\protect\citeauthoryear{Landau \& Lifshitz}{1959}]{Landau:1959}
Landau L.~D., Lifshitz E.~M., 1959, in ``Fluid Mechanics'' (Pergamon Press, Oxford)
\bibitem[\protect\citeauthoryear{Lazarian}{2006}]{Lazarian:2006}
Lazarian A., 2006, ApJ, 645, L25
\bibitem[\protect\citeauthoryear{Lee \& Burton}{2000}]{Lee:2000}
Lee J.-K., Burton M.~G., 2000, MNRAS, 315, 11
\bibitem[\protect\citeauthoryear{Levenson, Graham \& Walters}{Levenson et al.}{2002}]{Levenson:2002}
Levenson N.~A., Graham J.~R., Walters J.~L., 2002, ApJ, 576, 798
\bibitem[\protect\citeauthoryear{Mac Low \& Klessen}{2004}]{MacLow:2004}
Mac Low M.-M., Klessen R., 2004, Rev. Mod. Phys., 76, 125
\bibitem[\protect\citeauthoryear{Mac Low et al.}{1994}]{MacLow:1994}
Mac Low M.-M., McKee C.~F., Klein R.~I., Stone J.~M., Norman M.~L., 1994, ApJ, 433, 757
\bibitem[\protect\citeauthoryear{Marcolini et al.}{2005}]{Marcolini:2005}
Marcolini A., Strickland D.~K., D'Ercole A., Heckman T.~M., Hoopes C.~G., 2005, MNRAS, 362, 626
\bibitem[\protect\citeauthoryear{Martin et al.}{2007}]{Martin:2007}
Martin D.~C., et al., 2007, Nature, 448, 780
\bibitem[\protect\citeauthoryear{Matsuura et al.}{2007}]{Matsuura:2007}
Matsuura M., et al., 2007, MNRAS, 382, 1447
\bibitem[\protect\citeauthoryear{McKee \& Ostriker}{1977}]{McKee:1977}
McKee C.~F., Ostriker J.~P., 1977, ApJ, 218, 148
\bibitem[\protect\citeauthoryear{Melioli, de Gouveia Dal Pino \& Raga}{Melioli et al.}{2005}]{Melioli:2005}
Melioli C., de Gouveia Dal Pino E.~M., Raga A., 2005, A\&A, 443, 495
\bibitem[\protect\citeauthoryear{Mellema, Kurk \& R\"{o}ttgering}{2002}]{Mellema:2002}
Mellema G., Kurk J.~D., R\"{o}ttgering H.~J.~A., 2002, A\&A, 395, L13
\bibitem[\protect\citeauthoryear{Melnick, Tenorio-Tagle \& Terlevich}{Melnick et al.}{1999}]{Melnick:1999}
Melnick J., Tenorio-Tagle G., Terlevich R., 1999, MNRAS, 302, 677
\bibitem[\protect\citeauthoryear{Miceli et al.}{2006}]{Miceli:2006}
Miceli M., Reale F., Orlando S., Bocchino F., 2006, A\&A, 458, 213
\bibitem[\protect\citeauthoryear{Nakamura et al.}{2006}]{Nakamura:2006}
Nakamura F., McKee C.~F., Klein R.~I., Fisher R.~T., 2006, ApJSS, 164, 477
\bibitem[\protect\citeauthoryear{Narayan \& Medvedev}{2001}]{Narayan:2001}
Narayan R., Medvedev M.~V., 2001, ApJ, 562, L129
\bibitem[\protect\citeauthoryear{O'Dell, Henney \& Ferland}{O'Dell et al.}{2005}]{ODell:2005}
O'Dell C.~R., Henney W.~J., Ferland G.~J., 2005, AJ, 130, 172
\bibitem[\protect\citeauthoryear{Ohyama et al.}{2002}]{Ohyama:2002}
Ohyama Y., et al., 2002, PASJ, 54, 891
\bibitem[\protect\citeauthoryear{Orlando et al.}{2005}]{Orlando:2005}
Orlando S., Peres G., Reale F., Bocchino F., Rosner R., Plewa T., Siegel A.,
2005, A\&A, 444, 505
\bibitem[\protect\citeauthoryear{Orlando et al.}{2008}]{Orlando:2008}
Orlando S., Bocchino F., Reale F., Peres G., Pagano P.,
2008, ApJ, accepted (arxiv::0801.1403)
\bibitem[\protect\citeauthoryear{Parker}{1979}]{Parker:1979}
Parker E.~N., 1979, ``Cosmical Magnetic Fields'' (Clarendon Press, Oxford)
\bibitem[\protect\citeauthoryear{Patnaude et al.}{2002}]{Patnaude:2002}
Patnaude D.~J., Fesen R.~A., Raymond J.~C., Levenson N.~A., Graham J.~R., Wallace D.~J., 2002, AJ, 124, 2118
\bibitem[\protect\citeauthoryear{Pittard}{2007a}]{Pittard:2007a}
Pittard J.~M., 2007a, ApJ, 660, L141
\bibitem[\protect\citeauthoryear{Pittard}{2007b}]{Pittard:2007b}
Pittard J.~M., 2007b, in ``Diffuse Matter From Star Forming Regions to Active Galaxies'' (Springer-Verlag, Heidelberg)
\bibitem[\protect\citeauthoryear{Pittard et al.}{2005}]{Pittard:2005}
Pittard J.~M., Dyson J.~E., Falle S.~A.~E.~G., Hartquist, 2005, MNRAS, 361, 1077
\bibitem[\protect\citeauthoryear{Plewa \& M\"{u}ller}{2001}]{Plewa:2001}
Plewa T., M\"{u}ller E., 2001, Computer Physics Communications, 138, 101
\bibitem[\protect\citeauthoryear{Poludnenko, Frank \& Blackman}{Poludnenko et al.}{2002}]{Poludnenko:2002}
Poludnenko A.~Y., Frank A., Blackman E.~G., 2002, ApJ, 576, 832
\bibitem[\protect\citeauthoryear{Pope et al.}{2008}]{Pope:2008}
Pope E.~C.~D., Pittard J.~M., Hartquist T.~W., Falle S.~A.~E.~G., 2008, MNRAS, 385, 1779
\bibitem[\protect\citeauthoryear{Raymond et al.}{2007}]{Raymond:2007}
Raymond J.~C., Korreck K.~E., Sedlacek Q.~C., Blair W.~P., Ghavamian P., Sankrit R., 2007, ApJ, 659, 1257
\bibitem[\protect\citeauthoryear{Redman, Meaburn \& Holloway}{Redman et al.}{2002}]{Redman:2002}
Redman M.~P., Meaburn J., Holloway A.~J., 2002, MNRAS, 332, 754
\bibitem[\protect\citeauthoryear{Runacres \& Owocki}{2005}]{Runacres:2005}
Runacres M.~C., Owocki S.~P., 2005, A\&A, 429, 323 
\bibitem[\protect\citeauthoryear{Scalo \& Elmegreen}{2004}]{Scalo:2004}
Scalo J., Elmegreen B.~G., 2004, ARA\&A, 42, 275
\bibitem[\protect\citeauthoryear{Schultz et al.}{1999}]{Schultz:1999}
Schultz A.~S.~B., et al., 1999, ApJ, 511, 282
\bibitem[\protect\citeauthoryear{Serabyn, Lacy \& Achtermann}{Serabyn et al.}{1991}]{Serabyn:1991}
Serabyn E., Lacy J.~H., Achtermann J.~M., 1991, ApJ, 378, 557
\bibitem[\protect\citeauthoryear{Shin, Stone \& Snyder}{Shin et al.}{2008}]{Shin:2008}
Shin M.-S., Stone J.~M., Snyder G.~F., 2008, ApJ, accepeted (arXiv:0802.2708)
\bibitem[\protect\citeauthoryear{Spitzer}{1956}]{Spitzer:1956}
Spitzer L., 1956, ``Physics of Fully Ionized Gases'' (John Wiley and Sons, Inc., New York) 
\bibitem[\protect\citeauthoryear{Steffen \& L\'{o}pez}{2004}]{Steffen:2004}
Steffen W., L\'{o}pez J.~A., 2004, ApJ, 612, 319
\bibitem[\protect\citeauthoryear{Stone \& Norman}{1992}]{Stone:1992}
Stone J.~M., Norman M.~L., 1992, ApJ, 390, L17
\bibitem[\protect\citeauthoryear{Strickland \& Stevens}{2000}]{Strickland:2000}
Strickland D.~K., Stevens I.~R., 2000, MNRAS, 314, 511
\bibitem[\protect\citeauthoryear{Sutherland \& Bicknell}{2007}]{Sutherland:2007}
Sutherland R.~S., Bicknell G.~V., 2007, ApJSS, 173, 37
\bibitem[\protect\citeauthoryear{Tedds, Brand \& Burton}{Tedds et al.}{1999}]{Tedds:1999}
Tedds J.~A., Brand P.~W.~J.~L., Burton M.~G., 1999, MNRAS, 307, 337
\bibitem[\protect\citeauthoryear{Tenorio-Tagle et al.}{2006}]{Tenorio-Tagle:2006}
Tenorio-Tagle G., Mu\~{n}oz-Tu\~{n}\'{o}n, C., P\'{e}rez E., Silich S., Telles E., 2006, ApJ, 643, 186
\bibitem[\protect\citeauthoryear{Van Loo et al.}{2007}]{vanLoo:2007}
Van Loo S., Falle S.~A.~E.~G., Hartquist T.~W., Moore T.~J.~T., 2007, A\&A, 471, 213
\bibitem[\protect\citeauthoryear{Vikhlinin et al.}{2001}]{Vikhlinin:2001}
Vikhlinin A., Markevitch M., Forman W., Jones C., 2001, ApJ, 555, L87
\bibitem[\protect\citeauthoryear{Wagner et al.}{2006}]{Wagner:2006}
Wagner A.~Y., Falle S.~A.~E.~G., Hartquist T.~W., Pittard J.~M., 2006, A\&A, 452, 763
\bibitem[\protect\citeauthoryear{Weis, Duschl \& Chu}{Weis et al.}{1999}]{Weis:1999}
Weis K., Duschl W.~J., Chu Y.-H., 1999, A\&A, 349, 467
\bibitem[\protect\citeauthoryear{Westmoquette et al.}{2007a}]{Westmoquette:2007a}
Westmoquette M.~S., Exter K.~M., Smith L.~J., Gallagher J.~S., III, 2007a, MNRAS, 381, 894
\bibitem[\protect\citeauthoryear{Westmoquette et al.}{2007b}]{Westmoquette:2007b}
Westmoquette M.~S., Smith L.~J., Gallagher J.~S., III, O'Connell R.~W., Rosario D.~J., de Grijs R., 2007b, ApJ, 671, 358
\bibitem[\protect\citeauthoryear{Yirak et al.}{2008}]{Yirak:2008}
Yirak K., Frank A., Cunningham A., Mitran S., 2008, ApJ, 672, 996
\bibitem[\protect\citeauthoryear{Yusef-Zadeh \& Morris}{1991}]{Yusef-Zadeh:1991}
Yusef-Zadeh F., Morris M., 1991, ApJ, 371, L59
\end{thebibliography}
\end{document}